\documentclass[11pt,a4paper]{report}
\usepackage[inner=40mm, outer=30mm]{geometry}

\usepackage{authblk}
\usepackage{subfiles}

\usepackage{opendeduction}
\usepackage{virginialake}
\usepackage{hyperref}

\usepackage{hyperref}
\usepackage{mathtools}
\DeclarePairedDelimiter\bra{\langle}{\rvert}
\DeclarePairedDelimiter\ket{\lvert}{\rangle}
\DeclarePairedDelimiterX\braket[2]{\langle}{\rangle}{#1 \delimsize\vert #2}

\newcommand\collapse[1]{\lfloor #1\rfloor}

\usepackage{tikzit}

\tikzstyle{box}=[shape=rectangle, text height=1.5ex, text depth=0.25ex, yshift=0.5mm, fill=white, draw=black, minimum height=12.5mm, yshift=-0.5mm, minimum width=7.5mm, font={\small}]
\tikzstyle{Z dot}=[inner sep=0mm, minimum size=2mm, shape=circle, draw=black, fill={rgb,255: red,160; green,255; blue,160}]
\tikzstyle{Z phase dot}=[minimum size=5mm, font={\footnotesize\boldmath}, shape=rectangle, rounded corners=2mm, inner sep=0.2mm, outer sep=-2mm, scale=0.8, tikzit shape=circle, draw=black, fill={rgb,255: red,160; green,255; blue,160}, tikzit draw=blue]
\tikzstyle{X dot}=[Z dot, shape=circle, draw=black, fill={rgb,255: red,220; green,0; blue,0}]
\tikzstyle{X phase dot}=[Z phase dot, tikzit shape=circle, tikzit draw=blue, fill={rgb,255: red,220; green,0; blue,0}, font={\footnotesize\color{white}\boldmath}]
\tikzstyle{hadamard}=[fill=yellow, draw=black, shape=rectangle, inner sep=0.6mm, minimum height=1.5mm, minimum width=1.5mm]
\tikzstyle{small hadamard}=[hadamard]
\tikzstyle{vertex}=[inner sep=0mm, minimum size=1mm, shape=circle, draw=black, fill=black]
\tikzstyle{vertex set}=[inner sep=0mm, minimum size=1mm, shape=circle, draw=black, fill=white, font={\footnotesize\boldmath}]
\tikzstyle{new style 0}=[fill=none, draw=none, shape=rectangle, font={\large}]

\tikzstyle{simple}=[-]
\tikzstyle{hadamard edge}=[-, color=blue, dashed, dash pattern=on 2pt off 0.7pt]
\tikzstyle{brace edge}=[-, tikzit draw=blue, decorate, decoration={brace,amplitude=1mm,raise=-1mm}]
\tikzstyle{gray}=[-, draw={rgb,255: red,191; green,191; blue,191}]
\tikzstyle{arrow}=[<-, draw={rgb,255: red,128; green,128; blue,128}]
\tikzstyle{double-arrow}=[draw={rgb,255: red,128; green,128; blue,128}, <->]
\tikzstyle{new big text}=[-]

\newcommand\FMCV{\textup{!FMC}}
\newcommand\SLCV{\textup{!SLC}}
\newcommand\FMCBase{\textup{FMC}}

\usepackage{tikz}
	\usetikzlibrary{arrows.meta}
	\usetikzlibrary{calc}

\usepackage{csquotes}

\usepackage{xr-hyper}
\usepackage{amsmath,amssymb,amsthm}
\usepackage{accents}
\usepackage{cmll}
\usepackage{graphicx}
\usepackage{mathtools}
\usepackage{proof}
\usepackage{stmaryrd}
\usepackage{xcolor}
\usepackage{quiver}

\usepackage{opendeduction}
\usepackage{bussproofs}
\usepackage{virginialake}

\usepackage{tikz-cd}
	\usetikzlibrary{arrows.meta}
	\usetikzlibrary{positioning,intersections}

\usepackage{tikzit}

\tikzstyle{termbox}=[draw=term, fill={term!10}, rounded corners, minimum size=20pt]
\tikzstyle{tallbox}=[draw=term, fill={term!10}, rounded corners, minimum width=20pt, minimum height=40pt]
\tikzstyle{termpic}=[x=1pt, y=1pt, inner sep=0pt, outer sep=0pt, thick]
\tikzstyle{box}=[shape=rectangle, text height=1.5ex, text depth=0.25ex, yshift=0.5mm, fill=white, draw=black, minimum height=12.5mm, yshift=-0.5mm, minimum width=7.5mm, font={\small}]
\tikzstyle{Z dot}=[inner sep=0mm, minimum size=2mm, shape=circle, draw=black, fill={rgb,255: red,160; green,255; blue,160}]
\tikzstyle{Z phase dot}=[minimum size=5mm, font={\footnotesize\boldmath}, shape=rectangle, rounded corners=2mm, inner sep=0.2mm, outer sep=-2mm, scale=0.8, tikzit shape=circle, draw=black, fill={rgb,255: red,160; green,255; blue,160}, tikzit draw=blue]
\tikzstyle{X dot}=[Z dot, shape=circle, draw=black, fill={rgb,255: red,220; green,0; blue,0}]
\tikzstyle{X phase dot}=[Z phase dot, tikzit shape=circle, tikzit draw=blue, fill={rgb,255: red,220; green,0; blue,0}, font={\footnotesize\color{white}\boldmath}]
\tikzstyle{hadamard}=[fill=yellow, draw=black, shape=rectangle, inner sep=0.6mm, minimum height=1.5mm, minimum width=1.5mm]
\tikzstyle{small hadamard}=[hadamard]
\tikzstyle{vertex}=[inner sep=0mm, minimum size=6.5pt, shape=circle, draw=black, fill=black]
\tikzstyle{vertex set}=[inner sep=0mm, minimum size=1mm, shape=circle, draw=black, fill=white, font={\footnotesize\boldmath}]
\tikzstyle{new style 0}=[draw=term, fill={term!10}, rounded corners, minimum width=25pt, minimum height=30pt]
\tikzstyle{new style 2}=[draw=term, fill={term!10}, rounded corners, minimum width=25pt, minimum height=30pt]

\tikzstyle{new edge style 0}=[-, draw=black, line width=0.8pt]
\tikzstyle{blue 0}=[-, draw=blue, line width=0.8pt]
\tikzstyle{red 0}=[-, draw=red, line width=0.8pt]
\tikzstyle{black dash}=[-, color=black, dashed, dash pattern=on 1.5pt off 1.5pt, draw=black, line width=0.8pt]
\tikzstyle{red dash}=[-, color=red, dashed, dash pattern=on 1pt off 1.5pt, draw=red, line width=0.8pt, line cap=round]
\tikzstyle{blue dash}=[-, color=blue, dashed, dash pattern=on 1pt off 1.5pt, draw=blue, line width=0.8pt, line cap=round]
\tikzstyle{red term dash}=[-, dotted, draw={rgb,255: red,210; green,0; blue,0}, dash pattern=on 1pt off 2pt, line width=0.8pt, line cap=round]
\tikzstyle{brace edge}=[-, tikzit draw=blue, decorate, decoration={brace,amplitude=1mm,raise=-1mm}, line width=0.8pt]
\tikzstyle{gray}=[-, draw={rgb,255: red,191; green,191; blue,191}]
\tikzstyle{arrow}=[<-, draw={rgb,255: red,128; green,128; blue,128}]
\tikzstyle{double-arrow}=[draw={rgb,255: red,128; green,128; blue,128}, <->]
\tikzstyle{thick line}=[-, line width=0.8pt]

\makeatletter

\theoremstyle{plain}
\newtheorem{theorem}               {Theorem}    [section]
\newtheorem{corollary}   [theorem] {Corollary}
\newtheorem{lemma}       [theorem] {Lemma}
\newtheorem{proposition} [theorem] {Proposition}

\newtheorem*{notation*}  {Notation}

\theoremstyle{definition}
\newtheorem{definition}  [theorem] {Definition}
\newtheorem{example}     [theorem] {Example}
\newtheorem{notation}     [theorem] {Notation}
\newtheorem*{example*}             {Example}
\newtheorem{remark}      [theorem] {Remark}

\newcommand\black{\color{black}}

\colorlet{0}{black}
\colorlet{1}{red!80!black}
\colorlet{2}{blue!80!black}
\colorlet{3}{orange!90!black}
\colorlet{4}{green!50!black}
\colorlet{5}{violet!80!black}
\colorlet{6}{teal!90!white}
\colorlet{7}{brown!90!black}
\colorlet{8}{pink!70!blue}
\colorlet{9}{lime!80!black}

\colorlet{term}{1}
\colorlet{type}{2}
\colorlet{ctxt}{3}
\colorlet{simp}{4}
\colorlet{patn}{5}
\colorlet{ptpe}{6}

\newcommand\colorterm{\color{term}}
\newcommand\colortype{\color{type}}
\newcommand\colorctxt{\color{ctxt}}
\newcommand\colorsimp{\color{simp}}
\newcommand\colorpatn{\color{patn}}
\newcommand\colorptpe{\color{ptpe}}

\newcommand\colored{%
  \let\color@term\colorterm%
  \let\color@type\colortype%
  \let\color@ctxt\colorctxt%
  \let\color@simp\colorsimp%
  \let\color@patn\colorpatn%
  \let\color@ptpe\colorptpe%
}
\newcommand\uncolored{%
  \let\color@term\relax%
  \let\color@type\relax%
  \let\color@ctxt\relax%
  \let\color@simp\relax%
  \let\color@patn\relax%
  \let\color@ptpe\relax%
  \black
}
\uncolored

\newcommand\typecolor{\color{type}}

\newcommand\embedterm[1]{\embed{\term{#1}}}
\newcommand\embedint[1]{\termint{#1}}
\newcommand\embedintterm[1]{\termint{\term{#1}}}
\newcommand\collapint[1]{\ccc{#1}}
\newcommand\collapintterm[1]{\ccc{\term{#1}}}

\newcommand\result{\kern1pt{\Downarrow}\kern1pt}

\newcommand\machine[3]{%
\begin{array}{@{(~}l@{~,~}r@{~)}}%
#1 & \term {#2}%
\\\hline\hline\rule[-5pt]{0pt}{15pt}%
#3 & \term *\;%
\end{array}%
}

\newcommand\diagdots[2][1.5,3]{
  \node[fill=black,circle,inner sep=0pt,minimum size=1.5pt] at ($(#2) - (#1)$) {};
  \node[fill=black,circle,inner sep=0pt,minimum size=1.5pt] at (#2) {};
  \node[fill=black,circle,inner sep=0pt,minimum size=1.5pt] at ($(#2) + (#1)$) {};
}
\tikzstyle{termbox}=[draw=term,fill=term!10,rounded corners,minimum size=20pt]
\tikzstyle{tallbox}=[draw=term,fill=term!10,rounded corners,minimum width=20pt,minimum height=40pt]
\tikzstyle{termpic}=[x=1pt,y=1pt,inner sep=0pt,outer sep=0pt,thick]

\newcommand\wires[5]{
  \node[anchor=east] at (#2,#3) {$\type{#1}$};
  \node[anchor=west] at (#4,#3) {$\type{#5}$};
  \draw ($(#2,#3) + ( 2, 6)$) -- ($(#4,#3) + (-2, 6)$);
  \draw ($(#2,#3) + ( 8,-6)$) -- ($(#4,#3) + (-8,-6)$);
  \diagdots[-1.5,3]{$(#2,#3) + (10,0)$}
  \diagdots[ 1.5,3]{$(#4,#3) - (10,0)$}
}
\newcommand\wiresright[4]{
  \node[anchor=west] at (#3,#2) {$\type{#4}$};
  \draw ($(#1,#2) + ( 0, 6)$) -- ($(#3,#2) + (-2, 6)$);
  \draw ($(#1,#2) + ( 0,-6)$) -- ($(#3,#2) + (-8,-6)$);
  \diagdots[ 1.5,3]{$(#3,#2) - (10,0)$}
}
\newcommand\wiresleft[4]{
  \node[anchor=east] at (#2,#3) {$\type{#1}$};
  \draw ($(#2,#3) + ( 2, 6)$) -- ($(#4,#3) + ( 0, 6)$);
  \draw ($(#2,#3) + ( 8,-6)$) -- ($(#4,#3) + ( 0,-6)$);
  \diagdots[-1.5,3]{$(#2,#3) + (10,0)$}
}

\newcommand\vc[1]{\vcenter{\hbox{$#1$}}}

\newcommand\rvec[1]{\accentset{\rightharpoonup}{#1}}
\newcommand\lvec[1]{\accentset{\leftharpoonup}{#1}}

\newcommand\defeq{\stackrel{\scriptscriptstyle\Delta}=}
\newcommand*\coloneq{
 \mathrel{%
  \rlap{\raisebox{0.3ex}{$\m@th\cdot$}}%
        \raisebox{-0.3ex}{$\m@th\cdot$}%
 {=}}}
\newcommand*\coloneqq{
 \mathrel{%
  \rlap{\raisebox{0.3ex}{$\m@th\cdot$}}%
        \raisebox{-0.3ex}{$\m@th\cdot$}%
  \rlap{\raisebox{0.3ex}{$\m@th\cdot$}}%
        \raisebox{-0.3ex}{$\m@th\cdot$}%
 {=}}}

\newcommand\smallbin[1]{\mathchoice
      {\mathbin{\raise.2ex \hbox{$\scriptstyle      #1$}}}%
      {\mathbin{\raise.2ex \hbox{$\scriptstyle      #1$}}}%
      {\mathbin{\raise.12ex\hbox{$\scriptscriptstyle#1$}}}%
      {\mathbin{           \hbox{$\scriptscriptstyle#1$}}}}%

\newcommand\Con{\wedge}
\newcommand\Imp{\rightarrow}

\newcommand\con{\kern1pt{\smallbin\Con}\kern1pt}
\newcommand\imp{\kern1pt{\smallbin\Imp}\kern1pt}

\newlength{\ilength}
\newcommand\fv[1]{\mathsf{fv}(#1)}

\newcommand\seq@discard[1]{\seq@start}
\newcommand\seq@comma{\@ifnextchar,{{\black,\dots,}~\seq@discard}{{\black,}~\seq@start}}

\newcounter{@parens}

\newcommand\term[1]{{\colored\trm{#1}}}
\newcommand\trm[1]{%
  \setcounter{@parens}{0}%
  \vphantom(%
  \let\seq@start\term@start%
  \seq@start#1\@end%
}
\newcommand\term@start{\color@term\let\term@loop=\term@next\term@loop}
\newcommand\term@next[1]{%
  \ifx#1\@end\let\term@loop\term@end\else%
  \ifx#1`\let\term@loop\term@escape\else%
  \ifx#1_\let\term@loop\term@sub\else%
  \ifx#1^\let\term@loop\term@sup\else%
  \ifx#1?\let\term@loop\term@lvec\else%
  \ifx#1!\let\term@loop\term@rvec\else%
  \ifx#1"\let\term@loop\term@color\else%
  \ifx#1:\let\term@loop\term@colon\else%
  \ifx#1|\let\term@loop\term@bar\else%
  \ifx#1((\stepcounter{@parens}\else%
  \ifx#1))\addtocounter{@parens}{-1}\else%
  \ifx#1\{\{\stepcounter{@parens}\else%
  \ifx#1\}\}\addtocounter{@parens}{-1}\else%
  \ifx#1,\ifnum\value{@parens}=0\let\term@loop\seq@comma\else,\fi\else%
  \ifx#1*\star\else%
  \ifx#1l\lambda\else%
  \ifx#1<\langle\else%
  \ifx#1>\rangle\else%
  \ifx#1..\,\else%
  \ifx#1;\,{;}\,\else%
  \ifx#1=\kern1pt{\smallbin=}\kern1pt\else%
  \ifx#1G{{\black\Gamma}}\else%
  \ifx#1D{{\black\Delta}}\else%
  \ifx#1\sim{\,\black\sim\,}\else%
  #1%
  \fi\fi\fi\fi%
  \fi\fi\fi\fi\fi%
  \fi\fi\fi\fi\fi%
  \fi\fi\fi\fi\fi%
  \fi\fi\fi\fi\fi%
  \term@loop%
}
\newcommand\term@end{\color{black}\uncolored}
\newcommand\term@discard[1]{\term@start}
\newcommand\term@escape[1]{#1\term@start}
\newcommand\term@color[1]{\color{#1}\term@start}
\newcommand\term@sub[1]{_{#1}\term@start}
\newcommand\term@sup[1]{^{#1}\term@start}
\newcommand\term@rvec[1]{\rvecup{#1}\term@start}
\newcommand\term@lvec[1]{\lvecup{#1}\term@start}
\newcommand\term@bar{\@ifnextchar-{~{\black\vdash}~\term@discard}{|\term@start}}
\newcommand\term@colon{\@ifnextchar={\term@assign}{\term@type}}
\newcommand\term@assign{\mathbin{{:}{=}}\term@discard}
\newcommand\term@type{\black\colon\type@start}

\newcommand\type[1]{%
  \vphantom)%
  \let\seq@start\type@start%
  \seq@start#1\@end%
}
\newcommand\type@start{\typecolor\let\type@loop\type@next\type@loop}
\newcommand\type@next[1]{%
  \ifx#1\@end\let\type@loop\type@end\else%
  \ifx#1`\let\type@loop\type@escape\else%
  \ifx#1_\let\type@loop\type@sub\else%
  \ifx#1^\let\type@loop\type@sup\else%
  \ifx#1|\black\vdash\let\type@loop\seq@discard\else%
  \ifx#1,\let\type@loop\seq@comma\else%
  \ifx#1!\let\type@loop\type@rvec\else%
  \ifx#1?\let\type@loop\type@lvec\else%
  \ifx#1>\smallbin\Rightarrow\else%
  \ifx#1.\kern1pt{\cdot}\kern1pt\else%
  \typevar{#1}%
  \fi\fi\fi\fi%
  \fi\fi%
  \fi\fi\fi\fi%
  \type@loop%
}
\newcommand\type@end{\black}
\newcommand\type@escape[1]{#1\type@start}
\newcommand\type@sub[1]{_{#1}\type@start}
\newcommand\type@sup[1]{^{{\type@sup@color #1}}\type@start}
\newcommand\type@rvec[1]{\rvec{\typevar{#1}}\type@start}
\newcommand\type@lvec[1]{\lvec{\typevar{#1}}\type@start}

\newcommand\typevar[1]{\typevar@start#1\@end}
\newcommand\typevar@start{\let\typevar@loop\typevar@next\typevar@loop}
\newcommand\typevar@next[1]{%
  \ifx#1\@end\let\typevar@loop\typevar@end\else%
  \ifx#1a\alpha\else%
  \ifx#1b\beta\else%
  \ifx#1c\gamma\else%
  \ifx#1d\delta\else%
  \ifx#1n\nu\else%
  \ifx#1p\pi\else%
  \ifx#1k\kappa\else%
  \ifx#1r\rho\else%
  \ifx#1s\sigma\else%
  \ifx#1t\tau\else%
  \ifx#1u\upsilon\else%
  \ifx#1w\omega\else%
  #1%
  \fi\fi\fi\fi\fi\fi\fi\fi\fi\fi\fi\fi\fi%
  \typevar@loop%
}
\newcommand\typevar@end{}

\newcommand\rnd{\mathsf{rnd}} 
\newcommand\nd {\mathsf{nd}}  
\newcommand\inp{\mathsf{in}}  
\newcommand\out{\mathsf{out}} 

\tikzstyle{rwhead}=[>/.tip={Triangle[open,length=2.5pt,width=4.5pt]},|/.tip={Rectangle[length=.5pt,width=4.5pt]}]

\tikzstyle{lw} =[line width=.5pt,rwhead,->]
\tikzstyle{rw} =[line width=.5pt,rwhead,->]
\tikzstyle{rws}=[line width=.5pt,rwhead,->.>]
\tikzstyle{rwn}=[line width=.5pt,rwhead,->.>|]
\tikzstyle{rwp}=[line width=.5pt,rwhead,->,double]
\tikzstyle{rwps}=[line width=.5pt,rwhead,->.>,double]

\newcommand\rw  {\mathrel{\tikz\draw[rw]  (0,0)--(20pt,0pt);}}

\newcommand\TR[1]{\!\scriptstyle{\tr{#1}\vphantom{pb}}}%
\newcommand\tr[1]{%
  \ifx#1*\star\else%
  \ifx#1v\mathsf{fam}\else%
  \ifx#1p\mathsf{proj}\else%
  \ifx#1u\mathsf{unit}\else%
  \ifx#1q\mathsf{pair}\else%
  \ifx#1x\mathsf{var}\else%
  \ifx#1l\mathsf{abs}\else%
  \ifx#1a\mathsf{app}\else%
  \ifx#1c\mathsf{con}\else%
  \ifx#1f\mathsf{base}\else%
  \ifx#1<\mathsf{lcut}\else%
  \ifx#1>\mathsf{rcut}\else%
  \fi\fi\fi\fi\fi\fi\fi\fi\fi\fi\fi\fi%
}

\DeclareRobustCommand\widecheck[1]{{\mathpalette\@widecheck{#1}}}
\def\@widecheck#1#2{%
    \setbox\z@\hbox{\m@th$#1#2$}%
    \setbox\tw@\hbox{\m@th$#1%
       \widehat{%
          \vrule\@width\z@\@height\ht\z@
          \vrule\@height\z@\@width\wd\z@}$}%
    \dp\tw@-\ht\z@
    \@tempdima\ht\z@ \advance\@tempdima2\ht\tw@ \divide\@tempdima\thr@@
    \setbox\tw@\hbox{%
       \raise\@tempdima\hbox{\scalebox{1}[-1]{\lower\@tempdima\box
\tw@}}}%
    {\ooalign{\box\tw@ \cr \box\z@}}}

\newcommand{\red}{\textcolor{red}}
\newcommand{\gray}{\textcolor{black!70}}

\newcommand{\<}{\langle}

\newcommand\Z{\mathbb Z}

\protected\def\verythinspace{%
  \ifmmode
    \mskip0.5\thinmuskip
  \else
    \ifhmode
      \kern0.08334em
    \fi
  \fi
}

\newcommand\id{\mathsf{id}}

\newcommand\Tim{\times}
\newcommand\Ten{\otimes}
\newcommand\tim{\kern1pt{\smallbin\Tim}\kern1pt}
\newcommand\ten{\kern1pt{\smallbin\Ten}\kern1pt}

\newcommand\cat[1]{
  \cat@start#1\@end%
}
\newcommand\cat@start{\let\cat@loop\cat@next\cat@loop}
\newcommand\cat@next[1]{%
  \ifx#1\@end\let\cat@loop\cat@end\else%
  \ifx#1_\let\cat@loop\cat@sub\else%
  \ifx#1^\let\cat@loop\cat@sup\else%
  \ifx#1!\let\cat@loop\cat@rvec\else%
  \ifx#1?\let\cat@loop\cat@rvec\else
  \ifx#1-\let\cat@loop\cat@arrows\else%
  \ifx#1*\tim\else%
  \ifx#1A\!_A\else%
  \typevar{#1}%
  \fi\fi\fi\fi\fi\fi%
  \fi\fi%
  \cat@loop%
}
\newcommand\cat@sub[1]{_{#1}\cat@start}
\newcommand\cat@sup[1]{^{#1}\cat@start}
\newcommand\cat@rvec[1]{\rvecup{\typevar{#1}}\cat@start}
\newcommand\cat@lvec[1]{\lvecup{\typevar{#1}}\cat@start}
\newcommand\cat@arrows{%
  \@ifnextchar-{\longrightarrow\cat@discard@two}{%
  \@ifnextchar>{\imp\cat@discard}{-\cat@start}}}
\newcommand\cat@end{}
\newcommand\cat@discard[1]{\cat@start}
\newcommand\cat@discard@two[2]{\cat@start}

\newcommand\ac{\kern1pt{\smallbin\ggg}\kern1pt} 

\newcommand\B{\mathbb B}

\newcommand\ccc[1]{\llbracket#1\rrbracket}
\newcommand\sn[1]{\llbracket#1\rrbracket}

\newcommand\termint[1]{\{#1\}} 
\newcommand\termcat[1]{\textup{\textsf{!FMC}}(#1)}

\newcommand\ccceq{=_{\textsf{eqn}}}

\newcommand\concat[2]{#1 , #2}

\newcommand\rvecup[1]{\accentset{\rightharpoonup}{#1}}
\newcommand\lvecup[1]{\accentset{\leftharpoonup}{#1}}

\newcommand\get{\mathsf{get}}
\newcommand\set{\mathsf{set}}
\newcommand\print{\mathsf{write}}
\newcommand\rand{\mathsf{rand}}
\newcommand\rd{\mathsf{read}}

\newcommand\run[1]{\textsc{run}(\type{#1})}

\newcommand\arrimp{\kern1pt{\smallbin\rightsquigarrow}\kern1pt}

\newcommand\e{\epsilon}

\newcommand\rr[1]{%
  \ifx#1x\mathsf{var}\else%
  \ifx#1l\mathsf{abs}\else%
  \ifx#1a\mathsf{arg}\else%
  \ifx#1f\mathsf{fun}\else%
  \fi\fi\fi\fi%
}

\newcommand\cccterm[1]{\ccc{\term{#1}}}
\newcommand\ccctype[1]{\ccc{\type{#1}}}

\newcommand\uterm[1]{\underline{\term{#1}}}

\newcommand\cvdash{\vdash_{\textsf{c}}}
\newcommand\vvdash{\vdash_{\textsf{v}}}

\newcommand\stermcat[1]{\textup{\textsf{!SLC}}({#1})}
\newcommand\collap[1]{\lfloor #1 \rfloor}
\newcommand\embed[1]{\lceil #1 \rceil}
\newcommand\collapterm[1]{\collap{\term{#1}}}
\newcommand\popsterm[1]{\term{<}\collapterm{#1}\term{>.}}
\newcommand\popstermdot[1]{\term{<}\collapterm{#1}\term{>}}
\newcommand\pushsterm[1]{\term{[}\collapterm{#1}\term{]}}
\newcommand\pushkinvterm[1]{\term{[}\embed{\term{#1}}^*\term{]}}
\newcommand\pushkterm[1]{\term{[}\collapterm{#1}^*\term{]}}
\newcommand\ide{\textsf{id}}
\renewcommand\r{x}
\newcommand\s{y}

\renewcommand\u{w}

\newcommand\y{y}

\newcommand\step[4]{%
\begin{array}{@{(~}l@{~,~}r@{~)}}%
#1 & \term {#2}%
\\\hline%
#3 & \if#4*\term*\;\else\term{#4}\fi%
\end{array}%
}

\begin{document}

\title{On the Simply-Typed Functional Machine Calculus: Categorical Semantics and Strong Normalisation}
\author{\Large{Chris~Barrett}}
\affil{\small A thesis submitted for the degree of \\ \normalsize Doctor of Philosophy \\ \vspace{2cm} University of Bath \\ Department of Computer Science}
\date{October 2022}

\maketitle

\newpage

\begin{center}
\textbf{Copyright}
\\
\vspace{0.5cm}
Attention is drawn to the fact that copyright of this thesis rests with
the author and copyright of any previously published materials included may
rest with third parties. A copy of this thesis has been supplied on
condition that anyone who consults it understands that they must not copy it
or use material from it except as licenced, permitted by law or with the
consent of the author or other copyright owners, as applicable.
\\
\vspace{1cm}
\textbf{Declarations}
\\
\vspace{0.5cm}
The material presented here for examination for the award of a higher
degree by research has not been incorporated into a submission for
another degree.
\\
\textit{Chris Barrett}
\vspace{0.5cm}
\\
I am the author of this thesis, and the work described therein was carried
out by myself personally.
\\
\textit{Chris Barrett}
\end{center}

%

%
%

\newpage

\begin{abstract}
The Functional Machine Calculus (FMC) was recently introduced as a generalization of the lambda-calculus to include higher-order global state, probabilistic and non-deterministic choice, and input and ouput, while retaining confluence. The calculus can encode both the call-by-name and call-by-value semantics of these effects.  This is enabled by two independent generalizations, both natural from the perspective of the FMC's operational semantics, which is given by a simple (multi-)stack machine. 

The first generalization decomposes the syntax of the lambda-calculus in a way that allows for sequential composition of terms and the encoding of reduction strategies.  Specifically, there exist translations of the call-by-name and call-by-value lambda-calculus which preserve operational semantics. The second parameterizes application and abstraction in terms of `locations' (corresponding to the multiple stacks of the machine), which gives a unification of the operational semantics, syntax, and reduction rules of the given effects with those of the lambda-calculus.
The FMC further comes equipped with a simple type system which restricts and captures the behaviour of effects.
 
This thesis makes two main contributions, showing that two fundamental properties of the lambda-calculus are preserved by the FMC. 
The first is to show that the categorical semantics of the FMC, modulo an appropriate equational theory, is given by the free Cartesian closed category.
The equational theory is validated by a notion of observational equivalence. 
The second contribution is a proof that typed FMC-terms are strongly normalising. This is an extension (and small simplification) of Gandy's proof for the lambda-calculus, which additionally emphasizes its latent operational intuition.
\end{abstract}

\renewcommand{\abstractname}{Acknowledgements}
\begin{abstract}
First and absolutely foremost, I'd like to thank my supervisor, Willem Heijltjes.
It is perhaps a triviality to say that without the supervisor, the thesis would not exist. 
But this is doubly true, because I was lucky enough to study under Willem when he had the brilliant idea that is the Functional Machine Calculus, the topic of this thesis. 
Much more than that, I am grateful for his seemingly \textit{endless} patience throughout my PhD, and for his advice and guidance, which was \textit{always} insightful and has highly coloured my thinking. Finally, for giving me the {freedom to} follow my interests, take tangents, and develop as an independent mathematician.

I'd also to especially thank Alessio Guglielmi, who was my second supervisor during the first year of my PhD, and with whom I co-authored my first research paper. His unusual webpage brought me to Bath, and to a computer science, rather than mathematics, department: a choice that I have never since questioned. He founded Deep Inference, which was, in fact, the first proof formalism I understood -- before even the sequent calculus -- and has been formative in my thinking about proof theory. Guy McCusker also deserves a special mention, with whom Willem and I co-authored two papers,  for his support.  I'd like also to thank my examiners, Jim Laird and Ugo Dal Lago, for their careful attention and helpful suggestions. 

I'd like to additionally thank my colleagues -- Ben, Andrea, Alessio (Jr.), David, Georgi, Torie, Giulio, Tom, Vincent, Hollie, Cameron, Dan, and Keji for many interesting discussions, encouragement, and for putting up with more than a couple of rambles on whatever topic in proof theory I was most excited about that week. 
I have very much enjoyed the company of all of the above -- mentors and colleagues -- over the years. 

The wider community also deserves mention,  and I will single out Anupam Das for his encouragement.  I'd also like to thank my master's supervisor, Evgenios, who was the first person I had whiteboard discussions on mathematics with, and who helped me towards my current position in Bath.

 I also have to thank my family: my mother, Helen, and sister, Emma, and my friends, in particular, Edd and Elisa, who have supported me in every way I could have asked for when times were difficult, and for always believing that I would finish, even when I doubted myself. 
\end{abstract}

\tableofcontents

\listoffigures

\chapter{Introduction}

Since the 60’s, the $\lambda$-calculus has been regarded as the definitive model of higher-order
functional programming \cite{Barendregt-1984,Church-1941,Landin-1964,Landin-1965}. It has a remarkably compact syntax, a natural denotational semantics, and satisfies
confluence. That is, the order in which one chooses to evaluate sub-expressions makes no difference to
the eventual result of the computation. This means that each expression has one canonical `meaning', namely its result. 
Further, it comes equipped with a powerful type system that guarantees strong normalisation.

\section{The $\lambda$-calculus and Computational Effects}

Serious programming languages seem to demand  \textit{effectful} primitives be added to the $\lambda$-calculus: for example, to deal with input and output, mutable store, probabilistic choice, exceptions, continuations, or anything that isn't \textit{pure} computation. However, confluence is lost when one tries to extend the $\lambda$-calculus naively.  In an upcoming example, we will see how loss of confluence can mean for the programmer a confusing loss of the soundness of equational reasoning (one cannot always soundly substitute a function for its definition), and for the theoretician a concerning challenge to its canonical denotational
semantics (i.e., one independent of order of evaluation).

Thus, there is an important research program to be undertaken: how can we extend the $\lambda$-calculus to incorporate {computational effects} while maintaining the good properties listed above, which made it a centre of programming language research in the first place?

\subsection{The Problem}
Let us first explore the problem with the $\lambda$-calculus and effects. Take, as an example, a standard $\lambda$-calculus incorporating \textit{probabilistic choice} and \textit{state}.
\begin{align*}
       M \quad \Coloneqq \quad x ~\mid~ MN ~\mid~ \lambda x.M ~\mid~ M \oplus N ~\mid~ c := N;M ~\mid~ !c 
\end{align*}
The syntax of the $\lambda$-calculus is extended with new primitives: a term $M \oplus N$, which tosses a coin and evaluates as $M$ or $N$ based on the result; a term $c := N;M$, which assigns the value $N$ to the memory cell $c$, before continuing as $M$, and a term $!c$ which dereferences the cell $c$. We give two example $\lambda$-terms below, one for probabilistic choice, and one for state, demonstrating the failure of confluence. That is, each term gives different results depending on whether it is reduced by a call-by-name (CBN) or call-by-value (CBV) strategy \cite{Plotkin-1975}.  In the first term, $=$ is an equality-checking function, and $\top$ and $\bot$ are Booleans. 
\[
\begin{array}{crclc}
	\top \oplus \bot &  {}_{\textsf{cbn}}\!\leftarrow& (\lambda x. x = x)(\top \oplus \bot)  &\to_\textsf{cbv} & \top \\
	5 & _\textsf{cbn}\!\leftarrow& c := 3 ; (\lambda x. !c ) (c := 5 ; M) &\to_\textsf{cbv} & 3
\end{array}
\]
According to CBN and CBV, respectively, the case of probabilistic choice either \textit{duplicates first} its argument, and \textit{then tosses two distinct coins}, one for each resulting choice to be made, vs. \textit{tossing a coin first} to decide the result of its argument, and \textit{then duplicating} that result. For the case of state, the sub-term $c:= 5; M$ is either \textit{discarded before it has chance to run}, thus making no change to state, vs. \textit{being run before it is discarded}, thus updating the state.

The common approach of insisting on a fixed order of evaluation only vacuously preserves confluence.  
 Nevertheless, one would be forgiven for thinking that \textit{another} $\lambda$-calculus variant is not the way forward. 
Indeed, there is perhaps a problem of canonicity in the field of programming languages: to quote Abramsky, ``it is too easy to cook up yet another variant lambda calculus; there are too few constraints'' \cite{DBLP:journals/corr/Abramsky14a}.

We proceed in this vein of inquiry anyway.   
The remarkable work of developing a variant which is confluent in the presence of (an important subset of) effects, the \textit{Functional Machine Calculus (FMC)}, was recently carried out by Heijltjes \cite{Barrett-Heijltjes-McCusker-2022}, building on \cite{DalLago-Guerrieri-Heijltjes-2020}. However, to defend the calculus, we wish to show it preserves \textit{all} of the good properties of the $\lambda$-calculus.
The aforementioned work already equips the FMC with a system of simple types which restricts and captures the behaviour of effects.
The main body of the thesis contributes by showing important remaining properties of the $\lambda$-calculus are preserved: namely, a natural denotational (in this case, categorical) semantics and the strong normalisation of typed terms.  The former involves developing and justifying an appropriate equational theory on FMC-terms. Overall, the long-term aim of this program of research is to bring the powerful reasoning tools developed for the $\lambda$-calculus to bear on real-world, effectful {programming langugages}.

 Introduction and discussion of the Functional Machine Calculus is the main focus of the rest of this chapter, which then concludes with a summary of the remaining thesis. 
Before introducing the FMC from first principles, we begin, in the next section, with a motivating example.

\subsection{The Solution}\label{subsec:pmc-example}
Realizing that a fully general solution to confluence for effects is out of reach, at least for the moment, will restrict our focus to a subset of computational effects, which we call \textit{reader/writer effects}. Namely, they are input and output, global higher-order state, and probabilistic and non-deterministic choice.
This is an important subset, but we will also comment, briefly, in the final chapter on how the insights offered by the Functional Machine Calculus are suggestive of further extensions, \textit{e.g.} to local state, and even concurrency. 

The \textit{unique} insight of the Functional Machine Calculus is that the operational semantics of higher-order computation and reader/writer effects can be unified.
This leads further to a unification at the syntactic and algebraic level -- that is to say, both higher-order computation and effects are encoded by the same syntax, with their algebraic theory given by the same, familiar, \textit{beta reduction}.
In other words, instead of adding \textit{new} primitives for computational effects, the FMC \textit{decomposes} them into its existing syntax, which models higher-order computation in a standard way.
This approach results in the calculus retaining confluence, as proved by Heijltjes in \cite{Barrett-Heijltjes-McCusker-2022}, following standard techniques \cite{Huet-1994,Takahashi-1995}.

Before introducing a motivating example, let us take a step back. What \textit{is} a computational effect? Despite being an extremely active area of study, there is no generally accepted formal definition. We offer the following relevant quote from Filinski.
\begin{displayquote}
``Informally, an effect is any deviation from the intuitive characterization
of a program fragment as representing a simple function from inputs to outputs\ldots\,The challenge to the semanticist is thus to admit the possibility of effects, while
retaining as many as possible of the appealing properties of functional programming."\cite{Filinski-1996} 
\end{displayquote}
That is, a program has a side-effect if it causes some observable effect other than to consume and return its \textit{specified} inputs and output (author's emphasis). 
In other words,  one common perspective on side-effects is that they can be seen as inputs or outputs of a program that \textit{aren't} encoded in its type. \footnote{An example of the perspective under discussion is that of \textit{pre-monoidal} categories,  a discussion of which will follow. There do, however, exist various type systems which indicate the presence of effects.  }

\begin{figure}
	\[
	\begin{array}{cccc}
		\ \ \term{\rand}
		&\quad
		\term{\set}
		&\quad
		\term{\get}
		&
		 \term{\print}
	\\
		\ \ \term{\rnd<x>.[x]}
		&\quad
		\term{<x>.c<\_>.[x]c}
		&\quad
		\term{c<x>.[x]c.[x]}
		&
		 \term{<x>.[x]\out}
	\\
		\ \ \type{\rnd(\Z) > \Z}
		&\quad
		\type{\Z `c(\Z) > `c(\Z)}
		&\quad
		\type{`c(\Z ) > `c(\Z )\Z}
		&
		\type{\Z > \out(Z)}
	\\
		\begin{tikzpicture}[termpic, scale=2]
		\filldraw[opacity=0, black] (0,13) circle (3pt);
		\node (a2) at (-15,30) {$\textcolor{red}{\rnd}$};
		\draw[red] [densely dashed](-5,30) -- (30,30) ;
		\draw (20,30) -- (45,30) ;
		\filldraw[black] (20,30) circle (3pt);
		\node[opacity=0.2,draw=term,fill=term!10,rounded corners,minimum size=20pt] at (20,30) {};
		\node (a1) at (50,30) {$\ \ \lambda$};
		\end{tikzpicture}
		&
		\begin{tikzpicture}[termpic, scale=2]
		\node (a1) at (-5,20) {$\ \ \lambda$};
		\node (a1) at (-5,40) {$\ \ \textcolor{blue}{\textsf c}$} ;
		\draw[blue][densely dashed]  (5,40) -- (30,40) ;
		\draw (5,20) -- (20,20) ;
		\draw (20,20) -- (30, 25);
		\draw[blue][densely dashed] (30,25) -- (40,30) ;
		\draw[blue][densely dashed] (40,30) -- (55,30) ;
		\filldraw[black] (30,25) circle (3pt);
		\filldraw[black] (30,40) circle (3pt);
		\node[opacity=0.2,tallbox] at (30,30) {};
		\node (a1) at (55,30) {$\ \ \textcolor{blue}{\textsf c}$};
		\end{tikzpicture}
	&
		\begin{tikzpicture}[termpic, scale=2]
		\node (a1) at (-5,30) {$\ \ \textcolor{blue}{\textsf c}$};
		\draw[blue][densely dashed]  (5,30) -- (30,30) ;
		\draw[blue][densely dashed] (40,20) -- (55,20) ;
		\draw[blue][densely dashed] (30,30) -- (40,20) ;
		\draw(40,40) -- (55,40) ;
		\draw(30,30) -- (40,40) ;
		\filldraw[black] (30,30) circle (3pt);
		\node[opacity=0.2,tallbox] at (30,30) {};
		\node (a1) at (55,40) {$\ \ \  \lambda$};
		\node (a1) at (55,20) {$\ \ \textcolor{blue}{c}$};
		\end{tikzpicture}
		& \quad
		\begin{tikzpicture}[termpic, scale=2]
		\filldraw[opacity=0, black] (0,13) circle (3pt);
		\node (a2) at (-10,30) {$\lambda$};
		\draw(-5,30) -- (20,30) ;
		\draw [orange][densely dashed](20,30) -- (45,30) ;
		\filldraw[black] (20,30) circle (3pt);
		\node[opacity=0.2,draw=term,fill=term!10,rounded corners,minimum size=20pt] at (20,30) {};
		\node (a1) at (50,30){$\textcolor{orange}{\ \  \out}$};
		\end{tikzpicture}
	\end{array}
	\]
\[
\begin{array}{c}
	\term{~\rand~;\ ~\set~;~~\get~;~~\rand~;~~\set~;~~\get~;~+~;~~\print~ } \\ \\
	\vc{\begin{tikzpicture}[termpic, scale=2]
		\node (a1) at (-10,40) {$\ \ \textcolor{blue}{\textsf c}$};
		\node (a2) at (-10,20) {$\textcolor{red}{\rnd}$};
		\node (a3) at (-10,0) {$\textcolor{red}{\rnd}$};
		\draw[blue][densely dashed]  (0,40) -- (50,40) ;
		\draw[red] [densely dashed](0,20) -- (30,20) ;
		\draw (20,20) -- (40,20) ;
		\draw (40,20) -- (50,25) ;
		\draw[red][densely dashed] (0, 0) -- (110, 0) ;
		\draw[blue][densely dashed] (50,25) -- (60,30) ;
		\draw[blue][densely dashed] (60,30) -- (80,30) ;
		\draw (90,40) -- (210,40) ;
		\draw (80,30) -- (90,40) ;
		\draw[blue][densely dashed] (80,30) -- (90,20) ;
		\draw[blue][densely dashed] (90,20) -- (140,20) ;
		\draw (110,0) -- (130,0) ;
		\draw (130,0)-- (140,5)  ;
		\draw[blue][densely dashed] (140, 5) -- (150,10) ;
		\draw[blue][densely dashed] (150,10) -- (170,10) ;
		\draw (180,20) -- (210,20) ;
		\draw (170,10) -- (180,20) ;
		\draw[blue][densely dashed] (170,10) -- (180,0) ;
		\draw[blue][densely dashed] (180,0) -- (250,0) ;
		\draw (210,30) -- (230,30) ;
		\draw[orange][densely dashed] (230,30) -- (250,30) ;
		\node[opacity=0.2,draw=term,fill=term!10,rounded corners,minimum size=20pt] at (20,20) {};
		\filldraw[black] (20,20) circle (3pt);
		\node[opacity=0.2, tallbox] at (50,30) {};
		\filldraw[black] (50,40) circle (3pt);
		\filldraw[black] (50,25) circle (3pt);
		\node[opacity=0.2,tallbox] at (80,30) {};
		\filldraw[black] (80,30) circle (3pt);
		\node[opacity=0.2,termbox] at (110, 0) {};
		\filldraw[black] (110,0) circle (3pt);
		\node[opacity=0.2,tallbox] at (140,10) {};
		\filldraw[black] (140,20) circle (3pt);
		\filldraw[black] (140,5) circle (3pt);
		\node[opacity=0.2,tallbox] at (170,10) {};
		\filldraw[black] (170,10) circle (3pt);
		\node[tallbox] at (200,30) {$\textbf{\term+}$};
		\node[opacity=0.2,termbox] at (230,30) {};
		\filldraw[black] (230,30) circle (3pt);
		\node (a3) at (253,0) {$\textcolor{blue}{c}$};
		\node (a3) at (255,30) {$\textcolor{orange}{\ \  \out}$};
	\end{tikzpicture}}
	\end{array}
\]
\caption{String diagrams for effectful FMC-terms} 
\label{ex:effects-example}
\end{figure}
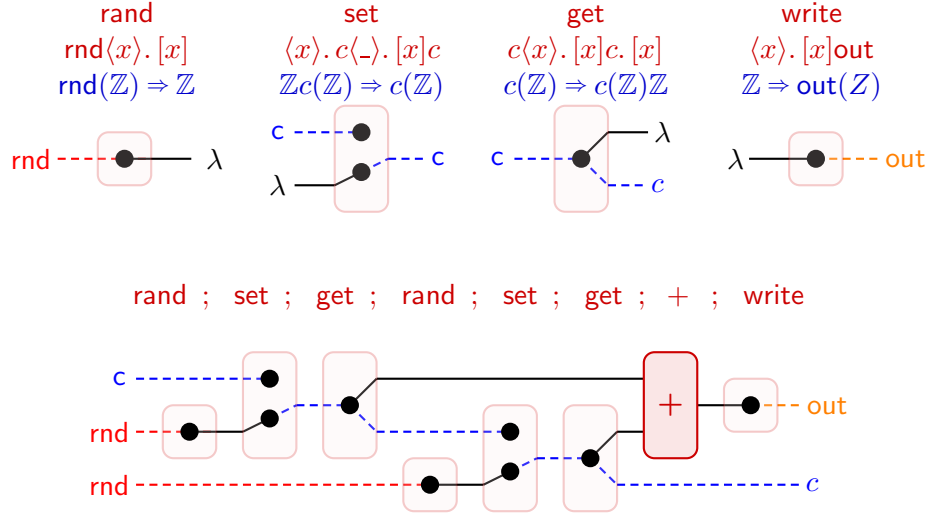

Consider that effectful primitves can be naively viewed as \textit{black boxes}. For example, to \textit{set} the value of a memory cell can be viewed as a black box which consumes the value to be written, performs the given side-effect, but has no return value. Similarly, to \textit{get} the value of a cell can be viewed as a black box which consumes no input, but returns the value held in state. In these cases, where the value is sent or where the value comes from, respectively, is left unaccounted for. 
Thus the black boxes have \textit{side-effects} which makes them sensitive to the order of operations, as in the previously given non-confluent $\lambda$-terms. This is because, despite there being no \textit{explicit} dependency between the operations of \textit{setting} and \textit{getting}, they nevertheless \textit{are} dependent on each other. 

From this perspective, the achievement of the FMC is to \textit{open} these boxes, and reveal their inner workings. Effectful primitives are \textit{decomposed} into the plain syntax of the FMC. FMC-terms can then be typed in a relatively standard way, allowing us to see previously \textit{hidden inputs and outputs}. 
This perspective is amenable to a graphical representation in via string diagrams. For an overview of these, see \cite{Selinger2011}, and for related literature, see \cite{DBLP:journals/corr/abs-2205-07664,Schweimeier2001CategoricalAG,JEFFREY199851,DBLP:conf/rta/Mellies14,JOYAL199155}.

For an example of this, consider Figure \ref{ex:effects-example}.\footnote{
The term for this diagram is given modulo permutations of locations and symmetries, where appropriate. This is all discussed later, and is justified by the Permutation Lemma \ref{lem:permutation-equivalence}  and Remark \ref{remark:permutation-ids}} Consider some standard effectful primitives: returning a random integer, setting and dereferencing a memery cell, and writing to output. These primitives, respectively, and their corresponding FMC-terms (in dark red), and FMC-types (blue) are given at the top of the figure. We delay discussion of the FMC-terms and -types.  An example of a larger effectful process given below the depiction of primitives. 
The rest of this section is devoted to explaining this example in detail. 

A key innovation of the FMC -- which is precisely what allows for the encoding of effects -- is that of parameterizing the application and abstraction of the $\lambda$-calculus in terms of \textit{locations}. 
In the example, we use black for the \textit{main}, location $\lambda$, which deals with the \textit{standard} arguments and return values of a function. Colours and dashed lines are then used to indicate the non-main locations \red{\textsf{rnd}}, \color{blue}{\textsf{c}, \color{black} and \color{orange}{\textsf{out}}\color{black}, which respectively track \textit{random inputs}, {\textit{the value held in a memory cell $c$}}, and \textit{output}, \textit{e.g.}, printing to the screen. We then allow the black dot to depict \textit{relocation}, as described below.
The primitives given in the figure can the be considered as follows. 
Note that the previously hidden inputs and outputs are exactly those on non-main locations, that is, the dashed wires. 
\begin{itemize}
\item The term $\term{\rand}$ consumes an integer from a random stream held on location $\term{\rnd}$, and relocates it to the \textit{main} location, on which it is returned;
\item The term $\term{\set~c}$ consumes an integer from the \textit{main} location $\lambda$, and one from a memory cell held at location $\color{blue}{c}$. The integer consumed from $\color{blue}{c}$ is \textit{deleted}, but the integer consumed from the main location is relocated to location $\color{blue}{c}$, taking its place;
\item The term $\term{\get~c}$ consumes an input from a memory cell held at $\term{c}$ and \textit{duplicates} it, returning one copy to $\color{blue}{c}$ and one copy to the main location as its standard return value. Thus, the cell can be read from again, returning same value;
\item The term $\term{\print}$ consumes an integer from the main location, and relocates it to an output stream held on  \color{orange}{\textsf{out}} \color{black}. 
\end{itemize}

 The expliciting of dependencies (\textit{i.e.}, dataflow) between operations tells us when we can safely reorder effectful operations, and when we cannot, achieving a \textit{properly monoidal} category. Further, we can now ``look inside'' the black box, revealing -- modulo relocation --  more familiar operations of duplication and deletion.

For example, it is clear from this diagram that the second call to $\term \rand$ may safely be made before the first calls to $\term \set$ and $\term \get$. This would be illustrated by `sliding' the $\term{\rand}$ operation along the wire, as in a monoidal category. We can also see that the second $\term \set$ is \textit{dependent} on the first $\term \get$: the value returned to $\color{blue}{c}$ by the first $\term \get$ is discarded by the second $\term \set$. Indeed, we will see later the composition ($\term{;}$) of the terms $\term \set$ and $\term \get$ creates such a beta-redex, corresponding to the expected interaction of duplication and deletion in string diagrams.

Note again that in the ``black box'' picture described previously, the dashed wires \textit{are not visible}, which means one cannot know when one can soundly rearrange the order of operations: one fears violating a hidden dependency. In the the study of \textit{pre-monoidal} categories \cite{power-robinson-1997} a strict ordering is enforced on operations, restricting completely their rearrangement.
In terms of string diagrams, this can be viewed as forgetting the dashed wires (non-main locations), and instead adding an additional \textit{control} wire \cite{Schweimeier2001CategoricalAG,JEFFREY199851} which is used to totally order the boxes.
The FMC offers another perspective, and 
the relationship between string diagrams and the FMC will be discussed further later in the introduction, and in detail in Chapter \ref{chapter:CCC-eqns}.


The syntax of the FMC generalizes and parameterizes that of the $\lambda$-calculus.
\begin{itemize}
\item First,  the variable construct is decomposed into \textit{variable-with-continuation} and \textit{skip} constructs, allowing a simple concatenation, or \textit{sequential composition}, operation to be defined on terms. This gives rise to translations of the CBN and CBV $\lambda$-calculus, each preserving operational semantics by expressing explicitly the appropriate sequencing of computation.
\item Second, the application and abstraction of the $\lambda$-calculus are parameterized in terms of \textit{locations} $\{\term{a, b, c,}\!\ldots\}$.  This allows the encoding of effects, and the unification of their operational semantics, syntax and reduction rules with those of the $\lambda$-calculus. 
\end{itemize} 
Importantly, the first point enables the expression of both the CBN and CBV semantics of effects, as well as giving rise to a notion of \textit{sequential composition}, familiar from imperative programming.\footnote{The notion of sequencing described here is syntactic and semantic, but does \textit{not} refer to reduction order.}
The second point can be seen in the example terms discussed earlier: abstraction is given by $\term{a<x>.M}$ and application by $\term{[N]a.M}$, with the argument written on the left. 
Both innovations comes from the consideration of the operational semantics of a simple variation of a standard stack Krivine machine which has \textit{multiple} stacks \cite{Krivine-2007}, and are considered in detail in the remaining chapter.

The results contained in this thesis contribute to the program of research outlined above by verifying that the FMC does indeed preserve two fundamental \textit{good properties} of the $\lambda$-calculus. 
The first is that the category of typed FMC-terms modulo a natural notion of observational equivalence is Cartesian closed, extending the ideas of the example described above.  In fact,  given an appropriate equational theory, it forms the \textit{free Cartesian closed category}. 
The second contribution is to show that the type system of the FMC guarantees \textit{strong normalisation} with respect to the (analogue of the) beta reduction relation.
Both results speak to the strength of the type system. 

Necessarily for the result of Cartesian closure,  the calculus under study here treats locations \textit{uniformly}: each location has an associated push- and pop- action, and its associated stack is of unbounded depth.  This is in contrast with the encoding of effects described above,  which requires certain special properties associated to effectful locations, \textit{e.g.}, read- or write-only streams to model input or output,  or bounded depth stacks to model memory cells.  Without such assumptions, we cannot claim to accurately model the given effects.  Indeed, there are more terms typeable than there would be if we were aiming to properly model the constraints associated to locations in effectful computation, and these extra terms are essential to recovering a Cartesian closed semantics: they result in there being more contexts with which to test for observational (contextual) equivalence, and in fact they
are needed to define the Cartesian closed equipment (e.g. the diagonal term at certain types).  
As such, the categorical semantics is intended as a semantics of the calculus itself, rather than of effects, aiming to show that the calculus is semantically well-behaved despite its novelties.  

With the assumption of uniformity, the inclusion of non-main locations into the calculus is semantically similar to the encoding of monadic state in simply-typed $\lambda$-calculus, where a stateful process  of type $A \to B$ can be considered a pure function with the larger type $A \to (S \to B \times S)$.  In the FMC, when a term accesses a second stack, this will be recorded in its type.  Thus, we can recover a Cartesian closed category of terms.  Together with uniformity,  it is to be understood that we achieve Cartesian closure because our notion of observation which determines our equivalence on terms is very strong: indeed, our type system alone tracks exactly how many times a term consults a random stream, which ought to be unobservable information. \footnote{One would expect that a term which reads one element from the random stream and then discards it ought to be observationally equivalent to a term which does nothing at all.}
Further investigation of the calculus as a model of computational effects is left as future work, however we note that tweaking the type system to account for read-only, write-only, or bounded depth stacks is trivial.

These results are a sanity-check for the FMC. Despite its novel perspective on effectful computation, it would appear that the calculus is very well-behaved: operationally (via the machine), semantically (via its cateogrical semantics), and algebraically (beta reduction is confluent and strongly normalising). 
These results are discussed in detail in the relevant chapters. 
For now, we proceed to a more detailed introduction to the FMC and its encoding of effects.



\label{sec:sequencing}
\section{The $\lambda$-calculus and the Stack Machine}

The Krivine Abstract Machine (KAM) \cite{Krivine-2007} is a standard call-by-name stack machine which gives an especially simple operational semantics to the $\lambda$-calculus. That is, the machine specifies \textit{how} to execute a $\lambda$-term; in particular it implements \textit{weak head reduction}. The machine is presented informally as follows.\footnote{For clarity, we present the machine with substitution rather than an environment, since we are not concerned with actual implementation. } 

A \textit{state} of the abstract machine is a pair $(S, M)$ consisting of a stack $S$ of $\lambda$-terms, and a $\lambda$-term $M$  to be executed.
The 
\textit{transitions}, or \textit{steps}, of the machine are given below,  where constructors of the $\lambda$-calculus are interpreted as intructions for the machine:
\begin{itemize}
\item
\textbf{Application}, $M(N)$, as \textbf{push}  $N$ onto the stack and continue as $M$; 
\item
\textbf{Abstraction}, $\lambda x.M$, as \textbf{pop} the head $N$ off the stack and continue as $M\{N/x\}$.
\end{itemize}
\[
\begin{array}{@{(~}l@{~,~}r@{~)}}
	S 	         & {M(N)}
\\\hline
	S \cdot N &       M
\end{array}
\qquad
\begin{array}{@{(~}l@{~,~}r@{~)}}
	S \cdot N &   {\lambda x.M}
\\\hline
	S               & {M\{N/x\}}
\end{array}
\]
A \textit{run}, or \textit{execution},  of the machine is a sequence of steps.
A redex $(\lambda x. M)N$ is thus executed by a \textit{push} followed by a \textit{pop}, and indeed the machine correctly evaluates such a term as $M\{N/x\}$, as shown below, left. 
Considering only closed terms, variables are never encountered and the machine terminates when it reaches a state $(\e, \lambda x.M')$, outputting the weak head normal form $\lambda x.M'$.\footnote{In general, we can run the machine with open terms, in which case it may also terminate on states of the form $(N_n \cdots N_1, x)$, corresponding to the weak head normal form $x N_1\ldots N_n$. However,  we need not consider open terms.}
A terminating run is shown below, right, where the double line indicates a sequence of steps. The symbol $\epsilon$ denotes the empty stack. 
\[
\begin{array}{@{(~}l@{~,~}r@{~)}}
	S 	         & {(\lambda x.M)N}
\\\hline
	S \cdot N 	         & {\lambda x.M}
\\\hline
	S         & {M\{N/x\}}
\end{array}
\qquad \qquad 
\begin{array}{@{(~}l@{~,~}r@{~)}}
	S  &   {M}
\\\hline \hline
	\e               & {\lambda x.M'}
\end{array}
\]
A call-by-name $\lambda$-term can thus be viewed as a sequence of \textit{push-} and \textit{pop-actions}, ending in a variable. In order to facilitate the reading of an application as \textit{push} $N$ \textit{and continue as} ${M}$, let us reimagine the syntax of $\lambda$-terms so that the argument $N$ is written as a prefix to $M$. 
\[
\begin{array}{r@{\quad\Coloneqq\quad}c@{~\mid~}c@{~\mid~}c}
	{M, N} & {x} &  MN & \lambda x.M  \\[5pt]
	\term{M, N} & \term{x} & \term{[N].M} & \term{<x>.M}  
\end{array}
\]
Two successive generalizations of the $\lambda$-calculus are presented, both based on the intuition given by the KAM. They are naturally presented using this new syntax. First, the \textit{Sequential $\lambda$-calculus (SLC)}, which incorporates a mechanism for \textit{sequential composition} into the $\lambda$-calculus, and then the \textit{Functional Machine Calculus (FMC)}, which further parameterizes actions in terms of \textit{locations}. Each is presented informally, in turn, delaying formal definitions until Chapter \ref{chapter:fmc-prelims} (see there for reference).
\section{The Sequential $\lambda$-Calculus}

An important issue when dealing with computational effects is that of \textit{evaluation order}. {We} saw in the introduction that a \textit{single} effectful $\lambda$-term can reduce to distinct normal forms, dependent on choice of reduction strategy. We do not wish to fix a reduction strategy,  as this entails a loss of expressivity -- instead, we wish to be able to express both the CBN and CBV semantics of effects within the same calculus, and with natural syntax. Ergo, we aim to refine the $\lambda$-calculus into a new language which satisfies the following.
\begin{displayquote}
Given a single $\lambda$-term, we aim to have \textit{distinct} call-by-name (CBN) \textit{and} call-by-value (CBV) translations into \textit{distinct} terms of some new, confluent language; each preserving the operational semantics of CBN and CBV, respectively. 
\end{displayquote}  Indeed, we will see in the review of related literature (Chapter \ref{chapter:related-lit}) that {something common to the various approaches to effectful $\lambda$-calculi is that they provide some way for the programmer to express the \textit{sequencing} of computations \textit{explicitly}, within syntax}. The notion of sequencing we refer to is semantic, and does not refer to a notion of reduction order. 
In the SLC, we also have a notion of explicit sequencing which we can use to define the desired translations. 

The Sequential $\lambda$-calculus is a novel approach towards this aim which is based on the stack machine intuition given previously. But it is not unique in its capabilities -- other languages (especially, Call-by-Push-Value \cite{Levy-2003}) provide similar capabilities. What is unqiue is the natural combination of our sequencing mechanism with a similarly machine-inspired perspective on effectful behaviour, which is introduced subsequently with the Functional Machine Calculus. This combination gives the programmer control over whether they want a CBN or CBV semantics for effects -- and for practical purposes, we need to be able to express both -- as well as resulting in a calculus which is confluent  \textit{in the presence of effects}.

\subsection{Syntax}
The Sequential $\lambda$-calculus allows for the sequencing of computations by taking seriously the operational intuition given by the Krivine machine. Observe that this perspective says that $\lambda$-terms are  sequences of instruction; push and pop actions and variables,  \textit{where variables occur exactly at the end of the sequence},  and this definition seems overly restrictive in the following way: \textit{the set of terms is not closed under concatenation of instruction sequences} (since this would result in a variable occuring in the middle of such a sequence).  However, such an operation is unproblematic from the operational perspective: if dealing with closed terms, the machine will never encounter a variable.\footnote{Every variable is bound, and thus will be substituted for another instruction sequence before it is reached. This justifies considering variables themselves as instructions. }  Further, including some notion of sequencing is well-motivated by the simple existence of imperative programming.

To allow for sequential composition, it suffices to generalize the syntax of the $\lambda$-calculus accordingly and, of course, provide a unit for composition. We thus achieve the \textit{Sequential $\lambda$-calculus (SLC)}.
\begin{align*}
\term{M,N}
  \quad\Coloneqq\quad \term *
       ~\mid~ \term{x.M}
       ~\mid~ \term{[N].M}
       ~\mid~ \term{<x>.M}
\end{align*}
the variable $\term x$ has been \textit{decomposed} into two constructs: a \textit{variable-with-continuation} $\term{x.M}$ and an \textit{identity} (or \textit{end-of-sequence}, or \textit{skip}), $\term *$, with the original variable construct recoverable as $\term{x.*}$. We will omit the trailing $\term{*}$ from terms to avoid unnecessary clutter. 
\textit{Sequential composition} is then given as a \textit{defined} operation $\term{N;M}$ on terms, which is capture-avoiding:\footnote{It would be possible to take the construct $\term{N;M}$ as primitive, but this would necessitate working modulo associativity of sequencing; by taking \textit{prefixing} as primitive, we have essentially chosen to associate to the right, considering this preferable.} 
\begin{align*}
		 \term{*;M} &=         \term M   &
	 \term{x.N;M} &=    \term{x.(N;M)} &
\\     \term{[P].N;M} &= \term{[P].(N;M)} &
	\term{<x>.N;M} &= \term{<x>.(N;M)} & (\term x\notin\fv{\term M})
\end{align*}
where the set of \textit{free variables} of a term $\term{M}$, $\fv{\term M}$, is defined as 
\[
	\begin{array}{lll@{\qquad}lll}
	\fv{\term *} &=& \emptyset & \fv{\term{x.M}} &=& \fv{\term{M}} \cup \{\term x\} \\
	\fv{\term{[N].M}} &=& \fv{\term{M}}\cup \fv{\term{N}} & \fv{\term{<x>.M}} &=& \fv{\term{M}} \setminus\, \{\term x\} 
	\end{array}
\]

There is an obvious inclusion of the set of $\lambda$-terms into the set of SLC-terms.
The following example gives some SLC-terms which are \textit{not} in the image of this inclusion. 
\begin{example}\label{ex:slc-terms}
Consider the terms below. 
The first term pops the top item off the stack and discards it. The second term pops the top item off the stack, but returns two copies of it to the stack. The third term makes \textit{no} change to the stack, while the fourth swaps the top two elements of the stack.  Note how the \textit{last-in first-out} nature of the stack is reflected in the form of these two terms. 
\[
	\term{<x>} \qquad \term{<x>.[x].[x]} \qquad \term{<x>.<y>.[y].[x]} \qquad \term{<x>.<y>.[x].[y]}  \qquad \term{<f>.f.f}
\]
The final term is a higher-order function, which pops the top item off the stack and executes it twice, using the remaining stack items as input. 
\end{example}
Beta reduction, shown below corresponds to an (operationally sound) optimization of the term for the machine\footnote{Note, we get the typical time/space tradeoff: a reduction leads to a larger term which evaluates more quickly on the machine. }, eliminating consecutive \textit{push} and \textit{pop} actions. 
\[
	\term{[N].<x>.M} \rw_\beta \term{\{N/x\}M} 
\]
{We formally define capture-avoiding substitution as follows.  In particular, we have $\term{\{N/x\}x.M} = \term{N;\{N/x\}M}$, but otherwise it is as expected for a $\lambda$-calculus.}
\[
\begin{array}{lll@{\qquad}llll}
	\term{\{N/x\}*}    &=& \term{*}  & \term{\{N/x\}x.M} &=& \term{N;M} \\
	\term{\{N/x\}[P].M}      &=& \term{[\{N/x\}P].\{N/x\}M}  &  \term{\{N/x\}<y>.M} &=& \term{<y>.\{N/x\}M}   
\end{array}
\]
where, in the abstraction case,  $\term y\notin\fv{\term N}$.  
Note, substitution is now written on the left, matching the \textit{pop} transition of the machine and the new syntax.

The beta law is applicable in any \textit{sequential} context, an appropriate notion of context for the SLC, which we define in Chapter \ref{chapter:fmc-prelims}. 
It is a result of \cite{Barrett-Heijltjes-McCusker-2022} that beta reduction remains confluent. 
Later, we will show that beta reduction is strongly normalising for well-typed terms.  
For our later semantic investigations, we will also give an appropriate equational theory on terms which includes that generated by the beta reduction relation, but also validates other equations such as $\term{*} =_\textsf{id} \term{<x>.[x]}$, which states that popping a term and pushing it back is the same as doing nothing.  

\subsection{The Sequential $\lambda$-Calculus as a Stack Transformer Language}
This generalized calculus facilitates a change in perspective from the $\lambda$-calculus. Previously, we saw that running a $\lambda$-term on the Krivine machine consumes its stack as input and terminates on the state $(\e, \lambda x.M)$, where the remaining term is considered the output of the machine. In the SLC, however, this is considered a \textit{failure} state: execution halts because not enough inputs have been provided. 
Instead, like in imperative languages, we can expect a successful computation to terminate in a \textit{skip} command, and outputs are given as the remaining stack items.
\[
	\machine SMT
\]

Thus, the SLC is a calculus of stack \textit{transformers}, whereas the $\lambda$-calculus is of one of stack \textit{consumers}.
Note that a certain input/output symmetry is thus recovered.
Then $\term{N;M}$ gives composition of runs, with the output stack of $\term{N}$ becoming the input stack of $\term{M}$, as follows, and $\term *$ gives the identity run (of zero steps). 
\[
	\text{if} \quad \machine RMS \quad \text{and} \quad \machine SNT \quad\text{then} \quad
    \begin{array}{@{(~}l@{~,~}r@{~)}}
    R & \term{M;N}
    \\\hline\hline\rule[-5pt]{0pt}{15pt}
    S & \term{N}
    \\\hline\hline\rule[-5pt]{0pt}{15pt}
    T & \term{*}
    \end{array}
\]
Following this observation, we will define in Chapter \ref{chapter:CCC-eqns} a natural category of \textit{closed} (typed) SLC-terms, with composition given by sequencing, in contrast with the $\lambda$-calculus, whose \textit{open} terms form a category with composition given by \textit{substitution}. As mentioned, closed terms are considered so that a machine run never encounters a variable. 

Note that, while we can hand a term too few inputs to succesfully complete a run, we cannot hand a term \textit{too many} inputs -- see below. 
We make later use of the \textit{expansion} of a stack, as this is called. 
\[
	\text{if} \quad \machine SMT \quad\text{then} \quad
    \begin{array}{@{(~}l@{~,~}r@{~)}}
    RS & \term{M}
    \\\hline\hline\rule[-5pt]{0pt}{15pt}
    RT & \term{*}
    \end{array}
\]

We emphasize again that beta reduction in the SLC corresponds not to \textit{evaluation} of a term, which is performed by the machine with respect to an input stack, but to \textit{compilation}, \textit{i.e.}, optimization of a term for the machine (with the usual time/space trade off - a larger term which runs more quickly, that is, in fewer steps). 

We proceed to discuss how the change in perspective from that of the $\lambda$-calculus to that of the SLC manifests in the type system.
%

\subsection{Simple Types for the Sequential $\lambda$-Calculus}
One of the aims of the Functional Machine Calculus is to bring the powerful reasoning tools developed for the $\lambda$-calculus to bear on effectful {languages}. Perhaps the foremost such tool is \textit{simple types} for the $\lambda$-calculus. In particular, the simple type system provides a guarantee of \textit{strong normalisation}, and is in  natural correspondence  with intuitionistic logic via the Curry-Howard isomorphism. Lambek further developed its categorical correspondence with the \textit{free Cartesian closed category}, giving it a natural denotational semantics and a characterization of its models. \cite{lambek2,Lambek1968, DBLP:conf/litp/Lambek85,Lambek1986-LAMITH-2}.
We present here an overview of the type system of the SLC, to be extended appropriately for the FMC, while the body of this thesis concentrates on proving analogues of the above results. 

Consider now the perspective of SLC-terms as stack transformers. The type of a term describes its net effect on the stack. Stacks are given a \textit{type vector} $\type{!t}$: a type for each element of the stack. Terms $\term{M}$  in a context $\Gamma$ are then given type $\type{t}$, which is either a base type $\type{a}$ or an implication between type vectors.
\[
	 \Gamma \vdash \term{M: t} \qquad \type{t} ~\Coloneqq~ \type{a} ~\mid~ \type{?s_A > !t_A} \qquad \type{!t} ~\Coloneqq~ \type{t_1 \ldots t_n}\, 
\]
On the left of the implication is recorded the type of each element that may be consumed from the stack, and on the right the type of each element left on the stack once the run is successfully completed. We \textit{reverse} a type vector on the left of an implication, indicated by a reversed arrow $\type{?t}$, reflecting the first-in last-out nature of the stack. Concatenation of vectors is denoted by juxtaposition. 
The typing rules for the SLC are shown in Figure \ref{fig:SLC-types-plain}, with some admissible rules shown in Figure \ref{fig:SLC-admissible-plain}.
\begin{figure}
\[
	\infer[\!\scriptstyle{\mathsf{id}}]{\Gamma \vdash \term{ *:?t>!t}}{}
	\qquad
	\infer[\!\scriptstyle{\mathsf{base}}]{\term{x: a}, \Gamma \vdash \term{x:a}}{}
\]
\\
\vspace{-\baselineskip}
\[
	\infer[\!\scriptstyle{\mathsf{app}}]
	{\Gamma \vdash \term{[N].M: ?s > !t}}
	{\Gamma \vdash \term{N: r} && {\Gamma \vdash \term{M:r?s > !t}}}
	\qquad
	\infer[\TR l]
	  {\Gamma \vdash \term{ <x>.M : r?s>!t}}
	  {\term{x:r, G} \vdash \term{ M:?s>!t}}
\]
\\
\vspace{-\baselineskip}
\[
	\infer[\!\scriptstyle{\mathsf{var}}]
	{\term{x:?r > !s},  \Gamma \vdash \term{x.M: ?r\,?t > !u}}
	{{\term{x:?r > !s}, \Gamma \vdash \term{M: ?s\,?t > !u}}}
\]
\caption{Typing rules for the Sequential $\lambda$-Calculus}
\label{fig:SLC-types-plain}
\end{figure}
%

\begin{figure}
\[
	\infer[\!\scriptstyle{\mathsf{seq}}]
	{\Gamma \vdash \term{M;N: ?r > !t}}
	{{\Gamma \vdash \term{M: ?r > !s}} & {\Gamma \vdash \term{N: ?s > !t}}}
	\qquad
	\infer[\!\scriptstyle{\mathsf{exp}}]
	{\Gamma \vdash \term{M: ?s\,?t > !t\,!u}}
	{{\Gamma \vdash \term{M: ?s > !u}}}	
\]
\\
\vspace{-\baselineskip}
\[	
	\infer[\!\scriptstyle{\mathsf{cut}}]
	{\Gamma \vdash \term{\{N/x\}M: ?s > !t}}
	{{\Gamma \vdash \term{N: r}} &{\term{x:r},  \Gamma \vdash \term{M: ?s > !t}}}
\]
\caption{Admissible rules for the Sequential $\lambda$-calculus}
\label{fig:SLC-admissible-plain}
\end{figure}
\begin{example}\label{ex:slc-typed-terms}
We described in Example \ref{ex:slc-terms} how the terms given below transform the stack. Here, we also give corresponding types.
Note how, in the final two terms, the last-in first-out nature of the stack is reflected in the types as well as the terms, so that the syntax of types and terms matches.
\begin{align*}
	\term{<x>:}&\type{t\, > } & \term{<x>.[x].[x]:}&\type{ t > tt} \\
	 \term{<x>.<y>.[y].[x]:}&\type{ st > ts} & \term{<x>.<y>.[x].[y]:}&\type{ st > st}
\end{align*}
For a higher-order example, consider the term $\term{<f>.f}$. It consumes a term from the stack, which then transforms the remaining stack in some way according to the type of that term. Thus, we can type it as below, left. The term below, right, is similar. 
\[
	\term{<f>.f: (?s > !t)?s > !t} \qquad \term{<f>.f.f: (?t > !t)?t > !t}
\]
\end{example}
The typing derivation for a term will in fact serve as a direct proof of termination of the machine, as is proved in Chapter \ref{chapter:fmc-prelims}, following the proof due to Heijltjes in \cite{Barrett-Heijltjes-McCusker-2022}.
This gives a new perspective on types, and in particular an \textit{operational intuition} as to their meaning. 
The result is that, for any stack $S: \type{!s}$ and any closed term $\term{M: ?s > !t}$, there exists a stack $T: \type{!t}$ and a successful run of the machine such that 
\[
	\machine {S}N{T}\, .
\]

Figure \ref{fig:string-equipment} describes how to represent first-order terms (that is, where the stack only holds base types) of the typed SLC in string diagrams. The notation $\term{!x}$ indicates a vector of variables, and $\term{?x}$ its reverse, and we accordingly let 
\[
	\term{<?x>.M} = \term{<x_n>\ldots <x_1>.M} \qquad \textup{and} \qquad \term{[!x].M} = \term{[x_1]\ldots [x_n].M}
\] denote a sequence of abstractions and applications on $\term{!x} = \term{x_1 \ldots x_n}$. The wires represent the input and output stacks, with the head of the stack at the top of the diagram. As is typical, where $\term{M}$ is the identity $\term *$, we simply draw the wires. The term $\term{M: ?r?t > !t!s}$ is the term $\term{M: ?r > !s}$ given a stack \textit{expanded} by an extra $\type{!t}$ elements.  
The term $\term{<?x>.M.[!x]: ?t?r > !s!t}$ pops the arguments for $\type{!t}$ from the stack as the variables $\term{?x}$, evaluates $\term M$ on the remaining stack of type $\type{!r}$, resulting in an output stack of type $\type{!s}$and then returns the values bound to  $\term{!x}$ to the stack. 
Some terms from Example \ref{ex:slc-typed-terms} are also represented in Figure \ref{fig:string-equipment}. 
 Note that we need a stronger equational theory than is generated by beta reduction alone  to get all the expected equivalences between diagrams, and this is developed in Chapter \ref{chapter:CCC-eqns}.

\begin{figure}
\[
\begin{array}{c@{\qquad\qquad\qquad}c}
\begin{tikzpicture}[termpic, scale=2]
	\draw
	  (-28,16)node[left]{$\type{r_1}\,$}--(28,16)node[right]{$\,\type{s_1}$}
	  (-22, 4)node[left]{$\type{r_m}$}--(22, 4)node[right]{$\,\type{s_n}$};
	\diagdots[-1.5,3]{-20,10}
	\diagdots[ 1.5,3]{ 20,10}
	\node[termbox] at (0,10) {$\term M$}; 
\end{tikzpicture}
&
{\begin{tikzpicture}[termpic, scale=2, baseline={(0,0.2)}]
	\wires{?r}{-50}{20}{50}{!t}
	\node[termbox] at (-20,20) {$\term M$};
	\node[termbox] at (20,20) {$\term N$};
	\diagdots[0,3]{0,20}
\end{tikzpicture}} 
\\
\term{M: ?r > !s} & \term{M;N: ?r > !t} 
\\
\\
	\vc{\begin{tikzpicture}[termpic, scale=2]
		\wires{?r}{-30}{20}{30}{!s}
		\wires{?t\,}{-20}{0}{20}{!t}
		\node[termbox] at (0,20) {$\term M$}; 
	\end{tikzpicture}}
	&
	\vc{\begin{tikzpicture}[termpic, scale=2]
		\node (a1) at (-28,36) {$\,\term{<x_1>}\,$};
		\node (b1) at ( 28,36) {$\,\term{[x_1]}\,$};
		\node (an) at (-22,24) {$\,\term{<x_n>}\,$};
		\node (bn) at ( 22,24) {$\,\term{[x_n]}\,$};
		\draw 
		  (-48,36)node[left] {$\type{t_1}\,$} -- (a1) (b1) -- ( 48,36)node[right]{$\,\type{t_1}$}
		  (-42,24)node[left] {$\type{t_n}\,$} -- (an) (bn) -- ( 42,24)node[right]{$\,\type{t_n}$};
		\draw[dotted,line cap=round,color=term] (a1)--(b1) (an)--(bn);
		\diagdots[-1.5,3]{-40,30}
		\diagdots[ 1.5,3]{ 40,30}
		\wires{?r}{-40}{10}{40}{!s}
		\node[termbox] at (0,10) {$\term M$}; 
	\end{tikzpicture}}
\\ \\
		\term{M:?r?t>!t!s}  &
		\term{<?x>.(M;[!x]) : ?t?r > !s!t}  
\end{array}
\]
\\
\\
\vspace{-0.25cm}
\vspace{-\baselineskip}
\[
\begin{array}{c@{\qquad\qquad}c@{\qquad\qquad}c}
\vc{\begin{tikzpicture}[termpic, scale=2]
		\node (a1) at (-28,36) {$\,\term{<x_1>}\,$};
		\node (an) at (-22,24) {$\,\term{<x_n>}\,$};
		\node (b1) at ( 34,36) {$\,\term{[x_1]}\,$};
		\node (bn) at ( 28,24) {$\,\term{[x_n]}\,$};
		\node (c1) at ( 22,12) {$\,\term{[x_1]}\,$};
		\node (cn) at ( 16, 0) {$\,\term{[x_n]}\,$};
		\draw 
		  (-48,36)node[left] {$\type{t_1}\,$} -- (a1) 
		  (-42,24)node[left] {$\type{t_n}\,$} -- (an) 
		  (b1) -- ( 54,36)node[right]{$\,\type{t_1}$}
		  (bn) -- ( 48,24)node[right]{$\,\type{t_n}$}
		  (c1) -- ( 42,12)node[right]{$\,\type{t_1}$}
		  (cn) -- ( 36, 0)node[right]{$\,\type{t_n}$};
		\draw[dotted,line cap=round,color=term] (a1)--(b1) (an)--(bn) (a1)--(c1) (an)--(cn);
		\diagdots[-1.5,3]{-40,30}
		\diagdots[ 1.5,3]{ 46,30}
		\diagdots[ 1.5,3]{ 34, 6}
	\end{tikzpicture}}
&
\vc{\begin{tikzpicture}[termpic, scale=2, baseline={(0,1.1)}]
		\node (a1) at (-28,36) {$\,\term{<x_1>}\,$};
		\node (an) at (-22,24) {$\,\term{<x_n>}\,$};
		\draw 
		  (-48,36)node[left] {$\type{t_1}\,$} -- (a1) 
		  (-42,24)node[left] {$\type{t_n}\,$} -- (an);
		\diagdots[-1.5,3]{-40,30}
	\end{tikzpicture}}
\\
\\
\term{<?x>.[!x].[!x]: ?s > !s!s} & \term{<?x>: ?s > } 
\end{array}
\raisebox{34pt}{
\begin{tikzpicture}
	\begin{pgfonlayer}{nodelayer}
		\node [style=none] (0) at (1.5, -0.5) {};
		\node [style=none] (1) at (1.5, -2.5) {};
		\node [style=none] (3) at (4.5, -0.5) {};
		\node [style=none] (4) at (4.5, -2.5) {};
		\node [style=none] (10) at (-0.75, -0.5) {$\type{?s}$};
		\node [style=none] (11) at (-0.5, -2.5) {$\type{?t}$};
		\node [style=none] (15) at (6.5, -2.5) {$\type{!s}$};
		\node [style=none] (16) at (7, -0.5) {$\type{!t}$};
		\node [style=none] (17) at (3, -4.5) {$\term{<?x>.<?y>.[!x].[!y]: ?s?t > !s!t}$};
		\node [style=none] (18) at (0.75, -0.5) {$\term{<?x>}$};
		\node [style=none] (19) at (1, -2.5) {$\term{<?y>}$};
		\node [style=none] (20) at (5, -2.5) {$\term{[!x]}$};
		\node [style=none] (21) at (5.5, -0.5) {$\term{[!y]}$};
		\node [style=none] (22) at (5.875, -0.5) {};
		\node [style=none] (23) at (5.375, -2.5) {};
		\node [style=none] (24) at (6, -2.5) {};
		\node [style=none] (25) at (6.5, -0.5) {};
		\node [style=none] (26) at (0.25, -0.5) {};
		\node [style=none] (27) at (0.5, -2.5) {};
		\node [style=none] (28) at (0, -2.5) {};
		\node [style=none] (29) at (-0.25, -0.5) {};
		\node [style=none] (30) at (1.25, -0.5) {};
		\node [style=none] (31) at (5, -0.5) {};
	\end{pgfonlayer}
	\begin{pgfonlayer}{edgelayer}
		\draw [style=red term dash, in=0, out=-180] (3.center) to (1.center);
		\draw [style=red term dash, in=-180, out=0] (0.center) to (4.center);
		\draw [style=thick line] (22.center) to (25.center);
		\draw [style=thick line] (24.center) to (23.center);
		\draw (29.center) to (26.center);
		\draw [style=thick line] (28.center) to (27.center);
		\draw [style=red term dash] (30.center) to (0.center);
		\draw [style=red term dash] (3.center) to (31.center);
	\end{pgfonlayer}
\end{tikzpicture}}
\]
\caption{String diagrams for Sequential $\lambda$-calculus}
\label{fig:string-equipment}
\end{figure}
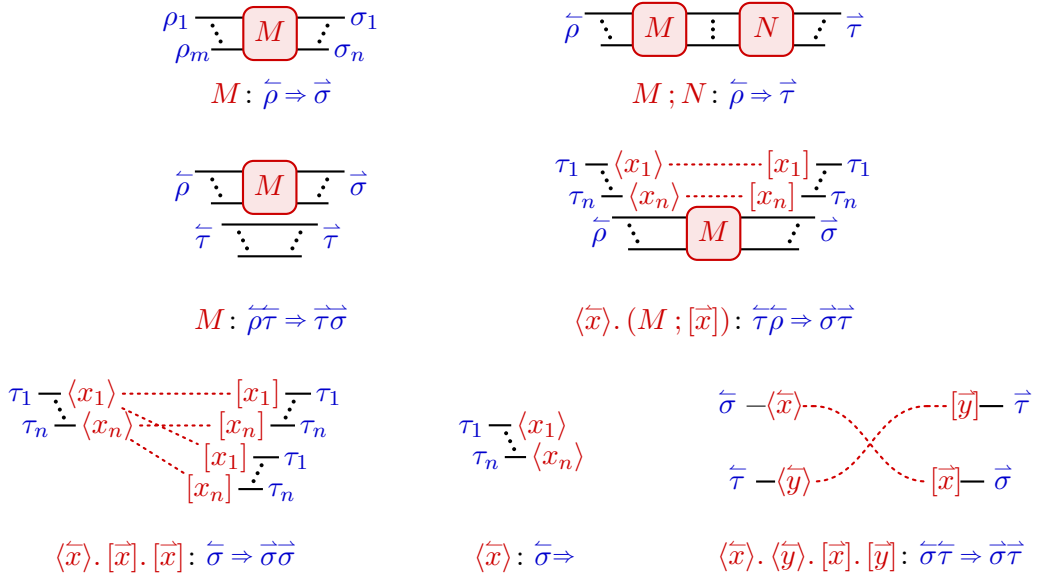

An important thing to note about the type system is that no new type constructors are added. In fact, the equational theory developed in Chapter \ref{chapter:CCC-eqns} makes the category of typed SLC-terms (and FMC-terms) into the free Cartesian closed category, giving a new computational interpretation of intuitionistic logic, in the style of the Curry-Howard-Lambek correspondence.\footnote{Although applying the full equational theory to the type derivations given here, which are in natural deduction style, may not seem a convincing Curry-Howard correspondence, considering intuitionistic proofs in \textit{deep inference} systems \cite{guglielmi2010proof,quasi,Tiu2006,Gundersen2013} makes for a more convincing analogy.} These results can suggest the canonicity of the SLC, if not the FMC itself.

\subsection{Translations of the CBN and CBV $\lambda$-calculus}\label{sec:intro:translations}
As promised, the CBN and CBV translations of the $\lambda$-calculus into the SLC, $(-)_n$ and $(-)_v$, respectively, are given below. 
 The translation $(-)_n$ is given by the obvious inclusion of terms. 
The image of $(-)_v$, however, is different. 
The CBV translation of a $\lambda$-term returns its resulting value (a variable or abstraction) to the stack. In the application case, the argument is executed first, storing the resulting value on the stack, before executing the function. The postfix $\term{<f>.f}$ takes the value resulting from execution of the function and executes it with the value returned from the argument as input.  
\[
\begin{array}{r@{}l@{\qquad}r@{}l@{\qquad}r@{}l}
	x_n &~\defeq~ x & (MN)_n &~\defeq~ \term{[}N_n\term{].}M_n & (\lambda x.M)_n &~\defeq \term{<x>.}M_n \\ 
	x_v &~\defeq~ \term{[x]} & (MN)_v &~\defeq~ N_v \term{;} M_v \term{; <f>.f} & (\lambda x.M)_v &~\defeq~ \term{[<x>.}M_v\term{]} \\
\end{array}
\]
The actions of the translations on types are given as follows:
if $\Gamma \vdash M:A$ then $\Gamma_n \vdash M_n: A_n$ and $\Gamma_v \vdash M_v: \type{>} A_v$,  where 
\[
\begin{array}{r@{}l@{\qquad}r@{}l}
	 \alpha_n ~&\defeq~ \type{> a} &  (A_1 \to \ldots \to A_k \to \alpha)_n ~&\defeq~ (A_1)_n \type{\ldots} (A_k)_n \type{> a}\\
	\alpha_v ~&\defeq~ \type{a} & (A \to B)_v ~&\defeq~ A_v \type{ > } B_v
\end{array}
\]
and $\Gamma_n$ and $\Gamma_v$ are given by the elementwise application of the respective translations. 

Note $(-)_n$ preserves types up to currying. 
In this translation, the variable which necessarily ends the term must thus consume the remaining stack and return a single element of base type $\type \alpha$.
Evaluation of the CBV translation of $\lambda$-term returns a value to the stack, so the overall type of its translation is $\type{> } A_{v}$, and types are preserved up to this final action (which is again an isomorphism). 

Running the translation of terms on the machine, we can see how they reflect the sequencing of CBN and CBV computation. In the former case, this is obvious, because we are working with a CBN machine. For the CBV translation, observe how running $((\lambda x.M)V)_v$ on the machine, where $V$ is a value (\textit{i.e.}, a variable or abstraction) evaluates the argument $V_v$ \textit{before} evaluating the function $(\lambda x.M)_v$. We show such a run below, where $V_v$ translates as $\term{[V]}$ for some $\term{V}$. The empty stack is given by $\epsilon$. 
\[
\begin{array}{@{(~}l@{~,~}r@{~)}}
	\e 	         & \term{[V]} \term{ ; [<x>.}M_v\term{] ; <f>.f}
\\\hline\hline
	\term{V} 	         & \term{ [<x>.}M_v\term{] ; <f>.f}
\\\hline
	\term{V} \cdot \term{<x>.}M_v         & \term{<f>.f}
\\\hline
	\term{V}          & \term{<x>.}M_v
\\\hline
	\e          & \term{\{V/x\}}M_v
\end{array}
\]
This is an example of how the CBN and CBV translations both preserve their respective operational semantics.\footnote{More formally,  a machine based operational semantics of the CBV $\lambda$-calculus is given by the SECD machine \cite{Plotkin-1975}.}
 Thus we can encode both behaviours of the $\lambda$-calculus in the SLC, \textit{even while working with the fixed reduction order determined by the machine}. The programmer can specify the operational behaviour they want by the choice of translation they use.  Compared to the simply-typed $\lambda$-calculus, we have gained the ability to express within syntax \textit{how} computation take place.  Again, this does \textit{not} mean that we \textit{need} to fix any particular reduction order for the SLC considered as a calculus: beta reduction remains confluent and applicable in any context, so we are free to reduce redexes in any order we choose, with the same resulting semantics.\footnote{In the untyped case, reduction order can still affect termination behaviour, but in the typed case, the strong normalisation result presented in Chapter \ref{chapter:SN} shows that the choice of reduction order really does not matter.} 


Nevertheless, it remains the case that adding effects like probabilistic choice or state to this calculus in the naive manner shown in the opening section will still break confluence.  However, we have gained the ability to express both the CBN and CBV semantics of such an operation \textit{if we were} to fix the reduction order. We proceed to explain how, in the Functional Machine Calculus, we can deal with this remaining issue.

\section{The Functional Machine Calculus}

We began with the operational intuition given to higher-order computation (embodied by the application and abstraction of the $\lambda$-calculus) by the Krivine machine. The key realization of the Functional Machine Calculus is the following:
\begin{displayquote}
Just as higher-order computation be given an operational semantics in terms of push and pop actions on the Krivine machine, \textit{so can reader/writer effects}.
A machine with \textit{multiple stacks} thus facilitates the unification of higher-order computation with these effects (and the effects with each other). 
\end{displayquote}
For example, consider the following effects, and their operational interpretation as actions on a stacks (or, more generally, streams):
\begin{itemize}
\item \textbf{Input and Output}: to \textbf{read} is a \textit{pop} from an input stream; to \textbf{write} is a \textit{push} to an output stream;
\item \textbf{Higher-Order Global State}: \textbf{set} pops the currently held term from a memory cell (modelled as a stack of depth at most one), discards it, and \textit{pushes} a new term; \textbf{get} \textit{pops} the currently held value, \textit{pushes} that same value back (for possible reuse later) and continues to subsequent use of that value;
\item \textbf{Probabilistic and Non-deterministic Choice}: a \textit{pop} from a stream of probabilistically (or non-deterministically) generated Church Booleans, and their subsequent application to a pair of terms to make a \textbf{choice} between them, following \cite{DalLago-Guerrieri-Heijltjes-2020}.
\end{itemize}

We proceed to define the Functional Machine Calculus, a natural extension of the SLC to work with multiple stacks, one for higher-order computation and one for each effect (and each memory cell). Then we \textit{decompose} the given effects into FMC-terms, so that the syntax and the equational theory of the effects is unified with that of higher-order computation given by the SLC.  Beta reduction \textit{remains confluent}, resulting in a confluent calculus which can naturally encode reader/writer effects. 

\subsection{Syntax}

\newcommand\SAouta{S_{A\setminus \{a\}}\,}

A natural language for a multiple-stack machine is given by a simple generalization of the SLC. 
Assigning to each stack a \textit{location} $\term{a, b, c, \ldots} \in A$, we can then \textit{parameterize} application and abstraction (push and pop-actions) in these locations, yielding the syntax of the \textit{Functional Machine Calculus (FMC)}.
\begin{align*}
\term{M,N}
  \quad\Coloneqq\quad 	
	\term{*}
	~\mid~ \term{x.M}
       ~\mid~ \term{[N]a.M}
       ~\mid~ \term{a<x>.M}
\end{align*}
The multi-stack machine transitions are given below, with constructors of the FMC interpreted as instructions for the multi-stack machine. 
We call an $A$-indexed \textit{family} of stacks a \textit{memory}.
Notation for memories is as follows. 
Formally, a memory $S_A$ can be regarded as a function from the set $A$ to the set of stacks. 
Where $B \subseteq A$, we write $S_B$ for the restriction of $S_A$ to $B$.  We write $S_a$ for $a \in A$ to access the stack held in memory $S_A$ at location $a$,  and we allow identification of $S_{\{a\}}$ with $S_a$. 
Let $S_{B};S_{C}$ denote the copairing of families (considered as functions) of stacks in sets $B$ and $C$.    
In particular this means we have $S_A = \SAouta;S_a$.
\begin{itemize}
\item
\textbf{Application on }$\term{a}$, $\term{a[N].M}$, as \textbf{push}  $\term N$ onto $\term a$; continue as $\term M$; 
\item
{\textbf{Abstraction on }$\term{a}$, $\term{a<x>.M}$, as \textbf{pop} the head $\term N$ off $\term a$; continue as $\term{\{N/x\}M}$.}
\end{itemize}
\[
\begin{array}{@{(~}l@{~,~}r@{~)}}
\SAouta ; S_a	         & \term{[N]a.M}
\\\hline
\SAouta ; S_a \cdot \term N &       \term M
\end{array}
\qquad
\begin{array}{@{(~}l@{~,~}r@{~)}}
\SAouta; S_a \cdot \term N &   \term{a<x>.M}
\\\hline
\SAouta ; S_a          & \term{\{N/x\}M}
\end{array}
\]
In addition to beta reduction, we extend the calculus with a new \textit{permutation} reduction step allowing push- and pop-actions on a location $\term{a}$ to interact even in the case there are intervening push- or pop-actions on \textit{other} locations. It is easy to verify this is sound with respect to the machine above. 
\begin{align*}
	\term{[N]a.a<x>.M} &\rw_\beta \term{\{N/x\}M} \\
	\term{[N]a.b<x>.M} &\rw_\pi \term{b<x>.[N]a.M} \qquad (\term{a} \neq \term b,  \term{x} \not\in \fv{\term{M}})
\end{align*}

By convention,  we will single out a particular stack as the \textit{main stack},  on which we consider the $\lambda$-calculus to operate.  That is,  we will always consider that the \textit{main location} $\term{l} \in A$,  and that this holds the main stack and then we consider effect operators that transfer terms between the main stack (where they may be made further use of) and other locations.  We omit the location $\term{l}$ from the syntax to avoid clutter.

What is remarkable here is that reader/writer effects and the higher-order mechanisms of the $\lambda$-calculus can be realized using same syntactic constructs. We proceed to show precisely how this is done. 	

\subsection{The FMC as a Multi-Stack Transformer Language}
We introduce by example some effectful terms in the FMC, and discuss later the differing CBN and CBV semantics of effects, and corresponding extensions of the previously given translations. 
As examples, consider the following operations for input and output, a memory cell $\term c$, and random and non-deterministic sums as defined constructs (``sugar'') into the FMC. 
\[
\begin{array}{r@{~}l}
	\term{\print} &~\defeq~\term{<x>.[x]\out}
\\	\term{\rd}    &~\defeq~\term{\inp<x>.[x]}
\end{array}
\quad
\begin{array}{r@{~}l}
	\term{\set ~c} &~\defeq~\term{<x>.c<\_>.[x]c}
\\	\term{\get ~c} &~\defeq~\term{c<x>.[x]c.[x]}
\end{array}
\quad
\begin{array}{r@{~}l}
	\term{N\oplus M} &~\defeq~ \term{[M].[N].\rnd<x>.x} 
\\	 \term{N+M} &~\defeq~ \term{[M].[N].\nd<x>.x} 
\end{array}
\]
where $(\term\_)$ represents a variable that does not occur elsewhere. These are explained in the following examples.

	\textbf{Input and Output:}
These make use of dedicated \emph{input} and \emph{outout} locations $\term\inp, \term\out \in A$, and the \emph{pop} and \emph{push} transitions, respectively, give the expected operational semantics.
For input, the machine is initialized with a stream $S_\inp=\cdots\term{N_3}\cdot\term{N_2}\cdot\term{N_1}$ (infinite to the left) at $\term\inp$. For output,
evaluation generates a stream $\term{N_1},\term{N_2},\dots$ (finite at any step) at $\term\out$.
\[
\begin{array}{@{}l@{}l@{~;~}l@{~;~}r@{~)}}
(~	S_{A\setminus\{\lambda, \inp\}} ~~;~&S_{\lambda}&S_\inp\cdot\term{N} & \term{\inp<x>.[x]}
\\\hline
(~	S_{A\setminus \{\lambda, \inp\}} ~~;~&S_{\lambda} &S_\inp              & \term{[N]}
\\\hline
(~	S_{A\setminus \{\lambda, \inp\}} ~~;~&S_{\lambda} \cdot \term{N}&S_\inp              & \term{*}
\\
\\
(~	S_{A\setminus \{\lambda, \out\}} ~;~& S_{\lambda} \cdot \term{N}&S_\out                & \term{<x>.[x]\out}
\\\hline
(~	S_{A\setminus \{\lambda, \out\}}  ~;~&S_{\lambda}&S_\out &           \term {[N]\out}
\\\hline
(~	S_{A\setminus \{\lambda, \out\}}   ~;~& S_{\lambda}&S_\out\cdot\term{N} &           \term {*}
\end{array}
\]

	\textbf{Higher-Order Global State:} A memory cell is modelled by a location $\term c\in A$. The associated stack is expected to hold at most one value, which is preserved by the encoding of the operators, and not enforced externally. 
In the machine, the stack for each cell is initialized with a (dummy) value, and the transitions then give the expected operational semantics.  Here, $\epsilon_c$ denotes the empty stack at location $c$.  
\[
\begin{array}{@{}l@{}l@{~;~}l@{~;~}r@{~)}}
    	 (~S_{A\setminus \{\lambda, c\}} ~;~& S_{\lambda} &\e_c\cdot \term N & \term{c<x>.[x]c.[x]}
\\\hline (~S_{A\setminus \{\lambda, c\}} ~;~& S_{\lambda} &\e_c              &       \term{[N]c.[N]}
\\\hline (~S_{A\setminus \{\lambda, c\}} ~;~& S_{\lambda} &\e_c\cdot \term N &            \term{[N]}
\\\hline (~S_{A\setminus \{\lambda, c\}} ~;~& S_{\lambda} \cdot \term N &\e_c\cdot \term N &            \term{*}
\\
\\
    	 (~S_{A\setminus \{\lambda, c\}} ~;~&  S_{\lambda} \cdot \term M &\e_c\cdot \term N &  \term{<x>.c<\_>.[x]c}
\\\hline (~S_{A\setminus \{\lambda, c\}} ~;~& S_{\lambda}&\e_c \cdot \term N    &       \term{c<\_>.[M]c}
\\\hline (~S_{A\setminus \{\lambda, c\}} ~;~& S_{\lambda} &\e_c &            \term{[M]c}
\\\hline (~S_{A\setminus \{\lambda, c\}} ~;~& S_{\lambda}&\e_c\cdot \term M &            \term{*}
\end{array}
\]

	\textbf{Probabilistic and non-deterministic sums:}
Following the probabilistic \linebreak{case~\cite{DalLago-Guerrieri-Heijltjes-2020},} probabilistic and non-deterministic sums are included via dedicated locations $\term\rnd,\term\nd\in A$,
where the machine is initialized with streams of  $\mathsf{\term{T}} \defeq \term{<x>.<y>.x}$ and $\mathsf{\term{F}} \defeq \term{<x>.<y>.y}$, generated probabilistically for $\term\rnd$ and non-deterministically for $\term\nd$.\footnote{We could equally generate a stream of Booleans and use a conditional statement to make the choice.} The probabilistic case is illustrated below, where, because the   input stream $\term \rnd$ is probabilistically generated, the result is the probabilistic mixture of the two machine runs. 
\[
\begin{array}{@{(~}l@{~;~}l@{~;~}l@{~;~}r@{~)}}
	S_{A \setminus \{\lambda, \rnd\}} & S_{\lambda} 			& S_\rnd \cdot \mathsf{\term{T}} & \term{[M].[N].\rnd<x>.x}
\\\hline
	S_{A \setminus\{\lambda, \rnd\}} & S_{\lambda} \cdot \term{M} & S_\rnd \cdot \mathsf{\term{T}} & \term{[N].\rnd<x>.x}
\\\hline
	S_{A \setminus\{\lambda, \rnd\}} & S_{\lambda} \cdot \term{M} \cdot \term{N} & S_\rnd \cdot \mathsf{\term{T}} & \term{\rnd<x>.x}
\\\hline
	S_{A \setminus\{\lambda, \rnd\}} & S_{\lambda} \cdot \term{M} \cdot \term{N} & S_\rnd  & \mathsf{\term{T}}
    \\\hline\hline\rule[-5pt]{0pt}{15pt}
	S_{A \setminus\{\lambda, \rnd\}} & S_{\lambda}  & S_\rnd              & \term{N}
\end{array}
\]
\[
\begin{array}{@{(~}l@{~;~}l@{~;~}l@{~;~}r@{~)}}
	S_{A \setminus\{\lambda, \rnd\}} & S_{\lambda} & S_\rnd \cdot \mathsf{\term{F}} & \term{[M].[N].\rnd<x>.x}
\\\hline
	S_{A \setminus\{\lambda, \rnd\}} & S_{\lambda} \cdot \term{M} & S_\rnd \cdot \mathsf{\term{F}} & \term{[N].\rnd<x>.x}
\\\hline
	S_{A \setminus\{\lambda, \rnd\}} & S_{\lambda} \cdot \term{M} \cdot \term{N} & S_\rnd \cdot \mathsf{\term{F}} & \term{\rnd<x>.x}
\\\hline
	S_{A \setminus\{\lambda, \rnd\}} & S_{\lambda} \cdot \term{M} \cdot \term{N} & S_\rnd  &  \mathsf{\term{F}}
    \\\hline\hline\rule[-5pt]{0pt}{15pt}
	S_{A \setminus\{\lambda, \rnd\}} & S_{\lambda}  & S_\rnd              & \term{M}
\end{array}
\]
Note that choice effects are dealt with by externalizing them: the calculus remains deterministic, but is evaluated in the context of a random (or non-deterministic) oracle. 

To see how beta reduction and permutation reduction allow for optimization of the above terms for the machine, consider the following example. 
\begin{example}
To see how $\term\set$ and $\term \get$ interact through beta reduction, consider the terms below.  
The first term can be verified to give the same result as $\term{[5].\set~c}$, as expected. Similarly, the second term can be verified to be the same result as $\term{[4].\set~c.[4]}$, as expected.
\begin{align*}
	\term{[3].\set~c;[5].\set~c} &= \underline{\term{[3].<x>}}\term{.c<\_>.[x]c.[5].<y>.c<\_>.[y]} \\
		&\rw_\beta \term{c<\_>.[3]c.}\underline{\term{[5].<y>}}\term{.c<\_>.[y]} \\
		&\rw_\beta \term{c<\_>.}\underline{\term{[3]c;c<\_>}}\term{.[5]} \\
		&\rw_\beta \term{c<\_>.[5]} \\
	\term{[4].\set~c;\get~c} &= \underline{\term{[4].<x>}}\term{.c<\_>.[x]c.c<y>.[y]c.[y]} \\
		 &\rw_\beta \term{c<\_>.}\underline{\term{[4]c.c<y>}}\term{.[y]c.[y]} \\
		 &\rw_\beta \term{c<\_>.[4]c.[4]} \, ,
\end{align*}
Observe further how permutation can unlock new redexes, for example in in the case of the term below with two memory cells $\term{c}$ and $\term{d}$, it allows interaction of $\term{\set~c}$ and $\term{\get~c}$ despite the intervening $\term{\set~d}$.
\begin{align*}
	\term{\set~c;\set~ d;\get~c} &= \term{<x>.c<\_>.}\underline{\term{[x]c.<y>}}\term{.d<\_>.[y]d.c<z>.[z]c.[z]}\\
		&\rw_\pi \term{<x>.c<\_>.}\term{<y>.}\underline{\term{[x]c.d<\_>}}\term{.[y]d.c<z>.[z]c.[z]}\\
		&\rw_\pi \term{<x>.c<\_>.}\term{<y>.}{\term{d<\_>.[x]c.}}\underline{\term{[y]d.c<z>}}\term{.[z]c.[z]}\\
		&\rw_\pi \term{<x>.c<\_>.}\term{<y>.}{\term{d<\_>.[x]c.}}\term{c<z>.[y]d}\term{.[z]c.[z]}\\
		&\rw_\pi \term{<x>.c<\_>.}\term{<y>.}{\term{d<\_>.}}\underline{\term{[x]c.c<z>}}\term{.[y]d.[z]c.[z]}\\
		&\rw_\beta \term{<x>.c<\_>.}\term{<y>.}{\term{d<\_>}}{\term{}}\term{.[y]d.[x]c.[x]}
\end{align*}
\end{example}
\subsection{Simple Types for the FMC}
The type system for the FMC extends that of the SLC with \textit{locations}. A \textit{memory} type $\type{!t_A}$ is given by an $A$-indexed family of type vectors, and computations track the input and output \textit{memories}, rather than the input and output \textit{stacks}. 
\[
	\type{t} ~\Coloneqq~ \type{a} ~\mid~ \type{?s_A > !t_A} \qquad \type{!t} ~\Coloneqq~ \type{t_1 \ldots t_n} \qquad \type{!t_A} ~\defeq~ \{\type{!t_a} ~\mid~ a \in A\}
\]
This means that the type system tracks, in addition to all function input and output, all the changes to state, all consumption from oracles, \textit{etc}. Therefore, from the denotational perspective, we can consider terms as a pure function between \textit{memories}, whereas by \textit{forgetting} all information about non-main stacks we could recover something more akin to the typical notion of a process with \textit{side-effects}. 

Typing rules are given in Figure \ref{fig:FMC-types-plain}, using the following notation. 
 A \emph{singleton} memory type $\type{`a(!t)}$ is empty at every location except $\term{a}$. We omit the main location $\term{\lambda}$ for singeton types, so $\type{\lambda(!t)}$ may be written as $\type{!t}$. \emph{Concatenation} extends to families point-wise, so $\type{!s_A !t_A}=\{\type{!s_a!t_a}\mid a\in A\}$, and similarly for the reverse of a family $\type{?t_A}$
.
An equivalent way to view computation types is as an implication between two vectors of singletons, considered modulo the permutation of types on different locations, as below. 
\[
	\type{`a_1(s_1)\dots `a_n(s_n)~>~`b_1(t_1) \dots `b_m(t_m)}
\qquad\qquad
	\type{`a(s)\,`b(t)}\sim\type{`b(t)\,`a(s)}
\]
Note that the notation $\type{`a(t)}$ isn't a novel type constructor, but notation for an indexed product. 
\begin{figure}
\[
	\infer[\!\scriptstyle{\mathsf{id}}]{\Gamma \vdash \term{ *:?t_A>!t_A}}{}
	\qquad
	\infer[\!\scriptstyle{\mathsf{base}}]{\term{x: a}, \Gamma \vdash \term{x:a}}{}
\]
\\
\vspace{-\baselineskip}
\[
	\infer[\!\scriptstyle{\mathsf{app}}]
	{\Gamma \vdash \term{[N]a.M: ?s_A > !t_A}}
	{\Gamma \vdash \term{N: r} && {\Gamma \vdash \term{M:`a(r)?s_A > !t_A}}}
	\qquad
	\infer[\TR l]
	  {\Gamma \vdash \term{ `a<x>.M : `a(r)\,?s_A>!t_A}}
	  {\term{x:r, G} \vdash \term{ M:?s_A>!t_A}}
\]
\\
\vspace{-\baselineskip}
\[
	\infer[\!\scriptstyle{\mathsf{var}}]
	{\term{x:?r_A > !s_A}, \Gamma \vdash \term{x.M: ?r_A?t_A > !u_A}}
	{{\term{x:?r_A > !s_A}, \Gamma\vdash \term{M: ?s_A?t_A > !u_A}}}
\]
\caption{Typing rules for the Functional Machine Calculus}
\label{fig:FMC-types-plain}
\end{figure}
\begin{example}
We can build up the type of the term given in in Figure \ref{ex:effects-example} as follows. 
\[
\begin{array}{rr@{}l}
 \term{\print}:&\type{ \Z } & \type{> \out(\Z)} \\
 \term{+; \print}:&\type{ \Z\Z } & \type{> \out(\Z)} \\
\term{\get ; + ; \print}:&\type{  \Z \,`c(\Z)} & \type{> `c(\Z)\,\out(\Z)} \\
\term{\set ; \get ; + ; \print}:&\type{ \Z\Z\,`c(\Z)  } & \type{> `c(\Z)\,\out(\Z)} \\
\term{\rand ; \set ; \get ; + ; \print}:&\type{ \Z\,\rnd(\Z)\,`c(\Z)  } & \type{> `c(\Z)\,\out(\Z)} \\
\term{\get ; \rand ; \set ; \get ; + ; \print}:&\type{ \rnd(\Z)\,`c(\Z)  } & \type{> `c(\Z)\,\out(\Z)} \\
\term{\set ; \get ; \rand ; \set ; \get ; + ; \print}:&\type{ \Z\,\rnd(\Z)\,`c(\Z)  } & \type{> `c(\Z)\,\out(\Z)} \\
\term{\rand ; \set ; \get ; \rand ; \set ; \get ; + ; \print}:&\type{ \rnd(\Z\Z)\,`c(\Z)  } & \type{> `c(\Z)\,\out(\Z)} 
\end{array}
\]
\end{example}

\subsection{Translations of CBN and CBV Effectful $\lambda$-Calculi}

It is now demonstrated, by example, how both CBN and CBV semantics for effects are expressible within the FMC, by extension of the CBN and CBV translations of the $\lambda$-calculus. We deal here with probabilistic choice and first-order state (that is, memory cells which hold base types only). 

\textbf{Probabilistic Choice:}
Consider the $\lambda$-calculus naively extended with probablistic choice and state.
\begin{align*}
       M \quad \Coloneqq \quad x ~\mid~ MN ~\mid~ \lambda x.M ~\mid~ M \oplus N ~\mid~ c := N;M ~\mid~ !c 
\end{align*}
Consider also the following non-confluent $\lambda$-term, and its CBN and CBV reductions respectively, given below. 
\[
	(\lambda f. \lambda x. f(f(x)))(M \oplus N) \to_\textsf{cbn} \lambda x.(M \oplus N)((M \oplus N)x) 
\]
\[
	(\lambda f. \lambda x. f(f(x)))(M \oplus N) \to_\textsf{cbv} \begin{cases} 50\% & (\lambda f.\lambda x.f(fx))M   \\  50\% & (\lambda f.\lambda x.f(fx))N     \end{cases}
\]
The translation of probabilistic choice follows the probabilistic $\lambda$-calculus from \cite{DalLago-Guerrieri-Heijltjes-2020}, which decomposes the choice operator into a \textit{generator} of a random value and a \textit{consumer}, which makes the choice depending on the value passed to it. Here, the role of the generator is played by an abstraction on a random stream of Booleans, and the role of consumer is played by the use of an abstracted Boolean. 

Recall that probabilistic choice in the $\lambda$-calculus is usually typed as below. 
\[
	\infer[]{\Gamma \vdash M \oplus N: A}
	{{\Gamma \vdash M: A} & {\Gamma \vdash N: A}}
\]
Recall the encoding of Booleans as $\mathsf{\term{T}} \defeq \term{<x>.<y>.x}$ and $\mathsf{\term{F}} \defeq \term{<x>.<y>.y}$ and let $\type{\B} = \type{tt > t}$ be the type of Booleans.
Note, we could similarly deal with \textit{actual} Booleans of base type, in combination with a conditional $\term{\if}$ for making the choice.
Then the translation of the types of probabilistic terms in CBN and CBV is given below.
We use a CBN-style encoding here, since even in CBV it is convention not to evaluate both branches of $M \oplus N$ before the choice is made. 
It is now recorded in the type system that each computation additionally reads some $n$ Booleans from the random stream. These are greyed out in the type system to represent that this information is \textit{missing} in the simply-typed $\lambda$-calculus. 
As before, the overall type of a CBV term is given by $\gray{\rnd(\B^n)} \type{> } A_{v}$. 
\[
\begin{array}{r@{}l@{\qquad}r@{}l}
	 \alpha_n ~&\defeq~ \gray{\rnd(\B^n)}\type{ > a} &  (A_1 \to \ldots \to A_m \to \alpha)_n ~&\defeq~ \gray{\rnd(\B^n)}(A_1)_n \type{\ldots} (A_m)_n \type{> a}\\
	\alpha_v ~&\defeq~ \type{a} & (A \to B)_v ~&\defeq~ \gray{\rnd(\B^n)}A_v \type{ > } B_v 
\end{array}
\]
We thus translate $M \oplus N$ in the same way for both CBN and CBV, but the resulting terms get different types. 
For $M \oplus N : A_1 \to \ldots A_m \to \alpha$ in CBN and $M \oplus N: A \to B$ in CBV, we have the following, where $n+1$ indicates we have read one more Boolean from the stream. 
\begin{align*}
(M \oplus N)_n &\defeq \term{\rnd<b>.[}M_n\term{].[}N_n\term{].b:} \gray{\rnd(\B^{n+1})}(A_{1})_n\type{\ldots} (A_m)_n\type{ > a} \\
(M \oplus N)_v &\defeq  \term{\rnd<b>.[}M_n\term{].[}N_v\term{].b:} \, \gray{\rnd(\B^{n+1})}\type{ > (}\gray{\rnd(\B^{m})}A_v \type{>} B_v\type{)}
\end{align*}
For $M \oplus N: (A \to \alpha) \to (A \to \alpha)$ and $x: A \to \alpha$ in CBN and $M \oplus N: A \to A$ and $x: A$ in CBV, we have the following translations (up to a small amount of simplification by beta reduction). 
We assume $M$ and $N$ to be pure terms (that is, which don't read from the probabilistic stream) below, in order to simplify the types. 
\begin{align*}
	((\lambda f. f(fx))(M \oplus N))_n &= {\term{[}(M \oplus N)_n\term{].<f>}}\term{.[[x].f].f}: \gray{\rnd(\B)}A_n\type{ > a} \\
	((\lambda f. f(fx))(M \oplus N))_v &= (M \oplus N)_v\term{ .<f>.[x].f.<y>.y.f.<z>.z:} \gray{\rnd(\B)}\type{  > } A_v 
\end{align*}
In the CBN case, the choice, including the \textit{generator} $\term{\rnd<x>}$ is duplicated, leading to a probabilistic sub-term $\term{[x].}(M \oplus N)_n$, which is then consumed by a second probabilistic sub-term $( M \oplus N)_n$. Note, the choice in the former sub-term is decided, then, until the latter subterm applies it to some argument (for example, consider $M,N = \lambda x.x$). 
In the CBV case, the choice between $M$ and $N$ is made once, at the start, and the result is duplicated (but not the generator). 
It is easy to verify that these terms give the expected distribution of results when evaluated on the machine. 

%
%

\textbf{First-Order State:}
Recall we will work with \textit{first-order} state here, that is where the memory cell can hold only terms of base type. That is, in $c:=N;M$, we require $N: \alpha$. We further make the simplifying assumption that such terms are not themselves effectful. We will elide much discussion of types in this section, but they can be dealt with similarly to the case of probabilistic choice. 
Consider now the non-confluent $\lambda$-term from the introduction.
\begin{align*}
	3 \ _\textsf{cbv}\leftarrow c := 3 ; (\lambda x. !c ) (c := 5 ; M) \to_\textsf{cbn} 5
\end{align*}
Recalling the $\term{\set ~c}$ and $\term{\get ~c}$ combinators from earlier, we translate the stateful constructors of the $\lambda$-calculus as follows.  
\begin{align*}
	\term{\set ~c} ~&\defeq~\term{<x>.c<\_>.[x]c}
&	\term{\get ~c} ~&\defeq~\term{c<x>.[x]c.[x]}\\
(!c)_n ~&\defeq~\term{\get~ c } & 
(c := N ; M)_n ~&\defeq~ N_n\term{;\set~ c; }M_n\\
(!c)_v ~&\defeq~\term{\get~ c } & 
(c := N ; M)_v ~&\defeq~ N_v\term{;\set~ c; }M_v
\end{align*}
Recalling that $\alpha_n = \alpha_v = \type{> \alpha}$, it can easily be checked that this gives the expected types (up to the actions on $\term{c}$, which do not appear in the $\lambda$-calculus), where the memory cell $term{c}$ holds a term of type $\type{alpha}$. 
We will now verify the following laws apply. 
\begin{align*}
	c: = M ; c := N ~\to~ c := N \qquad c := M ; !c ~\to~ c := M ; M
\end{align*}

%

These terms translate as follows, where we simplify by beta reduction where possible. 
We elide writing $M_n$, $N_n$, $M_v$ and $N_v$ here, and instead write simply $\term{M}$. 
\begin{align*}
	c: = M ; c := N ; P 
		&= \term{M.c<\_>.<x>.[x]c.N.c<\_>.<y>.[y]c;P}\\
	a := N ; P 
		&= \term{N.c<\_>.<y>.[y]c;P} \\
	a := M ; !a &= \term{M.a<\_>.<x>.}\underline{\term{[x]a.a<y>}}\term{.[y]a.[y]} \\
		&\rw_\beta \term{M.a<\_>.<x>.[x]a.[x]} \\
	a := M ; M &= \term{M.a<\_>.<x>.[x]a.M}
\end{align*}
To see these are equivalent, we can run them on the machine and see we achieve the same results.
In the following runs, let $\ iterm{M'}$ be the result, which is pushed to the stack, of running $\term{M}$.
We omit writing the full memory $S_A$ in the runs below, to save space. 
\[
\begin{array}{@{(~}l@{~;~}l@{~;~}r@{~)}}
    	 S_{\lambda} &\e_c\cdot \term Q & \term{M.c<\_>.<x>.[x]c.N.c<\_>.<y>.[y]c;P}
\\\hline S_{\lambda} \cdot \term{M'} &\e_c \cdot \term Q              &       \term{c<\_>.<x>.[x]c.N.c<\_>.<y>.[y]c;P}
\\\hline S_{\lambda}  \cdot \term{M'} &\e_c &          \term{<x>.[x]c.N.c<\_>.<y>.[y]c;P}
\\\hline S_{\lambda}   &\e_c &                \term{[M']c.N.c<\_>.<y>.[y]c;P}
\\\hline S_{\lambda}   &\e_c \cdot \term{M'} &                   \term{N.c<\_>.<y>.[y]c;P}
\end{array}
\]
\[
\begin{array}{@{(~}l@{~;~}l@{~;~}r@{~)}}
    	S_{\lambda} &\e_c\cdot \term Q & \term{M.c<\_>.<x>.[x]c.[x]}
\\\hline  S_{\lambda} \cdot \term{M'} &\e_c \cdot \term Q              &       \term{c<\_>.<x>.[x]c.[x]}
\\\hline  S_{\lambda}  \cdot \term{M'} &\e_c &          \term{<x>.[x]c.[x]}
\\\hline  S_{\lambda}   &\e_c &                \term{\term{[M']c.[M']}}
\\\hline S_{\lambda}   &\e_c \cdot \term{M'} &                \term{\term{[M']}}
\\\hline S_{\lambda} \cdot \term{M'}  &\e_c \cdot \term{M'} &                \term{\term{*}}
\end{array}
~~
\begin{array}{@{(~}l@{~;~}l@{~;~}r@{~)}}
    	 S_{\lambda} &\e_c\cdot \term Q & \term{M.a<\_>.<x>.[x]a.M}
\\\hline S_{\lambda} \cdot \term{M'} &\e_c \cdot \term Q              &       \term{c<\_>.<x>.[x]c.M}
\\\hline S_{\lambda}  \cdot \term{M'} &\e_c &          \term{<x>.[x]c.M}
\\\hline S_{\lambda}   &\e_c &                \term{\term{[M']c.M}}
\\\hline S_{\lambda}   &\e_c \cdot \term{M'} &                \term{\term{M}}
\\\hline S_{\lambda} \cdot \term{M'}  &\e_c \cdot \term{M'} &                \term{\term{*}}
\end{array}
\]
Note that, in the first case, we recover the expected term, although the memory now holds $\term{M'}$ instead of the original $\term{Q}$. However, since $a := N ; P$ discards the term in memory, this does not matter. 

In chapter \ref{chapter:CCC-eqns}, we will give an equational theory which extends the given reduction rules to capture more of \textit{machine equivalence} (also introduced formally in that chapter),  thus axiomatizing when two terms are equivalent with respect to the machine. 

We now return to the non-confluent $\lambda$-term we were aiming to translate.  Again, we have a CBN and a CBV translation, given below. Recall, again, that in the CBN translation $5: \Z$ is translated as $\term{5: > \Z}$, whereas in the CBV translation we have $\term{[5]: > \Z}$.
\begin{align*}
	(a := 3 ; (\lambda x. !a ) (a := 5 ; M))_n &= \term{ 3. \set ~a. }{\term{[5. \set ~a . }M_n\term{]. <x>}}\term{. \get ~a  } \\
		&\rw_\beta \term{3. \set ~a. \get ~a} \\
		&= a := 3 ; !a\\
		&= a := 3 ; 3\\
	(a := 3 ; (\lambda x. !a ) (a := 5 ; M))_v &=  \term{ [3]. \set ~a. [5]. \set ~a. }M_v\term{; }{\term{[<x>. \get ~a].<f>}}\term{.f}\\
		&\rw_\beta \term{[3]. \set ~a. [5]. \set ~a. }M_v\term{;<x>.\get~ a} \\
		&= a := 3; a := 5 ; M_v\term{;<x>.}!a\\
		&= a := 5 ; M_v\term{;<x>.}!a
\end{align*}
We can easily verify this second term is equivalent to $a:=5 ; 5$ on the machine. Note that the output of $M_v$ is deleted by the abstraction $\term{<x>}$.

The unification of the mechanism for effects (and higher-order functions) into beta reduction further facilitates the combination of effects within a single calculus, with the type system dealing with each effect (and with higher-oreder functions) in a uniform manner. While we have seen that our multi-stack machine gives the expected operational semantics of reader/writer effects, in fact it is the case that $\beta$-reduction captures the algebraic laws for \textit{e.g.}, state \cite{Barrett-Heijltjes-McCusker-2022}. Note that it is trivial to extend the type system with the expected restrictions used to model effects (that is, read- or write-only stacks for input and output, and one-place stacks for memory cells).

\section{Summary and Outline of Thesis}
The Functional Machine Calculus has been introduced and motivated, and we emphasize the following key points, before outlining the remaining content. 
\subsection{Summary}
\begin{itemize}
\item The FMC generalizes the $\lambda$-calculus to allow for {sequential composition}, by decomposing the variable construct. This allows translations of both the CBN {and} CBV $\lambda$-calculus into the FMC,  each preserving their respective operational semantics. 
\item Just as higher-order computation be given an operational semantics in terms of push and pop actions on the Krivine machine, so can reader/writer effects.
 The syntax and reduction rules of higher-order computation and effects are therefore unified via a paramterization of application (push) and abstraction (pop) in terms of locations.
\item Beta reduction remains confluent and applicable in any context \cite{Barrett-Heijltjes-McCusker-2022}, despite the presence of effects.  
\item
The calculus furthermore comes equipped with a type sytem which restricts and captures the behaviour of effects via their actions on non-main locations.
\end{itemize}
\subsection{Outline}

We now provide an overview of the {remaining} chapters and the main contributions of this thesis. 
\begin{itemize}
\item In Chapter \ref{chapter:cat}, the necessary categorical preliminaries for this thesis are given, namely, the Curry-Howard-Lambek correspondence in brief.
\item  We proceed to give formal definitions relating to the FMC in Chapter \ref{chapter:fmc-prelims}, in particular introducing a variant we name ``\FMCV''  (the Functional Machine Calculus \textit{with values}), which distinguishes between computations (which run on the machine) and values (which live on the stack), which will later allow for the most natural translation into the $\lambda$-calculus.  We conclude by reprising the result, due to Heijltjes \cite{Barrett-Heijltjes-McCusker-2022}, that type derivations provide a direct proof of machine termination for their corresponding terms. 
\item The first main contribution of the thesis is in Chapters \ref{chapter:CCC-eqns} and \ref{chapter:CCC-equiv}, where it is proved that the category of \textit{closed}, simply-typed \FMCV-terms taken modulo an appropriate equational theory,  with composition given by \textit{sequencing}, is equivalent to the category of simply-typed $\lambda$-terms. Given the ability of the Functional Machine Calculus to encode effects, this may be a surprising result. The nuance here is that we must assume that all locations are treated uniformly: in particular, we cannot restrict some locations to be read- or write-only, and we cannot limit the number of terms that can be held in a location, as was implicit in the encoding of effects described earlier in this chapter.  
In other words, we present an equational theory and denotational semantics for the \FMCV\  as a \textit{calculus}, and this is \textit{not} intended as a semantics of effects.
Chapter \ref{chapter:CCC-eqns} is structured as follows.
\begin{itemize}
\item The full equational theory is defined, extending the equivalence generated by beta and permutation reduction as introduced above. It is then proved that these equations are sufficient to form a Cartesian closed category. 
\item Then, a natural notion of observational equivalence of terms based on the machine,  called  \textit{machine equivalence}, is introduced, and it is shown to validate the equational theory. 
\end{itemize}
Chapter \ref{chapter:CCC-equiv} then proves the main result: equivalence of the category of \FMCV-terms and the free Cartesian closed category. The result is proved in two stages, using the simply-typed $\lambda$-calculus as a language for the latter. 
\begin{itemize}
\item First,  we prove the categorical equivalence of the $\lambda$-calculus and the \SLCV\  (the Sequential $\lambda$-calculus with values). This consists of considering the input stack to a closed \SLCV-term as the context of an open $\lambda$-term.
\item Second,  we prove the categorical equivalence of the \FMCV\  and the \SLCV. This consists of collapsing the memory of the \FMCV\  into a single large stack.
\end{itemize}
Note, the equivalence in the second point is up-to natural isomorphism. 
\item The second main contribution of the thesis is in Chapter $\ref{chapter:SN}$, where it is proved that simply-typed FMC-terms are strongly normalising with respect to beta and permutation reduction. The proof is an adaptation of Gandy's for the $\lambda$-calculus \cite{gandy}. Apart from the theorem itself, some adaptations made in the proof are also of note. First, it reveals a latent operational intuition, whereby the measure for strong normalisation is related to counting the number of steps remaining to be made by the machine on some given input. Second, there is a small refinement made to the structures of the proof, which arises naturally from the setting of the FMC. 
\item Finally, in Chapter \ref{chapter:related-lit} we review the related literature, and in Chapter \ref{chapter:conclusion}, we conclude and {note some of the many possible further avenues for research}. 
\end{itemize}

Neither of the two main contributions depend on each other, however, one point of similarity  to note is between the translation from the \SLCV\  to the $\lambda$-calculus in Chapter \ref{chapter:CCC-equiv} and the interpretation of FMC-terms (and therefore SLC-terms) into the semantic domain given in the strong normalisation proof of Chapter \ref{chapter:SN}. 

There are two publications associated with this thesis, the first laying out the foundations of the FMC \cite{Barrett-Heijltjes-McCusker-2022}, and the second \cite{barrett_et_al:LIPIcs.CSL.2023.10} containing early results from this thesis (amongst contributions by other authors).

\nocite{DBLP:journals/corr/abs-2105-01382}

\chapter{Preliminaries: Categorical Semantics of the Lambda Calculus}\label{chapter:cat}

This chapter presents the definition of a free Cartesian closed category, and recalls its equivalence with the category of terms of the simply-typed $\lambda$-calculus. This is preparation for Chapters \ref{chapter:CCC-eqns} and \ref{chapter:CCC-equiv}, where we investigate the denotational semanics of the FMC. In particular, in order to deal with constants, we define the free Cartesian closed category generated over a signature with products and arrows, rather than over some base category. Further, we introduce a slight variant of the $\lambda$-calculus (which is standard) which deals with elimination of products via \textit{pattern matching}. This proves to be a more natural variant to work with when relating to the FMC. Additionally, this $\lambda$-calculus has \textit{two} rules for introducing constants (which is not standard), for \textit{computation} and \textit{value} constants, which are introduced in a variant definition of its signature. This is to achieve the closest possible correspondence with the FMC with values,  \FMCV, which is introduced in Chapter \ref{chapter:fmc-prelims}. 

We will assume familiarity with basic category theory: the definition of categories, functors, natural isomorphisms, equivalence of categories, commuting diagrams and terminal objects. A basic introduction to category theory can be found in \cite{MacLane,BarrWell:90:Category:lh,leinster_2014}.
Our notation for composition is given in diagram order: $f;g$, instead of $g\circ f$. We use $\textsf{id}_\mathbb{C}$ to denote the identity functor on a category $\mathbb{C}$, and $\simeq$ to denote when two functors are naturally isomorphic. Given a terminal object $\top$, we write $!_A: A \to \top$ to be the unique such morphism. Sometimes we drop subscripts denoting the components of natural transformations when it is clear. We first recall the definition of adjoint functors here for convenience.

\section{Adjunctions}
We recall the definition of adjoint functor that will be useful for understanding the definition of the free Cartesian closed category. 
\begin{definition}[Adjoint Functors]
A functor $F: \mathbb{C} \to \mathbb{D}$ is a \textit{left adjoint functor} if for each object $A \in \mathbb{D}$, there exists an object $G(A) \in \mathbb{C}$ and a morphism $\epsilon_A: F(G(A)) \to A$, such that for every object $B \in \mathbb{C}$ and every morphism $f: F(A) \to B$ there exists a unique morphism $g: B \to G(A)$, such that the following diagram commutes:
\[\begin{tikzcd}
	{F(B)} \\
	{F(G(A))} & A
	\arrow["{F(g)}"', from=1-1, to=2-1]
	\arrow["f", from=1-1, to=2-2]
	\arrow["{\epsilon_A}"', from=2-1, to=2-2]
\end{tikzcd} \ .\]
A functor is  a \textit{right adjoint functor} if it satisfies the dual condition.
\end{definition}
\begin{remark}
It can be shown that $G$ can be turned into a functor $G: \mathbb{D} \to \mathbb{C}$ such that $\epsilon_D \circ F(G(f)) = f \circ \epsilon_C$, for all morphisms $f: C \to D$ in $\mathbb{C}$. In this case $F$ is called \textit{left adjoint} to $G$. Indeed, $F$ is left adjoint to $G$ if and only if $G$ is right adjoint to $F$. 
\end{remark}
\begin{remark}
It is the case that $F: \mathbb{C} \to \mathbb{D}$ is left adjoint to $G: \mathbb{D} \to \mathbb{C}$ if and only if there is a natural isomorphism
\[
	\alpha: Hom_{\mathbb{C}}(F(-), -) \simeq Hom_{\mathbb{D}}(-, G(-)), 
\]
which thus specifies a family of bijections
\[
	\alpha_{A,B}: Hom_{\mathbb{C}}(F(B), A) \to Hom_{\mathbb{D}}(B, G(A)), 
\]
for every $A \in \mathbb{C}$ and $B \in \mathbb{D}$.
\end{remark}

%

\section{Cartesian Closed Categories}

We define a Cartesian closed category, each example of which gives a model of the simply-typed $\lambda$-calculus. 

\begin{definition}
The \textit{(Cartesian) product} $(A_1 \times A_2, \pi_1, \pi_2)$ of two objects $A_1$ and $A_2$ in a category consists of an object $A_1 \times A_2$ together with two \textit{projections} $\pi_1: A_1 \times A_2 \to A_1$ and $\pi_2: A_1 \times A_2 \to A_2$ such that for every object $B$ and morphisms $f_1: B \to A_1$ and $f_2: B \to A_2$, there exists a unique morphism $\langle f_1, f_2 \rangle: B \to A_1 \times A_2$ such that the following diagram commutes:
\[\begin{tikzcd}
	& B \\
	& {} \\
	{A_1} & {A_1 \times A_2} & {A_2}
	\arrow["{\pi_1}", from=3-2, to=3-1]
	\arrow["{\pi_2}"', from=3-2, to=3-3]
	\arrow["{f_2}", from=1-2, to=3-3]
	\arrow[dashed, "{\exists !\langle f_1, f_2\rangle }"{description}, from=1-2, to=3-2]
	\arrow["{f_1}"', from=1-2, to=3-1]
\end{tikzcd}\ .
\] 
We will often write the product as simply $A_1 \times A_2$, omitting the equipment.
\end{definition}
\begin{remark}
If the Cartesian product of two objects exists, it is unique up to unique isomorphism, and so we will speak of \textit{the} Cartesian product. 
\end{remark}
\begin{proposition}\label{cartesian-equations}
The morphism $\langle f_1, f_2 \rangle $ is unique if for every $g: B \to A_1 \times A_2$, $\langle g ; \pi_1, h ; \pi_2 \rangle= g$.
\end{proposition}
\begin{proof}
For $\langle f_1, f_2 \rangle$ to be unique means that if $f_1 = u ; \pi_1$ and $f_2 = u ; \pi_2$ then $u = \langle f_1, f_2 \rangle$. 
Let $g: A_1 \times A_2 \to B$ be such that $g ; \pi_1 = f_1$ and $g ; \pi_2 = f_2$ . Then, by hypothesis, $g = \langle g ; \pi_1, g ; \pi_2\rangle = \langle f_1, f_2 \rangle$. 
\end{proof}
A Cartesian category is one which contains the product of any two objects and a terminal object; 
a Cartesian functor preserves this structure. This amounts to the following definition.
\begin{definition}
A \textit{Cartesian category} is a category $\mathbb{C}$, such that for any two objects $A, B \in \mathbb{C}$, there exists their product $A \times B$, and such that there exists a terminal object $\top$.
We denote a Cartesian category with its equipment as $(\mathbb{C}, \times, \top)$, but often write simply $\mathbb{C}$, omitting the equipment.
A \textit{Cartesian functor} $F: (\mathbb{C}, \times_{\mathbb{C}}, \top_{\mathbb{C}}) \to (\mathbb{D}, \times_{\mathbb{D}}, \top_{\mathbb{D}})$ between Cartesian categories is given by
\begin{itemize}
\item a functor $F: \mathbb{C} \to \mathbb{D}$, which additionally:
\item \textit{preserves products} in the sense that for every product $(A \times_\mathbb{C} B, \pi_1, \pi_2)$ in $\mathbb{C}$, $(F(A \times_\mathbb{C} B), F(\pi_1), F(\pi_2))$ is the product of $F(A)$ and $F(B)$ in $\mathbb{D}$, and 
\item \textit{preserves the terminal object} in the sense that $F(\top_\mathbb{C})$ is terminal in $\mathbb{D}$. 
\end{itemize} 
We call two Cartesian categories \textit{equivalent} when there exists a Cartesian functor between them which is an equivalence when considered as a plain functor.
\end{definition}

\begin{definition}
The  \textit{exponential} $(A \Rightarrow C, \epsilon)$ of two objects $A,C \in \mathbb{C}$ consists of an object $A \Rightarrow C$ together with an \textit{evaluation} morphism $\epsilon_{A,C}: A \times (A \Rightarrow C) \to C$ such that for every object $A \in \mathbb{C}$ and morphism $g: A \times B \to C$, there exists a unique morphism $\textsf{curry}(g): B \to (A \Rightarrow C)$, which we call \textit{currying}, such that 
\[\begin{tikzcd}
	{A \times B} \\
	{A \times (A \Rightarrow C)} & C
	\arrow[dashed, "{\textsf{id}_A \times \textsf{curry}(g)}"', from=1-1, to=2-1]
	\arrow["g", from=1-1, to=2-2]
	\arrow["\epsilon_{A,C}"', from=2-1, to=2-2]
\end{tikzcd}\]\ 
We will often write the exponential as simply $A \Rightarrow C$, omitting the equipment. 
\end{definition}
\begin{remark}
If the expoential of two objects exists, it is unique up to unique isomorphism, and so we will speak of \textit{the} exponential. 
\end{remark}
\begin{proposition}\label{ccc-equations}
The morphism $\textsf{curry}(g)$ is unique if for every $h: B \to (A \Rightarrow C)$, 
\[
	h = \textsf{curry}(( \textsf{id}_A \times h) ; \epsilon_{A,C})
\]
\end{proposition}
\begin{proof}
For $\textsf{curry}(g)$ to be unique means that if $g = \textsf{id}_A \times h ; ev_{B,C}$, then $h = \textsf{curry}(g)$. Let $g = \textsf{id}_A \times h ; ev_{B,C}$. Then, by hypothesis, $h = \textsf{curry}((\textsf{id}_B \times h) ; \epsilon_{A,C}) = \textsf{curry}(g)$. 
\end{proof}

A Cartesian closed category is one which has all exponentials, products, and a terminal object; a Cartesian closed functor is a functor which preserves this structure. This amounts to the following definition. 
\begin{definition}
A \textit{Cartesian closed category (CCC)}  is a Cartesian category \linebreak $(\mathbb{C}, \times, \top)$, such that for every two objects $A ,C \in \mathcal{C}$, there exists their exponential $A \Rightarrow C$. 
We denote a CCC with its equipment as $(\mathbb{C}, \Rightarrow, \times, \top)$ , but often write simply $\mathbb{C}$, omitting the equipment.
A \textit{Cartesian closed functor} $F: (\mathbb{C}, \Rightarrow_\mathbb{C}, \times_{\mathbb{C}}, \top_{\mathbb{C}}) \to (\mathbb{D}, \Rightarrow_\mathbb{D}, \times_{\mathbb{D}}, \top_{\mathbb{D}})$ is given by
\begin{itemize}
\item a Cartesian functor $F: (\mathbb{C}, \times_\mathbb{C}, \top_\mathbb{C}) \to (\mathbb{D}, \times_\mathbb{D}, \top_\mathbb{D})$ between the underlying Cartesian categories, which additionally:
\item \textit{preserves exponentials} in the sense that if  $(A \Rightarrow B, \epsilon)$ is an exponential in $\mathbb{C}$, $(F(A \Rightarrow B), F(\epsilon))$ is the exponential of $F(A)$ and $F(B)$ in $\mathbb{D}$. 
\end{itemize}
%
We call two Cartesian closed categories \textit{equivalent} if there exists a Cartesian closed functor between them, which is an equivalence when considered as a plain functor.
\end{definition}

\begin{remark}
The definition of a Cartesian closed category is equivalent to asking for right adjoints to the functor $A \times -$, for all objects $A$.
Equivalently this can be expressed as the existence of a bijection between the hom-sets
\[
	Hom(A \times B, C) \cong Hom(B, A \Rightarrow C),
\]
which is natural in $A$ and $B$. 
\end{remark}
Note that we could equivalently specify a CCC by asking for a right adjoint to $- \times A$, however it is important to use the formulation given in order to give the most natural correspondence with the FMC.

\begin{remark}
As expected, two Cartesian (closed) categories $\mathbb{C}$ and $\mathbb{D}$ are equivalent by the above definition(s), if and only if there exist Cartesian (closed)-functors $F: \mathbb{C} \to \mathbb{D}$ and $G: \mathbb{D} \to \mathbb{C}$ such that $F;G \simeq \textsf{id}_\mathbb{C}$ and $G;F \simeq \textsf{id}_\mathbb{D}$, using an appropriate definition of \textit{Cartesian (closed) natural isomorphisms}, which we do not give here. In particular, if we have a Cartesian closed functor $F: \mathbb{C} \to \mathbb{D}$ with a plain inverse functor $G: \mathbb{D} \to \mathbb{C}$, then $G$ is automatically a Cartesian closed functor. 
\end{remark}

\begin{example}

The category \textsf{Set} of sets and functions is a Cartesian closed category, with the product given by the Cartesian product of sets and the exponential object $B \Rightarrow C$ given by the set of all functions from $B$ to $C$.
\end{example}



%
%


\section{Signatures and The Free Cartesian Closed Category}

The free Cartesian closed category is, intuitively, a Cartesian closed category containing some \textit{generating} objects and morphisms, where two morphisms are equal if and only if their equality can by derived by the axioms of a CCC. That is, no more morphisms are identified than is strictly necessary in order for the category to be a CCC.
We generate the free CCC over a set of primitive morphisms of forms such as $f: (A \to B) \to (C \times D)$, in order to adequately deal with (higher-order) constants with similar types.  

We first define a notion of higher-order signature, which is the data over which we will generate a free Cartesian closed category. 
\begin{definition}
A \textit{(Cartesian closed) signature} is a tuple $\Sigma = (\Sigma_0, \Sigma_1, \textsf{dom}, \textsf{cod})$, where $\Sigma_0$ is a set of sorts, $\Sigma_1$ is a set of function symbols and $\textsf{dom},\textsf{cod}: \Sigma_1 \to \Sigma_0^{\times \rightarrow}$ are a pair of functions specifying the \textit{domain} and \textit{co-domain} of the function symbols, where $\Sigma_0^{\times\rightarrow}$ is freely generated by the grammar
\[
	A,B ~\Coloneqq~ \alpha ~|~ \top ~|~ A \times B ~\mid~  A \to B\, ,
\]
with $\alpha \in \Sigma_0$. A \textit{Cartesian} signature is defined similarly, except replacing $\Sigma_0^{\times\rightarrow}$  with $\Sigma_0^{\times}$, which is freely generated by the grammar
\[
	A,B ~\Coloneqq~ \alpha ~|~ \top ~|~ A \times B\, .
\]
We will sometimes write $f: A \to B$ when $f \in \Sigma_1$ to denote that $\textsf{dom}(f) = A$ and $\textsf{cod}(f) = B$. 
\end{definition}

\begin{example}\label{example:signature}
We can form a signature $\Sigma = (\Sigma_0, \Sigma_1, \textsf{dom}, \textsf{cod})$ with two base types $\Sigma_0 = \{\alpha, \beta\}$ and constants of type
\[
	a: 1 \to \alpha \qquad
	m: \alpha \times \alpha \to \alpha \qquad
	b: 1 \to \beta \qquad
	n: \beta \times \beta \to \beta
\]
by setting $\Sigma_1 = \{a,b,m,n\}$ and $\textsf{dom}, \textsf{cod}: \Sigma_1 \to \Sigma_0^{\times\rightarrow}$ such that
\begin{align*}
	\textsf{dom}(a) &= 1 & \textsf{cod}(a) &= \alpha\\
	\textsf{dom}(m) &= \alpha \times \alpha & \textsf{cod}(m) &= \alpha\\
	\textsf{dom}(b) &= 1 & \textsf{cod}(b) &= \beta\\
	\textsf{dom}(n) &= \beta \times \beta & \textsf{cod}(n) &= \beta\ . 
\end{align*}
\end{example}

\begin{definition}
Given signatures $\Sigma = (\Sigma_0, \Sigma_1, \textsf{dom}_\Sigma, \textsf{cod}_\Sigma)$ and \linebreak $\Pi = (\Pi_0, \Pi_1, \textsf{dom}_\Pi, \textsf{cod}_\Pi)$, a \textit{signature homomorphism} $f: \Sigma \to \Pi$ is defined as a pair of functions $f_0: \Sigma_0 \to \Pi_0$ and $f_1: \Sigma_1 \to \Pi_1$ such that the following diagrams commute:
\[\begin{tikzcd}
	{\Sigma_1} & {\Sigma_0^{\times\rightarrow}} & {\Sigma_1} & {\Sigma^{\times\rightarrow}_0} \\
	{\Pi_1} & {\Pi_0^{\times\rightarrow}} & {\Pi_1} & {\Pi_0^{\times\rightarrow}}
	\arrow["{\textsf{dom}_\Sigma}", from=1-1, to=1-2]
	\arrow["{\textsf{dom}_\Pi}"', from=2-1, to=2-2]
	\arrow["{f_1}"', from=1-1, to=2-1]
	\arrow["{f_0^{\times\rightarrow}}", from=1-2, to=2-2]
	\arrow["{f_1}"', from=1-3, to=2-3]
	\arrow["{f_0^{\times\rightarrow}}", from=1-4, to=2-4]
	\arrow["{\textsf{cod}_\Pi}"', from=2-3, to=2-4]
	\arrow["{\textsf{cod}_\Sigma}", from=1-3, to=1-4]
\end{tikzcd}\]
where $f_0^{\times\rightarrow}$ is the lifting of $f_0: \Sigma_0 \to \Pi_0$ to $f_0^{\times\rightarrow}: \Sigma_0^{\times\rightarrow} \to \Pi_0^{\times\rightarrow}$ defined in the obvious pointwise way.
\end{definition}

\begin{example}
Consider the signature $\Sigma$ from Example \ref{example:signature}, and a second signature $\Pi = (\Pi_0, \Pi_1, \textsf{dom}_\Pi, \textsf{co}_\Pi)$ with a single base type $\Sigma_0 = \{\gamma\}$ and constants of type
\[
	c: 1 \to \gamma \qquad
	p: \gamma \times \gamma \to \gamma \ ,
\]
that is, we have $\Pi_1 = \{c,p\}$ and $\textsf{dom}_\Pi, \textsf{cod}_\Pi: \Pi_1 \to \Pi_0^{\times \rightarrow}$ such that 
\begin{align*}
	\textsf{dom}_\Pi(c) &= 1 & \textsf{cod}_\Pi(c) &= \gamma \\
	\textsf{dom}_\Pi(p) &= \gamma \times \gamma & \textsf{cod}_\Pi(p) &= \gamma\ .
\end{align*}
Then an example of a signature homomorphism $f: \Sigma \to \Pi$ is given by a pair of functions $f_0: \Sigma_0 \to \Pi_0$ and $f_1: \Sigma_1 \to \Pi_1$ such that 
\begin{align*}
	f_0(\alpha) &= f_0(\beta) = \gamma \\
	f_1(a) &= f_1(b) = c \\
	f_1(m) &= f_1(n) = p \ ,
\end{align*}
where in particular the required commuting diagrams express that $f_1$ must respect the (co-)domain of constant symbols with respect to the transformation of base types described by $f_0$. 
\end{example}

Let \textbf{CCCSig} be the category of Cartesian closed signatures and their homomorphisms, and \textbf{CCCat} be the category of Cartesian closed categories and Cartesian closed functors. There is an evident \textit{forgetful} functor $U: \textbf{CCCat} \to \textbf{CCCSig}$, whose left-adjoint is the \textit{free} functor $F: \textbf{CCCSig} \to \textbf{CCCat}$. 
\[\begin{tikzcd}
	{\textbf{CCCSig}} & {\rotatebox{90}{$\vdash$}} & {\textbf{CCCat}}
	\arrow["F", curve={height=-18pt}, from=1-1, to=1-3]
	\arrow["U", curve={height=-18pt}, from=1-3, to=1-1]
\end{tikzcd}\]
For a signature $\Sigma$, we call $F(\Sigma)$ the \textit{free Cartesian closed category over $\Sigma$}. Note, the free CCC is indeed unique up to equivalence. Given such a signature and a CCC $(\mathbb{C}, \Rightarrow, \times, \top)$, the universal property of the adjunction then guarantees that for every signature homomorphism $f: \Sigma \to U(\mathbb{C})$, there exists a unique Cartesian closed functor $[-]: F(\Sigma) \to \mathbb{C}$ such that:
\[\begin{tikzcd}
	{\Sigma} & {UF(\Sigma)} \\
	& {U(\mathbb{C})}
	\arrow["{U([-])}", from=1-2, to=2-2]
	\arrow["", hook, from=1-1, to=1-2]
	\arrow["f"', from=1-1, to=2-2]
\end{tikzcd}\]
commutes.
This leads to the following definition.
\begin{definition}
A Cartesian (closed) category  $\mathbb{X}$ is the \textit{free Cartesian (closed) category over a signature $\Sigma$} if it is equipped with an \textit{interpretation} $i: \Sigma \to U(X)$  such that for every Cartesian (closed) category $\mathbb{C}$ and signature homomorphism $f: \Sigma \to U(\mathbb{C})$, there exists a unique Cartesian (closed) functor $\termint{-}: \mathbb{X} \to \mathbb{C}$ such that the following diagram commutes
\[\begin{tikzcd}
	{\Sigma} & {U(\mathbb{X})} \\
	& {U(\mathbb{C})}
	\arrow["{U(\termint{-})}", from=1-2, to=2-2]
	\arrow["i", hook, from=1-1, to=1-2]
	\arrow["f"', from=1-1, to=2-2]
\end{tikzcd}\]
We will denote the free Cartesian closed category over $\Sigma$ by $\textsf{CCC}(\Sigma)$. 
\end{definition}
Stated in other terms, the free CCC over $\Sigma$ is initial in the category of CCCs equipped with the extra structure of a specified set of morphisms $\Sigma$, and functors which respect this extra structure. As such `the free CCC over $\Sigma$' is to be read not as `the (free CCC) over $\Sigma$', where $\Sigma$ is some generating base category, but rather as `the free (CCC over $\Sigma$)'. 

\section{The Simply-Typed $\lambda$-Calculus with Patterns}
We define the simply-typed $\lambda$-calculus with pattern matching \cite{DBLP:journals/tcs/KlopOV08,DBLP:books/ph/Jones87}. 
We will use a signature with a little extra structure, in order to capture the distinction between values and computations  that will be made in the FMC, introduced in Chapter \ref{chapter:fmc-prelims}. In particular, we split the set of constants $\Sigma_1$ into the sets $\Sigma_c$ and $\Sigma_v$ of computation and value symbols, and require value constants to have the `empty' domain $\top$. Correspondingly, note that there are two typing rules for constants: one for values, and one for computations.
\begin{definition}
An \textit{$\lambda$-signature with values} is a signature $(\Sigma_0, \Sigma_c \uplus \Sigma_v, \textsf{dom}, \textsf{cod})$, where $\Sigma_c \uplus \Sigma_v$ is the disjoint sum of sets $\Sigma_c$ and $\Sigma_v$ 
and where $\textsf{dom}(v) = \top$ whenever $v \in \Sigma_v$.  We will write $f: A \to B \in \Sigma_c$ for a computation symbol $f \in \Sigma_1$ such that $\textsf{dom}(f) = A$ and $\textsf{cod}(f) = B$. For a value symbol $v$, we will write $v: A \in \Sigma_v$ when $\textsf{cod}(v) = A$. When writing a signature with values, we will often then omit the (co-)domain, so that we write $\Sigma = (\Sigma_0, \Sigma_c, \Sigma_v)$.
We will often refer to this as just a \textit{signature}, when there is no ambiguity.    
\end{definition}
\begin{definition}
The \textit{$\lambda$-calculus with pattern matching} generated by sets $\Sigma_v$ and $\Sigma_c$ of value and computation symbols is given by the following grammar:
\begin{align*}
	M, N &~\Coloneqq~ x ~\mid~ v~ \mid~ c@M ~\mid~ M @ N ~\mid~ \lambda p.M ~\mid~ (M_1, \ldots, M_n) \\
	p,q,r,s,t &~\Coloneqq~ x ~\mid~ (p_1, \ldots, p_n) 
\end{align*}
where from left to right the \textit{term} constructors are a \textit{variable}, a \textit{value constant} where $v \in \Sigma_v$ , a \textit{computation constant} where  $f \in \Sigma_c$, an \textit{application}, an \textit{abstraction} on the pattern $p$, which binds the variables of $p$ in $M$, and a \textit{n-ary tuple} of terms. \textit{Pattern} constructors are given by a \textit{variable} or a \textit{tuple of patterns}, respectively. {We freely allow coercion from patterns to terms.} Terms are considered modulo $\alpha$-equivalence. 
\end{definition}

\begin{definition}
Given a signature $\Sigma = (\Sigma_0,\Sigma_c, \Sigma_v)$, \textit{simple types}  are given by the following grammar\begin{align*}
A,B ~ \Coloneqq ~ \alpha  ~ \mid  \top ~\mid~  A \times B ~\mid~{A} \to {B}  
\end{align*}
for $\alpha \in \Sigma_0$.
The \textit{typing rules} for the \textit{simply-typed $\lambda$-calculus with pattern matching (STLC) over $\Sigma$}, for terms generated over $\Sigma_v$ and $\Sigma_c$, and types generated over $\Sigma_0$, are given in Figure \ref{fig:STLC-types}. A \textit{typing judgement} is of the form
\[
	\Pi \vdash M:A\, ,
\]
where a \textit{context} $\Pi$ is a finite sequence of variables, each associated with a type: ${x_1: A_1}, \ldots, {x_n: A_n}$. 
\end{definition}

\begin{definition}
The \textit{equational theory} of the STLC is the least equivalence generated by the following laws, closed under any context:
\[
\begin{array}{lrll}
	\textup{Beta (Function):} & (\lambda p.M) @ P &=_\beta M\{P/p\}&: A \\
	\textup{Eta (Function): } &\lambda p.M @ p &=_\eta M&: A \to B \quad  \\
	\textup{Eta (Product):} & (\pi_1(M), \ldots, \pi_n(M)) &=_{\pi} M &: A_1 \times \ldots \times A_n 
\end{array}
\]
where for $\eta$, $\fv{p} \cap \fv{M} = \emptyset$, and in the last case, we define $\pi_i = \lambda (x_1, \ldots, x_n).x_i$. Substitution \textit{pattern matches} in the sense that $\{(P_1, \ldots, P_n)/(p_1,\ldots, p_n), \ldots\} = \{P_1/p_1, \ldots, P_n/p_n, \ldots\}$. Otherwise, (simultaneous) substitution is capture-avoiding and defined as standard. 
\end{definition}
\begin{figure}
\[
	\infer[\!\scriptstyle{\mathsf{var}}]
	 {{\Pi, x:A, \Pi' \vdash {x}: {A}}}
	{}
\qquad 
	\infer[\!\scriptstyle{\mathsf{vconst}}]
	   {\Pi \vdash v: A}
	{ {v: A \in \Sigma_v}}
\qquad 
	\infer[\!\scriptstyle{\mathsf{cconst}}]
	   {\Pi \vdash c\, @M: B}
	{{\Pi \vdash M: A} & {c: A \to B \in \Sigma_c}}
\]
\newline
\vspace{-\baselineskip}
\[
\infer[\!\scriptstyle{\mathsf{tuple}}]
	  {{\Pi \vdash (M_1, \ldots, M_n): {A_1} \times \ldots \times A_n}}
	  {\{{\Pi \vdash M_i: {A_i}\}_{i \in I}} 
	  }
\qquad
\infer[\!\scriptstyle{\mathsf{pm}}]
	  {{(p_1, \ldots, p_n): {A_1} \times \ldots \times {A_n}, \Pi  \vdash M : {B}}}
	  {{ p_1: {A_1}, \ldots, p_n: {A_n}, \Pi \vdash M: {B}}}
\]	
\newline
\vspace{-\baselineskip}
\[
\begin{array}{cc}
	\infer[\TR a]
	  {{\Pi \vdash M@N: {B}}}
	  {{\Pi \vdash N: {A}} &&
	   {\Pi \vdash M: {A} \to {B}}
	  }
	&
	\infer[\TR l]
	  {{\Pi \vdash \lambda p. M : {A} \to {B}}}
	  {{p: {A}, \Pi \vdash M: {B}}}
\end{array}
\]
\caption{Typing rules for the $\lambda$-calculus with pattern matching}
\label{fig:STLC-types}
\end{figure}
\begin{figure}
\[
	\infer[\!\scriptstyle{\mathsf{cut}}]
	{\Pi \vdash M\{N/p\}: B}
	{{\Pi \vdash N: A} & {p: A, \Pi \vdash M:B}}
\]
\\
\vspace{-\baselineskip}
\[
	\infer[\!\scriptstyle{\mathsf{exch}}]
	{\Pi, q:B, p:A, \Pi' \vdash M: B}
	{{\Pi, p:A, q:B, \Pi' \vdash M:B}}
	\qquad 
	\infer[\!\scriptstyle{\mathsf{weak}}]
	{\Pi, p:A \vdash M: B}
	{{\Pi \vdash M:B}}
\]
\caption{Admissible rules for the the $\lambda$-calculus with pattern matching}
\label{STLC-admissible-rules}
\end{figure}
Note that in this calculus, the beta law for products $(\lambda (x_1,x_2).x_i)(M_1,M_2) = M_i$ is given by the pattern matching performed by the definition of substitution. Admissible typing rules are given in Figure \ref{STLC-admissible-rules}.

\section{The Category of Simply-Typed $\lambda$-Terms}

Finally, we define the category of simply-typed $\lambda$-terms and recall its equivalence with the free Cartesian closed category generated over the same signature, which is due to Lambek \cite{lambek2,Lambek1968,DBLP:conf/litp/Lambek85}.

We give more equipment than is necessary in the following definition, since this will make comparison with the FMC easier later on. Following the definition, we verify that the extra equipment defined in fact arises from the necessary equipment in the usual way. 
\begin{definition}\label{defn:stlc-equipment}
Given a signature $\Sigma = (\Sigma_0, \Sigma_v, \Sigma_c)$, the category of simply-typed $\lambda$-terms over $\Sigma$ $\Lambda(\Sigma)$ is defined with:
\begin{itemize}
\item Objects: simple types generated over $\Sigma_0$,
\item Morphisms: $Hom(A,B)$ is given by the set of simply-typed $\lambda$-terms $p: A \vdash M:B$ over $\Sigma$, modulo the equational theory $=$,
\item Composition: given morphisms $q: A \vdash N: B\in Hom(A,B)$ and $p: B \vdash M: C \in Hom(B,C)$, $N;M \in Hom(A,C)$ is given by substitution $q: A \vdash M\{N/p\}: C$,
\item Identity: given on every type $A$ by $p:A \vdash p:A$,
\item Products: given on types $A$ and $B$ by $A \times B$, with unit $\top$, and its action on morphisms and associated natural transformations:
\[
\begin{array}{rlcrl}
	M \times N :& A \times B \vdash A' \times B' &=& (p,q) &\vdash (M,N)\\
	!:& A \vdash \top &=& p &\vdash ()\\
	\Delta:& A \vdash A \times A &=& p &\vdash (p,p)\\
	{\langle M,N \rangle}:& A \vdash B \times C &=& r &\vdash (M\{r/p\},N\{r/q\})\\
	\pi_1:& A \times B \vdash A &=& (p,q) &\vdash p \\
	\pi_2:& A \times B \vdash B &=& (p,q) &\vdash q\\
	\textsf{sym}:& A \times B \vdash B \times A &=& (p,q) &\vdash (q,p) 
\end{array}
\]
for $p:A \vdash M:A'$ and $q:B \vdash N:B'$,
\item Exponents: on types $A$ and $B$ by $A \to B$, with its action on morphisms and associated natural transformations:
\[
\begin{array}{rlcrl}
	\epsilon&: A \times (A \to B) \vdash B &=& (p, f) &\vdash f@p \\
	\eta&: A \vdash B \to (B \times A) &=& p &\vdash \lambda q. (q,p) \\
	\textsf{curry}(M)&: B \vdash (A \to C) &=& q &\vdash \lambda p.M \\
	P \to Q&: (A \to B) \vdash (C \to D) &=& f &\vdash \lambda p. Q\{f(P) / q\}
\end{array}
\]
for $(p,q): A \times B \vdash M:C$ and 
$p: C \vdash P: A$ and $q: B \vdash Q: D$.
\item Associators and unitors:
\[
\begin{array}{rlcrl}
	\alpha &: A \times (B \times C) \vdash (A \times B) \times C &=& (p,(q,r)):  &\vdash ((p,q),r) \\
	\lambda &: \top \times A \vdash A  &=& ((),p):  &\vdash p \\
	\rho &: A \times \top \vdash A &=& (p,()) &\vdash p 
\end{array}
\]
\end{itemize}
We may sometimes omit the bracketing of the context, since it is unambiguous. 
\end{definition}

\begin{remark}\label{remark:overspec}
We have overspecified the necessary equipment, but the equipment given is consistent in the expected way: we can define the following equipment in terms of that required for the definition of a Cartesian closed category.
\[
\begin{array}{rlcrl}
	\Delta ~\defeq& \langle \textsf{id}, \textsf{id}\rangle \\
	M \times N ~\defeq& \langle \pi_1\,;\,M, \pi_2\,;\,N\rangle \\
\textsf{sym} ~\defeq&~ {\langle \pi_2, \pi_1\rangle} \\
	\eta ~\defeq&~  \textsf{curry}(\textsf{id}) \\
	M \to N ~\defeq&~ \textsf{curry}(\langle \pi_1 \,;\, M ,\pi_2\rangle  \,;\, \epsilon \,;\, N) 
\end{array}
\]
The corresponding terms given in Definition \ref{defn:stlc-equipment} can be seen to match this, up to a small amount of simplification using the equational theory.
Conversely, we can define the following equipment in terms of the natural transformations $\Delta, !, \eta, \epsilon$ and the bifunctors $\times$ and $\to$.
\[
\begin{array}{rlcrl}
	\langle {M}, {N}\rangle  ~\defeq& ~ \Delta \,;\, ({M} \times {N})\\
	    \pi_1 ~\defeq&~ (! \times \textsf{id}) \\
	    \pi_2 ~\defeq&~ (\textsf{id} \times \,!) \\
	\textsf{curry}({P})  ~\defeq&~ \eta \,;\, (\cat{\id} \xrightarrow{}  P) 
\end{array}
\]
\end{remark}

We will elide associativity and unitality in the remaining thesis, which will be justified by the following theorem. 

The following theorem is due to Lambek \cite{DBLP:conf/litp/Lambek85}, and extends the Curry-Howard correspondence between the simply-typed $\lambda$-calculus and intuitionistic logic to include their categorical counterpart, Cartesian closed categories. We sketch the proof, since we are working with a mostly standard definition of the $\lambda$-calculus.
We will deal with the \textit{strict} free CCC, where associativity and unitor isomorphisms are true identities. This is justifed by MacLane's \textit{coherence theorem}, which says that the free monoidal category  is (monoidally) equivalent to the \textit{strict} free monoidal category \cite{MacLane}.
Pay particular attention to the interpretation of computation constants in the following proof. 
\begin{theorem}[Curry-Howard-Lambek]\label{thm:curry-howard}
The category $\Lambda(\Sigma)$ is Cartesian closed and equivalent to $\textup{\textsf{CCC}}(\Sigma)$. 
\end{theorem}
\begin{proof}
We sketch the proof. 
The category $\Lambda(\Sigma)$ is Cartesian closed, as follows. 
To see the existence of products, let $q_1:A_1 \vdash M_1:A_1$ and $q_2: A_2 \vdash M_2: A_2$ in
\begin{align*}
	\langle M_1,M_2\rangle   ; \pi_i &= p \vdash (M_1\{p/q_1\}, M_2\{p/q_2\} ; (s_1,s_2) \vdash s_i \\
		&= p \vdash M_i\{p/q_i\} \\
		&=_\alpha q_i \vdash M_i,
\end{align*}
where we recall composition is substitution, which implements pattern-matching.
For uniqueness of products, let $s:A \vdash M: B \times C$, $M' = M\{r/s\}$ and $\pi'_i = \lambda (p_1,p_2).p_i$ in
\begin{align*}
\langle M;\pi_1, M;\pi_2\rangle &= r \vdash (M' ; (q_1,q_2) \vdash q_1, M' ; (q'_1,q'_2) \vdash q'_2) \\
					&=_\eta r \vdash ((\pi'_1@M', \pi'_2@M') ; (q_1,q_2) \vdash q_1, \\
					& \qquad\qquad\! (\pi'_1@M', \pi'_2@M')  ; (q'_1,q'_2) \vdash q'_2) \\
					&= r \vdash (\pi'_1@M' , \pi_2@M') \\
				&=_\eta r \vdash M' \\
			&=_\alpha s \vdash M
\end{align*}
For existence of exponents, let $(q,r): (A \times B) \vdash M:C$ in
\begin{align*}
	(\textsf{id} \times \textsf{curry}(M) ) ; \epsilon &= (p,r) \vdash (p, \lambda q.M) ; (p', f) \vdash f@p' \\
		&= (p,r) \vdash (\lambda q.M)@p \\
		&=_\beta (p,r) \vdash M\{p/q\}\\
		&=_\alpha (q,r) \vdash M
\end{align*}
For uniqueness of exponents, let $q: A \vdash N: B \to C$ in 
\begin{align*}
\textsf{curry}((\textsf{id} \times N) ; \epsilon) &= \textsf{curry}((p,q) \vdash (p,N) ; (p',f) \vdash f@p') \\
			&= \textsf{curry}((p,q) \vdash N@p) \\
			&= q \vdash \lambda p.N@p \\
			&=_\eta q \vdash N
\end{align*}
This defines a functor from $\textsf{CCC}(\Sigma)$ to $\Lambda(\Sigma)$ by freeness of $\textsf{CCC}(\Sigma)$.
We present the inverse functor $\termint{-}: \Lambda(\Sigma) \to \textsf{CCC}(\Sigma)$ below. We work with a \textit{strict} CCC, as justified by MacLane's strictification theorem \cite{MacLane}. We present the binary and nullary cases of tuples, with the $n$-ary case the obvious generalization. 
\begin{align*}
	\termint{\Gamma, x:A, \Delta \vdash x:A} &=~!_{\Gamma} \times \textup{\textsf{id}}_{A} \times\,  !_{\Delta}  \\
	\termint{\Pi \vdash v: A} &= ~!_{\Pi} ~;~ v   	\\
	\termint{\Pi \vdash c\, @M:B} &= \termint{\Pi \vdash M:A}~;~ c  	\\
	\termint{\Pi \vdash (M, N): A \times B} &= [\termint{\Pi \vdash M:A}, \termint{\Pi \vdash N:B}]\\
	\termint{(p, q): A \times B, \Pi \vdash M:C} &= \termint{p: A, q: B , \Pi \vdash M:C} \\
	\termint{(): \top, \Pi \times B \vdash M:C} &= \termint{\Pi \vdash M:C} \\
	\termint{\Pi \vdash (): \top} &= ~!_{\Pi}\\
	\termint{\Pi \vdash M @ N: B} &= [\termint{\Pi \vdash N: A}, \termint{\Pi \vdash M: A \to B}] ~;~ \epsilon\\
	\termint{\Pi \vdash \lambda p.M: A \to B} &= \eta ~;~ (\cat{\id_{A}} \to  \termint{\Pi, p:A \vdash M: B}) \qedhere
\end{align*}
\end{proof}
\begin{remark}
The translation of the admissible rules of cut, exchange and weakening into a Cartesian closed category are as follows.
\begin{align*}
	\termint{\Pi', \Pi \vdash M\{N/p\}: B} &= \Delta_{\Pi', \Pi} ; (\textsf{id}_{\Pi'} \times \termint{\Pi', \Pi \vdash N:A} \times \textsf{id}_{\Pi}) ;\\
		& \qquad  \termint{\Pi', p:A, \Pi \vdash M: B}\\
	\termint{\Pi, p:A, q:B, \Pi' \vdash M: C} &= (\textsf{id}_{\Pi} \times \textsf{sym}_{A,B} \times \textsf{id}_{\Pi'}) ; \termint{\Pi, q:B, p:A, \Pi' \vdash M: C}\\
 \termint{\Pi, p:A \vdash M:B} &= (\, !_A \times \textsf{id}_{\Pi}) ; \termint{\Pi \vdash M: C}
\end{align*}
\end{remark}
Note how the equational theory of the $\lambda$-calculus corresponds to the axiomatization of a CCC in terms of its univeral properties, as given earlier in the chapter. We will see later how this compares to the approach of the FMC.
\begin{remark}
By restricting the $\lambda$-calculus so that it doesn't include application or abstraction, we can similarly achieve a correspondence between the \textit{first-order} fragment of the $\lambda$-calculus and the free Cartesian category. 
\end{remark}

From now on, we can use the simply-typed $\lambda$-calculus as a convenient notation for the free Cartesian closed category. With this correspondence in mind, we will write $\textsf{CCC}(\Sigma)$ in place of $\Lambda(\Sigma)$, and, in particular, we will work modulo associativity and unitality of the $\lambda$-calculus. 

%


\chapter{The Functional Machine Calculus}
\label{chapter:fmc-prelims}

The material presented in this chapter was developed by Heijltjes and published in \cite{Barrett-Heijltjes-McCusker-2022}. The Functional Machine Calculus was amply motivated in the introduction, so we focus on concise definitions here for reference. 

There is one significant difference between the calculus presented in the introduction and the one presented here. 
Here, we make a distinction between \textit{values} (which live on the stack) and \textit{computations} (which act on the stack), following Call-by-Push-Value \cite{Levy-2003}.  We further introduce coercions \textit{force} and \textit{thunk} , taking values to computations and vice-versa. 
This variant was mentioned, but not developed, in previous publications \cite{Barrett-Heijltjes-McCusker-2022}.  The reason for the distinction is to achieve in Chapter \ref{chapter:CCC-eqns} the most natural correspondence with the $\lambda$-calculus. In particular, it turns out that \textit{force} and \textit{thunk} are naturally translated as the \textit{application} and \textit{abstraction}, respectively, of the $\lambda$-calculus. 
Furthermore, such a variant is natural for the inclusion of terms of base type: these highlight the difference between \textit{values}, which live on the stack, and \textit{computations}, which run on the machine, a natural operational distinction. \footnote{One suggestion that the original system is not quite right for the inclusion of base types is hidden in the typing rules of Figure \ref{fig:FMC-types-plain} in the introduction. Recalling that all terms defined in the original grammar end with a trailing $\term *$, which is usually omitted, we see that we have really typed $\term{x.*: a}$, thus giving a term of base type a continuation (albeit the trivial one).}

After introducing the \textit{Functional Machine Calculus with Values} -- which we will abbreviate \FMCV,  after the \textit{thunk} constructor $\term{`!M}$ -- the definitions relating to the ``classic'' FMC,  given in the introduction to this thesis, are collected.  The classic variant of the Functional Machine Calculus will be used in the proof of strong normalisation of Chapter \ref{chapter:SN}, where  the distinction between values and computations  is no longer useful. 

We fix a countable set of locations $A = \{\term a,\term b,\term c, \ldots\}$ throughout this chapter, and the remaining thesis, except where otherwise mentioned. We take $\term a, \term b$ to range over all locations.

\section{The FMC with Values: \FMCV}\label{sec:fmc-values}

\begin{definition}
\label{def:FMC}
The \textit{terms} of the \emph{Functional Machine Calculus with Values}, denoted \FMCV, generated by sets $\Sigma_v$ and $\Sigma_c$ of value and computation symbols, respectively, are given by the follwing grammar of  \emph{computations} and \emph{values}, respectively:
\begin{align*}
\term{M,N,P}
  \quad&\Coloneqq\quad \term *
	~\mid~ \term{c.M}
       ~\mid~ \term{[V]a.M}
       ~\mid~ \term{a<x>.M}
       ~\mid~ \term{`?V.M}\\
\term{V,W} \quad &\Coloneqq\quad \term x~\mid~ \term v ~\mid ~\term{`!M}
\end{align*}
where, from left to right, the \emph{computation} constructors are the \emph{identity}, a \emph{sequential constant} with $\term{c} \in \Sigma_c$,  \emph{application} or \emph{push action} on location $\term a$, an \emph{abstraction} or \emph{pop action} on  location $\term a$, which binds $\term x$ in $\term M$,  and a \emph{sequential execution} (or \emph{force}) of a value. The \emph{value} constructors are a \textit{variable}, a \textit{value constant} with $\term{v} \in \Sigma_v$, and a \emph{thunk} of a computation. Terms are considered modulo $\alpha$-equivalence.

Formally,  the set of \textit{free variables} of a computation $\term{M}$ or value $\term{V}$, $\fv{\term M}$ or $\fv{\term V}$, respectively,  is defined as follows.
\[
	\begin{array}{lll@{\qquad}lll}
	\fv{\term *} &=& \emptyset & 			\fv{\term{x}} &=& \{\term x\} \\
	\fv{\term{c.M}} &=& \fv{\term{M}} & 	 \fv{\term{v}} &=& \emptyset \\
	\fv{\term{[V]a.M}} &=& \fv{\term{V}}\cup \fv{\term{M}} &  \fv{\term{`!M}} &=& \fv{\term{M}} \\
	\fv{\term{a<x>.M}} &=& \fv{\term{M}} \setminus\, \{\term x\} & \\
	 \fv{\term{`?V.M}} &=& \fv{\term{V}} \cup \fv{\term M} &	
	\end{array}
\]
\end{definition} 
Note how the scoping of the $\term{`?}$ and $\term{`!}$ constructs is unambiguous.  
We will omit the trailing $\term{.*}$ from a computation, and use a \emph{main} location $\term l$, omitted from the notation. 
{We use Barendregt's variable convention, \textit{i.e.}, that all bound variables are chosen to be different from free variables.}

\begin{definition}\label{def:capture-comp}
Capture-avoiding \textit{composition}, or \textit{sequencing}, $\term{N;M}$ is given by
\[
\begin{array}{rcr@{\qquad}rrrrr}
		 \term{*;M} &=&         \term M   
& 	\term{c.N;M}  &=& \term{c.(N;M)} & 
\\
	\term{[V]a.N;M} &=& \term{[V]a.(N;M)}
&	\term{a<x>.N;M} &=& \term{a<x>.(N;M)} & 
\\     \term{`?V.N;M} &=&    \term{`?V.(N;M)} & 
\end{array}
\]
where, in the abstraction case, $\term{x} \not\in \fv{\term{M}}$.
Capture-avoiding \textit{substitution of values}  $\term{\{V/x\}M}$ and $\term{\{V/x\}W}$ into computations and values, respectively, is given by 
\[
\begin{array}{lll@{\qquad}llll}
	\term{\{V/x\}*}      &=& \term * & \term{\{V/x\}x} &=& \term{V}  \\
	\term{\{V/x\}c.M}    &=& \term{c.\{V/x\}M}   &  \term{\{V/x\}y}    &=& \term{y}       \\
	\term{\{V/x\}[W]a.M} &=& \term{[\{V/x\}W]a.\{V/x\}M} & \term{\{V/x\}v}    &=& \term{v}\\
	\term{\{V/x\}a<y>.M} &=& \term{a<y>.\{V/x\}M} & \term{\{V/x\}`!M} &=& \term{`!\{V/x\}M} \\
	\term{\{V/x\}`?W.M}    &=& \term{`?\{V/x\}W.\{V/x\}M} 
\end{array}
\]
%
where $\term{x} \neq \term{y}$ and, in the abstraction case,  $\term y\notin\fv{\term V}$. 
\end{definition}

\begin{remark}
Sequencing can easily be shown to be associative. It would be possible to take sequencing as a primitive constructor, but it would necessitate working modulo associativity. Instead, we take prefixing as the primitive constructor, which can be seen as a choice to associate to the right. 
\end{remark}

\begin{definition} \label{def:fmc-machine}
The \emph{functional abstract machine} is given by the following data. A \emph{stack} of terms $S$ is defined below left, and written with the top element to the right. A \emph{memory} $S_A$ is a family of stacks or streams in $A$, defined below right, and can formally be regarded as a function from the set of locations $A$ to the set of stacks. 
\[
	S ~\Coloneqq~\e~\mid~S{\cdot}\term V 
\qquad\qquad\qquad
	S_A~\Coloneqq~\{\,S_a\,\mid\,a\in A\}
\]
Where $B \subseteq A$, we write $S_B$ for the restriction of $S_A$ to $B$.  We write $S_a$ for $a \in A$ to access the stack held in memory $S_A$ at location $a$,  and we allow identification of $S_{\{a\}}$ with $S_a$. 
Let $S_{B};S_{C}$ denote the copairing of families (considered as functions) of stacks in disjoint sets $B$ and $C$.    
In particular this means we have $S_A = \SAouta;S_a$.
A \emph{state} is a pair $(S_A,\term M)$ of a memory and a term. 
The \emph{transitions} or \emph{steps} of the machine are given below. 
\[
\begin{array}{@{(~}l@{~,~}r@{~)}}
	S_{A\setminus\{a\}}~;~S_a 	         & \term{[V]a.M}
\\\hline
	S_{A\setminus\{a\}}~;~S_a{\cdot}\term V &      \term M
\end{array}
\qquad
\begin{array}{@{(~}l@{~,~}r@{~)}}
	S_{A\setminus\{a\}}~;~S_a{\cdot}\term V &   \term{a<x>.M}
\\\hline
	S_{A\setminus\{a\}}~;~S_a               & \term{\{V/x\}M}
\end{array}
\qquad 
\begin{array}{l@{,}r@{}}
(~S_A~& ~\term{`?`!N.M}~)
\\\hline
(~S_A~&~\term{N;M}~)
\end{array}
\]
A \emph{run} of the machine is a sequence of steps, written as $(S_A,\term M)\Downarrow(T_A,\term N)$ or with a double line as below.
\[
    \begin{array}{@{(~}l@{~,~}r@{~)}}
    S_A & \term{M}
    \\\hline\hline\rule[-5pt]{0pt}{15pt}
    T_A & \term{N}
    \end{array}~.
\]
A \emph{successful run} of the machine is a run $(S_A, \term{M})\Downarrow(T_A, \term{*})$, and in this case we will denote $T_A$ by simply $(S_A,\term{M})\Downarrow$.
\end{definition}

  We will introduce the reduction relation for the \FMCV, which will be studied throughout this thesis, but first we will need a notion of \textit{sequential context}. 
\begin{definition}\label{fmc-context}
A \textit{(sequential) context} is defined as an \FMCV-term with either a hole $\term{\{~\}_c}$ in place of a sequential {constant}, or a hole $\term{\{~\}_v}$ in place of a variable:
\begin{align*}
\term{K}
  \quad\defeq&\quad 
	\term{\{~\}_c.M} 
	~\mid~ \term{c.K}
	~\mid~ \term{[V]a.K} 
	~\mid~ \term{[`L]a.M}
	~\mid~ \term{a<x>.K}
	~\mid~ \term{`?V.K}
	~\mid~\term{`?L.M} 
\\
\term{L} \quad \defeq& \quad \term{\{~\}_v} ~\mid~ \term{`!K}
\end{align*}
We will write $\term{K\{~\}_c}$ or $\term{K\{~\}_v}$ to denote whether the context contains a hole for a computation or a value and refer to \textit{computation} or \textit{value} contexts, respectively. 
Given a  computation $\term{M}$ and a value $\term{V}$, define $\term{K\{M\}_c}$ and $\term{K\{V\}_v}$ to be the terms given by substituting $\term{M}$ and $\term{V}$ in for the holes $\term{\{~\}_c}$ and $\term{\{~\}_v}$, respectively. Substitution is defined as usual for values, except that in $\term{a<x>.K\{V\}}$, $\term x$ captures in $\term{V}$. \textit{Substitution of computations} is defined similarly, except with the case $\term{\{N\}_c.M} = \term{N;M}$.
\end{definition}
\begin{definition}\label{def:fmc-theory}
We define the \textit{beta}, \textit{force}, \textit{thunk} and \textit{permutation reductions} as the following rewrite rules on \FMCV-terms, respectively, applicable in all contexts: 
\begin{align*}
	\term{[V]a.a<x>.M} &\rw_\beta \term{\{V/x\}M} \\
	\term{`?`!N.M} &\rw_\phi \term{N;M}\\
	\term{`!`?V} &\rw_\tau \term{V} \\
	\term{[V]a.b<x>.M} &\rw_\pi \term{b<x>.[V]a.M},
\end{align*}
where in the final case, $\term x \not\in \fv{\term V}$ and $\term{a} \neq \term{b}$. We write $\rw$ to indicate the union of these reductions and $\rw^*$ to indicate the reflexive, symmetric, transitive closure of $\rw$. 
\end{definition}

\begin{remark}
Note that the definition of sequential (computation) context means each equation applies in the presence of some continuation $\term{M}$, as in $\term{\{N\}_c.M} = \term{N;M}$, and that the abstractions appearing in the equational theory will not bind in this continuation (because of the capture-avoiding definition of sequencing). Note, further, that the definition of sequential context means that a redex can be identified as being in more than one context. However, this does not matter for the purposes of the equational theory: we have
\[
\textup{if} ~  \term  M = \term N ~ \textup{then} ~\term{\{M\}_c.P;Q} = \term{\{N\}_c.P;Q} ~ \textup{if and only if} ~  \term{\{M;P\}_c.Q} = \term{\{N;P\}_c.Q} \, ,
\]
which can easily be seen to follow from the definitions and associativity of sequencing. 
\end{remark}

Beta reduction is analogous to that of the $\lambda$-calculus. 
Permutation reduction allows push- and a pop-action on the same location to interact (by beta reduction) despite the presence of intervening pushes and pops on \textit{other} locations. 

In the following chapters on categorical semantics, we will give an equational theory which extends the one generated by the reduction relation, for example with equations such as $\term{*} =_{\textsf{id}} \term{<x>.[x]}$ mentioned in the introduction.  
 In Chapter \ref{chapter:CCC-eqns}, we formalize an appropriate definition of \textit{machine equivalence} of terms (a kind of contextual equivalence based on the abstract machine), and show that each equation is valid with respect to this. 

We  introduce the following notation, which will be used throughout the thesis.
\begin{notation}\label{not:absapp}
The  \textit{empty memory} is denoted $\e_A$. 
A \textit{singleton} memory $a(S)$ is empty at every location except $\term a$, where is contains $S$, that is: $a(S)_a = S$ and $a(S)_b = \e$, for $\term{a} \neq \term{b}$. We omit the main location $\term{\lambda}$ for singleton types, so $\lambda(S)$ may be written as $S$. \textit{Concatenation} of stacks is denoted by $(\cdot)$, which we will sometimes omit, so it is denoted instead by juxtaposition, and this extends to families point-wise,  where it is denoted $(;)$, so $S_A ; T_A = \{S_a \cdot T_a ~|~ a \in A\}$. 

We use \textit{vector} notation to denote a stack of variables, and \textit{reverse} a vector by pointing the arrow left:
\[
\term{!x} = \term{x_1}\cdots\term{x_n} \qquad \term{?x} = \term{x_n}\cdots\term{x_1}\, .
\]
\textit{Concatenation} of vectors given by juxtaposition. 
Note that, although stacks grow to the right, we count elements from the left, in order to avoid reindexing when a new element is added.  
This notation is extended to families, so that $\term{!x_A} = \{\term{!x_a} ~|~ a \in A\}$, 
the \textit{reverse} of a family $\term{?x_A}$ and concatenation defined pointwise: \textit{e.g.}, $\term{!x_A!y_A} = \{\term{!x_a!y_a} ~|~ a \in A\}$.

We then lift the notation to sequences of abstractions and applications: given $\term{!x}$ and $\term{!x_A}$ as above, a stack $S = \term{V_1}\cdots \term{V_n}$ and memory $S_A = \{S_a ~|~ a \in A\}$,  let
\begin{align*}
	\term{<?x>.N} &= \term{<x_n>\dots<x_1>.N} & \term{[S].N} &= \term{[V_1]\dots[V_n].N} \\
	\term{<?x_A>.M} &= \term{a_1<?x_{a_1}>\ldots a_n<?x_{a_n}>.M} & \term{[S_A].M} &= \term{[S_{a_1}]a_1\ldots [S_{a_n}]a_n.M}\ 
\end{align*}
for $A = \{\term{a_1}, \ldots, \term{a_n}\}$. 
We will see later,  in Lemma \ref{lem:permutation-equivalence}, that any choice of ordering on the locations above gives terms equal in the equational theory. 
This notation is extended to simultaneous substitutions as 
\[
	\term{\{S/{}!x\}} = \term{\{V_1/x_1`,\dots`,V_n/x_n\}} \qquad \term{\{S_A / !x_A\}} = \term{\{S_{a_1}/!x_{a_1},\ldots,S_{a_n}/!x_{a_n}\}}\  ,
\]
so in particular it respects beta reduction in the sense that we have
\[
	\term{[S].<?x>.M} \rw^*_\beta \term{\{S/!x\}M} \qquad \term{[S_A].<?x_A>.M} \rw^*_\beta \term{\{S_A/!x_A\}M}\ .
\]
We let $\fv{\term{!x}}$ and $\fv{\term{!x_A}}$ denote the underlying sets of variables of $\term{!x}$ and $\term{!x_A}$, respectively. 
\end{notation}

\begin{remark}
Normal forms arrived at from the reduction rules are of the form
\[
	\term{<?x_A>.[S_A].(`?x.<?y_A>.[T_A])}^*\, ,
\]
where $(-)^*$ indicates some number of iterations of terms of that form. 
\end{remark}

\section{Simple Types for the Functional Machine Calculus}

We begin with the definition of \FMCV-types (which coincide with FMC-types).
Consider the elements which comprise an \FMCV\  machine state: we have \textit{values}, which comprise stack elements, the \textit{stack} itself, a \textit{memory} which is an indexed product of stacks. In the higher-order setting, values may of course be ({thunks} of) terms themselves. 
Correspondingly, we have the following grammar for types. 
\begin{definition}
Given a set $\Sigma_0$ of \textit{base types}, \textit{simple} \emph{value},  \emph{stack} and \emph{memory types over $\Sigma_0$} are given, respectively, by:
\[
		\type{t}      ~\Coloneqq~  \type{a}  ~|~ \type{?s_A>!t_A}
\qquad	\type{!t}	  ~\Coloneqq~ \type{t_1\dots t_n}
\qquad	\type{!t_A} ~\Coloneqq~ \{\type{!t_a}\mid a\in A\}
\]
where ${\type{a}}$ is drawn from $\Sigma_0$. We also refer to stack and memory types as type vectors and type families, respectively.
\end{definition}

We now define an \FMCV-signature, which holds the sets of constant value and compuation symbols and their types, over which the simply-typed \FMCV\  will be generated.
\begin{definition}
An \textit{\FMCV-signature} is a tuple $\Sigma_A = (\Sigma_0, \Sigma_c, \Sigma_v, \textsf{dom}, \textsf{cod}, \textsf{val})$, where $\Sigma_0$ is a set of sorts, $ \Sigma_c$ and $\Sigma_v$ are sets of function and value symbols, respectively, and $\textsf{dom},\textsf{cod}: \Sigma_c \to \Sigma_0^m$ are a pair of functions specifying the \textit{domain} and \textit{co-domain} of the computation function symbols, respectively, and $\textsf{val}: \Sigma_v \to \Sigma^v_0$ specifies the \textit{value} of the value function symbols, where $\Sigma^v_0$ and $\Sigma_0^m$ are given by the set of value and memory types, respectively, generated over $\Sigma_0$. We will write $\term{c: ?s_A > !t_A}  \in \Sigma_c$ for a computation symbol $\term{c} \in \Sigma_c$ such that $\textsf{dom}(\term{c}) = \type{!s_A}$ and $\textsf{cod}(\term{c}) = \type{!t_A}$. Similarly, we will write $\term{v: t}$ for a value symbol $\term{v} \in \Sigma_v$, such that $\textsf{val}(\term v) = \type{t}$. We often then denote an !FMC-signature as $\Sigma_A = (\Sigma_0, \Sigma_c, \Sigma_v)$, omitting the (co-)domain and value functions, and refer to this as just a \textit{signature}, when there is no ambiguity.    
\end{definition}

In addition to values, stacks and a memory, we have the computation term itself which is under execution: this receives the type of an implication between memories.  Overall, the type system will have two sorts of judgements: one for values, and one for computations. 
\begin{definition}\label{def:fmc-types}
Given an \FMCV-signature $\Sigma_A = (\Sigma_0, \Sigma_c, \Sigma_v)$, 
the typing rules for the \textit{simply-typed \FMCV\ over $\Sigma_A$} assign to terms generated over $\Sigma_v$ and $\Sigma_c$ types generated over $\Sigma_0$, and are given in Figure~\ref{fig:FMC-types} using notation given subsequently.
A \textit{computation type} is an implication between memory types: $\type{?s_A > !t_A}$. \textit{Computation} and \textit{value} judgements are of the form
\[
	\Gamma \vvdash \term{V:t} \qquad \Gamma \cvdash \term{M: ?s_A > !t_A},
\]
repsectively, where in each case the \textit{context} $\Gamma$ is a finite sequence of variables, each associated with a value type: $\term{x_1: t_1}, \ldots, \term{x_n: t_n}$. We will sometimes write $\Gamma \cvdash \term{M, N: ?s_A > !t_A}$ to indicate two terms of the same type, typed in the same context. 
\end{definition}
Admissible rules are shown in Figure \ref{fig:admissible rules}.
\begin{notation}
Similar to Notation \ref{not:absapp}, we introduce the following. 
The \emph{empty} type vector is $\type\e$, and the empty memory type $\type{\e_A}$. A \emph{singleton} memory type $\type{`a(!t)}$ is empty at every location except $\type{`a}$, where it has $\type{!t}$: that is, $\type{`a(!t)_a}=\type{!t}$ and $\type{`a(!t)_b}=\type{\e}$, for $\type{`a}\neq \type{`b}$. We omit the main location $\term{\lambda}$ for singeton types, so $\type{\lambda(!t)}$ may be written as $\type{!t}$. \emph{Concatenation} of type vectors is denoted by juxtaposition, and extends to families point-wise, $\type{!s_A !t_A}=\{\type{!s_a!t_a}\mid a\in A\}$. The \emph{reverse} of a type vector $\type{!t}=\type{t_1\dots t_n}$ is written $\type{?t}=\type{t_n\dots t_1}$, with the reverse of a type family also defined pointwise. For a context $\Gamma = \term{x_1:t_1,,x_n:t_n}$, we may write $\term{!x:!t}$.
\end{notation}

An equivalent way to view computation types is as an implication between two vectors of singletons:
\[
	\type{`a_1(s_1)\dots `a_n(s_n)~>~`b_1(t_1) \dots `b_m(t_m)}
\qquad\qquad
	\type{`a(s)\,`b(t)}=\type{`b(t)\,`a(s)}\ ,
\]
considered modulo the permutation of types on different locations, as above. 
\begin{remark}[Expansion and Sequencing]
Note that the \textit{strict sequencing} (\textsf{seq}) and \textit{expansion} (\textsf{exp}) rules can be combined to give the \textit{(general) 
{left- and right-}sequencing} (\textsf{lseq} and \textsf{rseq}, respectively). 
 Conversely, we can recover the expansion rule by left-sequencing a term $\term{M:?s_A>!t_A}$ with $\term{*:?s_A?r_A > !r_A!s_A}$. 
Note further that the in the \FMCV, the ``left'' or ``right'' orientation of a composition may differ on a per location basis. That is, one could type, for example, the composition of terms $\term{M: ?r_A > `a(!s)`b(!s'!t')}$ and $\term{N: `a(?s?t)`b(?t') > !u_A}$.
\end{remark}
We will often work with left sequencing as primary, which we will call simply \textit{sequencing}, rather than with strict sequencing and expansion.
%
\begin{figure}
\[
	\infer[\!\scriptstyle{\mathsf{id}}]{\Gamma \cvdash \term{ *:?t_A>!t_A}}{}
	\qquad
	\infer[\!\scriptstyle{\mathsf{var}}]{\Gamma, \term{x: t}, \Delta \vvdash \term{x: t}}{}
\]
\\
\vspace{-\baselineskip}
\[
	\infer[\!\scriptstyle{\mathsf{cconst}}]
	{\Gamma \cvdash \term{c.M: ?r_A?t_A > !u_A}}
	{{\Gamma \cvdash \term{M: ?s_A?t_A > !u_A}} & \term{c: ?r_A > !s_A} \in \Sigma_c}
	\qquad
	\infer[\!\scriptstyle{\mathsf{vconst}}]
	{\Gamma \vvdash \term{v: t}}
	{\term{v: t} \in \Sigma_v}
\]
\\
\vspace{-\baselineskip}
\[
	\infer[\!\scriptstyle{\mathsf{app}}]
	{\Gamma \cvdash \term{[V]a.M: ?s_A > !t_A}}
	{\Gamma \vvdash \term{V: r} && {\Gamma \cvdash \term{M:`a(r)?s_A > !t_A}}}
	\qquad
	\infer[\TR l]
	  {\Gamma \cvdash \term{ `a<x>.M : `a(r)\,?s_A>!t_A}}
	  {\term{x:r, G} \cvdash \term{ M:?s_A>!t_A}}
\]
\\
\vspace{-\baselineskip}
\[
	\infer[\!\scriptstyle{\mathsf{thunk}}]
	{\Gamma \vvdash \term{`!M: ?s_A > !t_A}}
	{\Gamma \cvdash \term{M: ?s_A > !t_A}} 
\qquad
	\infer[\!\scriptstyle{\mathsf{force}}]
	{\Gamma \cvdash \term{`?V.M: ?r_A?t_A > !u_A}}
	{{\Gamma \vvdash \term{V: ?r_A > !s_A}} & {\Gamma \cvdash \term{M: ?s_A?t_A > !u_A}}}
\]
\caption{Typing rules for the Functional Machine Calculus with Values}
\label{fig:FMC-types}
\end{figure}
%
\begin{figure}\label{fig:admissible rules}
\[
	\infer[\!\scriptstyle{\mathsf{seq}}]
	{\Gamma \cvdash \term{M;N: ?r_A > !t_A}}
	{{\Gamma \cvdash \term{M: ?r_A > !s_A}} & {\Gamma \cvdash \term{N: ?s_A > !t_A}}}
	\qquad
	\infer[\!\scriptstyle{\mathsf{exp}}]
	{\Gamma \cvdash \term{M: ?s_A?t_A > !t_A!u_A}}
	{{\Gamma \cvdash \term{M: ?s_A > !u_A}}}	
\]
\\
\vspace{-\baselineskip}
\[
	\infer[\!\scriptstyle{\mathsf{lseq}}]
	{\Gamma \cvdash \term{M;N: ?r_A?t_A > !u_A}}
	{{\Gamma \cvdash \term{M: ?r_A > !s_A}} & {\Gamma \cvdash \term{N: ?s_A?t_A > !u_A}}}
\]
\\
\vspace{-\baselineskip}
\[
	\infer[\!\scriptstyle{\mathsf{rseq}}]
	{\Gamma \cvdash \term{M;N: ?r_A > !s_A!u_A}}
	{{\Gamma \cvdash \term{M: ?r_A > !s_A!t_A}} & {\Gamma \cvdash \term{N: ?t_A > !u_A}}}
\]
\\
\vspace{-\baselineskip}
\[	
	\infer[\!\scriptstyle{\mathsf{ccut}}]
	{\Gamma \cvdash \term{\{V/x\}M: ?s_A > !t_A}}
	{{\Gamma \vdash_\textsf{v} \term{V: r}} &{\term{x:r},  \Gamma \vdash_\textsf{c} \term{M: ?s_A > !t_A}}}
	\qquad
	\infer[\!\scriptstyle{\mathsf{vcut}}]
	{\Gamma \vvdash \term{\{V/x\}W: t}}
	{{\Gamma \vdash_\textsf{v} \term{V: r}} &{\term{x:r},  \Gamma \vdash_\textsf{c} \term{W: t}}}
\]
\begin{align*}
	\infer[\!\scriptstyle{\mathsf{exch}}]
	{\Gamma, \term{y:r},\term{x:p}, \Gamma' \cvdash \term{M: ?s > !t}}
	{\Gamma, \term{x:p}, \term{y:r}, \Gamma' \cvdash \term{M: ?s > !t}}
\qquad
	\infer[\!\scriptstyle{\mathsf{weak}}]
	{\term{x:r}, \Gamma  \cvdash \term{M: ?s > !t}}
	{\Gamma \cvdash \term{M: ?s > !t}}
\end{align*}
\caption{Admissible rules for the Functional Machine Calculus with Values}
\end{figure}
\begin{remark}
It can be proved that
the \textit{computation cut} (\textsf{ccut}), \textit{value cut} (\textsf{vcut}) and \textit{sequencing} rules are admissible by applying the definitions of substitution and sequencing to type derivations. Further, from this, preservation of types during reduction can be proved \cite{Barrett-Heijltjes-McCusker-2022}. 
Additionally, it can be shown by a simple induction on type derivations that the usual structural rules of \textit{exchange} and \textit{weakening} are admissible in this system. 
\end{remark} 
We will from now on assume that all terms mentioned are well-typed.

\section{Variants}

Finally, we recall the definition of the original FMC (that discussed in the introduction), without constants, which we will use in Chapter \ref{chapter:SN} for the proof of strong normalisation. 
\begin{definition}
\label{def:FMC-plain}
The \textit{terms} of the \emph{(classic) Functional Machine Calculus (FMC)},  are given by the follwing grammar.
\begin{align*}
\term{M,N,P}
  \quad&\Coloneqq\quad \term *
       ~\mid~ \term{x.M}
       ~\mid~ \term{[N]a.M}
       ~\mid~ \term{a<x>.M}
\end{align*}
where the second term constructor is a \emph{sequential variable}. Terms are considered modulo $\alpha$-equivalence. 
Capture-avoiding \textit{sequencing}, $\term{N;M}$ and \textit{substitution} is defined similarly to Definition \ref{def:capture-comp}, except with  the following cases instead of the cases for values or the sequential execution construct.
\[
	  \term{x.N;M }=    \term{x.(N;M)} \qquad \term{\{N/x\}x.M} = \term{N;\{N/x\}M}
\]
\textit{Sequential contexts} are the obvious variation of Definition \ref{fmc-context}, and reductions are as in Definition \ref{def:fmc-theory}, except without the force or thunk reductions. 
The abstract machine is as in Definition \ref{def:fmc-machine}, except without the transition for terms $\term{`?`!M}$. 
Types are as in Definition \ref{def:fmc-types}, but with typing rules from Figure \ref{fig:FMC-types-plain}.  We do not consider constants for this calculus.  All notation is similar. 
\end{definition}
Note that there is an unambiguous translation between the two versions of the calculus without constants.

\begin{definition}\label{def:SLC}
The \textit{Sequential $\lambda$-calculus with values (!SLC)} is defined as the \FMCV\ with one location, $\term \lambda$. In particular, it is generated over an \FMCV-signature with one location, $\Sigma_{\{\term \lambda\}}$, which we call an \textit{\SLCV-signature} and denote simply $\Sigma$.  
We will freely confuse stack types and memory types generated over a single location, which are clearly isomorphic. 
Similarly, we can define the `classic' sequential $\lambda$-calculus, SLC, analogously as the FMC with one location. 
\end{definition}

We can define a first-order version of the FMC as follows. 
\begin{definition}
We can define the \emph{Machine Calculus (MC)}  as the \FMCBase\  without higher-order value types: 
\begin{align*}
	\term{M,N} ~&\defeq~ \term{*} ~\mid~ \term{c.M} ~\mid~ \term{[x]a.M} ~\mid~ \term{a<x>.M} 
\end{align*}
In particular, it makes use of a first-order variant of a signature $\Sigma_A$ generated over only \textit{products} of (and not arrows between) base types.
This is a first-order calculus in the sense that the only values are variables or value constants, and thus the argument of any application must be a variable or a constant. The only value types are base types $\type{\alpha}$, omitting higher-order types $\type{?s_A > !t_A}$, meaning a variable must be of base type and every computation (and sequential constant) must be an implication between stacks of base types. The type system for this calculus is then easily derived by restriction from that of the \FMCV. 
Similarly, we can define a first-order version of the SLC.
\end{definition}

This variant turns out be a natural term language for string diagrams, as detailed in Chapters \ref{chapter:CCC-eqns} and \ref{chapter:CCC-equiv}. 

\section{Machine Termination}\label{sec:termination}

We formalize the intuitive meaning of types as describing the initial and final stack of a run of the machine.
This is an adaptation from the SLC to the \FMCV\  of a proof from \cite{Barrett-Heijltjes-McCusker-2022}, which is due to Heijltjes. 
A similar proof also applies for the classic FMC. 
This proof is referred to later, in Chapter \ref{chapter:SN}. It is also a good demonstration of the operational intuition the Functional Machine Calculus can bring to old results of the $\lambda$-calculus. 


We consider here stacks to hold only closed terms, and work with the case of an empty computation signature, and a value signature which holds only base types. This is necessary, because the machine would otherwise halt on states $(S_A, \term{c.M})$ or $(S_A, \term{`?v.M})$. 
\begin{definition}
The set $\run{?s_A>!t_A}$ is the set of \FMCV\  computations $\term{M}$ such that for any memory $S_A\in\run{!s_A}$ there is a memory $T_A \in\run{!t_A}$ and a run of the machine
\[
\machine {S_A}M{T_A}
\]
where $\run{!t_A}$ is the set of memories $T_A$ such that each $T_a \in \run{!t_a}$ and \linebreak $\run{r_1\dots r_n}$ is the set of stacks $\term{V_1}\cdots\term{V_n}$ such that either $\type{r_i}$ is a base type or else $\term{`?V_i} \in\run{r_i}$. \footnote{Note that $\run{?s_A > !t_A}$ can denote the set of values (where $\type{?s_A > !t_A}$ denotes the type of a singleton stack), or the set of computations of type $\type{?s_A > !t_A}$, but which will be unambiguous from context.  }
\end{definition}

{
We will show that $\term{M:?s_A > !t_A}$ implies $\term M\in\run{?s_A > !t_A}$, thus proving termination. The proof is by a direct induction on type derivations, using the properties of machine runs, indicating that a type derivation constitutes a certificate of termination. The proof is analogous to a {Tait's} reducibility proof~\cite{Tait-1967}, with $\run{?s_A > !t_A}$ taking the role of the reducibility set. 
}
\begin{lemma}
\label{lem:run}
If $~\term{!w:!w} \cvdash \term{ M:?s_A > !t_A}$ then for any ${W}\in\run{!w}$, $\term{\{W/{}!w\,\}M}\in\run{?s_A > !t_A}$.
\end{lemma}

\begin{proof}
We proceed by induction on the type derivation of $~\term{!w:!w |- M:?s_A > !t_A}$. In each case, let $W$ be a stack in $\run{!w}$ and let $\term{\sigma} = \term{\{!W/!w\}}$. 
\begin{itemize}
	\item 
For the base case, 
\[
\term{!w:!w} \cvdash \term{M} \equiv \term{ *:?t_A>!t_A}, 
\]
observe that $\term{\sigma * }= \term{*}$ and that $(S_A, \term *)$ trivially terminates in zero steps, for any memory $S_A: \type{!t_A}$.
\item 
For the abstraction case, 
\[
	\term{!w:!w} \cvdash \term{M} \equiv \term{a<x>.M': r?s_A > !t_A},
\]
 the inductive hypothesis gives for any $\term{V}\in\run{r}$ and $S_A \in\run{!s_A}$ a run on $\term{\{V/x\}\sigma M'}$ as below, left, to some $T_A \in\run{!t_A}$. We then construct the run for $\term{\sigma (a<x>.M')}=\term{a<x>.\sigma M'}$ as below, right.
\[
    \vc{\begin{array}{@{(~}l@{~,~}r@{~)}}
      S_A               & \term{\{V/x\}\sigma M'}
    \\\hline\hline\rule[-5pt]{0pt}{15pt}
      T_A               &       \term{*}\;
    \end{array}}
\qquad
    \vc{\begin{array}{@{(~}l@{~,~}r@{~)}}
      \SAouta ; S_a \cdot \term{V} &    \term{a<x>.\sigma M'}
    \\\hline\rule[-5pt]{0pt}{15pt}
      S_A               & \term{\{V/x\}\sigma M'}
    \\\hline\hline\rule[-5pt]{0pt}{15pt}
      T_A               &       \term{*}\;
    \end{array}}
\]

\item For the application case, with $\term{V:r}$, 
\[
	\term{!w:!w}\cvdash \term{M} \equiv \term{[V]a.M': ?s_A > !t_A},
\]
the inductive hypothesis gives that $\term{\sigma V} \in \run{r}$, and for any $S_A \in \run{!s_A}$ a run on $\term{\sigma M'}$ as below, left, to some $T_A \in \run{!t_A}$. 
We then construct the run for $\term{\sigma[V]a.M'}$ = $\term{[\sigma V]a.\sigma M'}$ as below, right. 
\[
    \vc{\begin{array}{@{(~}l@{~,~}r@{~)}}
      \SAouta ; S_a\cdot\term{\sigma V} &      \term{\sigma M'}
    \\\hline\hline\rule[-5pt]{0pt}{15pt}
      T_A               &     \term{*}\;
    \end{array}}
\qquad
    \vc{\begin{array}{@{(~}l@{~,~}r@{~)}}
      S_A               & \term{[\sigma V]a. \sigma M'} 
    \\\hline\rule[-5pt]{0pt}{15pt}
      \SAouta ; S_a\cdot\term{\sigma V} &      \term{\sigma M'}
    \\\hline\hline\rule[-5pt]{0pt}{15pt}
      T_A               &     \term{*}\;
    \end{array}}
\]

\item For the sequential execution case, 
\[
	\term{!w:!w} \cvdash \term{M} \equiv \term{`?V.M': ?r_A?t_A > !u_A},
\]
Consider first that $\term{!w: !w} \vvdash \term{V} \equiv \term{x}$. Take $\term{U} \in \run{?r_A > !s_A}$, so that for $R_A \in \run{r_A}$ there is a run on $\term{`?U}$ as below, left, to some $S_A \in \run{s_A}$. 
The inductive hypothesis then gives for $T_A \in \run{!t_A}$, a run on $\term{\{U/x\}\sigma M'}$ as below, right, to some $U_A \in \run{u_A}$. 
\[
	\vc{\begin{array}{@{(~}l@{~,~}r@{~)}}
      R_A & \term{`?U}
    \\\hline\hline\rule[-5pt]{0pt}{15pt}
      S_A  &           \term*\;
    \end{array}}
\qquad \qquad
	\vc{\begin{array}{@{(~}l@{~,~}r@{~)}}
      T_AS_A & \term{\{U/x\}\sigma M'}
    \\\hline\hline\rule[-5pt]{0pt}{15pt}
      U_A  &           \term*\;
    \end{array}}
\]
We then construct the run for $\term{\{U/x\}\sigma `?x.M'}$ = $\term{`?U; \{U/x\}\sigma M'}$ by composing the two runs as follows, expanding the stack on the run for $\term U$ by T.
\[
	\vc{\begin{array}{@{(~}l@{~,~}r@{~)}}
      T_AR_A & \term{`?U;\{U/x\}\sigma M'}
    \\\hline\hline\rule[-5pt]{0pt}{15pt}
      T_AS_A &   \term{\{U/x\}\sigma M'}
    \\\hline\hline\rule[-5pt]{0pt}{15pt}
      U_A  &           \term*\;
    \end{array}}
\]

Consider second that $\term{!w: !w} \vvdash \term{V} \equiv \term{`!N}$, so that the inductive hypothesis gives for $R_A \in \run{r_A}$ a run on $\term{\sigma N}$ as below, left, to some $S_A \in \run{s_A}$. 
The inductive hypothesis also gives for $T_A \in \run{!t_A}$, a run on $\term{\sigma M'}$ as below, right, to some $U_A \in \run{u_A}$. 
\[
	\vc{\begin{array}{@{(~}l@{~,~}r@{~)}}
      R_A & \term{\sigma N}
    \\\hline\hline\rule[-5pt]{0pt}{15pt}
      S_A  &           \term*\;
    \end{array}}
\qquad \qquad
	\vc{\begin{array}{@{(~}l@{~,~}r@{~)}}
      T_AS_A & \term{\sigma M'}
    \\\hline\hline\rule[-5pt]{0pt}{15pt}
      U_A  &           \term*\;
    \end{array}}
\]
We then construct the run for $\term{\sigma `?V.M'} = \term{`?`! \sigma N. \sigma M'}$ by composing the two runs as follows, expanding the stack on the run for $\term{`?U}$ by $T_A$.
\[
	\vc{\begin{array}{@{(~}l@{~,~}r@{~)}}
      T_AR_A & \term{`?`!\sigma N.\sigma M'}
    \\\hline\rule[-5pt]{0pt}{15pt}
      T_AR_A & \term{\sigma N;\sigma M'}
    \\\hline\hline\rule[-5pt]{0pt}{15pt}
      T_AS_A &   \term{\sigma M'}
    \\\hline\hline\rule[-5pt]{0pt}{15pt}
      U_A  &           \term*\;
    \end{array}}\qedhere
\]
\end{itemize}
\end{proof}

\begin{theorem}[Termination]
\label{thm:run}
For any closed, typed \FMCV-term the machine terminates.
\end{theorem}
\begin{proof} 
Given some closed term $\cvdash\term{ M: ?s_A > !t_A}$, Lemma \ref{lem:run} tells us that $\term M \in \run{?s_A> !t_A}$. The same lemma gives us that any memory $R_A : \type{!r_A}$ is such that $R_A \in \run{!s_A}$ and thus that the machine terminates on $(R_A, \term M)$.
\end{proof}

\chapter{The Functional Machine Category}\label{chapter:CCC-eqns}
In this chapter, we define the Funtional Machine Category,  $\termcat{\Sigma_A}$, consisting of \textit{closed} simply-typed \FMCV-terms generated by a signature $\Sigma_A$ and taken modulo an appropriate equational theory, with composition given by \textit{sequencing}. A main contribution of this thesis is that this is \textit{equivalent} to the familiar category of simply-typed $\lambda$-terms. In other words, the Functional Machine Category forms the \textit{free} Cartesian closed category. Note that, as expected, the denotational perspective collapses all operational behaviour (\textit{i.e.}, the distinction between stacks and the expliciting of sequencing), which are exactly the refinements made of the $\lambda$-calculus by the \FMCV. In this chapter, we develop the necessary equational theory and show that the Functional Machine Category is \textit{a} Cartesian closed category. In Chapter \ref{chapter:CCC-equiv}, the main result is proved. 
Note that the \FMCV\  {remains} an \textit{operational} refinement of the $\lambda$-calculus, even if it is not a \textit{denotational} one. 

Throughout these two chapters we will refer to the Functional Machine Calculus with Values as simply `the Functional Machine Calculus'; however we will continue to write \FMCV\  in abbreviation and in formal statements to distinguish the two calculi.  It is, however, an easy exercise to transfer results between the \FMCV \ and the FMC.

The result stated above may be surprising, given the ability of the Functional Machine Calculus to encode effects, and requires some explanation. Most importantly, this is a semantics of the \textit{calculus itself}, and \textit{not} a semantics of effects. In particular, the result {requires} every location to be treated uniformly. So, for example, while we would encode random input in the \FMCV\  using a read-only stream of probabilistically generated values (with no associated \textit{push} action), for this result we must assume that every location is both read- and write-able. Similarly, we would encode state as a stack of depth (at most) one. However, we must further assume that every location can hold an arbitrary number of values.  Imposing that a location is read-only, write-only, or has some maximum depth breaks the Cartesian closed semantics, as we will demonstrate in a later example.\footnote{Note, however, that the uniform treatment of locations corresponds to the unifrom treatment of effects with each other, and with the higher-order machinery of the $\lambda$-calculus, in the operational semantics of the \FMCV. Hence it is the right assumption for modelling the \textit{calculus}, if not for modelling \textit{effects}.}

Given this consideration, there is precedent for calculi which can encode effects maintaining the semantics of a Cartesian closed category. For example, the encoding of monadic effects in simply-typed $\lambda$-calculus, where, for example, \emph{state} encodes as the monad $S\to (-\times S)$. Here, a stateful process  of type $A \to B$ can be considered a pure function with the larger type $A \to (S \to B \times S)$. 
Something analogous happens with the \FMCV: that is, a term which accesses a second stack will record this in its type.  

Now we have justified why the main result is \textit{plausible}, let us summarize the motivation \textit{for} the result of Cartesian closure.
\begin{itemize}
\item We can define a natural notion of {observational equivalence}  for typed \FMCV-terms based on the stack machine, which we call \textit{machine equivalence}, and prove the category of \FMCV-terms modulo this equivalence forms a Cartesian closed category (although, not the free one). This result is proved at the end of this chapter. 
 Note, we will see that it is the strength of the type system and of the corresponding notion of observation implicit in the definition of machine equivalence which allows us to  achieve a Cartesian closed semantics here.\footnote{In particular, the assumption of uniformity of locations gives us more terms, and thus more contexts with which to distinguish between terms, than we might expect if we were attempting to give a semantic account of effects. }
\item The result serves as a sanity check: neither the introduction of sequencing or locations introduces anything that is semantically unexpected.
\item We preserve one of the most important semantic properties of the $\lambda$-calculus: its categorical semantics.  In combination with the preservation of confluence and strong normalisation (the latter proved in Chapter \ref{chapter:SN}), we argue that the Funtional Machine Calculus is very well-behaved,  despite its novelties.
\end{itemize}

The content of the first section is as follows. 
A coarser equational theory than is generated by the rewrites in Definition \ref{def:fmc-theory} is indeed necessary to show the category of \FMCV-terms is Cartesian closed. 
This equational theory is first motivated using string diagrams, before being defined. 
Following the choice of equations, it is then proved that this equational theory is indeed sufficient to make the Functional Machine Category $\termcat{\Sigma_A}$ Cartesian closed.   

The second section shows the equational theory is validated by a natural notion of observational equivalence, which considers terms equivalent when the machine maps equivalent inputs to equivalent outputs. As a corollary, we achieve that the category of \FMCV-terms modulo machine equivalence is Cartesian closed. 

However, analogous to the case of the simply-typed $\lambda$-calculus, there are not enough typed contexts to separate all terms which ``should" be distinct (at least, without introducing some constants such as natural numbers and addition), so machine equivalence necessarily turns out to be coarser than the equational theory. In brief, by restricting inputs to the machine to be only well-typed terms, there are then too few possible inputs with which to `test' adequately for distinguishability. Thus,  we are justified in seeking a finer equivalence: namely, the aforementioned equational theory, which nevertheless captures more of our notion of observational equivalence than does the reduction relation presented previously. 

\section{String Diagrams}\label{sec:eqn-theory}
Before defining the full equational theory, we will motivate it step-by-step. We begin by considering the first-order calculus, and detail the first-order fragment of the equational theory using string diagrams.
The equations specific to locations are then discussed. Finally, we leave string diagrams behind to discuss the higher-order fragment, along with the full definition.
Note that the equational theory will subsume the reduction relation from Definition \ref{def:fmc-theory}.



\subsection{The First-Order Fragment}
In motivating the first-order fragment, let us simplify matters for now, by considering the \FMCV\  with just a single location $\term \lambda$, that is, the \textit{Sequential $\lambda$-calculus with Values} (\SLCV), returning to the string diagrammatic representation of terms with multiple locations afterwards. 

We begin with the identity equation $\term{<x>.[x]} =_{\textsf{id}}  \term{*} $ mentioned previously: it says that popping a value from the stack and then pushing it back is the same as doing nothing.  \footnote{This could have been presented as a rewrite,  from left to right, and strong normalisation would still hold in its presence, but confluence would not. Analogous to the $\lambda$-calculus, one would expect instead to \textit{expand} typed terms by this equation to retain confluence, but consideration of this is beyond the scope of the current work.}

Next, recall from Figure \ref{fig:string-equipment} the representation of the sequencing of \SLCV-terms as string diagams.
{Left-sequencing and right-}sequencing of computations is given below, for $\term{N:?r>!s}$ and $\term{M:?r?u>!t}$, and $\term{M:?r > !u!s}$ and $\term{N:?s > !t}$, respectively.
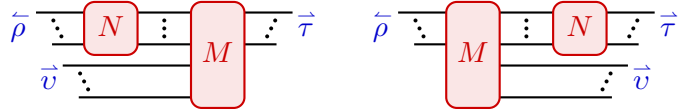
\[
\vc{\begin{tikzpicture}[termpic, scale=2]
	\wires{?r\,}{-50}{20}{50}{!t}
	\wiresleft{!u}{-40}{0}{20}
	\node[termbox] at (-20,20) {$\term N$};
	\node[tallbox] at ( 20,10) {$\term M$};
	\diagdots[0,3]{0,20}
\end{tikzpicture}}
\qquad
\vc{\begin{tikzpicture}[termpic, scale=2]
	\wires{?r\,}{-50}{20}{50}{!t}
	\wiresright{-20}{0}{40}{!u}
	\node[tallbox] at (-20,10) {$\term M$};
	\node[termbox] at ( 20,20) {$\term N$};
	\diagdots[0,3]{0,20}
\end{tikzpicture}}
\]
Further, recall from Figure \ref{fig:string-equipment} the representations of \textit{expansion} of a term $\term{M: ?r > !s}$ by a stack $\type{!t}$, and of a term $\term{<?x>.(M;[!x]): ?t?r > !s!t}$. This term pops the arguments for $\type{?t}$ from the stack as the variables $\term{?x}$, evaluates $\term M$, and then returns the values bound to  $\term{!x}$ to the stack. 
In combination with sequencing, this allows us to understand what will be the \textit{interchange equation} ($\iota$).
\[
	\begin{array}{c@{\qquad}c@{\qquad}c}
	\vc{\begin{tikzpicture}[termpic, scale=2]
		\node (a1) at (-28,36) {$\,\term{<y_1>}\,$};
		\node (b1) at ( 28,36) {$\,\term{[y_1]}\,$};
		\node (an) at (-22,24) {$\,\term{<y_m>}\,$};
		\node (bn) at ( 22,24) {$\,\term{[y_m]}\,$};
		\draw
		  (-60,36)-- (a1) (b1) -- ( 48,36)
		  (-60,24)-- (an) (bn) -- ( 42,24);
		\draw[dotted,line cap=round,color=term] (a1)--(b1) (an)--(bn);
		\wiresleft{?t\,}{-90}{30}{-60}
	    \node[anchor=west] at (50,30) {$\type{!u}$};
		\diagdots[-1.5,3]{-40,30}
		\diagdots[ 1.5,3]{ 40,30}
		\wires{?r}{-80}{10}{40}{!s}
		\node[termbox] at (  0,10) {$\term M$};
		\node[termbox] at (-60,30) {$\term N$};
	\end{tikzpicture}}
	&=_{\iota}&
	\vc{\begin{tikzpicture}[termpic, scale=2]
	    \node[anchor=east] at (-50,30) {$\type{?t}$\,};
		\node (a1) at (-28,36) {$\,\term{<x_1>}\,$};
		\node (b1) at ( 28,36) {$\,\term{[x_1]}\,$};
		\node (an) at (-22,24) {$\,\term{<x_n>}\,$};
		\node (bn) at ( 22,24) {$\,\term{[x_n]}\,$};
		\draw
		  (-48,36)-- (a1) (b1) -- ( 60,36)
		  (-42,24)-- (an) (bn) -- ( 60,24);
		\draw[dotted,line cap=round,color=term] (a1)--(b1) (an)--(bn);
		\wiresright{60}{30}{90}{!u}
		\diagdots[-1.5,3]{-40,30}
		\diagdots[ 1.5,3]{ 40,30}
		\wires{?r}{-40}{10}{80}{!s}
		\node[termbox] at ( 0,10) {$\term M$};
		\node[termbox] at (60,30) {$\term N$};
	\end{tikzpicture}}
	\\ \\
	\term{N;<?y>.(M;[!y])}
	&=_{\iota}&
	\term{<?x>.(M;[!x]);N}
	\end{array}
\]
Note, the action of the \textit{product} on terms, $\term{M} \times \term N$, in the category of closed \FMCV-terms will be given as above (in either form -- it does not matter).
As expected from the diagrams, the {interchange equation}  thus axiomatizes the \textit{bifunctoriality} of the product.  In the setting of the FMC, with its multiple locations, this equation becomes
\[
	\term{N;<?y_A>.(M;[!y_A])} =_\iota \term{<?x_A>.(M;[!x_A]);N}\ .
\]

\begin{remark}
Without the interchange equation, and in particular if we take the equational theory to be generated by \textit{just} {$\{\textsf{id}, \beta, \phi, \pi\}$}, the category of terms would only be \textit{pre-}monoidal. Indeed, this is a  category with the axioms of a monoidal category, except where bifunctoriality of the tensor product fails in general. Pre-monoidal categories are discussed later, in the chapter on related literature (Chapter \ref{chapter:related-lit}).
\end{remark}

Consider now that we can \textit{duplicate} and \textit{delete} elements from the top of the stack with the first two terms illustrated below, respectively.
The diagram corresponding to the term swapping the top two elements of the stack is also given below, right. 
\[
\begin{array}{c@{\qquad\qquad}c@{\qquad\qquad}c}
\vc{\begin{tikzpicture}[termpic, scale=2]
		\node (a1) at (-28,36) {$\,\term{<x_1>}\,$};
		\node (an) at (-22,24) {$\,\term{<x_n>}\,$};
		\node (b1) at ( 34,36) {$\,\term{[x_1]}\,$};
		\node (bn) at ( 28,24) {$\,\term{[x_n]}\,$};
		\node (c1) at ( 22,12) {$\,\term{[x_1]}\,$};
		\node (cn) at ( 16, 0) {$\,\term{[x_n]}\,$};
		\draw 
		  (-48,36)node[left] {$\type{t_1}\,$} -- (a1) 
		  (-42,24)node[left] {$\type{t_n}\,$} -- (an) 
		  (b1) -- ( 54,36)node[right]{$\,\type{t_1}$}
		  (bn) -- ( 48,24)node[right]{$\,\type{t_n}$}
		  (c1) -- ( 42,12)node[right]{$\,\type{t_1}$}
		  (cn) -- ( 36, 0)node[right]{$\,\type{t_n}$};
		\draw[dotted,line cap=round,color=term] (a1)--(b1) (an)--(bn) (a1)--(c1) (an)--(cn);
		\diagdots[-1.5,3]{-40,30}
		\diagdots[ 1.5,3]{ 46,30}
		\diagdots[ 1.5,3]{ 34, 6}
	\end{tikzpicture}}
&
\vc{\begin{tikzpicture}[termpic, scale=2, baseline={(0,1.1)}]
		\node (a1) at (-28,36) {$\,\term{<x_1>}\,$};
		\node (an) at (-22,24) {$\,\term{<x_n>}\,$};
		\draw 
		  (-48,36)node[left] {$\type{t_1}\,$} -- (a1) 
		  (-42,24)node[left] {$\type{t_n}\,$} -- (an);
		\diagdots[-1.5,3]{-40,30}
	\end{tikzpicture}}
\\
\\
\term{<?x>.[!x].[!x]: ?s > !s!s} & \term{<?x>: ?s > } 
\end{array}
\raisebox{34pt}{
\begin{tikzpicture}
	\begin{pgfonlayer}{nodelayer}
		\node [style=none] (0) at (1.5, -0.5) {};
		\node [style=none] (1) at (1.5, -2.5) {};
		\node [style=none] (3) at (4.5, -0.5) {};
		\node [style=none] (4) at (4.5, -2.5) {};
		\node [style=none] (10) at (-0.75, -0.5) {$\type{?s}$};
		\node [style=none] (11) at (-0.5, -2.5) {$\type{?t}$};
		\node [style=none] (15) at (6.5, -2.5) {$\type{!s}$};
		\node [style=none] (16) at (7, -0.5) {$\type{!t}$};
		\node [style=none] (17) at (3, -4.5) {$\term{<?x>.<?y>.[!x].[!y]: ?s?t > !s!t}$};
		\node [style=none] (18) at (0.75, -0.5) {$\term{<?x>}$};
		\node [style=none] (19) at (1, -2.5) {$\term{<?y>}$};
		\node [style=none] (20) at (5, -2.5) {$\term{[!x]}$};
		\node [style=none] (21) at (5.5, -0.5) {$\term{[!y]}$};
		\node [style=none] (22) at (5.875, -0.5) {};
		\node [style=none] (23) at (5.375, -2.5) {};
		\node [style=none] (24) at (6, -2.5) {};
		\node [style=none] (25) at (6.5, -0.5) {};
		\node [style=none] (26) at (0.25, -0.5) {};
		\node [style=none] (27) at (0.5, -2.5) {};
		\node [style=none] (28) at (0, -2.5) {};
		\node [style=none] (29) at (-0.25, -0.5) {};
		\node [style=none] (30) at (1.25, -0.5) {};
		\node [style=none] (31) at (5, -0.5) {};
	\end{pgfonlayer}
	\begin{pgfonlayer}{edgelayer}
		\draw [style=red term dash, in=0, out=-180] (3.center) to (1.center);
		\draw [style=red term dash, in=-180, out=0] (0.center) to (4.center);
		\draw [style=thick line] (22.center) to (25.center);
		\draw [style=thick line] (24.center) to (23.center);
		\draw (29.center) to (26.center);
		\draw [style=thick line] (28.center) to (27.center);
		\draw [style=red term dash] (30.center) to (0.center);
		\draw [style=red term dash] (3.center) to (31.center);
	\end{pgfonlayer}
\end{tikzpicture}}
\]
The following \textit{diagonal} $(\Delta)$ and \textit{terminal} $(!)$ equations axiomatize naturality of these operations in the FMC; that is, evaluating a computation and duplicating its output is the same as evaluating the term twice (in each case, given one of its duplicated inputs), and evaluating a term and discarding its output is the same as discarding its input and never evaluating the term. 
\begin{align*}
	\term{M;<?x_A>.[!x_A].[!x_A]} &=_\Delta \term{<?w_A>.[!w_A].M.[!w_A].M}\\
	\term{M;<?x_A>} &=_! \term{<?w_A>} 
\end{align*}
Note that naturality of symmetry is implicit in our $\Delta$ equation, and we detail this in a later remark 


Such equations seem well-motivated for terms without side-effects, as in the \SLCV, but may be surpising in the setting of the \FMCV, with its ability to encode effects.  Recall that we are not trying to give a semantics of effects here, but rather trying to capture a notion of observational equivalence for the calculus, which we have delayed introduction of. 

\subsection{The Locational Fragment}\label{subsec:nominal}
The locational fragment of the equational theory consists of the \textit{relocation} and the \textit{permutation} equations. 
The first axiomatizes the interaction of relocation isomorphisms with each other. Note, this is indeed subsumed by the beta rule, but we include it here for reasons that will become clear shortly. 
\[
	\term{a<x>.[x]b.b<y>.[y]c} =_\rho \term{a<x>.[x]c}
\]
This equation implies that each relocation term $\term{a<x>.b[x]}$ is indeed an isomorphism, with inverse $\term{b<x>.[x]a}$:
\[
	\term{a<x>.[x]b.b<y>.[y]a} =_\rho \term{a<x>.[x]a} =_\ide \term{*}\, .
\]
The role of the permutation equation, in contrast, is to \textit{collapse symmetries} on types at different locations into  \textit{identities}.
The following \textit{permutation lemma} shows that the equations generated from the identity and $\{\beta, \pi\}$ reductions suffices for permutation of applications and abstractions on different locations. 
We will refer to the equations $\pi'$ given the in the following lemma as the \textit{permutation equivalence}.
\begin{lemma}[Permutation]\label{lem:permutation-equivalence}
The following two equations are derivable: 
\[
	\term{[V]a.[W]b} =_{\pi'} \term{[W]b.[V]a} \qquad \term{a<x>.b<y>.M} =_{\pi'} \term{b<y>.a<x>.M} \ . 
\]
\end{lemma}
\begin{proof}
\begin{align*}
	\term{[V]a.[W]b} &=_{\textsf{id}} \term{[V]a.}\underline{\term{[W]b.a<x>}}\term{.[x]a} \\
		&=_\pi \underline{\term{[V]a.a<x>}}\term{.[W]b.[x]a} \\
		&=_{\beta} \term{[W]b.[V]a} \\
	\term{a<x>.b<y>.M} &=_{\textsf{id}} \term{b<z>.}\underline{\term{[z]b.a<x>}}\term{.b<y>.M} \\
		&=_\pi \term{b<z>.a<x>.}\underline{\term{[z]b.b<y>}}\term{.M} \\
		&=_\beta \term{b<z>.a<x>.\{z/y\}M} \\
		&=_\alpha \term{b<y>.a<x>.M} \qedhere
\end{align*}
\end{proof}
\begin{remark}\label{remark:permutation-ids}
The identity and symmetry at type $\type{`a(!s)`b(!t)}$ coincide.
First note, this is possible because, at the level of types, we have
\[
	\type{`a(!s)`b(!t) > `a(!s)`b(!t)} ~=~ \type{`a(!s)`b(!t) > `b(!t)`a(!s)}. 
\]
This is a result of the permutation equivalence, as follows. 
\[
	\term{a<?x>.b<?y>.[!x]a.[!y]b} ~=_{\pi}~ \term{a<?x>.[!x]a.b<?y>.[!y]b} ~=_\ide~ \term{*}\, 
\]
\end{remark}

We can represent terms of the \FMCV \ (as opposed to the \SLCV) using (a variant of) \textit{nominal} string diagrams \cite{DBLP:conf/calco/Balco019}. 
Whereas typical string diagrams can be used to represent actions on type \textit{vectors}, we can use nominal diagrams to represent actions on type \textit{families},
where the order between values in different families is no longer a consideration (due to the above remark). 
There are technical concerns to be considered when dealing with these diagrams (especially, composition of diagrams), but we elide a detailed discussion and instead refer the reader to the aforementioned citation. Given this, we make use of these diagrams only informally.

Such diagrams associate with each wire a label (its location) and include a \textit{renaming} generator (corresponding to the relocation isomorphism), which allows the changing of labels.
The interaction of two such generators corresponds to our \textit{relocation} $\rho$ equation, pictured below, left. 
We use colour to distinguish the wires with different labels (\textit{i.e.}, at different locations). 
\vspace{-\baselineskip}
\[
\input{nominal generator.tikz}
\]
\\
Assuming that we have at most one value held at each location, we can envisage these diagrams as being three-dimensional instead of planar, so that symmetries acting on wires become true identities, as above, right. In the general case (and ignoring {re}location) since wires on a single location are still ordered, we can imagine $n$ planes, one for each location, where symmetry of two wires on a given plane remains a true symmetry. 

We now reprise the example from the introduction, which uses such diagrams, aiming to illustrate why we would expect the \FMCV\  to form a Cartesian (closed) category, and how \textit{uniform} treatment of locations is crucial to achieving a CCC semantics.  This assumption of uniformity corresponds to our attempt to model the calculus itself, and \textit{not} computational effects. 
	\begin{example}
Consider the following diagram representing the term given below, recalling the definition of the combinators from Figure \ref{ex:effects-example}. Dashed lines here simply denote that a wire is on a non-main location. 
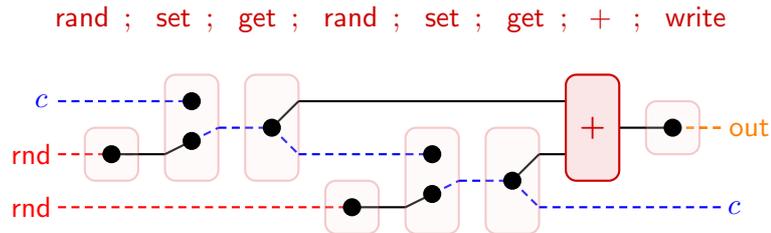
\[
\begin{array}{c}
	\term{~\rand~;\ ~\set~;~~\get~;~~\rand~;~~\set~;~~\get~;~+~;~~\print~ } \\ \\
	\vc{\begin{tikzpicture}[termpic, scale=2]
		\node (a1) at (-10,40) {$\ \ \textcolor{blue}{c}$};
		\node (a2) at (-10,20) {$\textcolor{red}{\rnd}$};
		\node (a3) at (-10,0) {$\textcolor{red}{\rnd}$};
		\draw[blue][densely dashed]  (0,40) -- (50,40) ;
		\draw[red] [densely dashed](0,20) -- (30,20) ;
		\draw (20,20) -- (40,20) ;
		\draw (40,20) -- (50,25) ;
		\draw[red][densely dashed] (0, 0) -- (110, 0) ;
		\draw[blue][densely dashed] (50,25) -- (60,30) ;
		\draw[blue][densely dashed] (60,30) -- (80,30) ;
		\draw (90,40) -- (210,40) ;
		\draw (80,30) -- (90,40) ;
		\draw[blue][densely dashed] (80,30) -- (90,20) ;
		\draw[blue][densely dashed] (90,20) -- (140,20) ;
		\draw (110,0) -- (130,0) ;
		\draw (130,0)-- (140,5)  ;
		\draw[blue][densely dashed] (140, 5) -- (150,10) ;
		\draw[blue][densely dashed] (150,10) -- (170,10) ;
		\draw (180,20) -- (210,20) ;
		\draw (170,10) -- (180,20) ;
		\draw[blue][densely dashed] (170,10) -- (180,0) ;
		\draw[blue][densely dashed] (180,0) -- (250,0) ;
		\draw (210,30) -- (230,30) ;
		\draw[orange][densely dashed, line width = 1pt] (230,30) -- (250,30) ;
		\node[opacity=0.2,draw=term,fill=term!10,rounded corners,minimum size=20pt] at (20,20) {};
		\filldraw[black] (20,20) circle (3pt);
		\node[opacity=0.2, tallbox] at (50,30) {};
		\filldraw[black] (50,40) circle (3pt);
		\filldraw[black] (50,25) circle (3pt);
		\node[opacity=0.2,tallbox] at (80,30) {};
		\filldraw[black] (80,30) circle (3pt);
		\node[opacity=0.2,termbox] at (110, 0) {};
		\filldraw[black] (110,0) circle (3pt);
		\node[opacity=0.2,tallbox] at (140,10) {};
		\filldraw[black] (140,20) circle (3pt);
		\filldraw[black] (140,5) circle (3pt);
		\node[opacity=0.2,tallbox] at (170,10) {};
		\filldraw[black] (170,10) circle (3pt);
		\node[tallbox] at (200,30) {$\textbf{\term+}$};
		\node[opacity=0.2,termbox] at (230,30) {};
		\filldraw[black] (230,30) circle (3pt);
		\node (a3) at (253,0) {$\textcolor{blue}{c}$};
		\node (a3) at (255,30) {$\textcolor{orange}{\ \  \out}$};
	\end{tikzpicture}}
	\end{array}
\]
\\
%
%
Due to the type system, we see all the dependencies between operations. 
For example, the second call to $\term \rand$ may safely be made before the first calls to $\term \set$ and $\term \get$. This would be illustrated by `sliding' the $\term{\rand}$ operation along the wire. We can also see that the second $\term \set$ is dependent on the first $\term \get$: the value pushed to $\term c$ by the first $\term \get$ is discarded by the second $\term \set$. Indeed, the sequential composition of the terms $\term \set$ and $\term \get$ creates such a beta-redex, corresponding to the expected interaction of $!$ with $\Delta$ in a Cartesian category.
Indeed, modulo renaming of wires, these effectful operations are encoded by the diagonal and terminal operations of a Cartesian category.

Note, one can see in the above example that applying the naturality of the diagonal would result in a duplication on the location $\term \rnd$. This relies on $\term \rnd$ being a location with no special status, and in particular, having an associated \textit{push} action, similar to the main location $\term \lambda$. If we were to enforce that $\term \rnd$ was a read-only stream, then this duplication would no longer be possible and the semantics can no longer be Cartesian.  Similar issues arise for memory cells, which ought to have depth (at most) one. We leave consideration of the particular properties of encoded effects for future work.  
\end{example}

\section{The Equational Theory}
We have seen in the previous section the two terms which allow for duplication and deletion.
For the higher-order calculus, we also expect the force and thunk equations, corresponding to the force and thunk reduction rules
\[
	\term{`?`!N.M} =_{\phi} \term{N;M} \qquad \term{`!`?V} =_{\tau} \term{V}
\]
We delay further discussion of these equations until the next chapter.  We will discuss the remaining two: \textit{beta} and \textit{eta}. 

There are two terms which are central to making the calculus higher-order, shown below. 
\[
	\term{<f>.`?f: (?s_A > !t_A)?s_A > !t_A} \qquad \term{<?x_A>.[`![!x_A].M]: ?s_A > (?r_A > !t_A)} \qquad 
\]
The first term allows the execution of a thunk held on the stack. The second term gives us the \textit{currying} of $\term{M: ?s_A?r_A > !t_A}$, by its partial application and thunking. It will be evident that the beta and eta equations detailed below relate to these terms. 

Remarkably, the beta law can be reduced to a more \textit{local} form: that is, we do not require full generality of $\term{[V].<x>.M} =_\beta \term{\{V/x\}M}$, but rather we can make do with the more restricted form below.
\[
	\term{[V].<x>.`?x} =_\beta \term{`?V}
\]
After presenting the full equational theory, it will be proved that the general, global beta law is in fact derivable. The idea behind the proof of this is to propogate an ``explicit substitution'' $\term{[V].<x>.M}$ along $\term{M}$ by applying the naturalities (and interchange) discussed in the first-order fragment. Duplication and deletion are handled solely by $\Delta$ and $!$, with the interchange rule also playing an essential role in the propogation.
When the redex meets a variable, the \textit{local} beta rule above can be applied, ``unboxing'' the value. However, there must also be an analogous rule for pushing the redex \textit{into} boxes (applications). This is the eta rule, given below.
\[
	\term{S;<?x>.[`![!x_A].M]} =_\eta \term{[`!S;M]}
\]
In other words, the global beta reduction rule is internalized in the calculus by the given set of naturalities.  

	We now introduce the equational theory we will work with. 
Recall that we fix a set of location $A = \{\term{a_1}, \ldots, \term{a_n}\}$ and we take $\term a, \term b$ to range over all locations. 

Note, some of these rules have been simplified from the form they were presented in above, but this is simply by an application of the identity law which is explained subsequently. Note, further, that   the equational theory implictly assumes each location is treated uniformly, as mentioned in the introduction of this chapter. 
	    
\begin{definition}\label{def:eqntheory}
We define the \textit{equational theory} $\ccceq$ of the \FMCV\  to be the least equivalence generated by the following laws, closed under all contexts: 
\begin{align*}
&\textup{Identity:} &\term{<?x_A>.[!x_A]} &=_\textsf{id} \term{*} & \type{?s_A} &\type{> !s_A} \\
&\textup{Local Beta:} & \term{[V]a.a<f>.`?f} &=_\beta \term{`?V} & \type{?s_A }&\type{> !t_A}\\ 
&\textup{Force:} & \term{`?`!M} &=_\phi \term{M} & \type{?s_A }&\type{> !t_A}\\ 
&\textup{Thunk:} & \term{`!`?V} &=_\tau \term{V} & \type{?s_A}&\type{> !t_A}\\ 
&\textup{Eta:} &\term{S;<?x_A>.[`![x_A].N]a} &=_\eta \term{[`!S;N]a} & &\type{> `a(?t_A > !u_A)}\\
&\textup{Diagonal:} &\term{S;<?x_A>.[!x_A].[!x_A]} &=_\Delta \term{S;S} &&\type{ > !s_A !s_A} \\ 
&\textup{Terminal:} &\term{S;<?x_A>} &=_! \term{*} & \phantom{\type{s}} &\type{>} \phantom{\type{s}} \\
&\textup{Interchange:} & \term{S;<?x_A>.(P;[!x_A])} &=_\iota \term{P;S}   & \type{?p_A} &\type{> !r_A!s_A} \\
&\textup{Relocation:} &\term{[V]a.a<y>.[y]b} &=_\rho \term{[V]b} && \type{> `b(t)} \\
&\textup{Permutation:}	&\term{[V]b.a<x>.M} &=_\pi \term{a<x>.[V]b.M} & \type{`a(r)?s_A }& \type{> !t_A} 
\end{align*}where, for eta, $\fv{\term{!x_A}} \cap \fv{\term{N}} = \emptyset$, for interchange, $\fv{\term{!x_A}} \cap \fv{\term{P}} = \emptyset$ and for permutation, $\term{x} \not\in \fv{\term{V}}$ and $\term{a} \neq \term{b}$, for relocation $\term{a} \neq \term{b}$, and  with $\term{x: !s_A, y: t, f: ?s_A > !t_A}$, $\term{M: ?s_A > !t_A}$, $\term{N: ?s_A?t_A > !u_A}$,  $\term{S: > !s_A}$, $\term{P: ?p_A > !r_A}$ throughout. We will often write simply $=$ for $=_{\textsf{eqn}}$,  and \textit{e.g.} $=_\beta$ for the least equivalence generated by $=_\beta$ closed under all contexts, and similarly for the other equations. 
\end{definition}

\begin{remark}\label{ren:expand-eqns}
Note, we could restrict identity to apply to a singleton type and recover the original by iteration.
By using the identity law, we could equivalently state the interchange equation in a more \textit{symmetric} manner, as follows: 
\[
	\term{Q;<?x_A>.(P;[!x_A])} =_\iota \term{<?w_A>.(P;[!w_A].Q)},
\]
 with $\term{P: ?p_A > !r_A}$, $\term{Q: ?s_A > !t_A}$, $\term{!w_A: s_A}$ and $\term{!x_A: t_A}$.\footnote{This is derived as $\term{Q;<?x_A>.(P;[!x_A])} =_\ide \term{<?y_A>.}\uterm{[!y_A].Q;<?x_A>.(P;[!x_A])} =_\iota \term{<?y_A>.(P;[!y_A].Q)}$.}
Similarly, we can apply the identity to terminality to recover 
\begin{align*}
	\term{M;<?x_A>.[`![!x_A].N] } &=_\eta \term{<?w_A>.[`![!w_A].M;N]}\\
	\term{M;<?x_A>.[!x_A].[!x_A]} &=_\Delta \term{<?w_A>.[!w_A].M.[!w_A].M}\\
	\term{M;<?x_A>} &=_! \term{<?w_A>} 
\end{align*}
for $\term{M: ?r_A > !s_A}$. We will sometimes refer to the above equations as just $\eta, !, \Delta$ or $\iota$.  
\end{remark}
It is now proved that the \textit{global} beta law is derivable from the given equational theory. 
We then proceed to show that the equational theory above is sufficient to make the \FMCV\  category Cartesian closed. 

\begin{proposition}[Soundness of beta]
The global beta equation (below) is derivable in $\ccceq$.
\[
\term{[V]a.a<x>.M} =_\beta \term{\{V/x\}M}
\] 
\end{proposition}
\begin{proof}
We prove a slightly stronger statement of the proposition, in conjunction with two appropriate statements on values. For all memories $S_A$, and all locations $\term{a}$, we claim the following are derivable:
\begin{align*}
	\term{[S_A].<?x_A>.M} &= \term{\{S_A/!x_A\}M} \\
	\term{[S_A].<?x_A>.[V]a} &=_1 \term{[\{S_A/!x_A\}V]a}\\
	\term{[S_A].<?x_A>.`?V} &=_2 \term{`?\{S_A/!x_A\}V}\, .
\end{align*}
Proceed by (mutual) induction on the type derivation of $\term{M}$ and $\term{V}$.
Let $\term{M'} = \term{\{!x'_A/!x_A\}M}$ so that in each case of the induction, we have $\term{M} = \term{[!x_A].<?x'_{\!A}>.M'}$. Similarly, let $\term{V'\,'} = \term{\{x'\,'/x\}V}$ so that in each case of the induction, we have $\term{[V]} = \term{[!x'\,'_{\!A}].<?x_A>.[V'\,']}$ and $\term{`?V} = \term{[!x'\,'_{\!A}].<?x_A>.`?V'\,'}$.
\begin{itemize}
\item Identity: 
\begin{align*}
	\term{[S_A].<?x_A>.*} &=_! \term{*} \\
	(\textsf{defn. subst.})	&= \term{\{S_A/!x_A\}*}
\end{align*}
\item Application:
\begin{align*}
\term{[S_A].<?x_A>.[V]a.M} ~(\textsf{I.H.})&=  \term{[S_A].<?x_A>.[!x_A].}\uterm{<?x'\,'_{\!\!\!A}>.[V'\,']a.[!x_A]}\term{.<?x'_{\!A}>.M'} \\
		&=_\iota  \term{[S_A].<?x_A>.[!x_A].}{\term{[!x_A].<?y_A>.<?x'\,'_{\!A}>.[V'\,']a.[!y_A]}}\term{.<?x'_{\!A}>.M'} \\
		&=_\Delta  \term{[S_A].}\uterm{[S_A].<?y_A>.<?x'\,'_{\!\!\!A}>.[V'\,']a.[?y_A]}\term{.<?x'_{\!A}>.M'} \\
		&=_\iota  \term{[S_A].<?x'\,'_{\!\!\!A}>.[V'\,']a.[S_A].<?x'_{\!A}>.M'} \\
		(\textsf{I.H.})&=   \term{[\{S_A/!x_A\}V].\{S_A/!x_A\}M} \\
	(\textsf{defn. subst.})	& = \term{\{S_A/!x_A\}[V].M} 
\end{align*}
\item Abstraction:
\begin{align*}
	\term{[S_A].<?x_A>.a<y>.M} &=_\textsf{id} \term{a<y'>.[y']a.[S_A].<?x_A>.a<y>.M}  \\
	(\textsf{I.H.})	&= \term{a<y'>.\{a(y')S_A/a(y)!x_A\}M} \\
				&=_\alpha \term{a<y>.\{S_A/!x_A\}M} \\
	(\textsf{defn. subst.})	&= \term{\{S_A/!x_A\}a<y>.M} 
\end{align*}

\item Sequential Execution:
\begin{align*}
	\term{[S_A].<?x_A>.`?V.M} 
	(\textsf{I.H.})	&= \term{[S_A].<?x_A>.}\uterm{[!x_{\!A}].<?x'\,'_{\!A}>.`?V'\,'.[!x_A]}\term{.<x'_{\!A}>.M'} \\
		&=_\iota \term{[S_A].<?x_A>.[!x_A].}{\term{[!x_A].}\term{<?y_A>.<?x'\,'_{\!A}>.?V'\,'.[!y_A]}}\term{.<?x'_{\!A}>.M'} \\
		&=_\Delta \term{[S_A].}\uterm{[S_A].<?y_A>.<?x'\,'_{\!A}>.`?V.[!y_A]}\term{.<?x'_{\!A}>.M'} \\
		&=_\iota \term{[S_A].<?x'\,'_{\!A}>.`?V'\,'}\term{.[S_A].<?x'_{\!A}>.M'} \\
	(\textsf{I.H.})	&= \term{`?\{!S_A/!x\}V.\{S_A/!x_A\}M}\\
	(\textsf{defn.\,subst.})	&= \term{\{S_A/!x\}`?V.M}
\end{align*}
\item Variable ($\term{y} \in \term{!x_A}$): let $\term{V}$ be the element of $\term{S_A}$ corresponding to $\term{y}$, and 
let $\term{[S_A]} = \term{[S'_A].[V]b.[S'\,'_A]}$ and $\term{<?x_A>.M} = \term{<?x'\,'_{\!A}>.b<y>.<?x'_A>.M}$.
\begin{itemize}
\item[$=_1:$]  
\begin{align*}
	\term{[S_A].<?x_A>.[y]a} &=_\rho \term{[S_A].<?x_A>.[y]b.b<z>.[z]a}\\
			&=_! \term{[S'_{\!A}].[V]b.b<y>.<?x'_{\!A}>.[y]b.b<z>.[z]a}\\
			&=_{\iota}  \term{[S'{\!_A}].<?x'_{\!A}>.[V]b.b<z>.[z]a}\\
			&=_! \term{[V]b.b<y>.[y]a}\\
			&=_\rho \term{[V]a}\\
		(\textsf{defn. subst.}) &= \term{[\{S_A/!x_A\}y]a}	
\end{align*}
\item[$=_2:$]  
\begin{align*}
	\term{[S_A].<?x_A>.`?y} &=_\beta \term{[S_A].<?x_A>.[y]b.b<z>.`?z} \\
		&=_\beta \term{[S'_{A}].[V]b.b<y>.<?x'_{\!A}>.[y]b.b<z>.`?z} \\
		&=_\iota \term{[S'_{\!A}].<?x'_{A}>.[V]b.b<z>.`?z} \\
		&=_! \term{[V]b.b<z>.`?z} \\
		&=_\beta \term{`?V}\\
	(\textsf{defn. subst.}) &= \term{`?\{S_A/!x_A\}y}
\end{align*}
\end{itemize}
\item Variable ($\term y \not\in \term{!x_A}$):
\begin{itemize}
\item[$=_1:$]  
\begin{align*}
	{\term{[S_A].<?x_A>}}\term{.[y]b} &=_! \term{[y]b}\\
	(\textsf{defn. subst.}) &= \term{[\{S_A/!x_A\}y]b}
\end{align*}
\item[$=_2:$]
\begin{align*}
	{\term{[S_A].<?x_A>}}\term{.`?y} &=_! \term{`?y} \\
	(\textsf{defn. subst.}) &= \term{`?\{S_A/!x_A\}y}
\end{align*}
\end{itemize}
\item Thunk:
\begin{itemize}
\item[$=_1:$]
\begin{align*}
	\term{[S_A].<?x_A>.[`!M]b} ~(\textsf{I.H.}) &= \term{[S_A].<?x_A>.[`![!x_A].<?x'_{\!A}>.M']b} \\
		&=_\eta \term{[`![S_A].<?x'_{\!A}>.M']b}\\
	(\textsf{I.H.})	&= \term{[`!\{S_A/!x_A\}M]b}\\
	(\textsf{defn. subst.}) &= \term{[\{S_A/!x_A\}`!M]b}	
\end{align*}
\item [$=_2:$]
\begin{align*}
	\term{[S_A].<?x_A>.`?`!M} &=_\phi \term{[S_A].<?x_A>.M}\\
	(\textsf{I.H.})	&= \term{\{S_A/!x_A\}M}\\
				&=_\phi \term{`?`!\{S_A/!x_A\}M}	\\
	(\textsf{defn. subst.}) &=_\phi \term{`?\{S_A/!x_A\}`!M}	\qedhere
\end{align*}
\end{itemize}
\end{itemize}
\end{proof}
We can now make use of this derived rule freely in our later calculations, and will henceforth use $=_\beta$ to refer to non-local beta equation, as usual. 
Technically, the admissibility of the global beta equation will allow us to bypass the usual work of showing that our translations between calculi (the STLC, \SLCV\  and \FMCV) respect substitution.

\begin{remark}\label{rem:sym-in-linear} 
Naturality of symmetry is implicit in the $\Delta$ equation. To see this, observe how $\Delta$ swaps the order of the central occurences of $\term{T}$ and $\term{S}$ in the right-hand side of the following equation. 
\begin{align*}
\term{T;S;<?x_A>.<?y_A>.[!y_A].[!x_A].[!y_A].[!x_A]} =_\Delta \term{T;S;T;S}
\end{align*}
More precisely, we can calculate the following.
\begin{align*}
&\term{T;S;<?x_A?y_A>.[!x_A!y_A]}\\
&=_\beta \term{T;S;<?x_A?y_A>.[!y_A!x_A!y_A!x_A].<?x'_A?y'_A?x'\,'_A?y'\,'_A>.[!x'\,'_A!y'_A]} \\
	&=_{\Delta} \term{T;S;T;S;<?x'_A?y'_A?x'\,'_A?y'\,'_A>.[!x'\,'_A!y'_A]}\\
	&=_{!}  \term{T;S;T;<?y'_A?x'\,'_A?y'\,'_A>.[!x'\,'_A!y'_A]}  \\
	&=_{\iota}  \term{T;<?y'\,'_A>.S;T}  \\
	& =_{!}\term{S;T} 
\end{align*}
For this reason, in the case of the \FMCV\  with a \textit{linear} variable policy, where we would drop the $\Delta$ and $!$ equations, we would have to add an equation axiomatizing the naturality of $\textit{symmetry}$: 
\[
\term{S.<?x_A>.<?w_A>.[!x_A].[!w_A]} =_{\sigma} \term{<?w_A>.S.[!w_A]: ?r_A > !s_A!r_A}\, , 
\]
where $\term{w_A: !r_A}, \term{x_A: !s_A}$. This would be natural for achieving a correspondence between symmetric monoidal (closed) categories and the \FMCV. 
\end{remark}

\section{The Functional Machine Category is Cartesian Closed}
Similar to our definition of the category of $\lambda$-calculus-terms, we will overspecify the equipment of the CCC. However, it is easy to verify that all the equipment given is consistent in the expected way. 

\begin{definition}\label{def:ccc}
Given an \FMCV\  signature $\Sigma_A = (\Sigma_0, \Sigma_c, \Sigma_v)$, the \textit{Functional Machine Category} of \textit{closed} \FMCV-terms $\termcat{\Sigma_A}$ is defined with
\begin{itemize}
\item Objects: families of type vectors $\cat{!\tau_A}= \type{!t_A}$ over $\Sigma_0$, 
\item Morphisms: $Hom(\cat{!\sigma_A}, \cat{!\tau_A})$ is given by the set of closed typed \FMCV\ computations $\term{M}: \type{?s_A > !t_A}$ over $\Sigma_A$, {modulo} $\ccceq$,
\item Composition: given morphisms ${M} \in Hom(\cat{!\rho_A}, \cat{!\sigma_A})$ and ${N} \in Hom(\cat{!\sigma_A}, \cat{!\tau_A})$, ${M} ; {N} \in Hom(\cat{!\rho_A}, \cat{!\tau_A})$ is given by sequencing $\term{M;N: ?r_A > !t_A}$,
\item Identity: given on every type $\cat{!\tau_A}$ by the term $\term{\star: ?t_A > !t_A}$. 
\item Products: given on objects by by concatenation: $\cat{!s_A} \times \cat{!t_A} = \type{!s_A!t_A}$, with the unit object given by $\type{\epsilon_A}$. The action on morphisms and the associated canonical natural transformations,  \emph{terminal}~$!$, the \emph{diagonal}~$\Delta$, \textit{pairing} $\langle P,Q \rangle$, the \emph{projections}~$\pi_1$ and $\pi_2$ and \emph{symmetry} $\textsf{sym}$, are respectively: 
\[
\begin{array}{@{}r@{~}llr@{\,}l@{}}
		\cat{M*N} :& \cat{!s_A*!r_A --> !t_A*!u_A} &=& \term{<?w_A>.P.[!w_A].Q} &: \type{?r_A?s_A > !t_A!u_A},
\\[5pt]     	        ! :& \cat{!t_A --> 1}        &=& \term{<?x_A>}           &: \type{?s_A\,>}
\\[5pt]	   \Delta :& \cat{!s_A --> !s_A * !s_A}  &=& \term{<?x_A>.[!x_A].[!x_A]} &: \type{?s_A>!s_A!s_A}
\\[5pt]	\langle P,Q\rangle : & \cat{!s_A --> !t_A * !u_A} &=& \term{<?x_A>.([!x_A].M;[!x_A].N)}  &: \type{ ?s_A > !t_A!u_A}
\\[5pt] 		\pi_1 : & \cat{!t_A * !s_A --> !s_A} &=& \term{<?x_A>} &: \type{?s_A ?t_A > !t_A}
\\[5pt] 		\pi_2 : & \cat{!t_A * !s_A --> !t_A} &=& \term{<?x_A>.<?y_A>.[!x_A]} &: \type{?s_A ?t_A > !s_A}
\\[5pt]	\textsf{sym}: & \cat{!t_A * !s_A --> !s_A * !t_A} &=& \term{<?x_A>.<?y_A>.[!x_A].[!y_A]} &: \type{?s_A?t_A > !s_A!t_A}
\end{array}
\]
where $\term{!w_A} :\type{!r_A},\term{!x_A: !s_A, !y_A: !t_A}$,  $\term{M:?s_A > !t_A}$, $\term{N:?s_A > !u_A}$, $\term{P: ?s_A > !t_A}$ and $\term{Q:?r_A > !u_A}$,
\item Exponential: given on objects by $\cat{!s_A} \xrightarrow{} \cat{!t_A} = \type{?s_A>!t_A}$. The action on morphisms and  associated canonical transformations, \emph{evaluation}~$\epsilon$, \emph{expansion}~$\eta$ and \emph{currying} $\textsf{curry}$, are respectively:
\[
\begin{array}{@{}r@{~}llr@{\,}l@{}}
		\epsilon   &: \cat{!s_A * }\cat{(!s_A->!t_A) --> !t_A}   &=& \term{<f>.`?f}       &: \type{(?s_A>!t_A)?s_A>!t_A} 
\\[5pt] 	\eta       &: \cat{!s_A --> }\cat{(!t_A -> !t_A*!s_A)}   &=& \term{<?x_A>.[`![!x_A]]} &: \type{?s_A>(?t_A>!t_A!s_A)} 
\\[5pt]	\textsf{curry}(\cat{M}) &:	\cat{!s_A --> }\cat{(!r_A --> !t_A)}		 &=& \term{<?x_A>.[`![!x_A].M]}  &: \type{?s_A > (?r_A > !t_A)}
\\[5pt] 	\cat{P} \xrightarrow{} \cat{Q} &: \cat{(!s_A->!t_A) -->} \cat{(!r_A->!u_A)}  &=& \term{<f>.[`!P;`?f.Q]} &: \type{(?s_A>!t_A)>(?r_A>!u_A)} 
\end{array}
\]
where $\term{f: ?s_A > !t_A, !x_A: !s_A}$ and $\term{M:?s_A?r_A>!t_A}$, $\term{P:?r_A>?s_A}$ and $\term{Q: ?t_A > !u_A}$,
\item the associators and unitors for each object are given by $\term{\star}$.
\end{itemize}
\end{definition}
We have that sequencing is associative and that $\term{*}$ is its identity, so we can indeed expect this to form a category. 

\begin{remark}
We have defined exponentials relative to the \textit{main} location $\term \lambda$ -- but there is nothing \textit{special} about this location.
Indeed, we can equivalently define an exponential $\type{!s_A} \xrightarrow{a} \type{!t_A}$ on \textit{any} location $\term{a}$ as $\type{`a(?s_A > !t_A)}$, with the following equipment
\[
\begin{array}{@{}r@{~}llr@{\,}l@{}}
		\epsilon^a   &: \cat{!s_A * }a\cat{(!s_A->!t_A) --> !t_A}   &=& \term{a<f>.`?f}       &: \type{`a(?s_A>!t_A)?s_A>!t_A} 
\\[5pt] 	\eta^a       &: \cat{!s_A --> }a\cat{(!t_A -> !t_A*!s_A)}   &=& \term{<?x_A>.[`![!x_A]]a} &: \type{?s_A>`a(?t_A>!t_A!s_A)} 
\\[5pt]	\textsf{curry}^a(\cat{M}) &:	\cat{!s_A --> }a\cat{(!r_A --> !t_A)}		 &=& \term{<?x_A>.[`![!x_A].M]a}  &: \type{?s_A > `a(?r_A > !t_A)}
\\[5pt] 	\cat{P} \xrightarrow{a} \cat{Q} &: a\cat{(!s_A->!t_A) -->} a\cat{(!r_A->!u_A)}  &=& \term{<f>.[`!P;`?f.Q]a} &: \type{`a(?s_A>!t_A)>`a(?r_A>!u_A)} 
\end{array}
\]
Any two of these bifunctors $\xrightarrow{a}$ and $\xrightarrow{b}$ are naturally isomorphic to each other, with the natural isomorphism given at type  type $\type{t}$ by 
the \textit{relocation} isomorphism.
\[
	\term{a<x>.[x]b: `a(t) > `b(t)}\, .
\]
This will be explained in detail by Chapter \ref{chapter:CCC-equiv}.  
\end{remark}

\begin{remark}
{Because symmetry isomorphisms on types at distinct locations are in fact a true identity, two different evaluation maps can apply to the same object: for example, given a stack of type $\type{?s_A`a(?s_A > !t_A)`b(?s_A > !u_A)}$, where $\type{!s_A}$ contains no elements at $\term{a}$ or $\term{b}$, we can apply either $\term{`a<f>.`?f}$ to get $\type{`b(?s_A > !u_A)!t_A}$ or $\term{`b<g>.`?g}$ to get $\type{`a(?s_A > !t_A)!u_A}$. Categorically, this is because a CCC typically has a left- and right-closure which are modulated by a symmetry; here, some symmetries are now identities. }
\end{remark}

The following lemmata give us that the necessary equations for uniqueness of products and exponents are indeed derivable from the equational theory, and we will use these in the subsequent proof that the category of \FMCV-terms is a CCC.
The following law will be used to give us to uniqueness of products.
\begin{lemma}\label{lem:producteqns}
The following equation is sound in $\ccceq$: 
\begin{align*}
	\term{P;<?x_A>;P;<?x`{'_A}>.<y_A>.[!x`{'_A}]} &=_{\times} \term{P: > !s_A!t_A} 
\end{align*}where  $\term{!x_A, !x_A': !t_A}$ and $\term{!y_A: !s_A}$.
\end{lemma}
\begin{proof}
We calculate: 
\begin{align*}
\term{P} 	&=_\beta \term{P.<?x_A?y_A>.[!y_A!x`{_A}]}\\
	&=_\beta \term{P.<?x_A?y_A>.[!y_A].[!y`{_A}!x`{_A}].<?x`{'''_A}?y`{''_A}>.[!x`{'''_A}]}\\
	&=_\beta \term{P.<?x_A?y_A>.[!y_A!x_A!y_A!x_A]} \term{.<?x`{'_A}?y`{'_A}?x`{''_A}>.[!y`{'_A}!x`{'_A}].<?x`{'''_A}?y`{''_A}>.[!x`{'''_A}]}\\
	&=_{\Delta} \term{P;}\uterm{P;<?x`{'_A}?y`{'_A}?x`{''_A}>.[!y`{'_A}!x`{'_A}]}\term{.<?x`{'''_A}?y`{''_A}>.[!x`{'''_A}]} \\
	&=_{\iota} \term{P;<?x`{''_A}>.P.<?x`{'''_A}>.<?y`{''_A}>.[!x`{'''_A}]}\, . \qedhere
\end{align*} 
\end{proof}
The following law will be used to give us uniqueness of exponents.
\begin{lemma}\label{rem:eta-equiv}
The following equation sound in $\ccceq$: 
\[
	\term{S} =_{\eta'} \term{[`!S;a<f>.`?f]a~: > `a(?s_A > !t_A)}\, ,
\]
where $\term{f: ?s_A > !t_A}$. 
\end{lemma}
\begin{proof}
We calculate:
\[
	 \term{[`!S;<f>.`?f]} =_\eta  \term{S;<g>.[`![g].<f>.`?f]} =_\beta \term{S;<g>.[`!`?g]} =_\tau \term{S;<g>.[g]} =_\ide  \term{S}\, . \qedhere 
\]
\end{proof}
\begin{remark}\label{rem:eta2}
In the presence of $\{\ide, \beta, \phi\}$, $\eta$ is derivable from $\eta'$: 
\begin{align*}
	\term{S;<?x_A>.[`![!x_A].M]a} &=_{\eta'} \term{[`!S;<?x_A>.[`![!x_A].M]a.a<f>.`?f]a} \\
		& =_\beta                \term{[`!S;<?x_A>.`?`![!x_A].M]a} \\
		&=_\phi \term{[`!S;<?x_A>.[!x_A].M]a} \\
		&=_\ide \term{[`!S;M]a} 
\end{align*}
so that in the presence of $\{\ide,\beta, \phi, \tau\}$, the two are equivalent. 
\end{remark}

It is now shown that the equational theory suffices to make the category of \FMCV-terms Cartesian closed. 
\begin{theorem}\label{thm:ccc}
The category $\termcat{\Sigma_A}$ is Cartesian closed, with equipment given by Definition \ref{def:ccc}.
\end{theorem}
\begin{proof}
Recalling Proposition \ref{cartesian-equations} and Proposition \ref{ccc-equations}, to prove we have a Cartesian closed category, we show existence of a terminal object and existence and uniqueness of products and existence and uniqueness of exponentials. 
Terminality of the empty type family follows from the terminality ($!$) equation. 
To verify existence of products, we show for $\term{M: ?s_A > !t_A, N: !s_A > ?u_A}$: 
\begin{align*}
	\langle \term{M}, \term{N} \rangle  \ ; \pi_1
		&=  \term{<?x_A>.([!x_A].M;}\term{[!x_A].N \ ;\  <?y_A>}\term{)} \\
		&=_{!} \term{<?x_A>.[!x_A]}\term{.M} \\
		&=_{\ide} \term{M}\\
	\langle \term{M}, \term{N}\rangle  \ ; \pi_2 &= \term{<?x_A>.(}\term{([!x_A].M;}\term{[!x_A].N\ ; \ <?y_A>.<?z_A>.[!y_A]}\term{)} \\
		 &=_\iota \term{<?x_A>.(}\term{[!x_A].M; <?y_A>.[!x_A].N}\term{)} \\
		&=_! \term{<?x_A>.[!x_A]}\term{.N} \\
		&=_{\ide} \term{N}
\end{align*}
where $\term{!x_A, !s_A, !y_A: !u_A}$ and $\term{!z_A: !t_A}$, respectively. 
To verify the uniqueness of the product, we show that for $\term{M: ?s_A > !t_A!u_A}$:
\begin{align*}
			\langle \term{M} ; \pi_1, \term{M} ; \pi_2 \rangle  &= \term{<?x_A>.(}\term{[!x_A].M;}\pi_1\term{;[!x_A].M;}\pi_2\term{)} \\
			&= \term{<?x_A>.[!x_A].M;<?z_A>.[!x_A].M;<?z`{'_A}>.<?y_A>.[!z`{'_A}]} \\
			&=_{\times} \term{<?x_A>.[!x_A]}\term{.M} \\
			&=_{!\times } \term{M}
\end{align*}
 where $\term{!x_A: !s_A}$.
To verify existence of exponentials, 
 we show that for  $\term{M: ?s_A?t_A>!u_A}$:
\begin{align*} 
	  (\textsf{id}_{\cat{!r_A}} \times \textsf{curry}(\term{M}))\  ;\  \epsilon &= \term{<?x_A>.[`![!x_A].M]} \term{;} \term{<f>.`?f} \\\
		& =_\beta \term{<?x_A>.`?`![!x_A].M} \\
		& =_\phi \term{<?x_A>.[!x_A].M} \\
		&=_{\ide} \term{M} 
\end{align*}
where $\term{!x_A : !s_A}$ and $\term{f : ?r_A > !t_A}$.
To verify uniqueness of exponentials, we show that for $\term{M: ?s_A > (?t_A > !u_A)}$:
\begin{align*}
	 \textsf{curry}((\textsf{id}_{\cat{!s_A}} \times \term{M}) \ ; \ \epsilon)) 
		 &= \term{<?x_A>.}\term{[`![!x_A].M;<f>.`?f]} \\
		&=_{\eta'} \term{<?x_A>.[!x_A].N} \\
		 &=_{\ide}  \term{M} 
\end{align*}
where $\term{!x_A : !s_A}$, $\term{f : ?r_A > !t_A}$.
\end{proof}
Note that the associators and unitors are  given by identities, which makes the category a \textit{strict} Cartesian closed category.

Let us proceed to show that the equational theory give above is valid with respect to the operational semantics given by the machine.

\section{Machine Equivalence}

The functional abstract machine gives the operational semantics of the \FMCV. The notion of a \textit{successful} run of the machine, where the run terminates in the state $(S_A, \term{*})$, is a novelty with respect to the $\lambda$-calculus. In Theorem \ref{thm:run}, it was proved that every typed term successfully runs on the machine when given appropriately typed inputs. Informally, we can thus consider a term $\term{M: ?s_A > !t_A}$ as generating an input/output function from memories of type $\type{!s_A}$ to memories of type $\type{!t_A}$,  giving a model of the \FMCV. Indeed, this model gives rise to a natural notion of equivalence on \FMCV-terms (\textit{i.e.}, by considering which terms generate the same function). We call this \textit{machine equivalence}, and it is a type of observational (or contextual) equivalence.

For the purposes of this section, we consider the computation signature to be empty, and the value signature to contain only values of base types.
This is because the machine cannot take any further steps if it were to reach the states $(S_A, \term{c.M})$ or $(S_A, \term{`?v.M})$, causing a premature failure state.
 Alternatively, we could consider constants as free variables, or include machine transitions for constants.
Indeed, adding transitions for \textit{e.g.} an addition function on a base type of natural numbers would be a sensible way to generate a finer equivalence on terms than is studied here. 

The following is a definition which captures {machine equivalence}.  Because the model we aim to capture is higher-order, the equivalence is naturally a \textit{logical relation}, as originally described by Plotkin \cite{DBLP:conf/icalp/PlotkinPST00}, and subsequently used in the study of program equivalence by Pitts and Stark \cite{Pitts-Stark-1998,Pitts-1996}.
We proceed to show this machine equivalence is a congruence on terms, and then show that the equational theory is valid with respect to it. 

Note that the definition of machine equivalence models a strong notion of observational equivalence, which would not be appropriate if we included \textit{e.g.} a location intended to model a stream of random variables.  
\begin{definition}
Closed computations $\term{M, N}: \type{ ?s_A > !t_A}$ are \textit{machine equivalent at type $\type{?s_A > !t_A}$} if for equivalent inputs the machine gives equivalent outputs:
\begin{align*}
	\term{M} \sim \term{M'}&: \type{?s_A > !t_A} &&\defeq&& \forall \ S_A \sim S'_A: \type{!s_A} . \ (S_A, \term{M})\Downarrow ~ \sim ~  (S_A', 
	\term{M'})\Downarrow\ : \type{!t_A},
\end{align*}
where, similarly, closed values $\term{V,V'}: \type{t}$ are machine equivalent at $\type{t}$ if their executions are equivalent (or, at base type, if they are identical constants)
\begin{align*}
	\term{V} \sim \term{V'}&: \type{?s_A > !t_A} ~ \defeq ~ \term{`?V} \sim \term{`?V'}: \type{?s_A > !t_A}
	\qquad
	\term{v} \sim \term{v'}: \type{a} ~\defeq~ \term{v} \equiv \term{v' : a}
\end{align*}
where two memories are equivalent if their values are pairwise equivalent. Equivalence extends to open terms as follows.
\begin{align*}
	\term{!u}: \type{!u} \cvdash \term{M} \sim  \term{M'} : \type{?s_A > !t_A} ~ &\defeq ~ \forall \ U \sim U': \type{!u}. \ \term{ \{U/!u\}M } \sim \term{\{!U'/!u\}M'} : \type{?s_A > !t_A} .
\\
	\term{?u}: \type{?u} \vvdash \term{V} \sim  \term{V'} : \type{t} ~ &\defeq ~ \forall \ U \sim U': \type{!u}. \ \term{ \{U/!u\}V } \sim \term{\{U'/!u\}V'} : \type{t} .
\end{align*}
\end{definition}
In the following, by congruence, we mean an equivalence relation $\sim$ on terms such that equivalent terms in equivalent contexts are equivalent. 
\begin{proposition}\label{congruence}
Machine equivalence $(\sim)$ is a congruence relation.
\end{proposition}
\begin{proof}
We must show that $(\sim)$ is symmetric, transitive, reflexive and closed under all contexts. Symmetry is immediate by the symmetry of the definition. Transitivity can be shown by a simple induction on the size of types. Reflexivity is not a priori clear, but follows trivially from closure under all contexts.
We proceed to show by induction on type derivations that $\sim$ is closed under all contexts. Throughout, let $W \sim W': \type{!w_A}$ and $\term{\sigma} = \term{\{W/!w\}}$ and $\term{\sigma'} = \term{\{W'/!w\}}$, being wary that we use $W$ here as a stack and not a value.
\begin{itemize}
\item Identity: $\term{*} \sim \term{*}$
is tautologous. 
\item Sequential Execution: given $\term{!w: !w} \vvdash \term{V} \sim \term{V': ?r_A > !u_A}$ and $\term{!w: !w} \cvdash \term{M} \sim \term{M': ?u_A?s_A > !t_A}$, we must show
\[
	\term{\sigma `?V'.M} \sim \term{\sigma' `?V'.M'}\ .
\]
Let $S_A \sim S'_A: \type{!s_A}$ and $R_A \sim R'_A: \type{!r_A}$. Applying the definition of substitution, we must thus show that $T_A \sim T'_A: \type{!t_A}$. 
\[
\begin{array}{l@{,}r@{}}
(~S_AR_A~& ~\term{`?\sigma V.\sigma M}~)
\\\hline\hline
(~S_AU_A~& ~\term{\sigma M}~)
\\\hline\hline
(~T_A~& ~\term{*}~)
\end{array}
\qquad 
\begin{array}{l@{,}r@{}}
(~S'_AR'_A~& ~\term{`?\sigma' V'.\sigma' M'}~)
\\\hline\hline
(~S'_AU'_A~& ~\term{\sigma' M'}~)
\\\hline\hline
(~T'_A~& ~\term{*}~)
\end{array}
\]
We have $U_A \sim U'_A : \type{!u_A}$ by $\term{\sigma V} \sim \term{\sigma' V'}$, and then the required result follows by $\term{\sigma M} \sim \term{\sigma' M'}$. 
\item Application: given $\term{!w: w} \cvdash \term{M} \sim \term{M'}: \type{`a(r)?s_A > !t_A}$ and $\term{!w: w} \vvdash \term{V} \sim \term{V'}: \type{r}$, we must show
\[
 	\term{\sigma [V]a.M} \sim \term{\sigma' [V']a.M': ?s_A > !t_A}\, .
\]
Let $S_A \sim S'_A : \type{!s_A}$. Applying the definition of substitution, we must thus show that  $T_A \sim T'_A: \type{!t_A}$ in
\[
\begin{array}{l@{,}r@{}}
(~S_A~& ~\term{[\sigma V]a.\sigma M}~)
\\\hline
(~\SAouta ; S_a \cdot \term{\sigma V}~& ~\term{\sigma M}~)
\\\hline \hline
(~T_A ~& ~\term{*}~)
\end{array}
\quad
\textup{and} 
\quad
\begin{array}{l@{,}r@{}}
(~S_{A}'~& ~\term{[\sigma' V']a.\sigma' M'}~)
\\\hline
(~S_{A\setminus \{a\}}'  ; S'_a \cdot \term{\sigma' V'}~& ~\term{\sigma' M'}~)
\\\hline \hline
(~T'_A ~& ~\term{*}~)
\end{array}
\]
This follows from $\term{\sigma M} \sim \term{\sigma' M'}$ and $\term{\sigma V} \sim \term{\sigma' V'}$. 
\item Abstraction: given $\term{!w: !w} \cvdash \term{M} \sim \term{M': ?s_A > !t_A}$, we must show
\[
	\term{\sigma a<x>.M} \sim \term{\sigma' a<x>.M': `a(r)?s_A > !t_A}
\]
Let $S_A \sim S'_A: \type{!s_A}$ and $\term{V} \sim \term{V': r}$. Applying the definition of substitution, we must thus show that $T_A \sim T'_A: \type{!t_A}$ in 
\[
\begin{array}{l@{,}r@{}}
(~\SAouta ; S_a \cdot \term{V}~& ~\term{a<x>.\sigma M}~)
\\\hline
(~S_A~& ~\term{\{V/x\}\sigma M}~)
\\\hline \hline
(~T_A ~& ~\term{*}~)
\end{array}
\quad
\textup{and} 
\quad
\begin{array}{l@{,}r@{}}
(~S_{A\setminus \{a\}}'; S_a \cdot \term{V'}~& ~\term{a<x>.\sigma' M'}~)
\\\hline
(~S_A'~& ~\term{\{V'/x\}\sigma' M'}~)
\\\hline \hline
(~T'_A ~& ~\term{*}~)
\end{array}
\]
The result follows from $\term{\{V/x\}\sigma M} \sim \term{\{V'/x\}\sigma' M': ?s_A > !t_A}$. 
\item Thunk:  given $\term{!w:!w} \cvdash \term{M} \sim \term{M: ?s_A > !t_A},$ we must show 
\[
	\term{\sigma`!M} \sim \term{\sigma'`!M': ?s_A > !t_A}\ . 
\]
Let $S_A \sim S'_A: \type{!s_A}$. Applying the definition of substitution, we must thus show that $T_A \sim T'_A: \type{!t_A}$ in 
\[
\begin{array}{l@{,}r@{}}
(~S_A~& ~\term{`?`!\sigma M}~)
\\ \hline
(~S_A~& ~\term{\sigma M}~)
\\\hline \hline
(~T_A ~& ~\term{*}~)
\end{array}
\quad
\textup{and} 
\quad
\begin{array}{l@{,}r@{}}
(~S'_A~& ~\term{`?`!\sigma' M}~)
\\ \hline
(~S'_A~& ~\term{\sigma' M}~)
\\\hline \hline
(~T'_A ~& ~\term{*}~)
\end{array}
\]
This follows from $\term{\sigma M} \sim \term{\sigma' M'}$. 
\qedhere
\end{itemize}
\end{proof}
The following result closes this section. We work here using the derived \textit{global} beta-equivalence, which clearly immediately implies the local beta and relocation equations which it replaces.  
\begin{proposition}[Soundness]\label{observational-soundness}
For all closed, typed computations $\term{M: ?s_A > !t_A}$ and $\term{N: ?s_A > !t_A}$,  we have that
\begin{align*}
	\term{M} =_{\textsf{eqn}} \term{N}\ \qquad \textup{implies} \qquad  \term{M} \sim \term{N: ?s_A > !t_A}. 
\end{align*}
\end{proposition}
\begin{proof}
We verify the statement for each equation generating $=_\textsf{eqn}$, which suffices since by Proposition \ref{congruence}, which says machine equivalence $\sim$ is closed under all contexts. 
In each case, we must show that the two sides of the equation, when evaluated on equivalent inputs, give equivalent outputs. Recall that by reflexivity of $\sim$, running a term $\term{M}$ on two equivalent inputs results in equivalent outputs. Throughout, let $S_A \sim S'_A: \type{!s_A}$ and let $(\e_A, \term{S}) \Downarrow S_A$, and that the evaluation of $\term{S: > !s_A}$ results a memory $S_A$ pushed to the stack, whereas $\term{M: ?s_A > !t_A}$ takes $S_A: \type{!s_A}$ to $T_A: \type{!t_A}$ and $\term{M: ?s_A > !u_A}$ takes $S_A: \type{!s_A}$ to $U_A: \type{!t_A}$, \textit{etc.}. 
\begin{itemize}
\item Identity: immediate from the following runs. 
\[
\begin{array}{l@{~,~}r}
(~S_A &\term{<?x_A>.[!x_A]}~)
\\\hline \hline
(~\e_A & \term{[S_A]}~)
\\\hline \hline
(~S_A & \term{*}~)
\end{array} \
\qquad \qquad
(~S'_A ~,~\term{*}~)
\]

\item Beta: we show for the global beta case, with local beta a special case. 
$T_A \sim T'_A$ follows from $S_A \sim S'_A$ and the following runs. 
\[
\begin{array}{l@{,}r@{}}
(~S_A~& ~\term{[ N]a.a<x>. M}~)
\\\hline
(~\SAouta ; S_a \cdot \term{ N}~& ~\term{a<x>. M}~)
\\\hline
(~S_A~& ~\term{\{ N/x\} M}~)
\\\hline \hline
(~T_A ~& ~\term{*}~)
\end{array}
\qquad\qquad
\begin{array}{l@{,}r@{}}
(~S'_A~& ~\term{\{ N/x\} M}~)
\\\hline \hline
(~T'_A ~& ~\term{*}~)
\end{array}
\]
\item Force: 
$T_A \sim T'_A$ follows from $S_A \sim S'_A$ and the following runs.
\[
\begin{array}{l@{,}r@{}}
(~S_A~&~\term{`?`! M}~)
\\\hline
(~S_A~&~\term{ M}~)
\\\hline \hline
(~T_A ~& ~\term{*}~)
\end{array}
\qquad\qquad
\begin{array}{l@{,}r@{}}
(~S'_A~&~ \term{M}~)
\\\hline \hline
(~T'_A ~& ~\term{*}~)
\end{array}
\]
\item Thunk: consider the following runs, assuming $S_A \sim S'_A$.
\[
\begin{array}{l@{,}r@{}}
(~S_A~&~\term{[`!`?V]}~)
\\\hline
(~S_A \cdot \term{`!`?V} ~&~\term{*}~)
\end{array}
\qquad\qquad
\begin{array}{l@{,}r@{}}
(~S'_A~&~ \term{[V]}~)
\\\hline \hline
(~S'_A \cdot \term{V} ~& ~\term{*}~)
\end{array}
\]
Then, we must prove $\term{`?`!`?V} \sim \term{`?V: ?r_A > !t_A}$. Indeed, $T_A \sim T'_A:\type{!t_A}$ follows from $R_A \sim R'_A: \type{!r_A}$, in the following runs.
\[
\begin{array}{l@{,}r@{}}
(~R_A~&~\term{`?`!`?V}~)
\\\hline
(~R_A  ~&~\term{`?V}~)
\\\hline\hline
(~T_A  ~&~\term{*}~)
\end{array}
\qquad\qquad
\begin{array}{l@{,}r@{}}
(~R'_A~&~ \term{`?V}~)
\\\hline \hline
(~T'_A  ~& ~\term{*}~)
\end{array}
\]
\item Eta: consider the following runs. 
\[
\begin{array}{l@{,}r@{}}
(~\e_A~& ~\term{S;<?x_A>.[`![!x_A].M]a}~)
\\\hline\hline
(~S_A & ~\term{<?x_A>.[`![!x_A].M]a}~)
\\\hline\hline
(~\e_A & ~\term{[`![!S_A].M]a}~)
\\\hline
(~\e_{A\setminus \{a\}} ; \e_a \cdot \term{`![!S_A].M}& ~\term{*}~)
\end{array} 
\]
\[
\begin{array}{l@{,}r@{}}
(~\e_A~& ~\term{[`!S;M]a}~)
\\\hline
(~\e_{A\setminus \{a\}} ; \e_a \cdot \term{S;M}& ~\term{*}~)
\end{array} 
\]
We are thus required to show that $\term{`?`![S_A].M} ~ \sim ~ \term{`?`!S;M: ?t_A > !u_A}$. Let $T_A \sim T'_A : \type{!t_A}$. Then the required result of $U_A \sim U'_A : \type{!u_A}$ is immediate from the following runs. 
\[
\begin{array}{l@{,}r@{}}
(~T_A~& ~\term{`?`![S_A];M}~)
\\\hline
(~T_A~& ~\term{[S_A];M}~)
\\\hline\hline
(~T_A \cdot S_A & ~\term{M}~)
\\\hline \hline
(~U_A ~& ~\term{*}~)
\end{array} 
\qquad\qquad
\begin{array}{l@{,}r@{}}
(~T'_A~& ~\term{`?`!S;M}~)
\\\hline
(~T'_A~& ~\term{S;M}~)
\\\hline\hline
(~T'_A \cdot S_A & ~\term{M}~)
\\\hline \hline
(~U'_A ~& ~\term{*}~)
\end{array} 
\]

\item Diagonal: immediate from the following runs. 
\[
\begin{array}{l@{,}r@{}}
(~\e_A~& ~\term{ S;<?x_A>.[!x_A].[!x_A]}~)
\\\hline \hline
(~S_A~& ~\term{<?x_A>.[!x_A].[!x_A]}~)
\\\hline \hline
(~\epsilon~& ~\term{[!S_A].[!S_A]}~)
\\\hline \hline
(~S_A~& ~\term{[!S_A]}~)
\\\hline \hline
(~S_A  S_A~& ~\term{*}~)
\end{array}
\qquad
\qquad
\begin{array}{l@{,}r@{}}
(~\e_A~& ~\term{ S; S}~)
\\\hline \hline
(~S_A~& ~\term{ S}~)
\\\hline \hline
(~S_A S_A~& ~\term{*}~)
\end{array}
\]
\item Interchange: 
Let $R_A\sim R_A':\type{!r_A}$, $T_A \sim T_A':\type{!t_A}$.
Then $S_A \sim S'_A: \type{!s_A}$ and $U_A \sim U'_A: \type{!u_A}$ follows by assumption and the following runs. 
\[
    \begin{array}{@{(~}l@{~,~}r@{~)}}%
    R_AT_A & \term{<?x_A>.M.[!x_A].N}%
    \\\hline\hline\rule[-5pt]{0pt}{15pt}%
    R_A  & \term{M.[T_A].N}%
    \\\hline\hline\rule[-5pt]{0pt}{15pt}%
    S_A  & \term{[T_A].N}%
    \\\hline\hline\rule[-5pt]{0pt}{15pt}%
    S_AT_A & \term{N}%
    \\\hline\hline\rule[-5pt]{0pt}{15pt}%
    S_AU_A & \term *\;%
    \end{array}%
\qquad\qquad
    \begin{array}{@{(~}l@{~,~}r@{~)}}%
    R'_AT_A' & \term{N.<?y_A>.M.[!y_A]}%
    \\\hline\hline\rule[-5pt]{0pt}{15pt}%
    R'_AU'_A & \term{<?y_A>.M.[!y_A]}%
    \\\hline\hline\rule[-5pt]{0pt}{15pt}%
    R'_A   & \term{M.[U'_A]}%
    \\\hline\hline\rule[-5pt]{0pt}{15pt}%
    S'_A   & \term{[U'_A]}%
    \\\hline\hline\rule[-5pt]{0pt}{15pt}%
    S'_AU'_A & \term *\;%
    \end{array}
\]

\item Terminal:
Note that both runs (by their type) must return the empty stack, which is trivially related to itself by $\sim$, the result follows immediately. We illustrate the two runs anyway.  
\[
\begin{array}{@{}l@{,}r@{}}
(~\e_A~&~\term{S;<?x_A>}~)
\\\hline
\hline
(~S_A~&~\term{<?x_A>}~)
\\\hline \hline
(~\e_A~&~\term{*}~)
\end{array}
\qquad\qquad
(~\e_A~,~ \term{ *}~)
\]

\item Permutation: Let $\term{W} \sim \term{W': r}$, then 
$T_A \sim T'_A: \type{!t_A}$ follows from this assumption, $S_A \sim S'_A$, and the following runs. 
\[
\begin{array}{l@{,}r@{}}
(~S_{A \setminus\{a,b\}} ; S_a \cdot \term{W}; S_b~& ~\term{[V]b.a<w>.M}~)
\\\hline
(~S_{A \setminus\{a,b\}} ; S_a \cdot \term{W} ; S_b \cdot \term{V}~& ~\term{a<w>.M}~)
\\\hline
(~S_{A \setminus\{a,b\}} ; S_a; S_b \cdot \term{V} ~& ~\term{\{W/w\}M}~)
\\\hline\hline
(~T_A  & ~\term{*}~)
\\ \multicolumn{2}{}{}
\\
(~S'_{A \setminus\{a,b\}} ; S'_a \cdot \term{W'} ; S_b~& ~\term{a<w>.[V]b.M}~)
\\\hline
(~S'_{A \setminus\{a,b\}} ; S'_a ; S'_b~& ~\term{[V]b.\{W'/w\}M}~)
\\\hline
(~S'_{A \setminus\{a,b\}} ; S'_a ; S'_b \cdot \term{V} ~& ~\term{\{W'/w\}M}~)
\\\hline\hline
(~T'_A  & ~\term{*}~)
\end{array}\qedhere
\]
\end{itemize} 
\end{proof}

\begin{remark}\label{rem:extensional}
In the case there are no constants, it trivally follows from the definition of $(\sim)$ that all terms of the same type are identified.
Consider then the \FMCV\  with one value constant $\term{c:a}$ for each base type $\type{a}$: every computation type is inhabited by the following term:
\begin{align*}
	(\type{?s_A > !t_A})^* = \term{<?x_A>.[}\type{!t_A}^*\term{]} \qquad 
	\type{a}^* = \term{c}\ .
\end{align*}
However, in this case, all terms of the same type are machine equivalent. 
For example, $\term{<x>.[c]} \sim \term{<x>.[x]: a > a}$, since for given any input (which must be the unique constant), the outputs are the same.
Given \textit{two} value constants, $\term{c, c': a}$, the machine equivalence becomes non-degenerate. Consider again the two terms above: they become distingushable by the input $\term{c'}$.
Nevertheless, for any number of value constants of base type, the machine equivalence is strictly coarser than the equational theory. Similar to the simply-typed $\lambda$-calculus, there are not enough contexts to distinguish between all equationally distinct terms. To see this, consider again that the only distinguishable terms at type $\type{a > a}$ are the identity and constant functions, as above. 
Then the Church numerals $\term{<f>.[f^n]: (a > a) > (a > a)}$, for $n \geq 1$, are not distinguishable by any of their possible inputs (namely, the identity and constant functions). 
\end{remark}

We can construct a category in the same way as $\termcat{\Sigma_A}$, except with terms taken modulo machine equivalence. In fact, this is a Cartesian closed category, too, although it is too degenerate to be the \textit{free} one. 
\begin{corollary}
 \FMCV-terms over a signature whose only constants are values of base type,  taken modulo machine equivalence, form a Cartesian closed category. 
\end{corollary}
\begin{proof}
Machine equivalence $(\sim)$ contains the equational theory $\ccceq$ (Theorem \ref{observational-soundness}), and terms modulo $\ccceq$ form a Cartesian closed category (Theorem \ref{thm:ccc}). 
\end{proof}
In fact, the category given by terms modulo machine equivalence is an extensional collapse of the category of terms modulo the equational theory, as commented in Remark \ref{rem:extensional}.
That is, we have that computations $\term{M} \sim \term{N: ?s_A > !t_A}$ if for every computation $\term{S: > !s_A}$, $\term{S;M} = \term{S;N: > !t_A}$. This follows immediately from the definition of machine equivalence and the observation that $\term{S;M} \sim \term{S;N: > !t_A}$ implies $\term{S;M} = \term{S;N: > !t_A}$.

\chapter{Categorical Semantics of the Functional Machine Calculus}\label{chapter:CCC-equiv}

This chapter contains the first main contribution of this thesis: the result that the category of \FMCV-terms is equivalent to the \textit{free} Cartesian closed category. 
This equivalence is naturally factorized into two parts: the equivalence between the free CCC (that is, the STLC) and the \SLCV, and that between the \SLCV\  and the \FMCV.
We give self-contained proofs for both.

We begin with the first equivalence, by giving an explicit construction of the free functor from the  category of STLC-terms to the category of \SLCV-terms. This is immediate from Theorem \ref{thm:ccc} of the previous chapter.
Its inverse functor must send the category of \textit{closed} \SLCV-terms, composed by sequencing, into the category of  \textit{open} $\lambda$-terms, composed by substitution. How to do this is explained in the upcoming section.

We expect there to be a denotational equivalence between these two categories, because, although the \SLCV\  gives us a sequencing mechanism on top of the $\lambda$-calculus, we expect its expressivity to remain the same: that is, to be able to express all the same higher-order functions. Further, we have chosen the equational theory so as to identify exactly what is needed to make the \SLCV\  a Cartesian closed category (and no more). 

The intuition behind the denotational equivalence of the \SLCV\  and the \FMCV\  is that of the isomorphism between an indexed product (\textit{i.e.}, a memory) and a plain product (\textit{i.e.}, a stack), given by fixing an ordering of indices. 

The functor from \SLCV-terms to \FMCV-terms is given by a simple \textit{embedding}, \textit{i.e.}, an inclusion of terms. This necessitates a choice of location on which to embed. 
For the inverse, we  collapse the memory of the \FMCV\  into single large stack (depending on a chosen of ordering on locations).
This collapsing of types (memories) is then lifted to terms, giving the desired functor.
The technical difficulty of this proof comes from necessarily working up-to (natural) isomorphism, and from the higher-order nature of the \FMCV, which means that we likewise will need appropriate higher-order isomorphisms. \footnote{Note, the embedding functor, which maps stacks to memories which make use of only one location,  is {essentially} surjective on objects, via the isomorphism exhibited in that section.} 

One perspective on this interpretation is that the \FMCV\  gives \textit{multiple} copies of the \SLCV, one for each location. This is confirmed by the observation that we have multiple isomorphic arrow types in the category of \FMCV-terms: one for each location. {
The interpretation of the \FMCV\  into the \SLCV\  can then be viewed as collapsing these multiple copies into one.} 
This perspective highlights how this result relies on the assumption that \textit{every location is equivalent}, and is reliant on the fact the type system accounts for \textit{everything} happening to the memory, not just actions on the main location. 

To summarize, to achieve the main result, we factorize the equivalence of the \FMCV\  into the free CCC into two steps, as below:
\[\begin{tikzcd}[ampersand replacement=\&]
	{\textsf{FMC}(\Sigma_A)} \& {\textsf{SLC}(\lfloor{\Sigma_A}\rfloor)} \& {\textsf{CCC}(\lfloor\Sigma_A\rfloor)} \\
	{}
	\arrow["{\llbracket - \rrbracket^<}", curve={height=-36pt}, from=1-1, to=1-2]
	\arrow["{\{-\}_a}", curve={height=-36pt}, from=1-2, to=1-1]
	\arrow["{\llbracket-\rrbracket'}", curve={height=-36pt}, from=1-2, to=1-3]
	\arrow["{\{-\}'}", curve={height=-36pt}, from=1-3, to=1-2]
\end{tikzcd}
\vspace{-0.4cm}
\]
where $\collap{-}$ is the collapsing of the types of an \FMCV\  signature, giving an \SLCV\  signature (and, isomorphically, an STLC signature).\footnote{Rightly, this should be parameterize in $<$ also. Note also that $\collap{-}$ is  surjective on signatures.}




Given these results, an obvious question is: if the simply-typed \FMCV\  is equivalent to the simply typed $\lambda$-calculus, what is to be gained from study of the \FMCV? We respond to this question in detail in the conclusion of the chapter. For now, we mention that, as expected, the denotational perspective collapses the all operational refinements made to the STLC by the \FMCV. 

\section{The Sequential $\lambda$-Calculus is Equivalent to the STLC}
%

We construct a map from the \SLCV \ to the STLC (rather than the free CCC) in order to compare calculi. We will, however, sometimes confuse the two when the differene is irrelevant. Note, this map will also bear a close resemblance to the quantitiative interpretation of terms given in Chapter \ref{chapter:SN}. The inverse map will be presented and discussed in the subsequent subsection. 

We begin with the translation from the simply-typed $\lambda$-calculus to the \SLCV. Indeed, knowing that the category of \FMCV-terms is Cartesian closed, we have in hand a functor from the free CCC (\textit{i.e.}, the simply-typed $\lambda$-calculus) to the category of \FMCV-terms. 
\subsection{The Free Functor: STLC to \SLCV}

We first define an interpretation of \SLCV\  types into the types of the $\lambda$-calculus. 
We then relate an \SLCV-signature and a signature with values used to generate the $\lambda$-calculus. 
\begin{definition}
The \textit{interpretation} of value types and stack types $\ccc{\type{!t}}$ and contexts are respectively defined by induction as:
\[
\ccc{\type{a}} = \alpha \qquad
\ccc{\type{?s > !t}} = \ccc{\type{!s}} \to \ccc{\type{!t}} \qquad
\ccc{\type{t_1 \ldots t_n}} = \ccc{\type{t_1}} \times \ldots \times \ccc{\type{t_n}} 
\]
\vspace{-\baselineskip}
\end{definition}

\begin{definition}
Given an \SLCV-signature $\Sigma = (\Sigma_0, \Sigma_c, \Sigma_v, \textsf{dom}, \textsf{cod}, \textsf{val})$, define its \textit{interpretation} $\ccc{\Sigma}$ as the $\lambda$-signature $(\Sigma_0, \Sigma_c, \Sigma_v, \ccc{\textsf{dom}}, \ccc{\textsf{cod}}, \ccc{\textsf{val}})$, where $\ccc{\textsf{dom}}(\term{c}) = \ccc{\textsf{dom}(\term{c})}$, $\ccc{\textsf{cod}}(\term{c}) = \ccc{\textsf{cod}(\term{c})}$ and $\ccc{\textsf{val}}(\term{v}) = \ccc{\textsf{val}(\term{v})}$. Since $\ccc{\Sigma}$ and $\Sigma$ are clearly isomorphic, and it will be clear from context if we are in the setting of the \SLCV or the $\lambda$-calculus, we confuse the two and write $\ccc{\Sigma}$ as $\Sigma$. 
\end{definition}

By the freeness of $\textsf{CCC}(\Sigma)$, knowing that $\stermcat{\Sigma}$ is a CCC (Theorem \ref{thm:ccc}), we can construct, for every function $j: \Sigma \to U(\stermcat{\Sigma})$, a unique CCC-functor $\termint{-}_j: \textsf{CCC}({\Sigma}) \to \stermcat{\Sigma}$. Clearly, there is a canonical choice of $j$ given by the inclusion of the signature.
Indeed, we will proceed show that the functor thus induced is in fact an equivalence of categories. 
\begin{definition}
Define $\termint{-}: \textup{\textsf{CCC}}(\Sigma) \to \stermcat{\Sigma}$ to be the unique CCC-functor making the following diagram commute:
\[\begin{tikzcd}
	{\Sigma} & {U({\textup{\textsf{CCC}}(\Sigma)})} \\
	& {U(\stermcat{\Sigma})}
	\arrow["{U(\termint{-})}", from=1-2, to=2-2]
	\arrow["i", hook, from=1-1, to=1-2]
	\arrow["j"', hook, from=1-1, to=2-2] \ 
\end{tikzcd}\]
where $i$ and $j$ are the obvious inclusions of signatures. 
\end{definition}
We can present this functor as a map from the STLC to closed, typed \SLCV-terms, defined inductively on type derivations. Note that, because we send \textit{open} $\lambda$-terms to \textit{closed} \SLCV-terms, we must map $\vdash$ into $\type{>}$, turning the free variables of a $\lambda$-term into the input stack of the corresponding \SLCV-term.

\begin{lemma}[The Free Functor]
The functor $\termint{-}: \textup{\textsf{CCC}}({\Sigma}) \to \stermcat{\Sigma}$ can equivalently be defined on $\lambda$-terms as follows. On types and contexts, respectively, define inductively:
\[
\begin{array}{cc}
	\begin{aligned}
	\termint{{\alpha}} &= \type{a}  \\
	\termint{{A} \to {B}} &= {}\termint{{A}}^{r} \type{>} \termint{{B}} \\
	\termint{A \times B} &= \termint{A}\termint{B} \\
	\termint{\top} &= \type{\epsilon_A} 
	\end{aligned}
	\quad&\quad
	\termint{A_1, \ldots, A_n} = \termint{A_1},{\ldots},\termint{A_n} 
\end{array}
\]
where $\termint{-}^{r}$ is defined as $\termint{-}$, except that $\termint{A ~\times~ B}^{r}  = \termint{B}^{r}\termint{A}^{r}$ and $\termint{A_1, \ldots, A_n}^r = \termint{A_n}^r\ldots\termint{A_1}^r$.
Define a closed \SLCV-term 
\[
\termint{\Gamma \vdash M: A} :  \termint{\Gamma}^r \type{>} \termint{A}
\]
inductively on the type derivation of $M$:
\begin{align*}
	\termint{\Pi', y:A, \Pi \vdash y:A} &= \term{<?x>.<?y>.<?z>.[!y]} \\
	\termint{\Pi \vdash v:A} &= \term{<?x>.[v]}\\
	\termint{\Pi \vdash c\, @M:B} &= \termint{\Pi \vdash M:A}\term{;c} \\
	\termint{(p, q): A \times B, \Pi \vdash M:C} &= \termint{p: A, q: B, \Pi \vdash M:C}\\
	\termint{(): \top, \Pi \vdash M:C} &= \termint{\Pi \vdash M:C}\\
	\termint{\Pi \vdash (M, N): A \times B} &= \term{<?x>.([!x].}\termint{\Pi \vdash M:A}\term{;[!x].}\termint{\Pi \vdash N: B}\term{)}\\
	\termint{\Pi \vdash (): \top} &= \term{<?x>} \\
	\termint{\Pi \vdash M @ N: B} &= \term{<?x>.([!x].}\termint{\Pi \vdash N: A}\term{;[!x].}\termint{\Pi \vdash M: A \to B}\term{);<f>.`?f}\\
	\termint{\Gamma \vdash \lambda p.M: A \to B} &= \term{<?x>.[`![!x];}\termint{p:A, \Pi \vdash M: B}\term{]}
\end{align*}
where in each case $\term{!x}: \termint{\Gamma}$, $\term{!y}: \termint{A}$, $\term{!z}: \termint{\Gamma'}$, $\term{f}: \termint{A}^r \type> \termint{B}$.
 Note that coherences are translated as identities, and we present the binary and nullary cases of $n$-ary tuples, and take the remaining cases to be clear. 
\end{lemma}
\begin{proof}
Recall the usual equivalence between the category of simply typed $\lambda$-terms and the free Cartesian closed category (with both generated by the signature $\Sigma$), given in Theorem \ref{thm:curry-howard}. Composing this with the definitions of the equipment of $\termcat{\Sigma}$ from Definition \ref{def:ccc} thus gives the functor described above (modulo a small amount of simplification via beta reduction).  
\end{proof}
\begin{remark}\label{rem:translate-cut}
The translation of the (admissible) cut, exchange and weakening rules are, respectively:
\begin{align*}
	\termint{\Pi', \Pi \vdash M\{N/x\}: B} &= \term{<?x?y>.[!y!y!x].}\termint{\Pi', \Pi \vdash N:A}\term{;[!x].}\termint{\Pi', p:A, \Pi \vdash M: B}\\
	\termint{\Pi', q:B, p:A, \Pi \vdash M: C} &= \term{<?x?y?z>.[!y!z!x].}\termint{\Pi', p:A, q:B, \Pi \vdash M: C}\\
 \termint{\Pi, p:A \vdash M:B} &=  \term{<?y>.}\termint{\Pi \vdash M:B},
\end{align*}
where, in the first case,  $\term{!x:} \termint{\Pi}, \term{!y:} \termint{\Pi'}$, in the following two cases case $\term{!x:} \termint{\Pi}, \term{!y}: \termint{A}$ and $\term{!z}: \termint{B}$, and in the last case, $p \not\in \fv{M}$.
\end{remark} 

\subsection{The Interpretation: \SLCV\  to STLC}

We construct an interpretation functor $\ccc{-}: \stermcat{\Sigma} \to \textup{\textsf{CCC}}(\Sigma)$, which we will show is inverse to $\termint{-}: \textup{\textsf{CCC}}({\Sigma}) \to \stermcat{\Sigma}$. 
\[
	\ccc{\Gamma \vdash_{\textsf{c}} \term{M: ?s > !t}} : \ccc{\type{!s}} \times \ccc{\Gamma} \vdash \ccc{\type{!t}}\\
\]
The top-level arrow $\type{>}$ becomes sequent entailment $\vdash$ with the type of the output stack becoming the type of the $\lambda$-term. 
The corresponding $\lambda$-term then has two ``inputs'', given by its context. The first corresponds to the input stack of the \SLCV, the second to the free variables of the \SLCV.

The interpretation can be considered via the operational intuition of a stack machine \textit{with closures}. The interpretation of a term can be considered as running the machine on its first set of inputs, with the closure dealing with the free variables (the second input) as usual. 
 The application of the \SLCV\  is interpreted as pushing a new element to the first input, while the abstraction of the \SLCV\  is interpreted as shifting from the first input to the second. 
The $\lambda$-term itself thus corresponds to the state of the stack at the end of a machine run, as a function of the (interpretation of the) input stack and the free variables of the \SLCV. 

The reason for the distinction between computations and values is to give the most natural correspondence with the $\lambda$-calculus. 
This now becomes clear: we see in the following definition that the force and thunk constructs of the \SLCV\  correspond to application and abstraction in the $\lambda$-calculus.
The soundness proof to follow shows clearly how the force and thunk \textit{equations} correspond to the beta and eta equations of the $\lambda$-calculus.
Thereby, the ability to \textit{force} a value to act as a computation, and to and \textit{thunk} a computation, considering it a value, is exactly what makes the \SLCV\  higher-order.

%

%

\begin{definition}
The \textit{interpretation} of value types and stack types $\ccc{\type{!t}}$ and contexts are respectively defined by induction as:
\[
\begin{array}{cc}
\begin{aligned}
\ccc{\type{a}} &= \alpha \\
\ccc{\type{?s > !t}} &= \ccc{\type{!s}} \to \ccc{\type{!t}} \\
\end{aligned}
\quad
&
\quad
\begin{aligned}
\ccc{\type{t_1 \ldots t_n}} &= \ccc{\type{t_1}} \times \ldots \times \ccc{\type{t_n}} \\
\ccc{\type{t_1}, \ldots, \type{t_n}} &= \ccctype{t_1} \times \ldots \times \ccctype{t_n}
\end{aligned}
\end{array}
\]
where $\alpha \in \Sigma_0$.
Note, the separating $(|)$ symbol in the following is just another notation for the standard product,  or comma.
Define the \textit{interpretation} of \SLCV-terms as open $\lambda$-terms
\begin{align*}
	\ccc{\Gamma \vdash_{\textsf{v}} \term{V: t}} : \ccc{{\Gamma}} \vdash \ccc{\type{t}}  \qquad \qquad
	\ccc{\Gamma \vdash_{\textsf{c}} \term{M: ?s > !t}} : \ccc{\type{!s}} ~|~ \ccc{\Gamma} \vdash \ccc{\type{!t}}
\end{align*}
by mutual induction on type derivations of computations and values, as follows.  
\begin{align*}
\ccc{\Gamma, \term{t: t}, \Delta \vdash_\textsf{v} \term{t:t}}&(v,t,w) &=&~ t \\
\ccc{\Gamma \vdash_\textsf{v} \term{v:t}}&(w) &=&~ v \\
\ccc{\Gamma \vdash_\textsf{v} \term{`!M: ?s > !t}}&(v) &=&  ~\lambda s. \ccc{\Gamma \vdash_\textsf{c} \term{M: ?s > !t}}(s~|~v)\\
\ccc{\Gamma \vdash_\textsf{c} \term{*: ?s > !s}}&(s~|~v)&=&~ s \\
\ccc{\Gamma \cvdash \term{c.M: ?r?s > !t}}&(s, r~|~v) &=&~ \ccc{\Gamma \cvdash \term{M: ?r?s > !t}}(s, c\, @r~|~v)\\
\ccc{\Gamma \vdash_\textsf{c} \term{<x>.M: r?s_A > !t_A}}&(s, r ~|~v)&=  &~ \ccc{\term{x: r}, \Gamma \vdash_\textsf{c} \term{M: ?s > !t}} (s~|~r,v) \\
\ccc{\Gamma \vdash_\textsf{c} \term{[V].M:?s > !t}}&(s~|~v)&=&~ \ccc{\Gamma \cvdash \term{M: r?s > !t}}(s, \ccc{\Gamma \vdash_\textsf{v} \term{V:r}}(v)~|~v) \\
\ccc{\Gamma \vdash_\textsf{c} \term{`?V.M: ?r?t > !u}}&(t,r~|~v)&= &\ \ccc{\Gamma \cvdash \term{M: ?s?t > !u}}(t,\ccc{\Gamma \vdash_\textsf{v} \term{V: ?r > !s}}(v) @ r ~|~ v)
\end{align*}
In each case, we have $r: \ccc{\type{r}}$ or $r: \ccc{\type{!r}}$, $s : \ccc{\type{!s}}$, $t: \ccc{\type{t}}$, $v : \ccc{\Gamma}$ and $w: \ccc{\Delta}$.
Note that the inputs to $\ccc{-}$ are open $\lambda$-terms to be substituted for free variables, and in particular having inputs corresponding to the context is another way to express a valuation.
We omit the context and/or types of terms inside the interpretation function when it is clear. 
\end{definition}

\subsection{Soundness of the Interpretation}
We proceed to show that the interpretation functor is well-defined. 
We first give the derived interpretations of our defined notation on $\term{!x}$.
\begin{lemma}[Application and Abstraction]
We have for every computation: 
\begin{gather*}
	\ccc{\term{!x: !r},\, \Gamma \cvdash \term{[!x].N: ?s > !t}}(s~|~r,v) = \ccc{\term{!x: !r},\,  \Gamma \cvdash \term{N:?r?s > !t}}(s,r~|~r,v)\\
	\ccc{\Gamma \cvdash \term{<?x>.M: ?r?s>!t}}( s,  r ~|~ v) = \ccc{\term{!x}: \type{!r},\, \Gamma \cvdash \term{M: ?s > !t}}(s ~|~ r,v) \, .
\end{gather*}
\end{lemma}
\begin{proof}
For each statement, proceed by induction on the size of $\term{!x}$. 
\end{proof}
\begin{remark}
Throughout the rest of this chapter, we will freely use the abstraction lemma without mention; however, we will still call attention to the use of the application lemma. 
\end{remark}

The following lemma gives the interpretation of the (admissible) exchange and weakening rules.

\begin{lemma}[Exchange and Weakening]
We have, for every computation:
\begin{align*}
	&\ccc{\Gamma, \term{y:p},\term{x:r}, \Gamma' \cvdash \term{M: ?s > !t}}(s~|~v,p,r,w) =\\
&\ccc{\Gamma, \term{x:r}, \term{y:p}, \Gamma' \cvdash \term{M: ?s > !t}}(s~|~v,r,p,w) 
\end{align*}
\[
\ccc{\Gamma \cvdash \term{ M:?s > !t}}(s~|~v) = \ccc{\term{x:r}, \Gamma \cvdash \term{ M:?s > !t}}(s~|~r,v) 
\]
where, in the latter case, $\term{r} \not\in \fv{\term{M}}$.
\end{lemma}
\begin{proof}
Induction on the type derivation of $\term M$. 
\end{proof}
%

The following lemma give the interpretation of the admissible sequencing rule,
which is necessary for proving that $\ccc{-}$ is a well-defined functor.

\begin{lemma}[Sequencing]
\label{lem:sequencing}\
For computations $\Gamma \cvdash \term{N: ?r > !u}$ and $\Gamma \cvdash \term{M: ?u?s > !t}$, we have that: 
\begin{align*}
 \ccc{\term{N;M}}({t} , {r}~|~v) &= \ccc{\term{M}} ({t} ,  \ccc{\term{N}}({r}~|~v)~|~v)
\end{align*}
\end{lemma}
\begin{proof}
We proceed by induction on the type derivation of $\term{N}$. In each case, we begin by unfolding the definition of sequencing. 
\begin{itemize}
\item Identity:
\begin{align*}
	\cccterm{*;M}(t,r~|~v) &= \ccc{\term{M}}(t,\underline{r}~|~v) \\
	(\textsf{defn.}\, \ccc{-}) &= \cccterm{M}(t,\cccterm{*}(r)~|~v) 
\end{align*}
\item Sequential Constant:
\begin{align*}
	\cccterm{c.N;M}(t,s,r~|~v) &= \cccterm{c.(N;M)}(t,s,r~|~v) \\
		(\textsf{defn.}\, \ccc{-}) &= \cccterm{N;M}(t,s,c\,@r,~|~v) \\
		(\textsf{I.H.}) &= \cccterm{M}(t, \underline{\cccterm{N}(s,c\,@r~|~v)}~|~v)\\
		(\textsf{defn.}\, \ccc{-}) &= \cccterm{M}(t, \cccterm{c.N}(s,r~|~v)~|~v)
\end{align*}
\item Application:
\begin{align*}
	\cccterm{[V].N;M}(t,r~|~v) &= \cccterm{[V].(M;N)}(t,r~|~v) \\	
		(\textsf{defn.}\, \ccc{-}) &= \cccterm{M;N}(t,r,\cccterm{V}(v)~|~v) \\
		(\textsf{I.H.}) &= \cccterm{M}(t, \underline{\cccterm{N}(r,\cccterm{V}(v)~|~v)}~|~v) \\
		(\textsf{defn.}\, \ccc{-}) &= \cccterm{M}(t, \cccterm{[V].N}(r~|~v)~|~v) 
\end{align*}
\item Abstraction: 
\begin{align*}
	\cccterm{<x>.N;M}(t,r,s~|~v) &= \cccterm{<x>.(N;M)}(t,r,s~|~v) \\
		(\textsf{defn.}\, \ccc{-})  &= \cccterm{N;M}(t,r~|~s, v) \\
		(\textsf{I.H.}) &= \cccterm{M}(t, \cccterm{N}(r~|~s,v)~|~s,v) \\
		(\textsf{Weak.}) &= \cccterm{M}(t, \underline{\cccterm{N}(r~|~s,v)}~|~v) \\
		(\textsf{defn.}\, \ccc{-})  &= \cccterm{M}(t, \cccterm{<s>.N}(r,s~|~v)~|~v)
\end{align*}
\item Sequential Excecution:
\begin{align*}
	\cccterm{`?V.N;M}(t,r,s~|~v) &= \cccterm{`?V.(N;M)}(t,r,s~|~v) \\
		(\textsf{defn.}\, \ccc{-}) &= \cccterm{N;M}(t, r, \cccterm{V}(v) @ s~|~v) \\
		(\textsf{I.H.})&= \cccterm{M}(t, \underline{\cccterm{N}(r, \cccterm{V}(v)@ s~|~v)}~|~v)\\
		(\textsf{defn.}\, \ccc{-})  &= \cccterm{M}(t, \cccterm{`?V.N}(r, s~|~v)~|~v) 
		\qedhere
\end{align*}
\end{itemize}

\end{proof}

We show the interpretation function is well-defined, \textit{i.e.}, that it preserves equality of terms. 
\begin{proposition}[Soundness]\label{lem:wd}
For any computations $\Gamma \vdash_{\textsf{c}} \term{M, N: ?s > !t}$, we have
\begin{align*}
	\term{M} =_{\textsf{eqn}} \term{N} \quad {implies} \quad \cccterm{M} = \cccterm{N}  . 
\end{align*}
\end{proposition}
\begin{proof}
We prove the statement in the case of each equation.
We begin each case by unfolding the definition of $\ccc{-}$, unless otherwise specified.
\begin{itemize}
\item Identity (\textsf{id}): 
\begin{align*}
	\cccterm{<?x>.[!x]}( s~|~v) &= \cccterm{[!x]}(\e~|~s,v) \\
	(\textsf{App.})	&= \cccterm{*}(s~|~s,v) \\
	(\textsf{Weak.}) &= 	\cccterm{*}(s~|~v)
\end{align*}
\item Beta ($\beta$): 
\begin{align*}
\cccterm{[V].<x>.`?x}( s~|~v) &= \cccterm{<x>.`?x}( s , \cccterm{V}(v) ~|~v) \\
	(\textsf{defn.}\, \ccc{-}) &=\cccterm{`?x}( s ~|~ \cccterm{V}(v),v) \\
	(\textsf{defn.}\, \ccc{-}) &=\cccterm{x}(\cccterm{V}(v),v) @ s \\
	(\textsf{defn.}\, \ccc{-})  &=\cccterm{V}(v) @ s \\
	(\textsf{defn.}\, \ccc{-}) &= \cccterm{`?V}(s~|~v) 
\end{align*}
\item Eta ($\eta$):
\begin{align*}
	\cccterm{S;<?x>.[`![!x].M]}(\e~|~v)~ (\textsf{Seq.}) &= \cccterm{<?x>.[`![!x].M]}(\cccterm{S}(\e~|~v), v) \\
	(\textsf{Abs.})	 &= \cccterm{[`![!x].M]}(\e~|~ \cccterm{S}(\e~|~v), v) \\
	(\textsf{defn.}\, \ccc{-})	 &= \cccterm{`![!x].M}(\cccterm{S}(\e~|~v), v) \\
	(\textsf{defn.}\, \ccc{-})	 &= \lambda s. \cccterm{[!x].M}(s~|~\cccterm{S}(\e~|~v), v) \\
	(\textsf{App.}\, \ccc{-})	 &= \lambda s. \cccterm{M}(s,\cccterm{S}(\e~|~v)~|~ v) \\
	(\textsf{Seq.})	 &= \lambda s. \cccterm{S;M}(s~|~ v) \\
	(\textsf{defn.}\, \ccc{-})	 &= \cccterm{`!S;M}(v) \\
	(\textsf{defn.}\, \ccc{-})	 &= \cccterm{[`!S;M]}(\e~|~v) \\
\end{align*}
\item Force ($\phi$): 
\begin{align*}
\cccterm{`?`!V}(s~|~v) &= \cccterm{`!V}(v) @ s \\
	(\textsf{defn.}\, \ccc{-}) &= (\lambda s'. \cccterm{V}(s'~|~v)) @ s \\
	&=_\beta \cccterm{V}(s~|~v)
\end{align*}
\item Thunk ($\tau$):
\begin{align*}
	\cccterm{`!`?x}(v) &= \lambda s. \cccterm{`?x}(s~|~v) \\
	(\textsf{defn.}\, \ccc{-}) 	&= \lambda s. \cccterm{x}(v) @ s \\
		&=_\eta \cccterm{x}(v) 
\end{align*}
\item Terminal ($!$): 
\begin{align*}
	\cccterm{S;<?x>}(\e~|~v)~
		(\textsf{Seq.}) &= \cccterm{<?x>}(\cccterm{S}(\e~|~v)~|~v)  \\
		(\textsf{defn.}\, \ccc{-})&= \cccterm{*}(\e~|~\cccterm{S}(\e~|~v),v) \\
		(\textsf{Weak.}) &= \cccterm{*}(\e~|~v) 
\end{align*}
\item Diagonal ($\Delta$): 
\begin{align*}
 \cccterm{S;<?x>.[!x].[!x]}(\e~|~v) ~(\textsf{Seq.}) &=  \cccterm{[!s].[!s]}(\cccterm{S}(\e~|~v)~|~v)  \\
	(\textsf{defn.}\, \ccc{-})  &= \cccterm{[!x].[!x]}(\epsilon~|~\cccterm{S}(\e~|~v),v)  \\
	(\textsf{App.})&= \cccterm{[!x]}(\cccterm{S}(\e~|~v)~|~\cccterm{S}(\e~|~v),v)  \\
	(\textsf{App.})&= \cccterm{*}(\cccterm{S}(\e~|~v),\cccterm{S}(\e~|~v)~|~\cccterm{S}(\e~|~v),v)  \\
	(\textsf{Weak.}) &= \cccterm{*}(\cccterm{S}(\e~|~v), \cccterm{S}(\e~|~v)~|~v) \\
	(\textsf{Seq.}) &= \cccterm{S;*}(\cccterm{S}(\e~|~v)~|~v) \\
	(\textsf{Seq.}) &= \cccterm{S;S}(\e~|~v)
\end{align*}
%
\item 
Interchange ($\iota$): 
\begin{align*}
 \cccterm{M;<?x>.(N;[!x])}({s}~|~v) &= \cccterm{N;[!x]}({s} ~|~ \cccterm{M}({\e~|~v}),v)  \\
	(\textsf{Seq.}) &= \cccterm{[!x]}(\cccterm{N}({s}~|~\underline{\cccterm{M}({\e~|~v})},v)~|~{\cccterm{M}({\e~|~v})},v) \\
	(\textsf{App.})&= \cccterm{*}(\cccterm{N}({s}~|~v) , \cccterm{M}( \e~|~v)~|~\underline{\cccterm{M}(\e~|~v)}, v) \\
	(\textsf{defn.}\, \ccc{-})&= (\cccterm{N}({s}~|~v) , \cccterm{M}( \e~|~v)) \\
(\textsf{defn.}\, \ccc{-})	 &=\cccterm{*}(\cccterm{N}({s}~|~v) , \cccterm{M}(\e~|~v)~|~ v) \\
	(\textsf{Seq.})&= \cccterm{M}(\cccterm{N}({s}~|~v) , v) \\
	(\textsf{Seq.})&= \cccterm{N;M}(s ~|~v) \qedhere
\end{align*}
\end{itemize}
\end{proof}

\begin{corollary}
The interpretation $\ccc{-}: \stermcat \to \textsf{CCC}(\Sigma)$ is a well-defined functor.
\end{corollary}
\begin{proof}
The interpretation preserves equality (\textit{i.e.}, is well-defined) by Propostion \ref{lem:wd}, preserves identity by definition and preserves composition by the Sequencing Lemma \ref{lem:sequencing}.
\end{proof}

\subsection{Equivalence}
In this section we demonstrate completeness of the free functor $\termint{-}: \textsf{CCC}(\Sigma) \to \stermcat{\Sigma}$ by proving that
\[
	\ccc{\termint{\Gamma \vdash M:A}} =_{\beta\eta\pi} \Gamma \vdash M:A,
\]
or, equivalently stated, that the free functor is faithful. Since the free functor sends equipment of the free CCC to the corresponding equipment in $\stermcat{\Sigma}$, the proof of this essentially amounts to showing that the interpretation $\ccc{-}: \stermcat{\Sigma} \to \textsf{CCC}(\Sigma)$ maps the equipment of $\stermcat{\Sigma}$ back to the corresponding equipment in $\textsf{CCC}(\Sigma)$. 

Further, it is then proved that $\termint{-}: \textsf{CCC}(\Sigma) \to \stermcat{\Sigma}$ in fact gives an equivalence of categories, by showing that
\[
	\termint{\ccc{ \term{ M: ?s > !t}}} =_\textsf{ccc} \term{ M: ?s > !t},
\]
for closed computations $\term{M}$, or, equivalently stated, that the free functor also full.

A reference for the two translations is presented in Figure \ref{fig:translations}. 
\begin{figure}
\[
\begin{array}{ll|ll}
M & \termint{M} & 						  \term{M} & \cccterm{M} \\
\hline
x,y,z \vdash y & \term{<?x>.<?y>.<?z>.[!y]}							& \term{*} & s~|~v \vdash s\\
\Pi \vdash c\,@M & \termint{M}\term{;c} 												& \term{c.M} & s,r~|~v \vdash \cccterm{M}(s,c\,@r~|~v) \\			
\Pi \vdash (M,N) & \term{<?x>.([!x].}\termint{M}\term{;[!x].}\termint{N}\term{)} 		& \term{[V].M} & s~|~v \vdash \cccterm{M}(s, \cccterm{V}(v)~|~v) \\					
\Pi \vdash () & \term{<?x>} 												& \term{<x>.M} & s, r~|~v \vdash \cccterm{M}(s~|~r,v)\\						
\Pi \vdash M@N & \term{<?x>.([!x].}\termint{M}\term{;[!x].}\termint{N}\term{);<f>.`?f} 	& \term{`?V.M} & t, r~|~v \vdash \cccterm{M}(t,\cccterm{V}(v)@r~|~v)\\
\Pi \vdash \lambda p.M & \term{<?x>.[`![!x].}\termint{M}\term{]}							& \term{`!M} & v \vdash \lambda s. \cccterm{M}(s~|~v) \\
\Pi \vdash v & \term{<?x>.[v]}								& \term{v} & w \vdash v
\end{array}
\]
\caption{Translations between the Sequential $\lambda$-calculus and the $\lambda$-calculus}
\label{fig:translations}
\end{figure}

\begin{theorem}[Completeness of the Free Functor]\label{thm:faithful}
The functor $\termint{-}: \textup{\textsf{CCC}}(\Sigma) \to \stermcat{\Sigma}$ is faithful.
\end{theorem}
\begin{proof}
We show that $\ccc{\termint{\Gamma \vdash M: A}} =_{\beta\eta\pi} \Gamma \vdash M: A$, or equivalently that 
\begin{align*}
	\ccc{\termint{s \vdash M}}(s'~|~\e) =_{\beta\eta\pi} M\{s'/s\}\ .
\end{align*}
We proceed by induction on the type derivation of $\Gamma \vdash M:A$. In each case (excepting for abstraction), we slightly abuse notation and choose for the fresh free variables $s' = s$ to avoid proliferation of $\alpha$-conversions. Although the $\lambda$-calculus has $n$-ary products, we show that binary and nullary cases and take the rest to be clear. 
 In each case, we begin by unfolding the definition of $\termint{-}$.
\begin{itemize}
\item Variable:
\begin{align*}
\ccc{\termint{z,y,x \vdash y:A}}(z,y,x~|~\epsilon) &= \ccc{\term{<?x>.\<?y>.<?z>.[!y]}}(z,y,x~|~\epsilon) \\
(\textsf{defn.}\, \ccc{-}) 	&= \ccc{\term{<?y>.<?z>.[!y]}}(z,y,x~|~x) \\
(\textsf{defn.}\, \ccc{-}) 	&= \ccc{\term{<?z>.[!y]}}(z~|~y,x) \\
(\textsf{defn.}\, \ccc{-}) 	&= \ccc{\term{[!y]}}(\e~|~z,y,x) \\
(\textsf{Exch.\,+\, App.}) &= y	
\end{align*}
\item Value Constant:
\begin{align*}
	\ccc{\termint{s \vdash v}}(s~|~\e) &= \ccc{\term{<?x>.[v]}}(s~|~\e) \\
	(\textsf{defn.}\, \ccc{-})	&= \ccc{\term{[v]}}(\e~|~s) \\
	(\textsf{defn.}\, \ccc{-})	&= \cccterm{v}(s) \\
	(\textsf{defn.}\, \ccc{-})	&= v \\
\end{align*}
\item Sequential Constant: 
\begin{align*}
	\ccc{\termint{s \vdash c\,@M}}(s~|~\e) &= \ccc{\termint{M}\term{;c}\,}(s~|~\e)  \\
	(\textsf{Seq.})	&= \ccc{\term{c}\,}(\ccc{\termint{M}}(s~|~\e)~|~\e)  \\
	(\textsf{I.H.})	&= \ccc{\term{c}\,}(M~|~\e)  \\
	(\textsf{defn.}\, \ccc{-}) &= c\,@M
\end{align*}
\item Pattern:
\begin{align*}
\ccc{\termint{s, (p, q) \vdash M}}(s,(p,q)~|~\e)  &=	\ccc{\termint{s, p, q \vdash M}}(s,(p,q)~|~\e) \\
&=_{\pi\eta}	\ccc{\termint{s, p, q \vdash M}}(s,\pi_1(p,q),\pi_2(p,q)~|~\e) \\
&=_{\beta}	\ccc{\termint{s, p, q \vdash M}}(s,p,q~|~\e) \\
(\textsf{I.H}) &= M 
\end{align*}
\item Binary Product:
\begin{align*}
	 \ccc{\termint{s \vdash (M, N)}}({s}~|~\e)  &= \ccc{\term{<?x>.([!x].}\termint{M}\term{;[!x].}\termint{N}\term{)}} (s~|~\e) \\
	  (\textsf{defn.}\, \ccc{-})  &=\ccc{\term{[!x].}\termint{M}\term{;[!x].}\termint{N}}({\epsilon}~|~s) \\
	 (\textsf{App.}) &=\ccc{\termint{M}\term{;[!x].}\termint{N}}(s~|~{s}) \\
	(\textsf{Seq.}) &=\ccc{\term{[!x].}\termint{N}} (\ccc{\termint{M}}({s}~|~s)~|~s) \\
	(\textsf{Weak.}) &=\ccc{\term{[!x].}\termint{N}} (\ccc{\termint{M}}({s}~|~\e)~|~s) \\
	(\textsf{I.H.}) &=\ccc{\term{[!x].}\termint{ N}}(M~|~s)  \\
	(\textsf{App.}) &=\ccc{\termint{N}} (M , s ~|~s)  \\
	(\textsf{Seq.})&=\cccterm{*}(M , \ccc{\termint{N}}(s~|~s)~|~s) \\
	(\textsf{defn.}\, \ccc{-}) &=(M , \ccc{\termint{N}}(s~|~s)) \\
	(\textsf{Weak.})&=(M , \ccc{\termint{N}}(s~|~\e)) \\
	(\textsf{I.H.})&= (M, N) 
\end{align*}
\item Nullary Product:
\begin{align*}
	\ccc{\termint{s \vdash ()}}(s~|~\e)&= \ccc{\term{<?x>}}(s~|~\e) \\
	(\textsf{defn.}\, \ccc{-}) 	&= \ccc{\term{*}}(\e~|~s) 	\\
	(\textsf{defn.}\, \ccc{-}) 	&= ()  
\end{align*}
\item Application:
\begin{align*}
	\ccc{\termint{{s: \Gamma} \vdash M @ N: B}}(s~|~\e) &= \ccc{\term{<?x>.([!x].}\termint{ N}\term{;[!x].}\termint{ M }\term{);<f>.`?f}}(s~|~\e) 
\\	(\textsf{Seq.}) &= \cccterm{<f>.`?f}(\ccc{\term{<?x>.([!x].}\termint{ N}\term{;[!x].}\termint{ M }\term{)}}(s~|~\e)~|~\e) 
\\	(\textsf{See Bin. Prod. Case}) &= \cccterm{<f>.`?f}(N,M~|~\e) 
\\	(\textsf{defn.}\, \ccc{-})&= \ccc{\term{`?f}}( N ~|~M )
\\	(\textsf{defn.}\, \ccc{-}) &= \ccc{\term{f}}(M) @ N
\\	(\textsf{defn.}\, \ccc{-})&= M @ N
\end{align*}
\item Abstraction:
\begin{align*}
	\ccc{\termint{{s} \vdash \lambda p.M}}({s}~|~\epsilon) &=	 \ccc{\term{<?x>.[`![!x].}\termint{ M}\term{]}} ({s}~|~\e)   \\
	(\textsf{defn.}\, \ccc{-})	&=\ccc{\term{[`![!x].}\termint{ M}\term{]}} (\epsilon~|~s) \\
	(\textsf{defn.}\, \ccc{-})	&=\ccc{\term{`![!x].}\termint{ M}}(s) \\
	(\textsf{defn.}\, \ccc{-})	&=\lambda p. \ccc{\term{[!x].}\termint{ M\term{}}}(p~|~s) \\
	(\textsf{App.})	 &=\lambda p.\ccc{\termint{ M}\term{}}(p,s~|~s) \\
	(\textsf{Weak.\,+\,I.H.})	&=\lambda p.M \qedhere
\end{align*}
\end{itemize}
\end{proof}

\begin{theorem}[Fullness of the Free Functor]\label{thm:full}
The functor $\termint{-}: \textup{\textsf{CCC}}(\Sigma) \to \stermcat{\Sigma}$ is full.
\end{theorem}
\begin{proof}
We require  that for closed terms $\term{M: ?s > !t}$, that $\termint{\ccc{ \term{ M}}} =_{\textsf{eqn}} \term{ M}$. We actually prove a stronger statement: for all (open) terms $\Gamma \cvdash \term{M: ?s > !t}$
\begin{align*}
	\termint{s: \ccc{\type{!s}} ~|~ v: \ccc{\Gamma} \vdash \ccc{\term{M}}(s~|~v): \ccc{\type{!t}}} &=_{\textsf{eqn}} \term{<?v>.M: ?v?s > !t}\\
	\termint{v: \ccc{\Gamma} \vdash \ccc{\term{V}}(v): \ccc{\type{t}}} &=_{\textsf{eqn}} \term{<?v>.[V]: ?v > t}.
\end{align*}
where $\term{!v}: \Gamma$
We proceed by mutual induction on the type derivations of $\term{M}$ and $\term{V}$. In each case, we begin by unfolding the definition of $\ccc{-}$. 
We will make use of the interpretation of the cut rule from Remark \ref{rem:translate-cut}.
\begin{itemize}
\item Variable: 
\begin{align*}
		\termint{z,y,x \vdash \cccterm{y}(z, y, x)} &=\termint{z,y,x \vdash y} \\
		(\textsf{defn.} ~\termint{-})  &=\term{<?x>.<y>.<?z>.[y]}
\end{align*}
\item Thunk: 
\begin{align*}
	\termint{v \vdash \cccterm{`!M}(v)}  &=\termint{v \vdash \lambda s. \cccterm{M}(s~|~v) }   \\
	(\textsf{defn.} \termint{-}) &=	\term{<?v>.[`![!v].}\termint{s, v \vdash \cccterm{M}(s~|~v)}\term{]} \\
	 (\textsf{I.H.})&=	\term{<?v>.[`!}\uterm{[!v].<?w>}\term{.\{!w/!v\}M]}\\ 
	&=_\beta	 \term{<?v>.[`!M]} \qedhere
\end{align*}
\item Value Constant:
\begin{align*}
	\termint{w \vvdash \cccterm{v}(w)} &= \termint{w \vvdash v} \\
		&= \term{<?w>.[v]}
\end{align*}
\item Identity: 
	\begin{align*}
	\termint{s,v \vdash \cccterm{*}(s~|~v)} &=	\termint{s,v \vdash s} \\
		(\textsf{defn.} ~\termint{-})&=	\term{<?v>.}\uterm{<?x>.[!x]} \\
		&=_{\textsf{id}}	\term{<?v>.*}
	\end{align*}
\item Sequential Constant:
\begin{align*}
	\termint{\cccterm{c.M}(s,r~|~v)}  &= \termint{\cccterm{M}(s,c\,@r~|~v)} \\
	(\textsf{Weak.\,+\,Cut.})	&= \term{<?v?w?x>.[!x].[!x!w!v].}\termint{s,r,v \vdash c@r}\term{.[!w!v].}\cccterm{M}(s,t~|~r,v)\\
	(\textsf{I.H}) &= \term{<?v?w?x>.[!x].[!x!w!v].}\termint{s,r,v \vdash c@r}\term{.}\uterm{[!r!v].<?v'?w'>}\term{.\{!v'/!v\}M} \\
		 &=_\beta \term{<?v?w?x>.[!x].[!x!w!v].}\termint{s,r,v \vdash c@r}\term{.}\term{M} \\
	(\textsf{defn.} ~\termint{-})	 &= \term{<?v?x>.[!x].[!x!w!v].}\termint{s,r,v \vdash r}\term{.c.}\term{M} \\
	(\textsf{defn.} ~\termint{-})	 &= \term{<?v?x>.[!x].}\uterm{[!x!w!v].<?v'?w'?x'>}\term{.[!w']}\term{.c.}\term{M} \\
		 &=_\beta \term{<?v?w?x>.[!x!w]}\term{.c.}\term{M} \\
		 &=_\ide \term{<?v>}\term{.c.}\term{M} 
\end{align*}
\item Application:
	\begin{align*}
		\termint{\cccterm{[V].M}(s~|~v)} &= \termint{\cccterm{M}(s, \cccterm{V}(v)~|~v)}\\
	(\textsf{Weak.\,+\,Cut.})	&= \term{<?v?x>.[!x].[!x!v].}\termint{\cccterm{V}(s,v)}\term{.[!v].}\cccterm{M}(s,t~|~v)\\
	(\textsf{I.H.})	&= \term{<?v?x>.[!x].}\uterm{[!x!v].<?v'?x'>}\term{.[\{v'/v\}V]}\term{.[!v].}\cccterm{M}(s,t~|~v)\\
		&=_\beta \term{<?v?x>.[!x].[V]}\term{.[!v].}\cccterm{M}(s,t~|~v)\\
	(\textsf{I.H.})	&= \term{<?v?x>.[!x].[V].}\uterm{[!v].<?v'\,'>}\term{.\{!v'\,'/!v\}M}\\
		&=_\beta \term{<?v?x>.[!x].[V]}\term{.M}\\
		&=_\ide \term{<?v>.[V]}\term{.M}
	\end{align*}
\item Abstraction: 
\begin{align*}
		\termint{s, r, v \vdash \cccterm{<x>.M}(s,r~|~ v)} &= \termint{s,r,v \vdash \cccterm{M}(s~|~r,v)} \\
		 (\textsf{I.H.}) &= \term{<?v>.<x>.M}
\end{align*}
\item Sequential Execution:
\begin{align*}
	\termint{\cccterm{`?V.M}(s,r~|~v)} &= \termint{\cccterm{M}(s, \cccterm{V}(v)@r~|~v)}\\
	(\textsf{Weak.\,+\,Cut.}) &=  \term{<?v?w?x>.[!x].[!x!w!v].}\termint{s,r,v \vdash \cccterm{V}(v)@r}\term{.[!w!v].}\cccterm{M}(s,t~|~r,v) \\
	(\textsf{I.H.}) &=  \term{<?v?w?x>.[!x].[!x!w!v].}\termint{s,r,v \vdash \cccterm{V}(v)@r}\term{.}\uterm{[!w!v].<?v'?r'>}\term{.\{!v'/!v\}M}\\
	 &=_\beta  \term{<?v?w?x>.[!x].[!x!w!v].}\termint{s,r,v \vdash \cccterm{V}(v)@r}\term{M}\\
	(\textsf{defn.} ~\termint{-}) &=  \term{<?v?w?x>.[!x].}\uterm{[!x!w!v].<?v'?w'?x'>}\term{.}\\
		& \qquad \quad \term{[!x'!w'!v'].}\termint{s,r,v \vdash r}\term{.[!x'!w'!v'].}\termint{s,r,v \vdash \cccterm{V}(v)}\term{M}\\
	 &=_\beta  \term{<?v?w?x>.[!x].[!x!w!v].}\termint{s,r,v \vdash r}\term{.[!x!w!v].}\termint{s,r,v \vdash \cccterm{V}(v)}\term{M} \\
	 (\textsf{defn.} ~\termint{-}) &=  \term{<?v?w?x>.[!x].}\uterm{[!x!w!v].<?v'\,'?w'\,'?x'\,'>}\term{.[!w'\,']}\term{.[!x!w!v].}\\ & \qquad\quad \termint{s,r,v \vdash \cccterm{V}(v)}\term{M} \\
	 &=_\beta  \term{<?v?w?x>.[!x].[!w].}\term{[!x!w!v].}\termint{s,r,v \vdash \cccterm{V}(v)}\term{M} \\
	(\textsf{I.H.}) &=  \term{<?v?w?x>.[!x].[!w].}\uterm{[!x!w!v].<?v'?w'?x'>}\term{.[\{v'/v\}V].}\term{M} \\
	 &=_\beta  \term{<?v?w?x>.[!x].[!w].}\term{[V].}\term{M} \\
	 &=_\ide  \term{<?v>}\term{.[V].}\term{M} 
\end{align*}
\end{itemize}
\end{proof}

\begin{corollary}\label{cor:stlc-equiv}
The categories $\stermcat{\Sigma}$ and $\textup{{\textsf{CCC}}}(\Sigma)$ are equivalent. 
\end{corollary}
\begin{proof}
We have shown that the free Cartesian closed functor $\termint{-}: \textsf{CCC}(\Sigma) \to \stermcat{\Sigma}$ is fully faithful (Theorems \ref{thm:faithful} and \ref{thm:full}), and it is clearly surjective on objects, thus we have an equivalence of categories.
\end{proof}
Indeed, we have that $\ccc{-}: \stermcat{\Sigma} \to \textsf{CCC}(\Sigma) $ is an inverse functor to $\termint{-}: \textsf{CCC}(\Sigma) \to \stermcat{\Sigma}$.
Note that it also follows that the interpretation $\ccc{-}$ is in fact a CCC-functor. Additionally, we note that, while the pattern $\lambda$-calculus is non-strict, we can consider instead the strict CCC \cite{MacLane}, as discused in Chapter \ref{chapter:cat}. The two functors described are indeed strict, in the sense that they map associators and unitors to the identity, and strictly preserve products and exponentials. 

The result above immediately gives us that the  first-order \SLCV, with an equational theory generated by the first-order equations, is a sound and fully complete language for \textit{string diagams} \cite{JOYAL199155}, in the way illustrated in Section \ref{sec:eqn-theory} of this chapter.
To be precise, the category of   first-order \SLCV-terms can be defined similarly to $\stermcat{\Sigma}$, and the same argument as made in this section (omitting the cases with higher-order constructs), then gives us that this is in fact the free Cartesian category.  Being cautious of Remark \ref{rem:sym-in-linear}, the above results should also extend easily to the \textit{linear} (higher-order) \SLCV\  and the free symmetric monoidal (closed) category.

\section{The \FMCV\  is Equivalent to the \SLCV}
Here, it is proved that the category of \SLCV-terms, which we now know to be the free CCC, is equivalent to the category of \FMCV-terms.
The structure of this section is as follows.  

We will give an \textit{embedding} $\embedint{-}_a$ of the \SLCV \ into the \FMCV, parameterized by the location $\term a \in A$ on which it is embedded. The inverse interpretation $\collapint{-}^<$ will be given by an extension to terms of the \textit{collapse} of a memory into a single stack. This is done by fixing an order $<$ on locations, and then concatenating the various stacks of the \FMCV into one large stack, in that order. This action on types must then be lifted to terms in an appropriate way, and the two functors arising from embedding and collapsing must be shown to be inverse. 
Although the roundtrip given by embedding stacks into memories and then collapsing memories into stacks is an obvious identity, the reverse composition is only equivalent to the identity functor up-to natural isomorphism. Thus, we require to prove that there exists a natural isomorphism $\kappa$ satisfying the following commutative diagram. 
\[
\scalebox{1.1}
{\begin{tikzcd}[ampersand replacement=\&]
	{\type{!s_A}} \&\& {\type{!t_A}} \\
	{\embedint{\collapint{\type{!s_A}}}} \&\& {\embedint{\collapint{\type{!t_A}}}}
	\arrow["{\term{M}}", from=1-1, to=1-3]
	\arrow["{\kappa}"', from=1-1, to=2-1]
	\arrow["{\kappa}", from=1-3, to=2-3]
	\arrow["{\embedint{\collapint{\term{M}}}}"', from=2-1, to=2-3]
\end{tikzcd}
}
\]
This isomorphism will be exhibited in the following subsection. It is given on base types by the relocation isomorphism, but requires an appropriate generalization for higher-order types.

\subsection{Embedding: \SLCV\  to \FMCV}
Recall that an \SLCV-signature is an \FMCV-signature with just this one \textit{main} location. Note, there is nothing special about the main location $\term \lambda$, except that we single it out by convention.
Recall also that we will freely confuse stack types and memory types generated over a single location, which are clearly isomorphic, \textit{i.e.}, the \SLCV \ is considered to have no memory and just a single stack.
\begin{definition}
Given an \SLCV-signature $\Sigma$, define the category $\stermcat{\Sigma}$ to be $\termcat{\Sigma_A}$, where $A = \{\term \lambda\}$.
\end{definition}

In order to relate the category of \FMCV-terms and \SLCV-terms, we first deal with the fact they are generated over different signatures. We will collapse the types of constants in the \FMCV-signature, which is generated over a set of locations $A$, into types of the \SLCV-signature, which is generated over a single location $\term{\lambda}$. Throughout this section, we fix $A =\{\term{a_1},\ldots,\term{a_m}\}$.  

Consider that an indexed product (\textit{i.e.}, a memory) is isomorphic to a standard product (\textit{i.e.}, a stack) by fixing a choice of ordering on its indices. 
Following this idea, we now define the \textit{collapse} of \FMCV-types into \SLCV-types.\footnote{Note that a significant amount of extra work goes into some of the definitions and proofs of this section in order to deal with arbitrary !FMC-signatures. A simpler presentation would be possible if we were to restrict the !FMC of this section to be generated by a signature with constants acting exclusively on the main location.}
We will later extend this definition to act on terms, which will then forms the desired equivalence.
 
\begin{definition}
Given a strict order $<$ on the set of locations $A =\{\term{a_1},\term{a_2}, \term{a_3}, \ldots\}$, such that $\term{a_i}<\term{a_j}$ when $i<j$, define the \textit{collapse interpretation} $\collapint{-}^<$ of value, stack and memory \FMCV-types generated over $A$ into \SLCV-types (generated over $\{\term \lambda\}$), respectively, by induction.
\[
	\begin{aligned}{}
	\collapint{\type{a}}^<&= \type{a}\\
	\collapint{\type{?s_A >!t_A}}^< &= \collapint{\type{?s_A}}^< \type{>} \collapint{\type{!t_A}}^<
	\end{aligned}
	\qquad
	\begin{aligned}{}
	\collapint{\type{t_1\ldots t_n}}^< &= \collapint{\type{t_1}}\type{\ldots} \collapint{\type{t_n}}^< \\
	\collapint{\type{!t_A}}^< &= \collapint{\type{!t_{a_1}}}\type{\ldots} \collapint{\type{!t_{a_m}}}^<
	\end{aligned}
\]
We will fix the order $<$ throughout this section, and omit it on $\collapint{-}^<$, writing just $\collapint{-}$.
\end{definition}
\begin{definition}
Given an \FMCV-signature $\Sigma_A = (\Sigma_0, \Sigma_c, \Sigma_v, \textsf{dom}, \textsf{cod}, \textsf{val})$, define its \textit{collapse} $\collap{\Sigma}$ as $(\Sigma_0, \Sigma_c, \Sigma_v, \collap{\textsf{dom}}, \collap{\textsf{cod}}, \collap{\textsf{val}})$, where $\collap{\textsf{dom}}(\term{c}) = \collapint{\textsf{dom}(\term{c})}$, $\collap{\textsf{cod}}(\term{c}) = \collapint{\textsf{cod}(\term{c})}$ and $\collap{\textsf{val}}(\term{v}) = \collapint{\textsf{val}(\term{v})}$. We will write $\collap{\term{c}}$ and $\collap{\term{v}}$ to distinguish the constant symbols of $\collap{\Sigma_A}$ and the constant symbols  $\term{c}$ and $\term{v}$ of $\Sigma_A$. 
\end{definition}

The aim, then,  is to prove that $\termcat{\Sigma_A} \simeq \stermcat{\collap{\Sigma_A}}$.
The desired equivalence of categories will indeed be up-to-isomorphism, and we exhibit (what will be proved to be) the relevant natural isomorphism $\kappa$ in the following definition. 
This will be given at base types by simple relocation, but must be appropriately lifted to higher-order types: we cannot simply return the variable since it has type $\type{?s_A > !t_A}$ and not type $\collapse{\type{?s_A > !t_A}}$ (in the case of $\kappa$, and vice-versa for $\kappa^{-1}$),  hence the conjugation by appropriate isomorphisms of smaller type. This then generalizes from singleton types to memories of any size in the obvious way.

\begin{definition}\label{def:isos}
For each memory type $\type{!t_A}$, we define terms 
\[
	\kappa(\type{!t_A}):  \type{?t_A} \type{>} \collapint{\type{!t_A}} \qquad  \kappa^{-1}(\type{!t_A}): \collapint{\type{?t_A}} \type{>} \type{!t_A}\, ,
\]
and \textit{iterated collapse} and \textit{iterated embed} functions $\collap{-}^*$ and $\embed{-}^*$,
 on (families of) variables, for each memory type, by mutual induction on types\footnote{Note that a type has the same size as its collapse.}. 
\[
\begin{array}{rlrl}
\kappa(\type{!t_A}) &=& \term{<?x_A>.[}\collap{{\term{!x_A}}}^*\term{]}~:& \type{?t_A} \type{>} \collapint{\type{!t_A}}\\ 
\collap{\term{!x_A: t_A}}^* &=& \collap{\term{!x_{a_m}}}_p^* \ldots \collap{\term{!x_{a_1}}}_p^* ~:& \collapint{\type{!t_A}} \\
\collap{\term{x: ?r_{\!A} > !u_{\!A}}}^* &=& \term{`!}\kappa^{-1}(\type{!r_{\!A}})\term{;`?x.}\kappa(\type{!u_{\!A}})~:& \collapint{\type{?s'_{\!A} > !t'_{\!A}}} \\
\collap{\term{x:a}}^* &=& \term{x}~:& \collapint{\type{a}}\\
\\
\kappa^{-1}(\type{!t_A}) &=&\term{<}\term{?x}\term{>.[}\embed{\term{!x}}^*\term{]} ~:&\collapint{\type{?t_A}} \type{>} \type{!t_A}\\
\embed{\term{!x_{a_m}} \ldots \term{!x_{a_1}}: \collapint{\type{!t_A}}}^* &=&  \{\embed{\term{!x_{b}}}_p\,^{\!*}~\mid~ b \in A\} ~:& \type{!t_A} \\ 
\embed{\term{x:} \collapint{\type{?r_{\!A} > !u_{\!A}}}}^* &=& \term{`!}\kappa(\type{!r_{\!A}})\term{;`?x.}\kappa^{-1}(\type{!u_{\!A}})~:& \type{?s'_{\!A} > !t'_{\!A}} \\
\embed{\term{x:}\collapint{\type{a}}}^* &=& \term{x}~:& {\type{a}}
\end{array}
\]
where $\collap{-}_p^*$ and $\embed{-}_p\,^{\!*}$ are the pointwise application of $\collap{-}^*$ and $\embed{-}^*$ on stacks, respectively, and we take $ \type{?s'_{\!A} > !t'_{\!A}}$ to be of smaller type than $\type{!t_A}$. 
We will omit the type associated with $\kappa(\type{!t_A})$ and $\kappa^{-1}(\type{!t_A})$ when it is clear. 
Note, the action of $\embed{-}^*$ on stacks is determined by $\type{!t_A}$.
\end{definition}
\begin{lemma}[Isomorphism]
For each type, $\kappa$ and $\kappa^{-1}$ are inverse to each other, as are $\embed{-}^*$ and $\collap{-}^*$.
\end{lemma}
\begin{proof}
Proceed by mutual induction on types. Note, it is clear that  $\embed{-}^*$ and $\collap{-}^*$ are inverses. We can then calculate the following.
\[
\kappa\term{;}\kappa^{-1} = \term{<?x_A>.[}\collap{{\term{!x_A}}}^*\term{].}\term{<}\term{?x}\term{>.[}\embed{\term{!x}}^*\term{]} = \term{<?x_A>.[}\embed{\collapterm{!x_A}^*}^*\term{]} = \term{<?x_A>.[!x_A]} = \term{*}
\]
\[
\kappa^{-1}\term{;}\kappa = \term{<}\term{?x}\term{>.[}\embed{\term{!x}}^*\term{].}\term{<?x_A>.[}\collap{{\term{!x_A}}}^*\term{]} = \term{<?x>.[}\collap{\embed{\term{!x}}^*}^*\term{]} = \term{<?x>.[!x]} = \term{*}\qedhere
\]
\end{proof}
Note that the definition of $\kappa$ relies on the strength of simple types: we need to know the size of the stacks in memory. 

The embedding of the \SLCV\  into the \FMCV\  is given by the obvious inclusion of terms, except for the case of sequential constants. Recalling the aim is to prove $
 \stermcat{\collap{\Sigma_A}} \simeq	\termcat{\Sigma_A}$, constants from the collapsed signature $\collap{\Sigma_A}$ must thus be mapped to those of the \FMCV\  signature $\Sigma_A$, and this is dealt with using the isomorphisms defined above.

Note that, similar to the previous section, the inputs to the following function are terms to be substituted for free variables, \textit{i.e.}, this is another way to express a valuation. 
The commas appearing in the inputs just represent concatenation, and could be omitted. This convention will appear throughout this section.
\begin{definition}
We define the \textit{embedding} $\embedint{-}_a$ of \SLCV-types into \FMCV-types as 
\[
	\begin{aligned}{}
	\embedint{\type{a}}_a&= \type{a}\\
	\embedint{\type{?s >!t}}_a &= \embedint{\type{?s}}_a \type{>} \embedint{\type{!t}}_a
	\end{aligned}
	\qquad
	\begin{aligned}{}
	\embedint{\type{t_1\ldots t_n}}_a &= \type{`a}(\embedint{\type{t_1}}_a \type{\ldots} \embedint{\type{t_n}}_a) \\
	\embedint{\type{t_1}, \ldots, \type{t_n}}_a &= \embedint{\type{t_1}}_a, \ldots, \embedint{\type{t_n}}_a
	\end{aligned}
\]
and the embedding of \SLCV-terms over $\collap{\Sigma_A}$ into \FMCV-terms over $\Sigma_A$,
\begin{align*}
	 \embedint{\Gamma}_a &\cvdash \embedint{\Gamma \cvdash \term{M: ?s > !t}}_a: \embedint{\type{?s}}_a \type{>} \embedint{\type{!t}}_a\\
	 \embedint{\Gamma}_a &\vvdash \embedint{\Gamma \cvdash \term{V: t}}_a: \embedint{\type{t}}_a
\end{align*}	
is given by mutual induction on type derivations of computations and values as follows.
\begin{align*}
\embedint{\Gamma, \term{x:t}, \Delta \vvdash \term{x: t}}_a(\term{!v, y, !w}) &= \term{y} \\
\embedint{\Gamma \vvdash \collap{\term{v}}: \type{a}}_a(\term{!v'})  &= {\term v} \\
\embedint{\Gamma \vvdash \collap{\term{v}}: \type{?s > !t}}_a(\term{!v'})  &= \term{`!}\kappa^{-1}{\term{;`? v.}}\kappa \\
\embedint{\Gamma \vvdash \term{`!M: ?s > !t}}_a(\term{!v})  &= \term{`!}\embedint{\Gamma \vvdash \term{M: ?s > !t}}_a(\term{!v}) \\
\embedint{\Gamma \cvdash \term{*:?t>!t}}_a(\term{!v})  &= \term{*} \\
\embedint{\Gamma \cvdash \term{`?V.M: ?r?s > !u}}_a(\term{!v}) &= \term{`?}\embedint{\Gamma \vvdash \term{V: ?r > !s}}_a(\term{!v}) \term{.}\embedint{\Gamma \cvdash \term{M: ?s?t > !u}}_a(\term{!v}) \\
\embedint{\Gamma \cvdash \term{<x>.M: r?s > !t}}_a(\term{!v}) &= \term{a<x'>.}\embedint{\term{x:r}, \Gamma \cvdash \term{M: ?s > !t}}_a(\term{x',!v}) \\
\embedint{\Gamma \cvdash \term{[V].M: ?s > !t}}_a(\term{!v})  &= \term{a[}\embedint{\Gamma \vvdash \term{V:r}}(\term{!v}) \term{].}\embedint{\Gamma \cvdash \term{M: ?r?s > !t}}_a(\term{!v})  \\
\embedint{\Gamma \cvdash \collapterm{c}\term{.M: ?r?t > !u}}_a(\term{!v})  &= \kappa^{-1}(\type{!r_A})\term{.}{\term{c}}\term{;}\kappa(\type{!s_A})\term{;}\embedint{\Gamma \cvdash \term{M: ?s?t > !u}}_a(\term{!v})  
\end{align*}
In each case, we have $\term{!v, !v':} \embedint{\Gamma}$, $\term{!w: } \embedint{\Delta}$,  $\term{y}: \embedint{\type{t}}$ and $\term{x'}: \termint{\type{r}}$.
We will often omit the input, or part of the input, when the meaning is unambiguous, especially when it is unchanging throughout a proof.
We omit the context and/or types of terms inside the interpretation function when it is clear. 
From now on, without loss of generality, we will choose to embed into the \textit{main} location, \textit{i.e.}, let $\term{a} = \term{\lambda}$ of the \FMCV, which will be omitted from terms. Similarly, we will omit the location $\term{a}$ from the embedding $\termint{-}_a$, which we write as simply $\termint{-}$.
\end{definition}
\begin{theorem}\label{lem:embed-faithful}
The category $\stermcat{\collap{\Sigma_A}}$ is a Cartesian closed category and 
the embedding $\embedint{-}_a: \stermcat{\collap{\Sigma_A}} \to \termcat{\Sigma_A}$ is a  CCC-functor. 
\end{theorem}
\begin{proof}
That $\stermcat{\Sigma_A}$ is a CCC is immediate from  Theorem \ref{thm:ccc}, applied in  the case of $A = \{\term \lambda\}$.
It is also immediate from the definition of $\stermcat{\Sigma_A}$ that the unit type, identity, products and exponentials are all preserved by embedding. Further, equality of terms is immediately preserved, as constants are not mentioned in any of the unit equations. 

We must further show that sequencing (composition) in preserved, \textit{i.e.}, that $\embedint{\term{M;N}} = \embedintterm{M}\term{;}\embedintterm{N}$. 
We proceed by induction on the type derivation of $\term{N}$, in the manner of Lemma \ref{lem:sequencing}, from the previous subssection. All cases are immediate here, though, except for the case of the sequential constant, since $\embedint{-}$ is simply inclusion of terms. The case of the sequential constant is detailed below. 
\begin{align*}
	\embedintterm{(c.M);N} &= \embedintterm{c.(M;N)}\\
	(\textsf{defn.}\, \embedint{-})	&= \kappa^{-1}\term{;c.}\kappa\term{.}\embedintterm{M;N} \\
	(\textsf{I.H.})	 &= \kappa^{-1}\term{;c.}\kappa\term{.}\embedintterm{M}\term{;}\embedintterm{N} \\
	(\textsf{defn.}\, \embedint{-}) &= \embedintterm{c.M}\term{;}\embedintterm{N} 
\end{align*}
 Overall, this gives that the embedding is a well-defined CCC-functor. 
\end{proof}
\begin{corollary}
The embedding $\embedint{-}_a$ maps the exponent $(\to, \epsilon)$ of the $\stermcat{\collap{\Sigma_A}}$ to $(\xrightarrow{a}, \epsilon^a)$ of $\termcat{\Sigma_A}$. 
\end{corollary}

In the next section, we construct what will prove to be the inverse functor. 

\subsection{Collapsing the Memory: \FMCV\  to \SLCV}

The definition of $\collapint{-}$, which is defined on \FMCV-types, is now extended to \FMCV-terms. 
We first extend this to our notation for families of variables. 
\begin{notation}
Given a memory $S_A = \{S_a ~|~ a \in A\}$, family of variables  $\term{!x_A} = \{\term{!x_{a_i}}~|~a_i \in A\}$ and memory type $\type{!t_A} = \{ \type{!t_{a_i}}~|~a_i \in A\}$, we use the following notation. 
\[
	\collap{S_A}= {S_{a_1}}\!\cdots {S_{a_m}} \qquad \collap{\term{!x_A}} = \term{!x_{a_1}\!\ldots !x_{a_m}} \qquad \collap{\type{!t_A}} = \type{!t_{a_1}\ldots !t_{a_m}}
\]
We will then use the following notation for applications and abstractions.
\[
	\pushsterm{S_A}\term{.M} = \term{[}{\term{S_{a_1}}} \term{\ldots}\, {\term{S_{a_m}}}\term{].M} \qquad \popsterm{?x_A}\term{M} = \term{<?x_{a_m} \ldots ?x_{a_1}>.M}\, .
\]	
{Note that, in the latter case, we reverse the vector $\collapterm{!x_A}$, rather than reversing $\term{!x_A}$ and \textit{then} applying $\collap{-}$. }
\end{notation}

The following \textit{collapse interpretation} can be understood as follows. 
The interpretation of application (and abstraction) must be given by an application (abstraction), except instead of acting on the head of one of the many stacks, they must now act on the same value, but which may now be somewhere deep in the single stack. We interpret application then as a sequence of abstractions which pick up the entire stack, followed by a sequence of applications, which replaces the stack, but with an extra application inserting the new value in the appropriate in the stack.  Similarly, an abstraction is interpreted as a sequence of abstractions which pick up the whole stack, followed by a sequence of applications, which replaces all but the variable to be bound. This variable is then the only one which remains free in the remaining term. 

The interpretation of sequencing $\term{N;M}$ of $\term{N: ?r_A > !u_A}$ and $\term{M: ?u_A?s_A > !t_A}$ is given below. This also illustrates how the cases of sequential constant and sequential execution are interpreted.
\[
\scalebox{1.2}{
\input{collapse sequence.tikz}
}
\vspace{-\baselineskip}
\vspace{-\baselineskip}
\vspace{-\baselineskip}
\]
Note that, unlike with the nominal string diagrams we used earlier, here coloured wires indicate what \textit{used to be} different locations, which have now been collapsed to a single stack. Note that the input $\collap{\type{!s_A!r_A}} = \type{!s_{a_m}!r_{a_m}\ldots !s_{a_1}!r_{a_1}}$. 
We will make the interpretation of sequencing formal in the next subsection, but give it here for illustrative purposes. 

\begin{definition}
 Define the \textit{collapse interpretation} $\collapint{-}$ on contexts as 
$
	\collapint{\type{t_1}, \ldots, \type{t_n}} = \collapint{\type{t_1}}, \ldots, \collapint{\type{t_n}}
$
and the \textit{collapse} of \FMCV-terms over $\Sigma_A$ into \SLCV-terms over $\collap{\Sigma_A}$
\begin{align*}
	  \collapint{\Gamma} &\cvdash \collapint{\Gamma \cvdash \term{M: ?s_A > !t_A}}:  \collapint{\type{?s_A}} \type{>} \collapint{\type{!t_A}}\\
	  \collapint{\Gamma} &\vvdash \collapint{\Gamma \cvdash \term{V: t}}: \collapint{\type{t}}
\end{align*}	
by mutual induction over type derivations of computatons and values, as follows:
\begin{align*}
	\collapint{\Gamma, \term{x:t}, \Delta \vvdash \term{x: t}}( \term{!u,  y,  !v} ) &= \term y\\	
	\collapint{\Gamma \vvdash \term{v:t}}(\,\term{!w}\,) &= \collap{\term{v}} \\
	\collapint{\Gamma \vvdash \term{`!M: ?s_A > !t_A}}(\term{!v})&= \term{`!}\collapint{\Gamma \cvdash \term{M: ?s_A > !t_A}}(\term{!v})\\
	\collapint{\Gamma \cvdash \term{*: ?s_A > !s_A}}(\term{!v}) &= \term{*}\\
	\collapint{\Gamma \cvdash \term{c.M: ?r_A?s_A > !s_A}}(\term{!v}) &=  \popsterm{?x_A?y_A}\pushsterm{!x_A}\term{.}\collapterm{c}\term{.}\term{}\popsterm{?z_A}\pushsterm{!y_A!z_A}\term{.}\\	& \qquad\quad \collapint{\Gamma \cvdash \term{M: ?r_A?s_A > !t_A}}(\term{!v})\\
	\collapint{\Gamma \cvdash \term{[V]a_{n}.M: ?s_A > !t_A}}(\term{!v}) &= \popsterm{?x_A}\pushsterm{!x_Aa_n(y)}\term{.}\collapint{\Gamma \cvdash \term{M: `a_n(r)?s_A > !t_A}}(\term{!v})\\
	\collapint{\Gamma \cvdash \term{a_n<x>.M: `a_{n}(r)?s_A > !t_A}}(\term{!v}) 
	&= \term{<}\collapterm{a_n(x')?y_A}\term{>.[}\collapterm{!y_A}\term{].}\collapint{\term{x}, \Gamma \cvdash \term{M: ?s_A > !t_A}}(\term{x',\!!v})\\
	\collapint{\Gamma \cvdash \term{`?V.M: ?r_A?s_A > !t_A}}(\term{!v}) &= \popsterm{?x_A?y_A}\pushsterm{!x_A}\term{.`?}\collapint{\Gamma \vvdash \term{V: ?r_A > !u_A}}(\term{!v})\term{.}\\		 
	&\qquad \quad\popsterm{?z_A}\pushsterm{!y_A!z_A}\term{.}\collapint{\Gamma \cvdash \term{M: ?u_A?s_A > !t_A}}(\term{!v}) 
\end{align*}
where, in the application case, $\term{y} = \collapint{\Gamma \vvdash \term{V: r}}(\term{!v})$, and throughout $\term{!v:} \Gamma$ and $\term{!w:} \Delta$.
We further have $\collap{\term{!y_A!z_A}}: \collapint{\type{!s_A!r_A}}, \collap{\term{!x_A}}: \collapint{\type{!s_A}}, \collap{\term{!y_A}}: \collapint{\type{!s_A}}$ and $\collap{\term{!y_A!z_A}}: \collapint{\type{!s_A!r_A}}$ in, respectively, the sequential constant, application, abstraction and sequential execution cases. 
We will sometimes omit the input, or part of the input, when the meaning is unambiguous, especially when it is unchanging throughout a proof.
We omit the context and/or types of terms inside the interpretation function when it is clear. Note, this map is implicitly parameterized in terms of an ordering $<$ on locations. 
\end{definition}

We proceed to show that this interpretation preserves equality of terms. 
\subsection{Soundness of the Collapse of the Memory}

The purpose of this subsection is to show that the collapse interpretation is a well-defined functor. 
We first give the derived interpretations of our defined notation $\collapterm{!x}$ for application and abstraction.

\begin{lemma}[Application and Abstraction]
We have for every computation:
\begin{align*}
&\collapint{\collap{\term{!y_A: !r_A}}, \Gamma \cvdash \term{[!y_A].M:  ?s_A > !t_A}}(\collapterm{!y\,'_{\!A}}, \term{!v} ) \\
&\qquad = \popsterm{?x_A}\pushsterm{!x_A!\y\,'_{\!A}}\term{.}\collapint{\collap{\term{!y_A: !r_A}}, \Gamma \cvdash \term{M: ?r_A?s_A > !t_A}}(\collapterm{!y\,'_{\!A}}, \term{!v} ) \, \\
&\collapint{\Gamma \cvdash \term{<?\r_A>.M: ?r_A?s_A > !t_A}}(\term{!v}) \\
&\qquad= \popsterm{?\r\,'_{\!A}?\s_A}\pushsterm{!\s_A}\term{.}\collapint{\collap{\term{!x_A:!r_A}}, \Gamma \cvdash \term{M: ?s_A > !t_A}}(\collapterm{!\r\,'_{\!A}}, \term{!v} )
\end{align*} 
\end{lemma}
\begin{proof}
In each case, we proceed by induction on the size of $\term{\r_A}$. 
\begin{itemize}
\item Abstraction:
For the base case, where $\term{\r_A} = \term{a_n(\r)}$  is a singleton, we have by definition
\begin{align*}
\collapint{\term{a_n<\r>.M}}(\term{!v}) &= \popsterm{a_n(\r')?\s_A}\pushsterm{!\s_A}\term{.}\collapint{\term{M}}(\term{\r', !v}) \ .
\end{align*}
For the inductive case, we calculate: 
\begin{align*}
&\collapint{  \term{a_n<\u>.<?\r_A>.M}}(\term{!v}) \\
(\textsf{defn.}\,\collapint{-}) &= \popsterm{a_n(\u)?\r\,'_{\!A}?\s_A}\pushsterm{!\s_A!\r\,'_{\!A}}\term{.}\collapint{\term{<?\r_A>.M}}(\term{\u,!v}) \\
(\textsf{I.H}) &= \popsterm{a_n(\u)?\r\,'_{\!A}?\s_A}\underline{\pushsterm{!\s_A!\r\,'_{\!A}}\term{.}\popstermdot{?\r\,'_{\!A}\!'?\s\,'_{\!A}}}\term{.}\pushsterm{!\s\,'_A}\term{.}\collapint{ \term{M}}(\collapterm{!\r\,'_{\!A}{\!'}},\term{\u,!v})  \\
&=_{\beta} \popsterm{a_n(\u)?\r\,'_{\!A}?\s_A}\pushsterm{!\s_A}\term{.}\collapint{ \term{M}}(\collapterm{!\r\,'_{\!A}},\term{\u,!v}) \\
(\textsf{Exch.}) &= \popsterm{a_n(\u)?\r\,'_{\!A}?\s_A}\pushsterm{!\s_A}\term{.}\collapint{ \term{M}}(\collapterm{!\r\,'_{\!A} `a_n(\u)},\term{!v})  \, .
\end{align*}
\item Application:
For the base case, where $\term{\r_A} = \term{a_n(\r)}$ is a singleton, we have by definition
\begin{align*}
\collapintterm{[y]a_n.M}(\term{\r',!v})&= \popsterm{?x_A}\pushsterm{!x_A a_n(y)}\term{.}\collapintterm{M}(\term{y'\,!v}) \ .
\end{align*}
For the inductive case, we calculate:
\begin{align*}
	&\collapintterm{[x]a_n.[!y_A].M}(\collapterm{!y\,'_{\!A}`a_n(x')}, \term{!v})\\
	(\textsf{defn.}\,\collapint{-}) &= \popsterm{?w_A}\pushsterm{!w_A a_n(x')}\term{.} \collapintterm{[!y_A].M}(\collapterm{!y\,'_{\!A}`a(x')}, \term{!v}) \\
	(\textsf{I.H})  &=   \popsterm{?w_A}\underline{\pushsterm{!w_A a_n(x')}\term{.} \popstermdot{a_n(x'\,')?w\,'_A}} \term{.}\pushsterm{!w\,'_{\!A} a_n(x'\,') !y\,'_{\!A}}\term{.}\\
		& \qquad \quad \collapint{\term{M}}(\collapterm{!y\,'_{\!A}`a_n(x')}, \term{!v})\\
	&=_\beta   \popsterm{?w_A}\pushsterm{!w_A!a_n(x') y\,'_{\!A} }\term{.}\collapint{\term{M}}(\collapterm{!y\,'_{\!A}`a_n(x')}, \term{!v})\, .  & \qedhere
\end{align*}
\end{itemize}
\end{proof}

The following lemma gives the interpretation of the (admissible) exchange and weakening rules.
\begin{lemma}[Exchange and Weakening]
For all computations, we have
\begin{align*}
	&\collapint{\Gamma, \term{x:p, y:r}, \Gamma' \cvdash \term{M: ?s_A > !t_A}}(\term{!u,x',y',!v}) = \\
	&\collapint{\Gamma, \term{y:r, x:p}, \Gamma' \cvdash \term{M: ?s_A > !t_A}}(\term{!u,y',x',!v})  
\end{align*}
\[
	\collapint{\Gamma \cvdash \term{M: ?s_A > !t_A}}(\term{!v}) = \collapint{\term{x:r}, \Gamma \cvdash \term{M: ?s_A > !t_A}}(\term{x',!v}) \, ,
\]
where, in the latter case, $\term{r} \not\in \fv{\term{M}}$
\end{lemma}
\begin{proof}
Induction on the type derivation of $\term{M}$.
\end{proof}

The following lemma will give the interpretation of the admissible sequencing rule,
which is necessary for proving that $\collap{-}$ is a well-defined functor.


\begin{lemma}[Sequencing]\label{lem:sequencing-collapse}
For computations $\term{N: ?r_A > !u_A}$ and $\term{M: ?u_A?s_A > !t_A}$, we have
\begin{align*}
	\collapintterm{N;M}(\term{!v}) = \term{<}\collapterm{?\r_A?\s_A}\term{>.[}\collapterm{!\r_A}\term{].}\collapintterm{N}(\term{!v})\term{;<}\collapterm{?z_A}\term{>.[}\collapterm{!\s_A!z_A}\term{].}\collapintterm{M}(\term{!v})\,.
\end{align*}
\end{lemma}
\begin{proof}
We proceed by induction on the type derivation of $\term{N}$.
We make use of the convention to omit inputs to $\collapint{-}$ where possible.
We begin each case by unfolding the definition of sequencing. 
\begin{itemize}
\item Identity:
\begin{align*}
	\collapintterm{*;M}  &= \collapintterm{M} \\
		&=_{\textsf{id}} \popsterm{?\r_A?\s_A}\pushsterm{!\s_A!\r_A}\term{.}\collapintterm{M} \\
		&=_{\beta} \popsterm{?\r_A?\s_A}\underline{\pushsterm{!\r_A}\term{.}\popstermdot{?\r'_A}}\term{.}\pushsterm{!\s_A!\r'_A}\term{.}\collapintterm{M} \\
		(\textsf{defn.}\,\collapint{-}) &= \popsterm{?\r_A?\s_A}\pushsterm{!\r_A}\term{.}\collapintterm{*}\term{;}\popsterm{?\r'_A}\pushsterm{!\s_A!\r'_A}\term{.}\collapintterm{M}
\end{align*}
\item {Sequential Constant}:
\begin{align*}
	\collapintterm{c.N;M} &= \collapintterm{c.(N;M)} \\
	(\textsf{defn.}\, \collapint{-}) &= \popsterm{?w_A ?x_A ?y_A}\pushsterm{!w_A}\term{.}\collapterm{c}\term{.}\popsterm{?u_A}\pushsterm{!y_A !x_A!u_A}\term{.}\collapintterm{N;M} \\
	(\textsf{I.H.}) &= \popsterm{?w_A?x_A?y_A}\pushsterm{!w_A}\term{.}\collapterm{c}\term{.}\popsterm{?u_A}\underline{\pushsterm{!y_A!x_A!u_A}\term{.}\popstermdot{?u'_{\!A}?x'_{\!A}?y'_{\!A}}}\term{.} \\
		&\qquad\quad \pushsterm{!x'_{\!A}!u'_{\!A}}\term{.}\collapintterm{N}\term{;}\popsterm{?z_A}\pushsterm{!y\,'_{\!A}!z_A}\term{.}\collapintterm{M} \\
	&=_\beta  \popsterm{?w_A?x_A?y_A}\pushsterm{!w_A}\term{.}\collap{\term c}\term{.}\popsterm{?u_A}\pushsterm{!x_A!u_A}\term{.}\collapintterm{N}\term{;}\\
		& \qquad\quad \popsterm{?z_A}\pushsterm{!y_A!z_A}\term{.}\collapintterm{M} \\
	&=_\beta \popsterm{?w_A?x_A?y_A}\underline{\pushsterm{!x_A!w_A}\term{.}\popstermdot{?w'_{\!A}?x'_{\!A}}}\term{.}\pushsterm{!w'_{\!A}}\term{.}\collapterm{c}\term{.}\popsterm{?u_A}\\
		& \qquad\quad \pushsterm{!x'_{\!A}!u_A}\term{.}\collapintterm{N}\term{;} \popsterm{?z_A}\pushsterm{!y_A!z_A}\term{.}\collapintterm{M}\\
	(\textsf{defn.}\,\collapint{-}) &= \popsterm{?w_A?x_A?y_A}\pushsterm{!x_A!w_A}\term{.}\collapintterm{c.N}\term{;}\popsterm{?z_A}\pushsterm{!y_A!z_A}\term{.}\collapintterm{M}
\end{align*}
\item Application: let $\term{w} = \collapintterm{V}$ in
\begin{align*}
		\collapintterm{[V]a_n.N;M}  &= \collapintterm{[V]a_n.(N;M)} \\
		(\textsf{defn.}\, \collapint{-}) &= \popsterm{?\r_A?\s_A}\pushsterm{!\s_A!\r_A `a_n(w)}\term{.} \collapintterm{N;M}  \\
		(\textsf{I.H.}) &= \popsterm{?\r_A?\s_A}\underline{\pushsterm{!\s_A!\r_A `a_n(w)}\term{.} \popstermdot{`a_n(w')?\r\,'_{\!A}?\s\,'_{\!A}}}\term{.} \pushsterm{!\r\,'_{\!_A}`a_n(w')}\term{.}\collapintterm{N} \term{;}\\
			&\qquad\quad \popsterm{?z_A}\pushsterm{!\s\,'_{\!A}!z_A}\term{.}\collapintterm{M} \\
		&=_\beta  \popsterm{?\r_A?\s_A} \pushsterm{!\r_A`a_n(w)}\term{.}\collapintterm{N} \term{;}\popsterm{?z_A}\pushsterm{!\s_A!z_A}\term{.}\collapintterm{M}  \\
		&=_\beta  \popsterm{?\r_A?\s_A} \underline{\pushsterm{!\r_A} \term{.}\popstermdot{?\r'_A}}\term{.} \pushsterm{!\r'_A `a_n(w)}\term{.}\collapintterm{N}\term{;}\\
			& \qquad\quad \popsterm{?z_A}\pushsterm{!\s_A!z_A}\term{.}\collapintterm{M}  \\
		(\textsf{defn.}\, \collapint{-}) &= \popsterm{?\r_A?\s_A} \pushsterm{!\r_A}\term{.}\collapintterm{[V]a_n.N}\term{;}\popsterm{?z_A}\pushsterm{!\s_A!z_A}\term{.}\collapintterm{M} 
\end{align*}
\item Abstraction:
\begin{align*}
	\collapintterm{a_n<u>.N;M} &= 	\collapintterm{a_n<w>.(N;M)}  \\
	(\textsf{defn.}\, \collapint{-}) &= \popsterm{`a_n(w')?\r_A?\s_A}\pushsterm{!\s_A!\r_A}\term{.} \collapintterm{N;M}(\term{w'}) \\
		(\textsf{I.H.}) &= \popsterm{`a_n(w')?\r_A?\s_A}\underline{\pushsterm{!\s_A!\r_A}\term{.}  \popstermdot{?\r\,'_{\!A}?\s\,'_{\!A}}}\term{.} \pushsterm{!\r'_A}\term{.}\collapintterm{N}(\term{w'})\term{;}\\
		& \qquad \popsterm{?z_A}\pushsterm{!\s\,'_{\!A}!z_A}\term{.}\collapintterm{M}(\term{w'}) \\
	 	&=_\beta \popsterm{`a_n(w')?\r_A?\s_A} \pushsterm{!\r_A}\term{.}\collapintterm{N}(\term{w'})\term{;} \popsterm{?z_A}\pushsterm{!\s_A!z_A}\term{.}\collapintterm{M}(\term{w'}) \\
	(\textsf{Weak.}) &= \popsterm{`a_n(w')?\r_A?\s_A} \pushsterm{!\r_A}\term{.}\collapintterm{N}(\term{w'})\term{;} \popsterm{?z_A}\pushsterm{!\s_A!z_A}\term{.}\collapintterm{M} \\
	&=_\beta \popsterm{`a_n(w')?\r_A?\s_A} \underline{\pushsterm{!\r_A`a_n(w'\,')} \term{.}\popstermdot{`a_n(w'\,')?\r\,'_{\!A}}}\term{.} \pushsterm{!\r\,'_{\!A}}\term{.} \\
		& \qquad \quad \collapintterm{N}(\term{w'\,'}) \term{;} \popsterm{?z_A}\pushsterm{!\s_A!z_A}\term{.}\collapintterm{M}\\
	(\textsf{defn.}\, \collapint{-})&= \popsterm{`a_n(w')?\r_A?\s_A} \pushsterm{!\r_A`a_n(w')} \term{.}\collapintterm{`a_n<w>.N} \term{;}\\
	& \qquad \quad \popsterm{?z_A}\pushsterm{!\s_A!z_A}\term{.}\collapintterm{M}
\end{align*}
\item Sequential Execution:
\begin{align*}
	\collapintterm{`?V.N;M} &= \collapintterm{`?V.(N;M)} \\
	(\textsf{defn.}\, \collapint{-}) &= \popsterm{?w_A ?x_A ?y_A}\pushsterm{!w_A}\term{.}\collapintterm{`?V}\term{.}\popsterm{?u_A}\pushsterm{!y_A !x_A!u_A}\term{.}\collapintterm{N;M} \\
	(\textsf{I.H.}) &= \popsterm{?w_A?x_A?y_A}\pushsterm{!w_A}\term{.}\collapintterm{`?V}\term{.}\popsterm{?u_A}\underline{\pushsterm{!y_A!x_A!u_A}\term{.}\popstermdot{?u'_{\!A}?x'_{\!A}?y'_{\!A}}}\term{.} \\
		&\qquad\quad \pushsterm{!x'_{\!A}!u'_{\!A}}\term{.}\collapintterm{N}\term{;}\popsterm{?z_A}\pushsterm{!y\,'_{\!A}!z_A}\term{.}\collapintterm{M} \\
	&=_\beta  \popsterm{?w_A?x_A?y_A}\pushsterm{!w_A}\term{.}\collapintterm{`?V}\term{.}\popsterm{?u_A}\pushsterm{!x_A!u_A}\term{.}\collapintterm{N}\term{;}\\
		& \qquad\quad \popsterm{?z_A}\pushsterm{!y_A!z_A}\term{.}\collapintterm{M} \\
	&=_\beta \popsterm{?w_A?x_A?y_A}\underline{\pushsterm{!x_A!w_A}\term{.}\popstermdot{?w'_{\!A}?x'_{\!A}}}\term{.}\pushsterm{!w'_{\!A}}\term{.}\collapintterm{`?V}\term{.}\popsterm{?u_A}\\
		& \qquad\quad \pushsterm{!x'_{\!A}!u_A}\term{.}\collapintterm{N}\term{;} \popsterm{?z_A}\pushsterm{!y_A!z_A}\term{.}\collapintterm{M}\\
	(\textsf{defn.}\,\collapint{-}) &= \popsterm{?w_A?x_A?y_A}\pushsterm{!x_A!w_A}\term{.}\collapintterm{`?V.N}\term{;}\popsterm{?z_A}\pushsterm{!y_A!z_A}\term{.}\collapintterm{M} & \qedhere
\end{align*}
\end{itemize}
\end{proof}
\begin{remark}
Note that it immediately follows from the above Lemma that $\collapint{-}$ respects \textit{strict} sequencing, \textit{i.e.},  where $\type{!s_A} = \type{\e_A}$, we have as below, left. Similarly, in the case of \textit{strict} sequential execution we have as below, right. 
\[
	\collapintterm{M;N}= \collapintterm{M}\term{;}\collapintterm{N}\qquad \collapint{\term{`?V.M}} = \term{`?}\collapint{\term{V}}\term{;}\collapint{\term{M}}
\]
To see this, applying the result above with $\type{!s_A} = \type{\e_A}$ an calculate
\begin{align*}
	\collapintterm{M;N} = \popsterm{?\r_A}\pushsterm{!\r_A}\term{.}\collapintterm{N}\term{;}\popsterm{?\u_A}\pushsterm{!\u_A}\term{.}\collapintterm{M} 		=_\textsf{id} \collapintterm{N}\term{;}\collapintterm{M}\, .
\end{align*}
with the result for strict sequential execution  similar.
We will use these results in future proofs, commenting only that they are applications of the Sequencing Lemma or the definition of $\collapint{-}$, respectively.  
\end{remark}

This section is concluded by proving that the collapse interpretation is a well-defined functor from $\termcat{\Sigma_A}$ to $\stermcat{\Sigma_A}$.
\begin{proposition}[Soundness] \label{lem:wd-collapse}
For any computations $\Gamma \cvdash \term{M, N: ?s_A > !t_A}$, we have
\[
	\term{M} = \term{N} \quad implies \quad \collapintterm{M} = \collapintterm{N}\, .
\]	
\end{proposition}
\begin{proof}
We prove the statement in the case of each equation.
In each case, we begin by unfolding the definition of $\collap{-}$, except where otherwise specified.
\begin{itemize}
\item Identity (\textsf{id}): 
\begin{align*}
	\collapintterm{<?x_A>.[!x_A]} ~(\textsf{Abs.})~&= \popsterm{?x_A}\collapintterm{!x_A}(\collapterm{!x_A})  \\
		(\textsf{App.}) &= {\popsterm{?x_A}\pushsterm{!x_A}}\term{.}\collapintterm{*}(\collapintterm{!x_A}) \\
		(\textsf{Weak.}) &= {\popsterm{?x_A}\pushsterm{!x_A}}\term{.}\collapintterm{*} \\
		&=_{\textsf{id}} \collapintterm{*} 
\end{align*}
\item Local Beta ($\beta$): let $\term{x'} = \collapintterm{V}$ in
\begin{align*}
	\collapintterm{[V]a_n.a_n<\r>.`?x} &= \popsterm{?\s_A}\pushsterm{!\s_A `a_n(x')}\term{.}\collapintterm{<\r>a_n.`?x} \\
	(\textsf{defn.}\, \collapint{-}) &= \popsterm{?\s_A}\underline{\pushsterm{!\s_A `a_n(x')}\term{.}\popstermdot{`a_n(x'\,')?\s\,'_{\!A}}}\term{.}\pushsterm{!\s\,'_{\!A}}\term{.}\collapintterm{`?x}(\term{x'\,'}) \\
	&=_\beta \popsterm{?\s_A}\pushsterm{!\s\,'_{\!A}}\term{.}\collapintterm{`?x}(\term{x'}) \\
	(\textsf{defn.}\, \collapint{-}) &=\popsterm{?\s_A}\pushsterm{!\s\,'_{\!A}}\term{.`?}\collapintterm{{x}}(\term{x'}) \\
	(\textsf{defn.}\, \collapint{-}) &=\popsterm{?\s_A}\pushsterm{!\s\,'_{\!A}}\term{.`?x'}\\
	(\textsf{defn.}\, \term{x'})	&= \term{`?}\collapintterm{V} \\ 
	(\textsf{defn.}\, \collapint{-})	&= \collapintterm{`?V} \\ 
\end{align*}
\item Eta ($\eta'$):  (Remark \ref{rem:eta2})
\begin{align*}
	&\collapintterm{[`!M;a<f>.`?f]a} \\
	(\textsf{defn.}\, \collapint{-})		&= \term{[`!}\collapintterm{M;a<f>.`?f}\term{]} \\
	(\textsf{Seq.})	&= \term{[`!}\popsterm{?s_A}\collapintterm{M}\term{;}\term{<g>.}\pushsterm{!s_A`a(g)}\term{;}\collapintterm{a_n<f>.`?f}\term{]} \\
	(\textsf{Abs.})	&= \term{[`!}\popsterm{?s_A}\collapintterm{M}\term{;}\term{<g>.}\underline{\pushsterm{!s_A`a(g)}\term{;}\popstermdot{`a(g')?s'_A}}\term{.}\pushsterm{!s_A}\term{.}\collapintterm{`?f}(\term{g'})\term{]} \\
		&=_\beta \term{[`!}\popsterm{?s_A}\collapintterm{M}\term{;}\term{<g>.}\pushsterm{!s_A}\term{.}\collapintterm{`?f}(\term{g})\term{]} \\
	(\textsf{defn.}\, \collapint{-})	&= \term{[`!}\popsterm{?s_A}\collapintterm{M}\term{;}\term{<g>.}\pushsterm{!s_A}\term{.`?}\collapintterm{f}(\term{g})\term{]} \\
	(\textsf{defn.}\, \collapint{-})		&= \term{[`!}\popsterm{?s_A}\collapintterm{M}\term{;}\term{<g>.}\pushsterm{!s_A}\term{.`?g}\term{]} \\
				&=\beta \term{[`!}\popsterm{?s_A}\collapintterm{M}\term{;}\term{<g>.}\pushsterm{!s_A}\term{.}\underline{\term{[g].<h>}}\term{.`?h}\term{]} \\
			&=_\iota \term{[`!}\popsterm{?s_A}\term{}\pushsterm{!s_A}\term{.}\collapintterm{M}\term{;<h>.`?h}\term{]} \\
			&=_\textsf{id}\term{[`!}\collapintterm{M}\term{.<h>.`?h]}\\
			&=_\eta' \collapintterm{M}
\end{align*}
\item Force ($\phi$): 
\begin{align*}
	\collapintterm{`?`!M} &= \term{`?}\collapintterm{`!M} \\
		(\textsf{defn.}\, \collapint{-}) &= \term{`?`!}\collapintterm{M} \\
		&=_\phi \collapintterm{M}
\end{align*}
\item Thunk ($\tau$): 
\begin{align*}
	\collapintterm{`!`?V} &=
		 \term{`!}\collapintterm{`?V}\term{} \\ 
		(\textsf{defn.}\, \collapint{-}) &= \term{`!`?}\collapintterm{V}\term{} \\ 
		 &=_\tau \term{}\collapintterm{V}\term{} \\ 
\end{align*}
\item Terminal ($!$): 
\begin{align*}
	 \collapintterm{S;<?x_A>} ~(\textsf{Seq.}) &= \collapintterm{S}\term{;}\collapintterm{<?x_A>} \\
		(\textsf{Abs.}) &= \collapintterm{S} \term{;} \popsterm{?x'_A} \collapintterm{*}(\collapterm{!x'_A}) \\
		(\textsf{Weak.}) &=  \collapintterm{S} \term{;} \popsterm{?x_A} \collapintterm{*} \\
		&=_{!} \collapintterm{*}
\end{align*}
\item Diagonal ($\Delta$):
\begin{align*}
	\collapintterm{S;<?x_A>.[!x_A].[!x_A]}~(\textsf{Seq.}) &= \collapintterm{S} \term{;} \collapintterm{<?x_A>.[!x_A].[!x_A]} \\
	(\textsf{Abs.})	&= \collapintterm{S} \term{;} \popsterm{?x'_A} \collapintterm{[!x_A].[!x_A]}(\collapterm{!x'_A}) \\
	(\textsf{App.})	&= \collapintterm{S} \term{;} \popsterm{?x'_A} \pushsterm{!x'_A} \term{.} \collapintterm{[!x_A]}(\collapterm{!x'_A}) \\
	(\textsf{App.})	&= \collapintterm{S}  \term{;} \popsterm{?x'_A} \pushsterm{!x'_A} \term{.} \pushsterm{!x'_A} \\
				&=_\Delta \collapintterm{S} \term{;} \collapintterm{S} \\
	(\textsf{Seq.}) &= \collapintterm{S;S}
\end{align*}
\item Interchange ($\iota$): 
\begin{align*}
	&\collapintterm{P;<?\s_A>.(Q;[!\s_A])}\\
	(\textsf{Seq.}) &= \popsterm{?\r_A}\collapintterm{P}\term{;}\popsterm{?\s_A}\pushsterm{!\r_A!\s_A}\term{.}\collapintterm{<?\s_A>.(Q;[!\s_A])}  \\
		&= \popsterm{?\r_A}\collapintterm{P}\term{;}\popsterm{?\s_A}\underline{\pushsterm{!\r_A!\s_A}\term{.}\popstermdot{?\s'_A?\r'_A}}\term{.}\pushsterm{!\r'_A}\term{.}\collapintterm{Q;[!\s_A]}(\collapterm{!\s'_A})\\
		&=_\beta \popsterm{?\r_A}\collapintterm{P}\term{;}\popsterm{?\s_A}\pushsterm{!\r_A}\term{.}\collapintterm{Q;[!\s_A]}(\collapterm{!\s_A})\\
		(\textsf{Seq.}) &= \popsterm{?\r_A}\collapintterm{P}\term{;}\popsterm{?\s_A}\pushsterm{!\r_A}\term{.}\collapintterm{Q}(\collapterm{!\s_A})\term{;}\collapterm{[!\s_A]}(\collapterm{!\s_A})\\
		(\textsf{Weak.}) &= \popsterm{?\r_A}\collapintterm{P}\term{;}\popsterm{?\s_A}\pushsterm{!\r_A}\term{.}\collapintterm{Q}\term{;}\collapterm{[!\s_A]}(\collapterm{!\s_A})\\
		&= \popsterm{?\r_A}\underline{\collapintterm{P}\term{;}\popsterm{?\s_A}\pushsterm{!\r_A}\term{.}\collapintterm{Q}\term{;}\pushsterm{!\s_A}}\\
		&=_\iota \popsterm{?\r_A}\pushsterm{!\r_A}\term{.}\collapintterm{Q}\term{;}\collapintterm{P}\\
		&=_\textsf{id} \collapintterm{Q}\term{;}\collapintterm{P}\\
		(\textsf{Seq.})&= \collapintterm{Q;P}
\end{align*}
\item Permutation ($\pi$), where $\term{a_n} < \term{a_k}$, with the $\term{a_k} < \term{a_n}$ case similar: 
\begin{align*}
	\collapintterm{[V]a_k.a_n<x>.M}&= \popsterm{`a_n(x')?y_A}\pushsterm{!y_A`a_n(x')`a_k(w)}\term{.}\collapintterm{a_n<x>.M} \\
	(\textsf{defn.}\, \collapint{-})	&= \popsterm{`a_n(x') ?y_A}\underline{\pushsterm{!y_A `a_n(x') `a_k(w)}\term{.}\popstermdot{`a_k(w') `a_n(x'\,') ?\s'_A}}\term{.}\\
			& \qquad \pushsterm{!y'_A`a_k(w')}\term{.}\collapintterm{M}(\term{x'\,'}) \\
		&=_\beta \popsterm{`a_n(x') ?y_A}\pushsterm{!y_A`a_k(w)}\term{.}\collapintterm{M}(\term{x'}) \\
		&=_\beta \popsterm{`a_n(x')?y_A}\underline{\pushsterm{!y_A}\term{.}\popstermdot{?\s'_A}}\term{.}\pushsterm{!\s'_A`a_k(w)}\term{.}\collapintterm{M}(\term{x'}) \\
	(\textsf{defn.}\, \collapint{-})	&= \popsterm{`a_n(x')?y_A}\pushsterm{!y_A}\term{.}\collapintterm{[V]a_k.M}(\term{x'})  \\
	(\textsf{defn.}\, \collapint{-})	&= \collapintterm{a_n<x>.[V]a_k.M}
\end{align*}
where $\term{w} = \collapintterm{V} = \collapintterm{V}(\term{r}) $ (by the Weakening Lemma).
\item Relocation ($\rho$):
\begin{align*}
	\collapintterm{[V]a.a<x>.[x]b} &= \term{[}\collapintterm{V}\term{].}\collapintterm{a<x>.[x]b} \\
	(\textsf{defn.}\, \collapint{-})	&= \term{[}\collapintterm{V}\term{]}\term{.}\term{<x'>.}\collapintterm{[x]b}(\term x') \\
	(\textsf{defn.}\, \collapint{-})	&= \term{[}\collapintterm{V}\term{]}\term{.}\term{<x'>.[x]}\\
		&=_\beta  \term{[}\collapintterm{V}\term{]}\\
	(\textsf{defn.}\, \collapint{-})	&= {\collapintterm{[V]b}}\qedhere
\end{align*}
\end{itemize}
\end{proof}

\begin{corollary}
The interpretation $\collapint{-}: \termcat{\Sigma_A} \to \stermcat{\collap{\Sigma_A}}$ is a well-defined functor.
\end{corollary}
\begin{proof}
The interpretation preserves equality of terms by Propostion \ref{lem:wd-collapse}, preserves identity by definition and preserves composition by the Sequencing Lemma \ref{lem:sequencing-collapse}.
\end{proof}
We proceed to show that this functor is the inverse of $\embedint{-}: \stermcat{\collap{\Sigma_A}} \to \termcat{\Sigma_A}$

\subsection{Equivalence}
In this section, we prove the equivalence of $\termcat{\Sigma_A}$ and $\stermcat{\collap{\Sigma_A}}$. That is, we show that the embedding functor $\embedint{-}_a: \stermcat{\collap{\Sigma_A}} \to \termcat{\Sigma_A}$ is full and faithful. That is, we prove
\[
	\collapint{\embedint{\term{M}}_a} = \term{M} \qquad \embedint{\collapint{{-}}}_a \simeq \textsf{id} \, .
\]
We begin with the faithfulness, which is nearly trivial, since the embedding is essentially just the inclusion of terms: except for the case of (computation) constants. 
For fullness, we require working up-to a natural isomorphism, which will be given by $\kappa$. 

The following lemma will allow us to deal with the case of constants in the proof of faithfulness. To see this, recalling the definition of the embedding functor on constants applies the isomorphisms $\kappa$. We show that the collapse interpretation maps these back to the identity. Note, it is easy to see the following type-checks, by noting that collapsing a type is an idempotent operation. 
\begin{lemma}\label{lem:collapse-isos}
For every $\type{!t_A}$, we have $\collapint{\kappa(\type{!t_A})}(\term{!v}) = \term*$ and $ \collapint{\kappa^{-1}(\type{!t_A})}(\term{!v})  = \term{*}$.
\end{lemma}
\begin{proof}
For each statement, we proceed by induction on the size of the (collapse of the) memory type $\type{!t_A}$. For each statement, the base case, the empty memory, is trivial. 
Note that the following statement holds. 
\[
	(\ast) \qquad  \collapint{\collapterm{x}^*}(\term{x'}, \term{!v}) = \term{x'}  \qquad \collapint{\embedterm{x}^*}(\term{x'}, \term{!v}) = \term{x'} 
\]
This is true by definition in the case that $\type{s} = \type{a}$. In the case that $\type{r} = \type{?p_A > !u_A}$ (where, in particular, $\type{!p_A}$ and $\type{!u_A}$ are smaller than $\type{!t_A}$), see that the first statement of $(\ast)$ is given by the following calculation.
\begin{align*}
	\collapint{\collapterm{x}^*}(\term{x'}, \term{!v}) &= \collapint{\term{`!}\kappa^{-1}(\type{!p_A})\term{;`?x;}\kappa(\type{!u_A})}(\term{x'}, \term{!v})\\
	(\textsf{defn.}\, \collapint{-})	&= \term{`!} \collapint{\kappa^{-1}(\type{!p_A})\term{;`?x;}\kappa(\type{!u_A})}(\term{x'}, \term{!v})\\
	(\textsf{Seq.})	&= \term{`!} \collapint{\kappa^{-1}(\type{!p_A})}(\term{x'}, \term{!v})\term{;}\collapint{\term{`?x}}(\term{x'}, \term{!v})\term{;}\collapint{\kappa(\type{!u_A})}(\term{x'}, \term{!v})\\
	(\textsf{I.H.})	&= \term{`!} \collapint{\term{`?x}}(\term{x'}, \term{!v})\\
	(\textsf{defn.}\, \collapint{-})	&= \term{`!`?} \collapint{\term{x}}(\term{x'}, \term{!v})\\
						&=_\phi  \collapint{\term{x}}(\term{x'}, \term{!v})\\
	(\textsf{defn.}\, \collapint{-})	&= \term{x'}
\end{align*}
The second statement of  $(\ast)$ follows from a similar argument. 
With this setup, we now prove the two statements of the lemma.

For the inductive case of the first statement of the lemma,
let $\term{x}$ be the left-most element of $\collap{\term{!z_A}}$, so that for some $\term{!\y_{A}}$, we have
\[
	\collap{\term{!z_A}} = \term{x}\cdot\collap{\term{!y_{A}}} \, .
\]
Note that this implies $\collapterm{!z_A}^* = \collap{\collapterm{!z_A}}^*_p = \collap{\term x \cdot \collapterm{!y_A}}_p^* = \collapterm{x}^* \cdot \collap{\collapterm{!y_A}}^*_p = \collapterm{x}^* \cdot \collapterm{!y_A}^* $.
\begin{align*}
		\collapint{\kappa(\type{!t_A})} &= \collapint{\term{<?z_A>.[}\collap{{\term{!z_A}}}^*\term{]}} \\
	(\textsf{Abs.})	 &= \popsterm{?z\,'_{\!\!A}}\term{}\collapint{\term{[}\collap{{\term{!z_A}}}^*\term{]}}(\collapterm{!z\,'_{\!\!A}}) \\
	(\textsf{defn.}\, \collapterm{!z_A})	 &= \popsterm{?y\,'_{\!A}}\term{<x'>.}\collapint{\term{[}\collap{\term{x}}^*\term{].[} \collap{\term{!y_{\!A}}}^*\term{]}}(\term{x'},\collapterm{!y\,'_{\!\!A}}) \\
	(\textsf{defn.}\, \collapint{-})	&= \popsterm{?y\,'_{\!\!A}}\term{<x'>.[}\collapint{\collapterm{x}^*}(\term{x'},\collapterm{!y\,'_{\!\!A}})\term{]}\term{.}\collapint{\term{[}\collap{{\term{!y'_{\!A}}}}^*\term{]}}(\term{x'},\collapterm{!y\,'_{\!\!A}}) \\
	({\ast})	&= \popsterm{?y\,'_{\!\!A}}\term{<x'>.[x']}\term{.}\collapint{\term{[}\collap{{\term{!y'_{\!A}}}}^*\term{]}}(\term{x'},\collapterm{!y\,'_{\!\!A}}) \\
		&=_\ide \popsterm{?y\,'_{\!\!A}}\term{}\collapint{\term{[}\collap{{\term{!y'_{\!A}}}}^*\term{]}}(\term{x'},\collapterm{!y\,'_{\!\!A}})\\	
	(\textsf{Weak.})	&= \popsterm{?y\,'_{\!\!A}}\term{}\collapint{\term{[}\collap{{\term{!y'_{\!A}}}}^*\term{]}}(\collapterm{!y\,'_{\!\!A}})\\	
	(\textsf{Abs.})	&= \collapint{\term{<?y'_{\!A}>.[}\collap{{\term{!y'_{\!A}}}}^*\term{]}} \\
	(\textsf{defn.}\, \kappa)	&= \collapint{\kappa(\type{!s_A})} \\
	(\textsf{I.H.})	&= \term{*}
\end{align*}

For the inductive case of the second statement of the lemma, let $\term{x:}\collapint{\type{r}}$ be the leftmost element of $\term{!z:}\collapint{\type{ !t_A}}$, so that for some $\term{!y:}\collapint{\type{!s_A}}$, we have
\[
	\term{!z:}\collapint{\type{ !t_A}} = \term{x:}\collapint{\type{r}} \cdot \term{!y:}\collapint{\type{!s_A}}\, .
\]
Note that this implies $\embed{\term{!z}}^* = \embed{\term{x!y}}^* =  b(\embed{\term{x}}^*) \cdot \embed{\term{!y}}^*$, where $\term b$ is the first (with respect to the order $<$ on locations) non-empty location in $\type{!t_A}$.
\begin{align*}
		\collapint{\kappa^{-1}(\type{!t_A})} &= \collapint{\term{<?z>.[}\embed{{\term{!z}}}^*\term{]}} \\
	(\textsf{Abs.})	 &= \term{<?z>.}\collapint{\term{[}\embed{{\term{!z}}}^*\term{]}}(\term{!z\,'{}}) \\
	(\textsf{defn.}\, \collapterm{!z_A})	 &= \term{<?y'>.<x'>.}\collapint{\term{[}\embed{\term{x}}^*\term{]b.[}\embed{\term{!y}}^*\term{]}}(\term{x'},\term{!y\,'}) \\
	(\textsf{defn.}\, \collapint{-})	&= \term{<?y'>.}\term{<x'>.[}\collapint{\embedterm{x}^*}(\term{x'},\term{!y\,'})\term{]}\term{.}\collapint{\term{[}\collap{{\term{!y'}}}^*\term{]}}(\term{x'},\term{!y\,'}) \\
	({\ast})	&= \term{<?y>.}\term{<x'>.[x']}\term{.}\collapint{\term{[}\embed{{\term{!y'}}}^*\term{]}}(\term{x'},\term{!y\,'}) \\
		&=_\ide \term{<?y>.}\term{}\collapint{\term{[}\embed{{\term{!y'}}}^*\term{]}}(\term{x'},\term{!y\,'})\\	
	(\textsf{Weak.})	&= \term{<?y>.}\term{}\collapint{\term{[}\embed{{\term{!y'}}}^*\term{]}}(\term{!y\,'})\\	
	(\textsf{Abs.})	&= \collapint{\term{<?y'>.[}\embed{{\term{!y'}}}^*\term{]}} \\
	(\textsf{defn.}\, \kappa^{-1})	&= \collapint{\kappa^{-1}(\type{!s_A})} \\
	(\textsf{I.H.})	&= \term{*} \qedhere
\end{align*}
\end{proof}

\begin{proposition}[Completeness of Embedding]\label{prop:embed-faith}
The embedding functor  $\embedint{-}_a: \stermcat{\collap{\Sigma_A}} \to \termcat{\Sigma_A}$ is faithful.
\end{proposition}
\begin{proof}
We have for any closed \SLCV-term $\term{M: ?s > !t}$, that 
\[
	\collapint{\embedint{\term{M}}_a} = \term{M}\,.
\]
Proceed by induction on the type derivation of $\term{M}$. The only non-trivial cases are for higher-order value constants,  sequential constants , and sequential execution, which we detail below, taking $\term{a} = \term{\lambda}$ without loss of generality. 
\begin{itemize}
\item Value constant ($\term{v: ?s_A > !t_A}$):
\begin{align*}
	\collapint{\embedint{\collap{\term{v}}}} &= \collapint{\kappa^{-1}\term{;}{\term{v}}\term{.}\kappa} \\
		(\textsf{Seq.})	&= \collapint{\kappa^{-1}}\term{;} \collapint{{\term{v}}}\term{;} \collapint{\kappa} \\
		(\textsf{Lem.}\, \ref{lem:collapse-isos})	&= \collapint{{\term{v}}}\\
		(\textsf{defn.}\, \collapint{-})	&= \collap{{\term{v}}}\\
\end{align*}
\item Sequential constant: 
\begin{align*}
	\collapint{\embedint{\term{c.M}}} &= \collapint{\kappa^{-1} \term{;}\embed{\term{c}}\term{.} \kappa \term{;}\embedint{\term{M}}}\\
		(\textsf{defn.}\, \collapint{-})	&= \term{<?x?y>.[!x].}\collapint{\kappa^{-1} \term{;}\embed{\term{c}}\term{.} \kappa}\term{.<?z>.[!y!z].}\collapint{\embedint{\term{M}}}\\
		(\textsf{Seq.})	&= \term{<?x?y>.[!x].}\collapint{\kappa^{-1}} \term{;}\collapint{\embed{\term{c}}}\term{.} \collapint{\kappa}\term{.<?z>.[!y!z].}\collapint{\embedint{\term{M}}}\\
		(\textsf{Lem.}\, \ref{lem:collapse-isos})	&= \term{<?x?y>.[!x].}\collapint{\embed{\term{c}}}\term{.<?z>.[!y!z].}\collapint{\embedint{\term{M}}} \\
		(\textsf{defn.}\, \collapint{-})	&= \term{<?x?y>.[!x].}{{\term{c}}}\term{.<?z>.[!y!z].}\collapint{\embedint{\term{M}}} \\
			&=_\iota \term{<?x?y>.[!x!y].c.}\collapint{\embedint{\term{M}}} \\
		(\textsf{I.H.})	&= \term{c.M} 
\end{align*}
\item Sequential Execution:
\begin{align*}
\collapint{\embedint{\term{`?V.M}}} &= \collapint{{\term{`?}\embedint{\term V}\term{.}\embedint{\term M}}} \\
	(\textsf{defn.}\, \collapint{-}) &= \term{<?x?y>.}\term{[!x].`?}\collapint{\embedint{\term{V}}}\term{.<?z>.[!y!z]}\term{.}\collapint{\embedint{\term M}}\\
 	&=_\iota \term{<?xy>.[!x!y].`?}\collapint{\embedint{\term V}}\term{.}\collapint{\embedint{\term M}} \\
	&=_{\textsf{id}} \term{`?}\collapint{\embedint{\term V}}\term{.}\collapint{\embedint{\term M}}   \\
	(\textsf{I.H.}) &= \term{`?}{{\term V}}\term{.}{{\term M}}   \qedhere
\end{align*}
\end{itemize}
\end{proof}

\begin{theorem}[Fullness of Embedding]\label{prop:embed-full}
The embedding functor  $\embed{-}_a:$ \linebreak $ \stermcat{\collap{\Sigma_A}} \to \termcat{\Sigma_A}$ is full.
\end{theorem}
\begin{proof}
We aim to show that $\embedint{\collapint{-}}_a \simeq \textsf{id}: \termcat{\Sigma_A} \to \termcat{\Sigma_A}$, with the natural isomorphism given by $\kappa$ and its inverse $\kappa^{-1}$ at each memory type. 
In other words, we must show that the following diagram commutes, for every closed \FMCV-term $\term{M: ?s_A > !t_A}$.
We will omit the location on $\embed{-}_a$, taking, without loss of generality, $\term a = \term \lambda$, so we can also omit the location on terms.
\[
\scalebox{1.1}
{\begin{tikzcd}[ampersand replacement=\&]
	{\type{!s_A}} \&\& {\type{!t_A}} \\
	{\embedint{\collapint{\type{!s_A}}}} \&\& {\embedint{\collapint{\type{!t_A}}}}
	\arrow["{\term{M}}", from=1-1, to=1-3]
	\arrow["{\kappa}"', from=1-1, to=2-1]
	\arrow["{\kappa}", from=1-3, to=2-3]
	\arrow["{\embedint{\collapint{\term{M}}}}"', from=2-1, to=2-3]
\end{tikzcd}
}
\]
We will prove a stronger statement: for \textit{open} \FMCV-terms, where $\term{!u,!w:!w}$ and $\term{!v}: {\collapint{\type{!w}}}$
and $\term {\sigma} = \term{\{}\embed{\,\term{!u}\,}_p^*\term{/!w\, \}}$, 
\begin{align*}
	\kappa\term{;}\embedint{\term{!v} \cvdash \collapint{\term{!w} \cvdash \term{M: ?r_A > !s_A}}(\term{!v})}(\term{!u})&\term{;}\kappa^{-1} = \term{\sigma M}\\
	\embedint{\term{!v} \vvdash \collapint{\term{!w} \vvdash \term{V: ?t_A > !u_A} }(\term{!v})}(\term{!u})&= \term{`!}\kappa^{-1}(\type{ !t_A})\term{;}\term{`?\sigma V;}\kappa(\type{!t_A})\\ 
	\embedint{\term{!v} \vvdash \collapint{\term{!w} \vvdash \term{V: a} }(\term{!v})}(\term{!u})&= {\term{\sigma V}} 
\end{align*}
and in particular, we take the valuation (input) to hold only variables. 
Note that when applied to closed computations, this gives the required result. 
We proceed by induction on type derivations.  
We begin each case by unfolding the definitions of $\collapint{-}$, and we immediately apply $\embedint{-}$ whenever possible (when its action is just the obvious inclusion of terms). 
Note that we make use of the fact that $\collapterm{!x_A!y_A}^* = \collap{\collap{\term{!x_A}}_p^*\collap{\term{!y_A}}_p^*}$, {where $\collap{-}^*_p$ is defined to be the pointwise application of $\collap{-}^*$ to a family of variables. }
\begin{itemize}
\item Variable ($\term{z: a}$): 
\begin{align*}
	\embedint{\term{!v}, \term{y},\term{!v'} \vvdash \collapintterm{z}(\term{!v}, \term{y},\term{!v'} )}(\term{!u, x, !u'}) &= \embedint{\term{!v}, \term{y},\term{!v'} \vvdash \term{y}}(\term{!u, x, !u'}) \\
	(\textsf{defn.}\,\embedint{-}) &= \term{x} \\
	(\textsf{defn. subst.})	&= {\term{\{}\embed{\term{!u, x, !u'}\, }_p^* / \term{!w, z, !w'\, \} z}}
\end{align*}
\item Variable ($\term{z: ?s_A > !t_A}$):  
\begin{align*}
	\embedint{\term{!v}, \term{y},\term{!v'} \vvdash \collapintterm{z}(\term{!v}, \term{y},\term{!v'} )}(\term{!u, x, !u'}) &= \embedint{\term{!v}, \term{y},\term{!v'} \vvdash \term{y}}(\term{!u, x, !u'}) \\
	(\textsf{defn.}\,\embedint{-}) &= \term{x} \\
	&=_{\tau} \term{`!`?`!`?x} \\
	(\textsf{Iso.}) &= \term{`!}{\kappa^{-1}}\term{`?`!;}\kappa\term{;`?x.}\kappa^{-1}\term{;}\kappa\\
	(\textsf{defn. subst}) &= \term{`!}{\kappa^{-1}}\term{;`?\{`!}\kappa\term{;`?x.} \kappa^{-1} / z\term{\}z.}\term{;}\kappa\\
	(\textsf{defn.}\,\collap{-}^*)	&=  \term{`!}{\kappa^{-1}}\term{;}{\term{`?\{}\embed{\term{x}\, }^* / \term{z\, \} z}}\term{;}\kappa\\
	(\textsf{defn. subst})		&=  \term{`!}{\kappa^{-1}}\term{;}{\term{`?\{}\embed{\term{!u, x, !u'}\, }_p^* / \term{!w, z, !w'\, \} z}}\term{;}\kappa
\end{align*}
\item Value Constant ($\term{v:a}$):
\begin{align*}
	\embedint{\term{!v'} \vvdash \collapintterm{v}(\term{!v'} )}(\term{!u}) &= \embedint{\term{!v'} \vvdash \term{v}}(\term{!u}) \\
	(\textsf{defn.}\,\embedint{-}) &= \term{v} \\
	(\textsf{defn. subst.})	&= {\term{\{}\embed{\term{!u}\, }_p^* / \term{!w\, \} v}}
\end{align*}
\item Value Constant  ($\term{v:?s_A > !t_A}$):
\begin{align*}
	\embedint{\term{!v'} \vvdash \collapintterm{v}(\term{!v'})}(\term{!u}) &= \embedint{\term{!w} \vvdash \collap{\term v}}(\term{!u}) \\
	(\textsf{defn.}\,\embedint{-})	&= \term{`!}{\kappa^{-1}}\term{;}\term{`?v}\term{;}{\kappa}\\
	(\textsf{defn. subst.})	&= \term{`!}{\kappa^{-1}}\term{;}{\term{`?}\term{\{}\embed{\term{!w}\, }^* / \term{!w\, \} v}}\term{;}{\kappa}
\end{align*}
\item Thunk:
\begin{align*}
	\embedint{\term{!v} \vvdash \collapintterm{`!M}(\term{!v})}(\term{!u}) &= \embedint{\term{!v}  \vvdash \term{`!}\collapintterm{M}(\term{!v})}(\term{!u}) \\
	(\textsf{defn.}\,\embedint{-})	&=  \term{`!}\embedint{\term{!v}  \vvdash \collapintterm{M}(\term{!v})}(\term{!u}) \\
		\textsf{I.H.} &= \term{`!}\,\kappa^{-1} \term{;\{}\embed{\, \term{!v}\, }_p^*\term{/!w\, \}M}\term{;}\kappa \\
		&=_\phi \term{`!}\,\kappa^{-1} \term{;`?`!\{}\embed{\, \term{!v}\, }_p^*\term{/!w\, \}M}\term{;}\kappa 
\end{align*}
\item Identity: 
\begin{align*}
	\kappa\term{;}\embedint{\collapint{\term{*}}(\term{!v})}( \term{!u} )\term{;}\kappa^{-1} &= \kappa\term{;}\embedint{\term *}( \term{!u} )\term{;}\kappa^{-1}\\
	(\textsf{defn.}\,\embedint{-}) &= \kappa\term{;}\term{*}\term{;}\kappa^{-1}  \\
	(\textsf{defn. subst.})	& = \kappa^{-1}\term{;\{}\embed{\,\term{!v}\,}_p^*\term{/!w\, \}*;}\kappa
\end{align*}
\item Sequential Constant: let $\term \sigma = \term{\{}\embed{{\term{!v}}\, }_p^*\term{/!w\, }\term{\}}$ in 
\begin{align*}
&\kappa\term{;}\embedint{\collapint{\term{c.M}}(\term{!v})}(\term{!u})\term{;}\kappa^{-1} \\
(\textsf{defn.}\, \collapint{-}, \embedint{-})	&= \kappa\term{;}\popsterm{?x_A?y_A}\pushsterm{!x_A}\term{.`?}\embedint{\collap{{\term{c}}}(\term{!v})}(\term{!u})\term{.}\popsterm{?z_A}\\&\qquad\quad\pushsterm{!y_A!z_A}\term{.}
		       \embedint{\collapint{\term{M}}(\term{!v})}(\term{!u})\term{;}\kappa^{-1} \\
(\textsf{defn.}\, \collapint{-}, \embedint{-}) &= \kappa.\popsterm{?x_A?y_A}\pushsterm{!x_A}\term{.}\kappa^{-1}\term{;}{{\term{c}}}\term{.}\kappa\term{;}\popsterm{?z_A}\\&\qquad\quad\pushsterm{!y_A!z_A}\term{.}\kappa^{-1}
		       \term{;}{{\term{\sigma M}}}\term{;}\kappa\term{;}\kappa^{-1}  \\
(\textsf{Iso.}) &= \kappa.\popsterm{?x_A?y_A}\pushsterm{!x_A}\term{.}\kappa^{-1}\term{;}{{\term{c}}}\term{.}\kappa\term{;}\popsterm{?z_A}\\&\qquad\quad\pushsterm{!y_A!z_A}\term{.}\kappa^{-1}
		       \term{;}{{\term{\sigma M}}} \\
(\textsf{defn.}\, \kappa^{-1})	&= \kappa.\popsterm{?x_A?y_A}\pushsterm{!x_A}\term{.}\kappa^{-1}\term{;}{{\term{c}}}\term{.}\kappa\term{;}\popsterm{?z_A}\\&\qquad\quad\underline{\pushsterm{!y_A!z_A}\term{.<?s>}}\term{.[}\embed{\term{!s}}^* \term{].}
		       {{\term{\sigma M}}} \\
	&=_\beta \kappa.\popsterm{?x_A?y_A}\pushsterm{!x_A}\term{.}\kappa^{-1}\term{;}{{\term{c}}}\term{.}\kappa\term{;}\popsterm{?z_A}\term{[}\embed{\collap{\term{!y_A!z_A}}}^* \term{].}
		       {{\term{\sigma M}}} \\
	&=_\alpha \kappa.\popsterm{?x'_{\!A}?y'_{\!A}}\pushsterm{!x'_{\!A}}\term{.}\kappa^{-1}\term{;}{{\term{c}}}\term{.}\kappa\term{;}\popsterm{?z_A}\term{[}\embed{\collap{\term{!y'_{\!A}!z_A}}}^* \term{].}
		       {{\term{\sigma M}}} \\
(\textsf{defn.}\, \kappa)	&=\term{<?x_A?y_A>.}\underline{\term{[}\collap{\term{!y_A!x_A}}^*\term{].}\popstermdot{?x'_{\!A}?y'_{\!A}}}\term{.}\pushsterm{!x'_{\!A}}\term{.}\kappa^{-1}\term{;}{{\term{c}}}\term{.}\kappa\term{;}\popsterm{?z_A}\\&\qquad\quad\term{[}\embed{\collap{\term{!y'_{\!A}!z_A}}}^* \term{].}
		       {{\term{\sigma M}}} \\
	&=_\beta \term{<?x_A?y_A>.[}\collap{\term{!x_A}}^*\term{]}\term{.}\kappa^{-1}\term{;}{{\term{c}}}\term{.}\kappa\term{;}\popsterm{?z_A}\term{[}\embed{\collap{\collap{\term{!y_A}}_p^*\term{!z_A}}}^* \term{].}
		       {{\term{\sigma M}}} \\
(\textsf{I.H})	&= \term{<?x_A?y_A>.[}\collap{\term{!x_A}}^*\term{].}\term{}\kappa^{-1}\term{;}{{\term{c}}}\term{.}\kappa\term{;}\popsterm{?z_A}\term{[}\embed{\collap{\collap{\term{!y_A}}_p^*\term{!z_A}}}^* \term{].}
		       {{\term{\sigma M}}} \\
	&=_\phi \term{<?x_A?y_A>.[}\collap{\term{!x_A}}^*\term{];}\term{.}\kappa^{-1}\term{;}{{\term{c}}}\term{.}\kappa\term{;}\term{}\popsterm{?z_A}\term{[}\embed{\collap{\collap{\term{!y_A}}_p^*\term{!z_A}}}^* \term{].}
		       {{\term{\sigma M}}} \\
(\textsf{defn.}\, \kappa^{-1})	&= \term{<?x_A?y_A>.}\underline{\term{[}\collap{\term{!x_A}}^*\term{].}\popstermdot{?x'_{\!A}}}\term{.}\pushkinvterm{!x'_{\!A}}\term{.}{{\term{c}}}\term{.}\kappa\term{;}\\&\qquad\quad\popsterm{?z_A}\term{[}\embed{\collap{\collap{\term{!y_A}}_p^*\term{!z_A}}}^* \term{].}
		       {{\term{\sigma M}}} \\
	&=_\beta \term{<?x_A?y_A>.[}\embed{\collap{\term{!x_A}}^*}^*\term{]}\term{.}{{\term{c}}}\term{.}\kappa\term{;}\popsterm{?z_A}\term{[}\embed{\collap{\collap{\term{!y_A}}_p^*\term{!z_A}}}^* \term{].}
		       {{\term{\sigma M}}} \\
(\textsf{Iso.}) &= \term{<?x_A?y_A>.[}{{\term{!x_A}}}\term{]}\term{.}{{\term{c}}}\term{.}\kappa\term{;}\term{}\popsterm{?z_A}\term{[}\embed{\collap{\collap{\term{!y_A}}_p^*\term{!z_A}}}^* \term{].}
		       {{\term{\sigma M}}} \\
(\textsf{defn.}\, \kappa) &= \term{<?x_A?y_A>.[}{{\term{!x_A}}}\term{]}\term{.c.}\term{<?z'_{\!A}>.}\underline{\pushkterm{!z'_{\!A}}\term{.}\popstermdot{?z_A}}\term{.[}\embed{\collap{\collap{\term{!y_A}}_p^*\term{!z_A}}}^* \term{].}
		       {{\term{\sigma M}}} \\
&=_\beta \term{<?x_A?y_A>.[}{{\term{!x_A}}}\term{]}\term{.c.}\term{<?z'_{\!A}>.}\term{[}\embed{\collap{\collap{\term{!y_A}}_p^*\collap{\term{!z'_{\!A}}}_p^*}}^* \term{].}
		       {{\term{\sigma M}}} \\
(\textsf{Iso.}) &= \term{<?x_A?y_A>.[}{{\term{!x_A}}}\term{]}\term{.c.}\term{<?z'_{\!A}>.}\term{[}{{\term{!y_A}\term{!z'_{\!A}}}} \term{].}
		       {{\term{\sigma M}}} \\
(\textsf{Perm.})	 &=_{\pi'\ide} \term{<?y_A>.}\term{c.}{{\term{[!y_A}}} \term{].}
		       {{\term{\sigma M}}} \\
	&=_\iota \term{ c.\sigma M}\\
(\textsf{defn. subst.})	&= \term{\sigma c.M}
\end{align*}
\item Application: let $\term \sigma = \term{\{}\embed{\,\term{!v}\,}^*\term{/!w\, \}}$ in
\begin{align*}
& {\kappa}\term{;} \embedint{\collapint{\term{[V]a_n.M}}(\term{!v})}(\term{!u})\term{;}{\kappa^{-1}} \\
(\textsf{defn.}\,  \collapint{-},\embedint{-})	&={\kappa} \term{;}\popsterm{?x_A}\pushsterm{!x_A`a(y)}\term{.}\embedint{\collapint{\term{M}}(\term{!v})}(\term{!u})\term{;}{\kappa^{-1}} \\
(\textsf{defn.}\, \kappa)	&=\term{<?x_A>.}\underline{\term{[}\collap{{\term{!x_A}}}^*\term{]} \term{.}\popstermdot{?x_A}}\term{.}\pushsterm{!x_A`a(y)}\term{.}\embedint{\collapint{\term{M}}(\term{!v})}(\term{!u}){\term{;}\kappa^{-1}} \\
	&=_\beta\term{<?x_A>} \term{.}\term{[}\collap{\collap{\term{!x_A}}_p^*\term{`a_n(y)}}\term{].}\embedint{\collapint{\term{M}}(\term{!v})}(\,\term{!u}\,)\term{;}\kappa^{-1} \\
(\textsf{I.H.})	&=\term{<?x_A>} \term{.}\term{[}\collap{\collap{\term{!x_A}}_p^*\term{`a_n(y)}}\term{].}\kappa^{-1}{{\term{.\sigma M}}}\term{;}\kappa\term{;}\kappa^{-1}\\
(\textsf{Iso.}) 	&=\term{<?x_A>} \term{.}\term{[}\collap{\collap{\term{!x_A}}_p^*\term{`a_n(y)}}\term{].}\kappa^{-1}{{\term{.\sigma M}}}\\
(\textsf{defn.}\, \kappa^{-1}) &=\term{<?x_A>} \term{.}\underline{\term{[}\collap{\collap{\term{!x_A}}_p^*\term{`a_n(y)}}\term{].<?z>.[}}\embed{\term{!z}}^*\term{]}{{\term{.\sigma M}}}\term{}\\
(\ast) 	&=_\beta 
	 \term{<?x_A>} \term{.[\{\sigma V/y'\}}\embed{\collap{\term{!x_A`a_n(y')}}^*}^*\term{]}{{\term{.\sigma M}}}\term{}\\
(\textsf{Iso.})  &=\term{<?x_A>} \term{.[!x_A`a_n(\sigma V)]}{{\term{.\sigma M}}}\term{}\\
	&=_\ide \term{[\sigma V]a_n.\sigma M}\\
(\textsf{defn. subst.})	&= \term{\sigma [V]a_n.M}
\end{align*}
where $\term{y} = \embedint{\collapint{\term V}}(\term{!u})$ and $(\ast)$ requires to note that, by inductive hypothesis, we have  $\term{y} = \term{\{\sigma V/y'\}}\collapterm{y'}^*$. To see this, consider the two cases of the IH on values, and the two cases of the definition of $\collapterm{y'}$.
\item Abstraction: let $\term \sigma = \term{\{}\embed{{\term{!v}}\, }_p^*\term{/!w\, }\term{\}}$ in
\begin{align*}
	&\kappa \term{;}\embedint{\collapint{\term{a_n<x>.M}}( \term{!v} )}( \term{!u} )\term{;} \kappa^{-1}  \\
(\textsf{defn.}\, \collapint{-},\embedint{-}) 	&=\kappa \term{;} \popsterm{`a_n(x')?y_A}\pushsterm{!y_A}\term{.}\embedint{\collapint{\term{M}}( \term{y', !v} )}(\term{x',!u} )\term{;} \kappa^{-1}\\
(\textsf{defn.}\, \kappa) 		&=\term{<`a_n(x'\,')?y_A>.}\underline{\term{[}\collap{{\term{!y_A`a_n(x'\,')}}}^*\term{]} \term{.} \popstermdot{`a_n(x')?y\,'_{\!A}}}\term{.}\pushsterm{!y\,'_{\!A}}\term{.}\\
	& \qquad \quad\embedint{\collapint{\term{M}}( \term{y, !v} )}( \term{x',!u} )\term{;} \kappa^{-1} \\
	&=_\beta \term{<`a_n(x'\,')?y_A>.[}\collap{{\term{!y_A}}}^*\term{].}\embedint{\collapint{\term{M}}( \term{y', !v} )}( \collap{\term{ x'\,'}}^*,\term{!u} )\term{;} \kappa^{-1} \\
(\textsf{I.H.})	&= \term{<`a_n(x'\,')?y_A>.[}\collap{{\term{!y_A}}}^*\term{];}\kappa^{-1} \term{;}\term{\{}\embed{\collap{\term{ x'\,'}}^*,{\term{!v}}\, }_p^*\term{/x,!w\, }\term{\}}\term{M} \term{;} \kappa^{} \term{;} \kappa^{-1} \\
(\textsf{Iso.})	&= \term{<`a_n(x'\,')?y_A>.[}\collap{{\term{!y_A}}}^*\term{];}\kappa^{-1} \term{;}\term{\{x'\,'},\embed{{\term{!v}}\, }_p^*\term{/x,!w\, }\term{\}}\term{M} \term{;} \kappa \term{;} \kappa^{-1} \\
&=_\alpha \term{<`a_n(x)?y_A>.[}\collap{{\term{!y_A}}}^*\term{];}\kappa^{-1} \term{;}\term{\sigma M} \term{;} \kappa \term{;} \kappa^{-1} \\
(\textsf{Iso.})	&= \term{<`a_n(x)?y_A>.[}\collap{{\term{!y_A}}}^*\term{];}\kappa^{-1} \term{;}\term{\sigma M} \\
(\textsf{defn.}\,  \kappa^{-1})	&= \term{<`a_n(x)?y_A>.}\underline{\term{[}\collap{{\term{!y_A}}}^*\term{].}\term{<?z>}}\term{.[}\embed{\term{!z}}^*\term{].}\term{\sigma M} \\
	&=_\beta \term{<`a_n(x)?y_A>.}\term{[}\embed{\collap{\term{!y_A}}^*}^*\term{].}\term{\sigma M}  \\
(\textsf{Iso.})	&= \term{<`a_n(x)?y_A>.[}{{\term{!y_A}}}\term{]}\term{.}\term{\sigma M}  \\
	&=_{\pi',\ide} \term{a_n<x>.} \term{\sigma M} \\
(\textsf{defn.}\, \term{\sigma},\,  \textsf{subst.})	&= \term{\{}\embed{\, \term{!v}\, }^*\term{/!w\, \} } \term{a_n<x>.M} 
\end{align*}
\item Sequential Execution: let $\term \sigma = \term{\{}\embed{{\term{!v}}\, }_p^*\term{/!w\, }\term{\}}$ in
\begin{align*}
	&\kappa\term{;}\embedint{\collapint{\term{`!V.M}}(\term{!v})}(\term{!u})\term{;}\kappa^{-1} \\
(\textsf{defn.}\, \collapint{-}, \embedint{-})	&= \kappa\term{;}\popsterm{?x_A?y_A}\pushsterm{!x_A}\term{.`?}\embedint{\collapint{\term{V}}(\term{!v})}(\term{!u})\term{.}\popsterm{?z_A}\\&\qquad\quad\pushsterm{!y_A!z_A}\term{.}
		       \embedint{\collapint{\term{M}}(\term{!v})}(\term{!u})\term{;}\kappa^{-1} \\
(\textsf{I.H.}) &= \kappa.\popsterm{?x_A?y_A}\pushsterm{!x_A}\term{.`?}\embedint{\collapint{\term{V}}(\term{!v})}(\term{!u})\term{.}\popsterm{?z_A}\\&\qquad\quad\pushsterm{!y_A!z_A}\term{.}\kappa^{-1}
		       \term{;}{{\term{\sigma M}}}\term{;}\kappa\term{;}\kappa^{-1}  \\
(\textsf{Iso.}) &= \kappa.\popsterm{?x_A?y_A}\pushsterm{!x_A}\term{.`?}\embedint{\collapint{\term{V}}(\term{!v})}(\term{!u})\term{.}\popsterm{?z_A}\\&\qquad\quad\pushsterm{!y_A!z_A}\term{.}\kappa^{-1}
		       \term{;}{{\term{\sigma M}}} \\
(\textsf{defn.}\, \kappa^{-1})	&= \kappa.\popsterm{?x_A?y_A}\pushsterm{!x_A}\term{.`?}\embedint{\collapint{\term{V}}(\term{!v})}(\term{!u})\term{.}\popsterm{?z_A}\\&\qquad\quad\underline{\pushsterm{!y_A!z_A}\term{.<?s>}}\term{.[}\embed{\term{!s}}^* \term{].}
		       {{\term{\sigma M}}} \\
	&=_\beta \kappa.\popsterm{?x_A?y_A}\pushsterm{!x_A}\term{.`?}\embedint{\collapint{\term{V}}(\term{!v})}(\term{!u})\term{.}\popsterm{?z_A}\term{[}\embed{\collap{\term{!y_A!z_A}}}^* \term{].}
		       {{\term{\sigma M}}} \\
	&=_\alpha \kappa.\popsterm{?x'_{\!A}?y'_{\!A}}\pushsterm{!x'_{\!A}}\term{.`?}\embedint{\collapint{\term{V}}(\term{!v})}(\term{!u})\term{.}\popsterm{?z_A}\term{[}\embed{\collap{\term{!y'_{\!A}!z_A}}}^* \term{].}
		       {{\term{\sigma M}}} \\
(\textsf{defn.}\, \kappa)	&=\term{<?x_A?y_A>.}\underline{\term{[}\collap{\term{!y_A!x_A}}^*\term{].}\popstermdot{?x'_{\!A}?y'_{\!A}}}\term{.}\pushsterm{!x'_{\!A}}\term{.`?}\embedint{\collapint{\term{V}}(\term{!v})}(\term{!u})\term{.}\popsterm{?z_A}\\&\qquad\quad\term{[}\embed{\collap{\term{!y'_{\!A}!z_A}}}^* \term{].}
		       {{\term{\sigma M}}} \\
	&=_\beta \term{<?x_A?y_A>.[}\collap{\term{!x_A}}^*\term{]}\term{.`?}\embedint{\collapint{\term{V}}(\term{!v})}(\term{!u})\term{.}\popsterm{?z_A}\term{[}\embed{\collap{\collap{\term{!y_A}}_p^*\term{!z_A}}}^* \term{].}
		       {{\term{\sigma M}}} \\
(\textsf{I.H})	&= \term{<?x_A?y_A>.[}\collap{\term{!x_A}}^*\term{].`?}\collapint{\termint{\term{\sigma V}}}^*\term{.}\popsterm{?z_A}\term{[}\embed{\collap{\collap{\term{!y_A}}_p^*\term{!z_A}}}^* \term{].}
		       {{\term{\sigma M}}} \\
(\textsf{defn.}\, \collap{-}^*)	&= \term{<?x_A?y_A>.[}\collap{\term{!x_A}}^*\term{]}\term{.`?}\term{`!}\kappa^{-1}\term{;`?\sigma V;}\kappa\term{.}\popsterm{?z_A}\term{[}\embed{\collap{\collap{\term{!y_A}}_p^*\term{!z_A}}}^* \term{].}
		       {{\term{\sigma M}}} \\
	&=_\phi \term{<?x_A?y_A>.[}\collap{\term{!x_A}}^*\term{];}\kappa^{-1}\term{;`?\sigma V;}\kappa\term{.}\popsterm{?z_A}\term{[}\embed{\collap{\collap{\term{!y_A}}_p^*\term{!z_A}}}^* \term{].}
		       {{\term{\sigma M}}} \\
(\textsf{defn.}\, \kappa^{-1})	&= \term{<?x_A?y_A>.}\underline{\term{[}\collap{\term{!x_A}}^*\term{].}\popstermdot{?x'_{\!A}}}\term{.}\pushkinvterm{!x'_{\!A}}\term{.`?\sigma V;}\kappa\term{.}\popsterm{?z_A}\\&\qquad\quad\term{[}\embed{\collap{\collap{\term{!y_A}}_p^*\term{!z_A}}}^* \term{].}
		       {{\term{\sigma M}}} \\
	&=_\beta \term{<?x_A?y_A>.[}\embed{\collap{\term{!x_A}}^*}^*\term{]}\term{.`?\sigma V;}\kappa\term{.}\popsterm{?z_A}\term{[}\embed{\collap{\collap{\term{!y_A}}_p^*\term{!z_A}}}^* \term{].}
		       {{\term{\sigma M}}} \\
(\textsf{Iso.}) &= \term{<?x_A?y_A>.[}{{\term{!x_A}}}\term{]}\term{.`?\sigma V;}\kappa\term{.}\popsterm{?z_A}\term{[}\embed{\collap{\collap{\term{!y_A}}_p^*\term{!z_A}}}^* \term{].}
		       {{\term{\sigma M}}} \\
(\textsf{defn.}\, \kappa) &= \term{<?x_A?y_A>.[}{{\term{!x_A}}}\term{]}\term{.`?\sigma V.}\term{<?z'_{\!A}>.}\underline{\pushkterm{!z'_{\!A}}\term{.}\popstermdot{?z_A}}\term{.[}\embed{\collap{\collap{\term{!y_A}}_p^*\term{!z_A}}}^* \term{].}
		       {{\term{\sigma M}}} \\
&=_\beta \term{<?x_A?y_A>.[}{{\term{!x_A}}}\term{]}\term{.`?\sigma V.}\term{<?z'_{\!A}>.}\term{[}\embed{\collap{\collap{\term{!y_A}}_p^*\collap{\term{!z'_{\!A}}}_p^*}}^* \term{].}
		       {{\term{\sigma M}}} \\
(\textsf{Iso.}) &= \term{<?x_A?y_A>.[}{{\term{!x_A}}}\term{]}\term{.`?\sigma V.}\term{<?z'_{\!A}>.}\term{[}{{\term{!y_A}\term{!z'_{\!A}}}} \term{].}
		       {{\term{\sigma M}}} \\
(\textsf{Perm.})	 &=_{\pi'\ide} \term{<?y_A>.}\term{;`?\sigma V.}{{\term{[!y_A}}} \term{].}
		       {{\term{\sigma M}}} \\
	&=_\iota \term{`?\sigma V.\sigma M}\\
(\textsf{defn. subst.})	&= \term{\sigma `?V.M}
 \qedhere
\end{align*}
\end{itemize}
\end{proof}

\begin{theorem}\label{cor:fmc-equiv}
The embedding functor $\embed{-}_a: \stermcat{\collap{\Sigma_A}} \to \termcat{\Sigma_A}$ gives an equivalence of categories. 
\end{theorem}
\begin{proof}
We have shown that embedding functor $\embed{-}:  \stermcat{\collap{\Sigma_A}} \to \termcat{\Sigma_A}$ is Cartesian closed, fully faithful (Proposition \ref{prop:embed-faith} and Theorem \ref{prop:embed-full}), and is essentially surjective on objects via the isomorphisms given in Definition \ref{def:isos}, thus we have an equivalence of categories. 
\end{proof}

\begin{corollary}\label{thm:ccc-equiv}
The categories $\termcat{\Sigma_A}$ and $\textup{{\textsf{CCC}}}(\collap{\Sigma_A})$ are equivalent. 
\end{corollary}
\begin{proof}
Compose the equivalences: $\termint{-}: \textsf{CCC}(\Sigma_A) \to \stermcat{\collap{\Sigma_A}}$ and $\embed{-}_a: \stermcat{\collap{\Sigma_A}} \to \termcat{\collap{\Sigma_A}}$ from Corollary \ref{cor:stlc-equiv} and Corollary \ref{cor:fmc-equiv}.
\end{proof}

Given these results, an obvious question is: if the simply-typed \FMCV\  is equivalent to the simply typed $\lambda$-calculus, what is to be gained from study of the \FMCV? We respond to this in two ways. 

First, the results of this chapter tell us that the \FMCV\  is \textit{denotationally} equivalent to the $\lambda$-calculus: however, it is a useful \textit{operational} refinement. The difference between the calculi lies in the fact we have gained the ability to express \textit{how} computation takes place: how computation is sequenced, whether inputs are passed as function arguments or as mutable state, when the random generator is consulted, \emph{etc}.

The achievement of these results is evidence of the strength of the \FMCV's type system. Simple types record all information about execution: all input and output, all changes to state, and all consultation of oracles. Indeed, strong type sytems for effects is a major research objective, and the strength of the \FMCV's type system is what allows for the result. Note how, unlike other approaches to sequencing and effects, and to decompositions of CBN and CBV $\lambda$-calculi,  no new type constructors are involved: we remain in the realm of pure intuitionistic logic. 

However, as emphasized in the introduction to Chapter \ref{chapter:CCC-eqns}, it is emphasized that what is presented here is not in fact a semantics \textit{of effects}, but of a calculus which can encode effects naturally. Relatedly, the type system presented here is \textit{too strong} for dealing with effects in reality: for example, when reading from a probabilistic generator, the programmer does not typically want to know \textit{how many times} the generator was consulted, which is currently information recorded in the type system. Indeed, such information is \textit{intensional} and not even observable in general.\footnote{For example, reading from a random generator and discarding the result is observationally equivalent to doing nothing at all!} 

Note that, although the \textit{current} type system is too strong, and the current semantics does not account for effects, we believe the \FMCV\  provides an excellent basis for future variants which address properly this major research objective. In particular, the \FMCV\  provides valuable intuitions about how to usefully think about effectful progams, and how (certain) effects relate to higher-order mechanisms which are well-understood. The ultimate research aim is to be able to extend reasoning techniques familiar from the $\lambda$-calculus to real-world (especially, imperative) languages. The type system allows us to ``see" side-effects, and this makes an excellent basis for future research. 

\newcommand\sntype[1]{\sn{\type{#1}}}
\newcommand\snterm[1]{\sn{\term{#1}}}
\newcommand\N{\mathbb N}

\newcommand\floor[1]{\lfloor#1\rfloor}
\newcommand\ceil[1]{\lceil#1\rceil}

\renewcommand\labelenumi{(\theenumi)}

\newtagform{roman}[\renewcommand{\theequation}{\roman{equation}}]()

\chapter{Strong Normalisation}
\label{chapter:SN}

As a second main contribution of this thesis, we show that simply typed terms  are strongly normalising with respect to beta and permutation reduction by extending Gandy's proof for the $\lambda$-calculus \cite{gandy}. 
This proof interprets terms in a domain of monotone functionals, equipped with a partial order. One may further collapse a functional to a natural number to give an overestimate of the longest reduction sequence of a term. 
The literature has several variants on Gandy's proof, including one by De Vrijer that calculates longest reduction sequences exactly~\cite{DeVrijer-1987}; see also~\cite{Schwichtenberg-1982,VanDePol-Schwichtenberg-1995,VanDePol-1995}

Our proof is a variant of Gandy's, which emphasizes its latent operational intuition and makes use of simpler structures -- rather than interpreting terms in domains of \textit{strictly} ordered, \textit{strict} monotone functionals, as Gandy does, we interpret terms in the more standard domain of \textit{non-strict} monotone functionals. The author is not aware of other proofs in the literature that
 are formulated in this way. We discuss later the small technical difference that makes this possible. 

Now, we will give an overview of this section; we will forward-reference various constructs, but will complete the formal definitions later. 
We will work with the variant of the FMC \textit{without base types or values}, given in Definition \ref{def:FMC-plain}, since base types and the distinction between computations and values is of no interest here.
Note, this means that type $\type{t}$ refers to a type of form $\type{?s_A > !t_A}$, and $\type{(>)}$ is the base case of an induction on types.  We now recapitulate the notion of reduction we are concerned with is the following.

\begin{definition}
Beta and permutation reduction for the FMC are respectively defined by the following rules, applicable in all contexts:
\[
	\term{[N]a.a<x>.M} \rw_\beta \term{\{N/x\}M} \qquad \term{[N]a.b<x>.M} \rw_\pi \term{b<x>.[N]a.M},
\]
where,  in the second case, $\term x \not\in \fv{\term N}$ and $\term a \neq \term b$.  We will often write simply $\rw$ for $\rw_\beta$. 
\end{definition}

We will begin by showing beta reduction is strongly normalising,  with strong normalisation for both beta and permutation reduction together following from a further (simple) argument in Section \ref{sec:perm}.   Note, by a similar argument to that section, we could also include the rewrite $\term{<x>.[x]} \rw_\textsf{id} \term{*}$ in our reduction relation, and prove it strongly normalising. We choose not to do this, as the property of confluence would be lost.\footnote{We expect that the right way to include the identity equation in the reduction relation is to \textit{expand} typed terms, similar to the case of $\eta$-expansion in the $\lambda$-calculus -- however, consideration of this is beyond the scope of the current work. }
It is an easy exercise to transfer the results to the \FMCV\ and its reduction relation, which includes reductions for \textit{force} and \textit{thunk}.
\begin{definition}\label{def:sn}
A term $\term{M}$ is \textit{strongly normalising} with respect to a reduction relation if there exists no infinite sequence of reductions beginning from $\term{M}$.
\end{definition}

Similar to Gandy's proof, we exhibit a measure function
which strictly reduces with every reduction step $\term{G |- M`{\black\rw_\beta}`M':t}$. This is done by defining a quantitative semantic interpretation:
we interpret the type $\type{ ?s_A > !t_A }$ as the set of monotone functions
 \begin{align*}
 	\sn{\type{?s_A > !t_A}}  = \sn{\type{!s_A}} \to \mathbb{N} \times \sn{\type{!t_A}}, \ 
 \end{align*}
 equipped with the extensional partial ordering (generated from $\mathbb{N}$, carrying its usual ordering). 
We note that each partial order has a least  element, which we denote $0_{\type{t}}$.\footnote{We prefer to avoid the use of $\bot$ for these least elements because of its connotations of undefinedness and non-termination; the least elements of these posets are better regarded as indicating immediate termination in zero steps.}


 The interpretation then associates to each term a function taking the
 interpretation of its inputs to an overestimation of the length of
 its reduction (this is the numeric output) and the interpretation of
 its outputs. This interpretation of outputs is one small generalization needed from the standard Gandy proof in order to account for sequential composition of terms.

The measure on terms is extracted from the interpretation by a \textit{collapse} function\footnote{No relation to the similarly named collapse functor of Chapter \ref{chapter:CCC-equiv}} $\collapse{-}_{\type{t}}{:}\,\sntype{t} \to \N$, defined at each type $\type{t}$, which provides a least element as input and discards the non-numeric output:    
\begin{alignat*}{3}
&\collapse{f}_{\type{t}} &\quad= \quad & \pi_1(f(0_{\type{!r_A}})), 
\end{alignat*}
where $f: \type{t} = \type{?r_A > !s_A}$ (with the definition extended element-wise for type vectors and families). We omit the subscript $\type{t}$ whenever possible.
The \textit{measure function} will then be  given by the composition of the interpretation and the collapse function, $\collapse{\sn{-}}: \sntype{t} \to \N$, and we aim to prove that 
\[
	\term{M} \rw_\beta \term{M': t} \quad \textup{implies} \quad \collapse{\snterm{M}} >_{\N} \collapse{\snterm{M'}}, 
\]
from which strong normalisation immediately follows.

 The problem of strong normalisation is more difficult than that of weak normalisation due to the possibility of reduction within discarded arguments. 
To illustrate this, consider a machine-based intuition for the quantitative interpretation, where the numeric output of $\snterm{M}(\sn{S_A})$, with $\sn{S_A}$ defined pointwise on memories, measures the length of the machine run $(S_A, \term{M})$.
Then, given a naive interpretation function which measures only the length of the machine run (see Section \ref{sec:WN}), we run into the following problem:
pushing an argument to the stack takes only one transition step, but that leaves redexes inside the argument unaccounted for. Indeed, if the argument is then discarded or left unused, reductions inside it will not reduce the resulting measure at all. 

By adding the measure of the argument to the count as it is pushed to the stack, we can achieve a strict decrease in the measure. This ``trick'' of accounting for discarded terms in order to move from a proof of weak to strong normalisation is typical of Gandy-style proofs.
This results in an overestimation of the number of machine steps remaining, but it means that reductions inside the argument result in reductions in the measure, as required.

The proof of strong normalisation for the FMC described above directly specializes to one for
 simply-typed lambda calculus, but the proof so obtained differs from
 previous formulations in the way in which it enforces strict
 monotonicity. 

As stated, Gandy's proof interprets terms in domains of \textit{strictly} ordered, \textit{strict} monotone functions. 
These domains are well-founded (because they are strict) and the interpretation of terms is such that it decreases on reduction, giving strong normalisation.
This means that in the interpretation of $\lambda x.M$, we must then include the measure of the argument supplied to $x$, 
so that it always contributes to the overall interpretation, even if it's discarded. This is necessary in order for the interpretation of the abstraction to be strictly monotone. In other words, the measure of an argument is added to the count not as it is pushed to the stack, but as it is popped from the stack. 
 The literature has several further such constructions~\cite{Gortz-Reuss-Sorensen-2003}. However, this approach somewhat obscures the operational intuition.   
%


Our proof avoids the domain of \emph{strict} functionals and instead interprets terms in the domain of (non-strict) monotone functionals. 
One benefit of this is that the usual domain of monotone functions is much easier to work with than the strict domains. Non-strict domains are not well-founded, but our interpretation of terms ensures that when functionals are collapsed to a natural number, this strictly decreases upon reduction, again giving strong normalisation. 

Working with monotone functions is made possible because, in our proof,  we account for the possibility of reduction within arguments \textit{when they are supplied as arguments}, as opposed to when the arguments are consumed by an abstraction (as in Gandy's proof). That is, in the interpretation of $\term{[N].M}$, we increment the overall measure with that for $\term N$, measuring potential reduction in $\term N$ even if it will be discarded by $\term M$. 
An abstraction $\term{<x>.M}$ may then be given its natural interpretation as a standard monotone function, avoiding strictness. In fact, for the FMC, this variation is especially natural, because the problem of reduction inside arguments lies not just in arguments which are discarded, but also in certain terms not seen in the $\lambda$-calculus, such as $\term{[N].*}$, where an argument is merely never consumed.

We now formally define the interpretation of typed terms described above. From now on, when we say `monotone function', we mean \textit{non-strict} monotone function. 

\section{A Quantitative Interpretation}

In this section, we define the \textit{interpretation} of the type $\type{ ?s_A > !t_A }$ as the set of monotonic functions $\sn{\type{!s_A}} \to \mathbb{N} \times \sn{\type{!t_A}}$, equipped with the extensional partial ordering on functions and the interpretation of a term $\term{M:t}$ as $\snterm{M} \in \sntype{t}$.
We take the base domain $\N$ to have the usual ordering $\leq$.

The following work culminates in Section \ref{sec:meas}, where we prove that if $\term M\rw\term N$ at type $\type t$ then 
$\collapse{\snterm M} >_\N\collapse{\snterm N}$, to give strong normalisation.
We will prove that top-level beta reduction reduces the measure, and that reduction inside any context reduces the measure. 

Indeed, most of the following work is taken up by proving that top-level beta reduction -- the base case -- reduces the measure, which requires proving substitution commutes with the interpretation (Section \ref{sec:subst}) and that the interpretation respects the permutation of actions (Section \ref{sec:perm}).
Proving that reduction in any context  -- the inductive case -- reduces the measure is straightforward, excepting the one interesting case of reduction in an argument. 
This case necessitates proving that reduction of a term is monotonic, \textit{i.e.}, that
$\term{N} \rw \term{N'}$ {implies} $\snterm{N} \geq_{\sntype{t}} \snterm{N'}$ ,
which is the content of Section \ref{sec:monotonic}. 

\begin{notation}
We write elements of product domains as vectors $(t_1,\dots,t_n)$, and will elide the isomorphisms for associativity and unitality so that concatenation of $s$ and $t$ may be written $(s,t)$.
If ${t} \in \{T_a \ | \ a \in A\}$ denotes an $A$-indexed family of vectors, we denote by $t_a \in T_a$ the $a$'th projection of $t$. 
A \textit{singleton} $a({t})$, where $t$ is a vector and $a \in A$, is a family of vectors over $A$, such that $a({t})_a = {t}$ and  $a({t})_b$ is empty for $a \neq b$. Concatenation lifts to indexed products pointwise: $(s,t)_a = (s_a,t_a)$.
\end{notation}

\begin{definition}
The interpretation of each type $\type{t}$ of the FMC is given by the set $\sn{\type{t}}$, equipped with the binary relation $\leq_{\sntype{t}}$. We define these inductively on types by:
\begin{gather*}
	\sntype{ ?s_A > !t_A } = 
		\{f \in \sntype{!s_A} \to \N \times \sntype{!t_A} \mid  \forall s \leq_{\sntype{!s_A}} s'.~ f(s) \leq_{\mathbb{N}\times\sntype{!t_A}} f(s')~ \} \\
	f \leq_{\sntype{?s_A > !t_A}} g \quad \textup{iff} \quad \forall s \in \sntype {!s_A} . \ f(s) \leq_{\mathbb{N} \times\sntype{!t_A}} g(s) 
\end{gather*}
with vectors and families interpreted as products and dependent products, respectively,
\begin{gather*}
	\sntype{t_1 \ldots t_n} = \sntype{t_1} \times \ldots \times \sntype{t_n}
\qquad\textup{and}\qquad
	\sntype{!t_A} = \Pi_{a\in A}\,\sntype{!t_a}
\end{gather*}
and given the element-wise ordering.
\end{definition}

\begin{remark}
Note that,  because we are working without base types,  the base case of induction on types is  $\type{>}$, whose interpretation is isomorphic to $\N$. 
\end{remark}

It is worth observing that for the simple types of the $\lambda$-calculus, as embedded in the FMC, these domains are the natural ones. Briefly, a simple type $\type{t_1 \imp \dots \imp t_n \imp o}$ embeds as the FMC-type $\type{t_1\dots t_n>}$ with the domain $\sntype{t_1}\times\dots\times\sntype{t_n}\to\N$, which is the expected one modulo Currying.

\begin{proposition}
For all types $\type{t}$, the relation $\leq_{\sntype{t}}$ on $\sntype{t}$ is a partial order. 
\end{proposition}
\begin{proof}
Reflexivity, transitivity and anti-symmetry of $\leq_{\sntype{t}}$ follow by induction on types. 
\end{proof}
%

For every type $\type{t}$, we give a least inhabiting element $0_{\type{t}}$ in $\sntype{t}$.  These elements will be used in the definition of the measure, in order to extract the bound on the length of the machine run. 

\begin{definition}\label{def:inhabitants}Define $0_{\type{t}} \in \sntype{t}$ inductively on types:
\begin{alignat*}{3}
	&0_{\type{?s_A > !t_A}}     &\quad = \quad & s \in \sntype{!s_A} \mapsto (0 , 0_{\type{!t_A}}),
\end{alignat*}
where $0_{\type{!t_A}}$ is defined element-wise for vectors and families. 
\end{definition}
\begin{remark}
We can see by induction on types that $0_{\type{t}}$ is well-defined: it defines a constant function, which is thus monotonic. 
In fact, for all types $\type{t}$, $0_{\type{t}}$ is a least element of $\sn{\type{t}}$: the constant function which outputs a least element is itself a least element in the extensional ordering on functions. 
\end{remark}



\newcommand\p {\kern1pt{\smallbin+}\kern1pt}

We now define the interpretation function. First, we require to define the collapse function and a \textit{valuation} to interpret the free variables of a term. 
\begin{definition}
The \textit{collapse function} $\collapse{-}_{\type{t}}{:}\,\sntype{t} \to \N$ at each type $\type{t}$ is given by:
\begin{alignat*}{3}
&\collapse{f}_{\type{t}} &\quad= \quad & \pi_1(f(0_{\type{!r_A}})) ,
\end{alignat*}
for $f: \type{t} = \type{?r_A > !s_A}$. We omit the subscript ${\type{t}}$ whenever the type is clear. 
\end{definition}
\begin{remark}\label{rem:measmono}
For all types $\type{t}$, $\collapse{-}$ is monotonic.
\end{remark}

\begin{definition}
A \textit{valuation $v$} is a function assigning to each variable $\term{x:t}$ a value $v(\term{x}) \in \sntype t$. 
Given a valuation $v$, let $v\{\term{x}\leftarrow t\}$ denote the valuation on $\Gamma \cup \{\term{x: t}\}$ which assigns $t: \type{t}$ to $\term x$ and otherwise behaves as $v$. 
\end{definition}



The interpretation is then defined as follows. 
For the interpretation to be well-defined, the construction for each term must be shown to preserve monotonicity. We will do so in a subsequent lemma (Lemma \ref{increasing}).
\begin{definition}
For each term $\term{G |- M: ?s_A > !t_A}$ and valuation $v$, we inductively define the \textit{interpretation} function
\begin{align*}
	\sn{\Gamma \vdash \term{M: t}}_v \in \sntype{t}.
\end{align*}
We omit the context and/or types of terms inside an interpretation when it is clear. 
\begin{alignat*}{3}
		      \snterm{*: ?t_A > !t_A}_v(t)                  &= (0, t)
\\       \snterm{a<x>.M: `a(r)\,?s_A > !t_A}_v(s, a(r)) &= (1+n, t)
\\ \qquad \textup{where } (\concat nt) &= \snterm{M: ?s_A > !t_A}_{v\{\term{x} \leftarrow r\}} (s)
\\       \snterm{[N]a.M: ?s_A > !t_A}_v (s)  &= (1+n + \collapse{\snterm{N}_v}, t)
\\		\qquad \textup{where } (n, t) &= \snterm{M: `a(r)\,?s_A >!t_A}_v (s, a(\snterm{N: r}_v))
\\      \snterm{x.M: ?r_A?s_A > !t_A}_v(\concat sr) &= (\concat{n+m}t)
\\      \qquad \textup{where } (\concat mt) {}&=\snterm{M: ?u_A ?s_A > !t_A}_v (\concat su)
    \\  \qquad \textup{and } (\concat nu) {}&= v(\term{x: ?r_A > !u_A})(r)
\end{alignat*}
We define  $\sn{\Gamma \vdash \term{M: t}}$ to be $\sn{\Gamma \vdash \term{M: t}}_w$, for the \textit{least valuation} $w = \{\term{x}\smallbin \leftarrow 0_{\type{s}}\mid\term{x:s}\}$ given by the minimal elements. 
\end{definition}

\begin{remark}
In this definition, the application case $\snterm{[N]a.M}_v(s)=$ \linebreak$(1\p m\p \collapse{\snterm{N: r}_v}, t)$ adds the measure $\collapse{\snterm{N: r}_v}$ to account for reduction inside the argument $\term N$. Further, both it and the abstraction case $\snterm{a<x>.M}_v(s,a(r))=(1\p m,t)$ add $1$ to count redexes, so that a reduction step reduces the overall measure by (at least) $2$. It would suffice to count only abstractions or only applications, but the choice to count both is so that we count steps of the stack machine. We observe the following: for the alternative interpretation that omits to count $\collapse{\snterm{N: r}_v}$, and instead has $\snterm{[N]a.M}_v(s)=(1\p m,t)$, then $\snterm{M}(\snterm{S_A})$ measures the exact length of the machine run on $(S_A, \term M)$, where $\snterm{S_A}$ is defined element-wise. This is formally proven later, in Section \ref{sec:WN}, and provides the proof with an operational intuition.
\end{remark}

Note that in giving the above definition, technically, we must simultaneously prove its well-definedness at each inductive step. {That is, each level relies on the previous level being well-defined. For example, in the application case, to define $\snterm{[N].M: ?s_A > !t_A}_v$, we assume that $\snterm{N}_v \in \sntype{r}$ and $\snterm{M}_v \in \sntype{`a(r)\,?s_A >!t_A}$, (\textit{i.e.} that they are monotonic) in order to construct  $\snterm{[N].M} \in \sntype{?s_A > !t_A}$, and then prove that $\snterm{[N].M}$ is indeed monotonic.

We prove well-definedness in the lemma below. However, we need to strengthen the inductive hypothesis, as proving monotonicity in the abstraction case necessitates proving that ``increasing'' the valuation increases the interpretation. Thus, we add a second such clause to the statement of the lemma, which relies on the following definition. 
\begin{definition}
Given two valuations $v$ and $w$, write
\begin{align*}
	v \leq w \quad \textup{if} \quad \forall \term{x: t} \in \Gamma, \ v(\term{x}) \leq_{\sntype{t}} w(\term{x}) \ .
\end{align*}
\end{definition}
\begin{lemma}\label{increasing}
For all terms $\Gamma \vdash \term{M: t}$ and valuations $v$ and $w$, we have that:
\begin{align*}
{(i)}\  \sn{\term{M}}_v \in \sn{\type{t}} \quad {and} \quad (ii) \ v \leq w \ \  {implies} \ \   \sn{\term{M}}_v \leq_{\sntype{t}} \sn{\term{M}}_w\ .
\end{align*}
\end{lemma}
\begin{proof}
We prove both statements simultaneously by induction on the type derivation of $\term{G |- M: t}$. 
Recall that the first item is equivalent to claiming $\sn{\term{M: t}}_v$ is monotonic. When we write `increasing' here, we mean non-strictly. 
\begin{itemize}
\item
For the base case 
\begin{align*}
	\Gamma \vdash \term{M} \equiv \term{*: ?t_A > !t_A},
\end{align*}
\begin{enumerate}
\item
Observe that $\sn{\term{*}}_v$(t) = (0, t) and so is monotonic. 
\item
Observe that
$\sn{\term{*	}}_v
  = \sn{\term{*}}_w$
for every $v$ and $w$. 
\end{enumerate}
\item
For the abstraction case, where $\term{x: r}$ and
\begin{align*}
	\Gamma \vdash \term{M} \equiv \term{a<x>.M': `a(r)?s_A > !t_A},
\end{align*}
\begin{enumerate}
\item
We must show the function 
\begin{align*}
	\snterm{a<x>.M}_v(s, a(r)) &= (1+n, t)
\\ \qquad \textup{where } (\concat nt) &= \snterm{M}_{v\{\term{x} \leftarrow r\}} (s)\ 
\end{align*} is monotonic. Indeed, we have that
$ (s, a(r)) \leq_{\sntype{!s_A`a(r)}} (s', a(r')) $
 implies, by inductive hypothesis (ii) on $\term{M'}$, that 
\begin{align*}
 \sn{\term{M'}}_{v\{\term{x} \leftarrow r\}} \leq_{\sntype{?s_A > !t_A}} \sn{\term{M'}}_{v\{\term{x} \leftarrow r'\}} \ ,
\end{align*}
and then by inductive hypothesis $(i)$ on $\term{M'}$, this implies
\begin{align*}
 \sn{\term{M'}}_{v\{\term{x} \leftarrow r\}}(s) \leq_{\sntype{?s_A > !t_A}} \sn{\term{M'}}_{v\{\term{x} \leftarrow r'\}}(s') \ .
\end{align*}
Thus, increasing the input of the function increases its output. 
\item 
We wish to show, for arbitrary $(s, a(r)) \in \sntype{!s_A`a(r)}$, that
\[
v \leq w \quad \textup{implies} \quad \snterm{a<x>.M'}_v(s, a(r)) \leq_{\N \times \sntype{!t_A}} \snterm{a<x>.M'}_w(s, a(r))\ .
\]
Unfolding definitions, we see we must show that $v \leq w$ implies
\begin{align*}  (1+n, t) &\leq_{\N \times \sntype{!t_A}} (1+n', t')
\\  \textup{where }  (\concat nt) &= \snterm{M}_{v\{\term{x} \leftarrow r\}} (s)\ ,
\\  \textup{and }   (\concat n't') &= \snterm{M}_{w\{\term{x} \leftarrow r\}} (s)\ 
\end{align*}
By assumption, we have that, for all $r \in \sntype{r}$, $v\{\term{x} \leftarrow r\} \leq w\{\term{x} \leftarrow r\}$. Applying inductive hypothesis $(ii)$ on $\term{M'}$, we thus have that 
\[
	\snterm{M'}_{v\{\term{x} \leftarrow r\}} \leq_{\sntype{?s_A > !t_A}} \snterm{M'}_{w\{\term{x} \leftarrow r\}}\ .
\]
and consequently $(n, t) \leq_{\N \times \sntype{!t_A}} (n', t')$. 
Indeed, the required result
immediately follows. 
\end{enumerate}
\item
For the application case, with $\Gamma \vdash \term{N: r}$ and
\begin{align*}
	\Gamma \vdash \term{M} \equiv \term{[N]a.M': ?s_A > !t_A}, 
\end{align*}
\begin{enumerate}
\item
We have to show that 
\begin{align*}
 	\snterm{[N]a.M'}_v (s)  &= (1+n + \collapse{\snterm{N}_v}, t)
	\\		 \textup{where } (n, t) &= \snterm{M'}_v (s, a(\snterm{N}_v))\ .
\end{align*}
is monotonic. Applying inductive hypothesis $(i)$ on $\term{M'}$, we achieve that  $\snterm{M'}_v$ is monotonic. Thus, increasing the input $s$ increases $(n, t)$, which therefore increases the output of the entire function.  
\item
We have to show, for arbitrary $s \in \sntype{!s_A}$, that
\[
v \leq w \qquad \textup{implies} \qquad \snterm{[N]a.M'}_v (s) \leq_{\N \times \sntype{!t_A}} \snterm{[N]a.M'}_w (s)
\]
Unfolding definitions, we see we must show that $v \leq w$ implies
\begin{align*}
 	(1+n + \collapse{\snterm{N}_v}, t) & \leq_{\N \times \sntype{!t_A}} (1+n' + \collapse{\snterm{N}_w}, t')
	\\		 \textup{where } (n, t) &= \snterm{M'}_w (s, a(\snterm{N}_v))\ .
	\\		 \textup{and } (n', t') &= \snterm{M'}_w (s, a(\snterm{N}_w))\ .
\end{align*}
Applying inductive hypothesis $(ii)$ on $\term{M'}$ and $\term{N}$, we achieve 
\[
	\sn{\term{M'}}_v \leq_{\sntype{`a(r)?s_A > !t_A}} \sn{\term{M'}}_w \quad \textup{and} \quad \sn{\term{N}}_v \leq_{\sntype{r}} \sn{\term{N}}_w. 
\]
The conjunction of both statements implies that 
$(n, t) \leq_{\N \times \sntype{!t}} (n', t')$ .
Using Remark \ref{rem:measmono}, we aditionaly observe  $\collapse{\snterm{N}_v} \leq_{\N} \collapse{\snterm{N}_w}$, and the result follows.
\end{enumerate}
\item
For the variable case, 
\begin{align*}
	\Gamma, \term{x: ?r_A > !u_A} \vdash \term{M} \equiv \term{x.M': ?r_A?s_A > !t_A},
\end{align*}
\begin{enumerate}
\item
We have to show that 
\begin{align*}
 \snterm{x.M'}_v(\concat sr) &= (\concat{n+m}t)
\\       \textup{where } (\concat mt) {}&=\snterm{M'}_v (\concat su)
    \\   \textup{and } (\concat nu) {}&= v(\term{x})(r)\ .
\end{align*}
is monotonic. 
Observe that $v(\term{x}) \in \sn{\type{?r_A > !u_A}}$ and so is monotonic. Thus, increasing the input $(s, r)$ increases $(n, u)$. We have from  inductive hypothesis (i) on $\term{M'}$ that  $\sn{ \term{M'}}$ is monotonic.  Altogether, this results in an increase in $(m, t)$ , which therefore increases the output of the entire function. 
\item 
We wish to show, for arbitrary $(s, r) \in \type{!s_A!r_A}$, that
\[
	v \leq w \qquad \textup{implies} \qquad \snterm{x.M'}_v(s, r) \leq_{\N \times \sntype{t_A}} \snterm{x.M'}_w(s,r)\ .
\]
Unfolding definitions, we see we must show that $v \leq w$ implies
\begin{align*}
 (\concat{n+m}t) &\leq_{\N \times \sntype{t_A}} (\concat{n'+m'}t')
\\       \textup{where } (\concat mt) {}&=\snterm{M'}_v (\concat su)
    \\   \textup{and   }\ \  (\concat nu) {}&= v(\term{x})(r)
\\       \textup{and } (\concat m't') {}&=\snterm{M'}_w (\concat su')
    \\   \textup{and } (\concat n'u') {}&= w(\term{x})(r)\ .
\end{align*}
By assumption, we have $v(\term{x}) \leq_{\sntype{?r_A > !u_A}} w(\term{x})$, which implies that 
\linebreak $(n, u) \leq_{\N \times \sntype{!u_A}} (n', u')$. Applying inductive hypothesis (ii) on $\term{M'}$, we have that 
\[
	\sn{\term{M'}}_v \leq_{\sntype{?u_A?s_A>!t_A}}  \sn{\term{M'}}_w. 
\]
Altogether, this implies $(m, t) \leq_{\N \times \sntype{!t_A}} (m', t')$.
Thus, we achieve the required result. \qedhere
\end{enumerate} 
\end{itemize}
 \end{proof}

\begin{remark}\label{weak-is-monotonic}
Note that monotonicity does not rely on the measure of the argument, $\collapse{\snterm{N}_{v}}$, being added to the count in the application case. 
\end{remark}

We will see later that monotonicity alone suffices to prove that reduction inside a term non-strictly decreases the measure (Lemma \ref{non-increasing}). Adding $\collapse{\snterm{N}_{v}}$ to the count is, however, necessary for achieving a strict inequality.

\section{The Substitution Lemma}\label{sec:subst}
To prove the base case of the main proposition  (Proposition \ref{prop:onestep}, that beta reduction reduces the measure), we need to prove that substitution commutes with the interpretation in an appropriate way. In order to prove that, we must first show that weakening and sequencing behave well with respect to the interpretation. Each of these lemmata is proved by a simple induction on the type derivation of terms.

\begin{lemma}[Weakening]\label{sn:lem:weakening}
For terms $\Gamma \vdash \term{M: t}$ and valuation $v$ and $s \in \sntype{s}$, we have that
\begin{align*}
\sn{\Gamma \vdash \term{M:t }}_{v} = \sn{\Gamma, \term{x:s} \vdash \term{M:t}}_{v\{\term{x} \leftarrow s\}} ,
\end{align*}
where $\term{x} \not\in \fv{\term{M}}$.
\end{lemma}
\begin{proof}
Induction on the type derivation of $\Gamma \vdash \term{M: t}$.
\end{proof}
As a consequence of this lemma, we may speak of $\snterm{M:t}$ independent of a valuation $v$ when $\term M$ is closed. 

We must show that sequencing behaves well with respect to the interpretation because substitution for a variable results in a sequencing operation, and this is required in the variable case of the Substitution Lemma. 

\begin{lemma}[Sequencing] \label{sn:lem:sequencing}
For terms $\Gamma \vdash \term{ M: ?s_A?t_A > !u_A}$ and $\Gamma \vdash \term{N: ?r_A > !s_A}$  and valuation $v$, we have:
\begin{align*}
 \sn{\term{N;M}}_v  ({t}, {r}) &= (i+j, {u}) \\
    \textup{where } (i, {s}) &=  \sn{\term{N}}_{v} ({r}) \\
     \textup{and } (j, {u}) &= \sn{\term{M}}_{v} ({t}, {s}) .
\end{align*}
\end{lemma}

\begin{proof}
We proceed by induction on the type derivation of $\term{G |- N:t}$.
\begin{itemize}

	\item
For the base case, 
\[
	\term{G |- N }\equiv \term{* : ?s_A > !s_A},
\]
let $\snterm{M}_v(t,s)=(m,u)$. Since $\snterm{*}_v(s)= (0,s)$, we need to show that \linebreak $\snterm{*;M}_v(t,s)=(0+m,u)$, but this is immediate since $\term{*;M}=\term M$.

	\item
For the abstraction case,
\[
	\term{G |- N }\equiv\term{a<x>.N': a(p)\,?r_A > !s_A},
\] 
let 
\[
	\snterm{N'}_{v\{\term x\leftarrow p\}}(r)=(i,s), \qquad \snterm{M}_v(t,s)=(j,u),
\]
so that $\snterm{a<x>.N'}_v(r,p)=(1+i,s)$. 
Because $\term x$ is not free in $\term M$, by Lemma~\ref{sn:lem:weakening} we have $\snterm{M}_{v\{\term x\leftarrow p\}}(t,s)=(j,u)$. The inductive hypothesis gives \linebreak $\snterm{N';M}_{v\{\term x\leftarrow p\}}(t,r)=(i+j,u)$, so that $\snterm{a<x>.(N';M)}_v(t,r,p)=(1+i+j,u)$. By definition, $\term{a<x>.N';M}=\term{a<x>.(N';M)}$, so this is the required result.

	\item
For the application case, with $\term{P:p}$,
\[
	\term{G |- N }\equiv \term{[P]a.N': ?r_A > !s_A},
\]
let 
\[
	\qquad \snterm {N'}_v(r,\snterm P_v)=(i,s), \qquad \snterm M_v(t,s) = (j,u),
\]
so that $\snterm{[P]a.N'}_v(r)=(\floor {\snterm P_v}+1+i,s)$. 
 The inductive hypothesis gives $\snterm{N';M}_v(t,r,\snterm P_v)=(i+j,u)$, so that $\snterm{[P]a.(N';M)}_v(t,r)=(\floor {\snterm P_v}+1+i+j, u)$. By definition, $\term{[P]a.N';M}=\term{[P]a.(N';M)}$, so this is the required result.

	\item
For the variable case,
\[
	\term{G , x: ?n_A > !p_A |- N }\equiv \term{x.N' : ?n_A\,?r_A > !s_A},
\]
let 
\[
	v(\term x)(n)=(i,p) \qquad \snterm {N'}_v(r,p) = (j,s) \qquad \snterm M_v(t,s) = (k,u)
\]
so that $\snterm{x.N'}_v(r,n)=(i+j,s)$. 
The inductive hypothesis gives \linebreak $\snterm{N';M}_v(t,r,p)=(j+k,u)$, so that $\snterm{x.(N';M)}_v(t,r,n)=(i+j+k,u)$. By definition, $\term{x.N';M}=\term{x.(N';M)}$, so this is the required result. \qedhere
\end{itemize}
\end{proof}

We finally come to prove the Substitution Lemma. This is required in the base case of the main Proposition \ref{prop:onestep} to follow. 
\begin{lemma}[Substitution]\label{sn:lem:substitution}
For terms $\Gamma, \term{x:w} \vdash \term{M: t}$ and $\Gamma \vdash \term{N: w}$ and valuation $v$, we have
\begin{align*}
	\sn{\term{\{N/x\}M}}_{v} = \sn{\term{M}}_{v\{\term{x} \leftarrow \sn{\term{N}}_v\} } \ .
\end{align*}
\end{lemma}
\begin{proof}
We proceed by induction on the type derivation of $\Gamma, \term{x:w} \vdash \term{M: t}$. 
\begin{itemize}
\item For the base case,  
\begin{align*}
	\Gamma, \term{x:w} \vdash \term{M} \equiv \term{*: ?t_A > !t_A}, 
\end{align*}
we have 
\begin{align*}
\snterm{G |- \{N/x\}*}_v (t) = \snterm{G |- *}_v(t) = (0,t) = \snterm{G, x |- *}_{v\{\term x \leftarrow \snterm{N}_v\}}(t)\ ,
\end{align*}
as required.
\item
For the abstraction case, 
\begin{align*}
	\Gamma, \term{x:w} \vdash \term{M} \equiv \term{a<y>.M': `a(r)?s_A > !t_A}, 
\end{align*}
we need to show that 
\begin{align*}
\snterm{\{N/x\}a<y>.M'}_v(s, a(r)) = \snterm{a<y>.M'}_{v\{\term x \leftarrow \snterm{N}_v\}}(s, a(r)) \ ,
\end{align*}
where $\term x \neq \term y$. For the left-hand side, we have
\begin{align*}
 \sn{\term{\{N/x\}a<y>.M'}}_{v} ({s}, a(r)) &= \\
\sn{\term{a<y>.\{N/x\}M'}}_{v} ({s}, a(r)) & = (1+n, {t})\\
\textup{where } (n, {t}) &= \sn{\term{\{N/x\}M'}}_{v\{\term{y} \leftarrow r\}} ({s})\ ,
\end{align*}
and, for the right-hand side, we have
\begin{align*}
\sn{\term{a<y>.M'}}_{v\{\term x \leftarrow \snterm{N}_v\}} ({s}, a(r)) &= (1+n', {t'}) \\
\textup{where } (n', {t'}) &= \sn{\term{M'}}_{u\{\term x \leftarrow \snterm{N}_v\}} ({s}) \ .
\end{align*} 
Let $u = v\{\term{y} \leftarrow r\}$. 
Applying the inductive hypothesis on $\term{M'}$ gives the first equality below. 
\[
	\snterm{\{N/x\}M}_{v\{\term y \leftarrow r\}} = \snterm{M'}_{u\{\term x \leftarrow \snterm{N}_u\}} =  \snterm{M'}_{u\{\term x \leftarrow \snterm{N}_v\}} 
\]By the Weakening Lemma \ref{sn:lem:weakening},
and since $\term y \not\in \fv{\term{N}}$, we have $\snterm{N}_u = \snterm{N}_v$. This gives the second equality. 
Thus, we have that $(n, t) = (n', t')$ and the required result follows.
\item
For the application case, where $\Gamma, \term{x: w} \vdash  \term{P: r}$ and
\begin{align*}
	\Gamma, \term{x:w} \vdash \term{M} \equiv \term{[P]a.M': ?s_A > !t_A}, 
\end{align*}
we need to show that
\[
	\snterm{\{N/x\}[P]a.M'}_v(s) = \snterm{[P]a.M'}_{v\{\term x \leftarrow \snterm{N}_v\}}(s)\ .
\]
For the left-hand side, we have
\begin{align*}
 \sn{\term{\{N/x\}[P]a.M'}}_{v}({s}) &= \\
 \sn{\term{[\{N/x\}P]a.\{N/x\}M'}}_{v}({s}) &= (1+n+\collapse{\snterm{\{N/x\}P}_v}, t)  \\
 \textup{where } (n, t) &=  \sn{\term{\{N/x\}M'}}_{v}  ({s}, a(\sn{\term{\{N/x\}P}}_{v})) \ ,
\end{align*}
and, for the right-hand side, we have
\begin{align*}
\sn{\term{[P]a.M'}}_{v\{\term x \leftarrow \snterm{N}_v\}}({s}) &= (1+n'+\collapse{\snterm{P}_{v\{\term x \leftarrow \snterm{N}_v\}}}, t') \\
 \textup{where } (n', t') &= \sn{\term{M'}}_{v\{\term x \leftarrow \snterm{N}_v\}}  ({s}, a(\sn{ \term{P}}_{v\{\term x \leftarrow \snterm{N}_v\}})) \ .
\end{align*}
Applying the inductive hypothesis on  $\term{M'}$ and $\term{P}$ achieves 
\begin{align*}
 \snterm{\{N/x\}M'}_{v} &= \snterm{M'}_{v\{\term x \leftarrow \snterm{N}_v\}} \\
 \snterm{\{N/x\}P}_{v} &= \snterm{P}_{v\{\term x \leftarrow \snterm{N}_v\}} \  .
\end{align*}
Thus, we have that $(n, t) = (n', t')$ and indeed $1+n'+\collapse{\snterm{\{N/x\}P}_{v}} = 1+n'+\collapse{\snterm{P}_{v\{\term x \leftarrow \snterm{N}_v\}}}$ as required.
\item
For the variable case, where $\term{y} \neq \term{x}$,
\begin{align*}
	\Gamma, \term{x:w}, \term{y: ?r_A > !u_A} \vdash \term{M} \equiv \term{y.M': ?r_A?s_A> !t_A}, 
\end{align*}
 we need to show that 
\[
	\snterm{\{N/x\}y.M'}_v(s, r) = \snterm{y.M'}_{v\{\term x \leftarrow \snterm{N}_v\}}(s,r)\ .
\]
For the left-hand side, 
\begin{align*}
 \sn{\term{\{N/x\}y.M'}}_{v}({s}, {r}) &= \\
 \sn{\term{y.\{N/x\}M'}}_{v}({s}, {r}) &= (n+m, {t}) \\
  \textup{where } (m, {t}) &= \sn{\term{\{N/x\}M'}}_{v} ({s}, {u}) \\
  \textup{and }  (n, {u}) &= v(\term{y})({r})\ ,
\end{align*}
and for the right-hand side, 
\begin{align*}
\sn{\term{y.M'}}_{v\{\term x \leftarrow \snterm{N}_v\}} ({s}, {r}) &= (n'+m', {t'}) \\
  \textup{where } (m', {t'}) &= \sn{\term{M'}}_{v\{\term x \leftarrow \snterm{N}_v\}} ({s}, {u'}) \\
   \textup{and } (n', {u'}) &= v\{\term x \leftarrow \snterm{N}_v\}(\term{y})({r})  \ .
\end{align*}
Observe that $v\{\term x \leftarrow \snterm{N}_v\}(\term{y}) = v(\term{y})$, which implies  $(n, u) = (n', u')$. 
Application of the inductive hypothesis on $\term{M'}$ achieves
\begin{align*}
 \snterm{\{N/x\}M'}_{v} &= \snterm{M'}_{v\{\term x \leftarrow \snterm{N}_v\}} \ .
\end{align*}
Thus, we have $(m, t) = (m', t')$ and the result follows. 
\item
For the variable case,
\begin{align*}
	\Gamma, \term{x:?r_A > !u_A} \vdash \term{M} \equiv \term{x.M': ?r_A?s_A > !t_A}, 
\end{align*}
we need to show that 
\[
	\snterm{\{N/x\}x.M}_v(s, r) = \snterm{x.M}_{v\{\term x \leftarrow \snterm{N}_v\}}(s,r)\ .
\]
For the left-hand side, by application of the Sequencing Lemma \ref{sn:lem:sequencing}, we have
\begin{align*}
 \sn{\term{\{N/x\}x.M'}}_{v}  ({s}, {r}) &=\\
 \sn{\term{N;\{N/x\}M'}}_{v}  ({s}, {r}) &= (n+m, {t}) \\
    \textup{where } (m, {t}) &= \sn{ \term{\{N/x\}M'}}_{v} ({s}, {u}) \\
    \textup{and } (n, {u}) &=  \sn{\term{N}}_v ({r})  \ ,
\end{align*}
and, for the right-hand side,
\begin{align*}
 \sn{ \term{x.M'}}_{v\{\term x \leftarrow \snterm{N}_v\}} ({s}, {r}) &= (n'+m', {t'})  \\
    \textup{where }   (m', {t'}) &= \sn{\term{M'}}_{v\{\term x \leftarrow \snterm{N}_v\}} ({s}, {u'})\\
  \textup{and } (n', {u'}) &= v\{\term x \leftarrow \snterm{N}_v\}(\term{x})({r}) \ .
\end{align*}
Observe that  $v\{\term x \leftarrow \snterm{N}_v\}(\term{x}) = \sn{\term{N}}_{v}$, which implies $(n, u) = (n', u')$.
Application of the inductive hypothesis on $\term{M'}$ achieves
\begin{align*}
 \snterm{\{N/x\}M'}_{v} &= \snterm{M'}_{v\{\term x \leftarrow \snterm{N}_v\}} \ .
\end{align*}
Thus, we have $(m, t) = (m', t')$ and the result follows.\qedhere
\end{itemize}
\end{proof}

\section{Beta Reduction is Monotonic}\label{sec:monotonic}
We now prove that reduction inside a term non-strictly decreases the interpretation of that term, which is necessary for the argument case of the main lemma, which claims that reduction \textit{strictly} decreases the measure. 
\begin{lemma}\label{non-increasing}
For every valuation v, 
\begin{align*}
	\term{G |- M}  \rw  \term{M': t} \quad {implies} \quad \snterm{M}_v \geq_{\sntype{t}} \snterm{M'}_v \ .
\end{align*}
\end{lemma}
\begin{proof}
We proceed by induction on the derivation of one-step reductions. 
\begin{itemize}
\item The base case,
\begin{align*}
	\term{[N]a.a<x>.M}~\rw~\term{\{N/x\}M} : \type{?s_A > !t_A},
\end{align*}
where $\term{x, N: r}$ requires showing, for arbitrary $s \in \sntype{!s_A}$, that 
\begin{align*}
	\snterm{[N]a.a<x>.M}_v(s)~\geq_{\N \times \sntype{!t_A}}~\snterm{\{N/x\}M}_v(s) \ .
\end{align*}
Unfolding definitions, we must show
\begin{align*}
(2 + n + \collapse{\snterm{N}_{v}}, t) & \geq_{\N \times \sntype{!t_A}}~\snterm{\{N/x\}M}_v(s)
\\		 \textup{where } (n, t) &= \snterm{M}_{v\{x \leftarrow \snterm{N}_v \}}(s)\ .
\end{align*}
Applying the Substitution Lemma \ref{sn:lem:substitution}, 
\[
	\snterm{M}_{v\{x \leftarrow \snterm{N}_v \}} =  \snterm{\{N/x\}M}_{v},
\]
 we see $(n, t) = \snterm{\{N/x\}M}_{v}(s)$.
Observing that $(2 + n + \collapse{\snterm{N}_{v}}, t) \geq_{\N \times \sntype{!t_A}} (n, t)$, the result follows.
\item The abstraction case,
\begin{align*}
	\term{a<x>.M}~\rw ~\term{a<x>.M'} : \type{ `a(r)?s_A > !t_A},
\end{align*}
requires showing, for arbitrary $(s, a(r)) \in \sntype{!s_A`a(r)}$, that
\[
	\snterm{a<x>.M}_v(s, a(r))~ \geq_{\N \times \sntype{!t_A}} ~\snterm{a<x>.M'}_v(s, a(r)) \ .
\]
Unfolding definitions, we must show
\begin{align*}
	(1+n, t) &\geq_{\N \times \sntype{!t_A}} (1+n', t')
\\  \textup{where } (\concat nt) &= \snterm{M}_{v\{\term{x} \leftarrow r\}} (s) 
\\  \textup{and } (\concat {n'}t') &= \snterm{M'}_{v\{\term{x} \leftarrow r\}} (s)\ .
\end{align*}
Applying the inductive hypothesis on $\term{M} \rw \term{M'}$, we achieve
\[
	\snterm{M}_w(s) \geq_{\N \times \sntype{!t_A}} \snterm{M'}_{w}(s)  \ ,
\]
for any valuation $w$. In particular, we can set $w = \{\term{x} \leftarrow r\}$
and thus we have that $(n, t) \geq_{\N \times \sntype{!t_A}} (n', t')$. The result follows. 
\item The application, function case,
\begin{align*}
	\term{[N]a.M}~\rw ~\term{[N]a.M'} : \type{ ?s_A > !t_A},
\end{align*}
requires showing, for arbitrary $s \in \sntype{!s_A}$, that
\[
	\snterm{[N]a.M}_v(s)~ \geq_{\N \times \sntype{!t_A}} ~\snterm{[N]a.M'}_v(s) \ .
\]
Unfolding definitions, we must show
\begin{align*}
	(1+n+\collapse{\snterm{N}_v}, t) &\geq_{\N \times \sntype{!t_A}} (1+n'+\collapse{\snterm{N}_v}, t')
\\  \textup{where } (\concat nt) &= \snterm{M}_{v}(s, a(\snterm{N}_v)) 
\\  \textup{and } (\concat n't') &= \snterm{M'}_{v}(s, a(\snterm{N}_v)) \ .
\end{align*}
Applying the inductive hypothesis on   $\term{M} \rw \term{M'}$,  we achieve 
\[
	\snterm{M}_w(s, a(\snterm{N}_v)) \geq_{\N \times \sntype{!t_A}} \snterm{M'}_{w}(s, a(\snterm{N}_v))  \ ,
\]
and thus that $(n, t) \geq_{\N \times \sntype{!t}} (n', t')$. The result follows. 
\item The application, argument case, where $\term{N: r}$,
\begin{align*}
	\term{[N].M}~\rw ~\term{[N'].M} : \type{ ?s_A > !t_A},
\end{align*}
requires showing, for arbitrary $s \in \sntype{!s_A}$, that
\[
	\snterm{[N]a.M}_v(s)~ \geq_{\N \times \sntype{!t_A}} ~\snterm{[N']a.M}_v(s) \ .
\]
Unfolding definitions, we must show
\begin{align*}
(1+n+\collapse{\snterm{N}_v}, t) &\geq_{\N \times \sntype{!t_A}} (1+n'+\collapse{\snterm{N'}_v}, t')
\\  \textup{where } (\concat nt) &= \snterm{M}_{v}(s, a(\snterm{N}_v)) 
\\  \textup{and } (\concat n't') &= \snterm{M}_{v}(s, a(\snterm{N'}_v)) 
\end{align*}
Applying the inductive hypothesis on $\term{N} \rw \term{N'}$, we achieve
\[
	\snterm{N}_v \geq_{\sntype{r}} \snterm{N'}_v
\]
which allows us to apply monotonicity of $\snterm{M}_v$ to see that $(n, t) \geq_{\N \times \sntype{!t_A}} (n', t')$. Remark \ref{rem:measmono}, stating monotonicity of $\collapse{-}$, then implies that $ (1+n+\collapse{\snterm{N}_v}, t)\geq_{\N \times \sntype{!t_A}}  (1+n'+\collapse{\snterm{N'}_v}, t')$, as required.
\item The variable case,
\begin{align*}
	\term{x.M}~\rw~ \term{x.M'} : \type{?r_A?s_A > !t_A},
\end{align*}
with $\term{x: ?r_A > !u_A}$
requires showing, for arbitrary $(s,r) \in \sntype{!s_A!r_A}$, that
\[
	\snterm{x.M}_v(s, r)~ \geq_{\N \times \sntype{!t_A}} ~\snterm{x.M'}_v(s, r) \ .
\]
Unfolding definitions, we must show
\begin{align*}
	 (n+m, t) &\geq_{\N \times \sntype{!t_A}} (n+m', t') \\
  \textup{where } (\concat mt) &= \snterm{M}_{v}(s, u) \\
  \textup{and } (\concat m't') &= \snterm{M'}_{v}(s, u) \\
  \textup{and } (n, u) &= v(\term{x})(r) 
\end{align*}
Applying the inductive hypothesis on   $\term{M} \rw \term{M'}$,  we achieve, for any $u \in \sntype{!u_A}$,
\[
	\snterm{M}_w(s, u) \geq_{\N \times \sntype{!t_A}} \snterm{M'}_{w}(s, u)  \ ,
\]
and thus that $(m, t) \geq_{\N \times \sntype{!t_A}} (m', t')$ and consequently $(n+m, t) \geq_{\N \times \sntype{!t_A}} (n+m', t')$, as required. \qedhere
\end{itemize}
\end{proof}
\section{The Measure for Strong Normalisation}\label{sec:meas}

The next lemma immediately implies the strong normalisation result: if $\term{G |- M }\rw \term{N: t}$ then $\floor{\snterm M} >_\N \floor{\snterm N}$, so that $\floor{\snterm M}$ gives a bound for the length of any reduction path from $\term M$. The following proof relies essentially on monotonicity of the interpretation of a term, in particular for the application, argument case, as well as the fact the measure of the argument is added to the count as it is pushed to the stack 

\begin{proposition}\label{prop:onestep}
For every valuation $v$ and for all $s \in \sn{\type{!s_A}}$, we have that
\begin{align*}
	\term{G |- M} \rw \term{M': ?s_A > !t_A} \quad {implies} \quad \pi_1(\snterm{M}_v(s)) >_{\N} \pi_1(\snterm{M'}_v(s))\ .
\end{align*}
\end{proposition}
\begin{proof}
We proceed by induction on the derivation of one-step reductions.
\begin{itemize}
\item 

The base case,
\begin{align*}
	\term{[N]a.a<x>.M}~\rw~\term{\{N/x\}M} : \type{?s_A > !t_A},
\end{align*}
where $\term{x, N: r}$ requires showing, for arbitrary $s \in \sntype{!s_A}$, that 
\begin{align*}
	\pi_1(\snterm{[N]a.a<x>.M}_v(s))~ >_{\N}~ \pi_1(\snterm{\{N/x\}M}_v(s)) \ .
\end{align*}
Unfolding definitions, we must show
\begin{align*}
2 + n + \collapse{\snterm{N}_{v}} & >_{\N} ~\pi_1(\snterm{\{N/x\}M}_v(s))
\\		 \textup{where } (n, t) &= \snterm{M}_{v\{x \leftarrow \snterm{N}_v \}}(s)\ .
\end{align*}
Applying the Substitution Lemma \ref{sn:lem:substitution}, 
\[
	\snterm{M}_{v\{x \leftarrow \snterm{N}_v \}} =  \snterm{\{N/x\}M}_{v},
\]
 we see $(n, t) = \snterm{\{N/x\}M}_{v}(s)$.
Indeed, the required result follows immediately from observing that  $n = \pi_1( \snterm{\{N/x\}M}_{v}(s))$.
\item The abstraction case, 
\begin{align*}
	\term{a<x>.M}~\rw ~\term{a<x>.M'} : \type{ `a(r)?s_A > !t_A},
\end{align*}
requires showing, for arbitrary $(s, a(r)) \in \sntype{!s_A`a(r)}$, that
\[
	\pi_1(\snterm{a<x>.M}_v(s, a(r)))~ >_{\N} ~\pi_1(\snterm{a<x>.M'}_v(s, a(r))) \ .
\]
Unfolding definitions, we must show
\begin{align*}
	1+n &>_{\N} 1+n'
\\  \textup{where } (\concat nt) &= \snterm{M}_{v\{\term{x} \leftarrow r\}} (s) 
\\  \textup{and } (\concat {n'}t') &= \snterm{M'}_{v\{\term{x} \leftarrow r\}} (s)\ .
\end{align*}
Applying the inductive hypothesis on $\term{M} \rw \term{M'}$, we achieve
\[
	\pi_1(\snterm{M}_w(s)) >_{\N} \pi_1(\snterm{M'}_{w}(s))  \ ,
\]
for any valuation $w$. In particular, we can set $w = \{\term{x} \leftarrow r\}$ and thus we have that $n >_{\N} n'$, as required. 
\item The application, function case,
\begin{align*}
	\term{[N]a.M}~\rw ~\term{[N]a.M'} : \type{ ?s_A > !t_A},
\end{align*}
requires showing, for arbitrary $s \in \sntype{!s_A}$, that
\[
	\pi_1(\snterm{[N]a.M}_v(s))~ >_{\N} ~\pi_1(\snterm{[N]a.M'}_v(s)) \ .
\]
Unfolding definitions, we must show
\begin{align*}
	1+n+\collapse{\snterm{N}_v} &>_{\N} 1+n'+\collapse{\snterm{N}_v}
\\  \textup{where } (\concat nt) &= \snterm{M}_{v}(s, a(\snterm{N}_v)) 
\\  \textup{and } (\concat n't') &= \snterm{M'}_{v}(s, a(\snterm{N}_v)) \ .
\end{align*}
Applying the inductive hypothesis on   $\term{M} \rw \term{M'}$,  we achieve 
\[
	\pi_1(\snterm{M}_w(s, a(\snterm{N}_v)))  >_{\N} \pi_1(\snterm{M'}_{w}(s, a(\snterm{N}_v)))  \ ,
\]
and thus that $n >_{\N} n'$, as required. 
\item The application, argument case, where $\term{N: r}$,
\begin{align*}
	\term{[N]a.M}~\rw ~\term{[N']a.M} : \type{ ?s_A > !t_A},
\end{align*}
requires showing, for arbitrary $s \in \sntype{!s_A}$, that
\[
	\pi_1(\snterm{[N]a.M}_v(s))~ >_{\N} ~\pi_1(\snterm{[N']a.M}_v(s)) \ .
\]
Unfolding definitions, we must show
\begin{align*}
1+n+\collapse{\snterm{N}_v} &>_{\N} 1+n'+\collapse{\snterm{N'}_v}
\\  \textup{where } (\concat nt) &= \snterm{M}_{v}(s, a(\snterm{N}_v)) 
\\  \textup{and } (\concat n't') &= \snterm{M}_{v}(s, a(\snterm{N'}_v)) 
\end{align*}
Applying the inductive hypothesis on $\term{N} \rw \term{N'}$, we achieve
\[
	\pi_1(\snterm{N}_v(s)) >_{\N} \pi_1(\snterm{N'}_v(s))
\]
In the special case where $s$ is the minimal element, we recover $\collapse{\snterm{N}_v} >_{\N} \collapse{\snterm{N'}_v}$.
Additionally, by Lemma \ref{non-increasing}, we have that $\snterm{N} \geq_{\N \times \sntype{!t}} \snterm{N'}$. This allows us to apply monotonicity of $\snterm{M}$ to deduce that $(n, t) \geq_{\N \times \sntype{!t}} (n', t')$. 
Altogether, this suffices for the result. 
%
\item
The variable case,
\begin{align*}
	\term{x.M}~\rw~ \term{x.M'} : \type{?r_A?s_A > !t_A},
\end{align*}
with $\term{x: ?r_A > !u_A}$
requires showing, for arbitrary $(s,r) \in \sntype{!s_A!r_A}$, that
\[
	\pi_1(\snterm{x.M}_v(s, r))~ >_{\N} ~\pi_1(\snterm{x.M'}_v(s, r)) \ .
\]
Unfolding definitions, we must show
\begin{align*}
	 n+m &>_{\N} n+m' \\
  \textup{where } (\concat mt) &= \snterm{M}_{v}(s, u) \\
  \textup{and } (\concat m't') &= \snterm{M'}_{v}(s, u) \\
  \textup{and } (n, u) &= v(\term{x})(r) 
\end{align*}
Applying the inductive hypothesis on   $\term{M} \rw \term{M'}$,  we achieve, for any $u \in \sntype{!u_A}$,
\[
	\pi_1(\snterm{M}_w(s, u)) >_{\N} \pi_1(\snterm{M'}_{w}(s, u))  \ ,
\]
and thus that $m >_{\N} m'$, as required. \qedhere
\end{itemize}
\end{proof}

%

\begin{theorem}
All typed FMC-terms are strongly normalizing with respect to beta reduction. 
\end{theorem}
\begin{proof}
For a term $\term{G |- M:t}$, Proposition~\ref{prop:onestep} gives that any reduction path from $\term M$ gives a strictly decreasing sequence in $\N$, by application of $\collapse{\sn{-}}$, which must therefore have finite length. 
\end{proof}

\section{Permutation Reduction}\label{sec:perm}
We show that the measure is invariant under permutation reduction, giving rise to a simple argument proving strong normalisation for the combined reduction relation given by both beta and permutation reduction.

\begin{lemma}[Permutation]\label{sn:lem:permutation}
For  terms $\Gamma \vdash \term{M: t}$ and $\Gamma \vdash \term{N: t}$ and valuations $v$, we have 
\begin{align*}
	\term{M} \rw_\pi \term{N} \quad {implies} \quad \sn{\term{M}}_v = \sn{\term{N}}_v\ .
\end{align*}
\end{lemma}
\begin{proof}
We recall the permutation equivalence:
\begin{align*}
	\term{[N]b.a<x>.P} \rw_\pi \term{ a<x>.[N]b.P: `a(r)?s_A > !t_A} \ ,
\end{align*}
where $\term{x} \not\in \fv{\term{N}}$.
To see this, observe the following are equivalent:
\begin{align*}
 \sn{\term{[N]b.a<x>.P}}_v ({s}_A, a(r)) &= (n+2+\collapse{\snterm{N}_{v}}, {t}), \\
   \textup{where } (\concat{n}{{t}}) &= \sn{\term{P}}_{v\{\term{x} \leftarrow r\}}  ({s}, b(\sn{\term{N}}_{v} ), \\
 \sn{\term{a<x>.[N]b.P: }}_v ({s}, a(r)) &= (n+2+\collapse{\snterm{N}_{v\{\term{x} \leftarrow r\}}}, {t}), \\
 \textup{where } (\concat{n}{{t}}) &= \sn{\term{P}}_{v\{\term{x} \leftarrow r\}} ({s}, b(\sn{\term{N}}_{v\{\term{x} \leftarrow r\}}), 
\end{align*}
using the Weakening Lemma \ref{sn:lem:weakening} ($\term{x} \not\in \fv{\term{N}}$). 
\end{proof}

\begin{corollary}
All typed FMC-terms are strongly normalising with respect to beta reduction and permutation reduction.
\end{corollary}
\begin{proof}
It is easy enough to show that permutation reduction alone is strongly normalising.
\[
	\term{[N]a.b<x>.M} \rw_\pi \term{b<x>.[N]a.M}
\]
Consider the multi-set measure given by, for each application $\term{[N]a.M}$ occurring in the term, the number of abstractions \textit{not} on location $\term a$ that are in $\term{M}$. Then permutation reduction strictly decreases this measure. Since permutation reduction makes no change to the measure for strong normalisation of beta  reduction (Lemma \ref{sn:lem:permutation}), we can interleave beta reduction with any number of permutation reduction steps (which will necessarily be a finite number) without affecting strong normalisation. 
\end{proof}

\section{The Weak Interpretation}\label{sec:WN}
In order to clarify the operational intuition of the proof above, we inductively define the \textit{weak interpretation} $\sn{-}^W$, which associates to each term a function taking the interpretation of its inputs to the \textit{exact} length of its machine run on those inputs and the interpretation of its outputs. It is defined as the interpretation in the previous section, except without adding the measure of the argument to the count in the application case. Following the definition, we prove that $\snterm{M}^W(\snterm{N_1}^W\cdots\snterm{N_n}^W)$ measures the exact length of the machine run on $(\term{N_1}\cdots \term{N_n}, \term M)$, for closed terms. 

Recalling Remark \ref{weak-is-monotonic}, we note that a similar proof to \ref{increasing} shows that the weak interpretation is indeed well-defined -- that is, the interpretation of a term is indeed a monotonic function. 
\begin{definition}\label{def:wsninterp}
For each term $\term{G |- M: ?s_A > !t_A}$ and valuation $v$, we inductively define a \textit{weak interpretation} function
\begin{align*}
	\sn{\Gamma \vdash \term{M: t}}^W_v : \sn{\type{t}} .
\end{align*}
We omit the context and/or types of terms inside an interpretation when it is clear. Each case is defined in the same way as the strong interpretation $\sn{-}$ in the previous section, except for the application case, which is given by
\begin{alignat*}{3}
       \snterm{[N]a.M: ?s_A > !t_A }^W_v (s)  &= (1+n, t)
\\		\qquad \textup{where } (n, t) &= \snterm{M: `a(r)\,?s_A >!t_A}^W_v (s, a(\snterm{N: r}^W_v))
\end{alignat*}
We define $\sn{\Gamma \vdash \term{M: t}}^W$ to be $\sn{\Gamma \vdash \term{M: t}}^W_w$, for the \textit{least valuation} $w = \{\term{x}\smallbin \leftarrow 0_{\type{s}}\mid\term{x:s}\}$ given by the minimal elements. 
The interpretation of a stack of terms is defined element-wise, and the interpretation of a memory $S_A$ is given by $\Pi_{a\in A} \sn{S_a}$.
\end{definition}
We now recall the substitution lemma from the previous section, applied instead to the \textit{weak} interpretation. The proof of this lemma is similar to the previous one. 
\begin{lemma}[Substitution]\label{lem:weak-substitution}
For terms $\Gamma, \term{x:w} \vdash \term{M: t}$ and $\Gamma \vdash \term{N: w}$ and valuations $v$, we have
\begin{align*}
	\sn{\term{\{N/x\}M}}^W_{v} = \sn{\term{M}}^W_{v\{\term{x} \leftarrow \sn{\term{N}}^W_v\} } \ .
\end{align*}
\end{lemma}
\begin{proof}
We omit the full proof here, but reference the Substitution Lemma \ref{sn:lem:substitution} proved in detail for the strong interpretation, which differs only in the extra count included in the application case (as do the lemmata of Weakening and Sequencing, which it relies on). 
\end{proof}

The following proof verifies that the weak interpretation counts exactly the number of machine steps required to evaluate a term. It relies essentially on machine termination.
\begin{proposition}\label{prop:countsteps}
For any closed term $\term{M: ?s_A > !t_A}$ and memories of closed terms $S_A: \type{!s_A}$ and $T_A: \type{!t_A}$, we have that 
\begin{align*}
	(S_A, \term{M})\Downarrow (T_A, \term{*})  \quad {implies} \quad \sn{\term{M}}^W (\sn{S_A}^W ) = (n, \sn{T_A}^W)\ ,
\end{align*}
where $n$ is the number of machine transitions until $(S_A, \term{M})$ terminates. 
\end{proposition}
\begin{proof}
We proceed by induction on the length $n$ of the machine run 
\begin{align*}
\machine {S_A}M{T_A} \ .
\end{align*}
The base case, a run of length $0$, implies that $\term{M} \equiv \term{*: ?t_A > !t_A}$ and $S_A \equiv T_A$. Thus, it suffices to observe that 
\begin{align*}
\sn{ \term{*}}^W ( \sn{S_A}^W ) = (0, \sn{S_A}^W) \ .
\end{align*}

Now, let us consider such a machine run of length $n+1$. Because $\term{M}$ is closed, there are just two cases: where $\term{M} \equiv \term{a<x>.M': `a(r)?s_A > !t_A}$ and where $\term{M} \equiv \term{[N]a.M': ?s_A > !t_A}$ (with $\term{N: r}$). We prove the statement for each case in turn.
\begin{itemize}
\item
The application case: let
\begin{align*}
	(S_A, \term{[N]a.M'}) \Downarrow (T_A, \term{*})
\end{align*}
have run length $n + 1$. Then the run $(\SAouta ; S_a \cdot \term{N}, \term{M'}) \Downarrow (T_A, \term{*})$ has length $n$, so we can apply the inductive hypothesis to achieve
\begin{align*}
\sn{ \term{M'}}^W (\sn{S_A}^W, a(\sn{ \term{N}}^W)) = (n, \sn{T_A}^W) \ . 
\end{align*}
By definition, we have  
\begin{align*} 
\sn{ \term{[N]a.M'}}^W (\sn{S_A}^W )&= (m+1, {t}) && \\
	\textup{where } (m, {t}) &= \sn{ \term{M'}}^W (\sn{S_A}^W,  a(\sn{ \term{N}}^W)) .&& 
\end{align*}
so altogether we have $(m,t) = (n, \sn{T_A}^W)$ and thus the required result
\begin{align*}
\sn{ \term{[N]a.M'}}^W (\sn{S_A}^W )&= (n+1, \sn{T_A}^W)  .
\end{align*}
\item
The abstraction case, let
\begin{align*}
	(\SAouta ; S_a \cdot \term{N},\term{a<x>.M'}) \Downarrow (T_A, \term{*})
\end{align*}
have run length $n+1$. Then the run $(S_A,\term{\{N/x\}M'}) \Downarrow (T_A, \term{*})$ has length $n$, so we can apply the inductive hypothesis to achieve
\begin{align*}
\sn{ \term{\{N/x\}M'}}^W (\sn{S_A}^W) = (n, \sn{T_A}^W) \ . 
\end{align*}
By definition, we have 
\begin{align*}
\sn{ \term{a<x>.M'}}^W (\sn{S_A}^W, a(\sn{ \term{N}})) &= (m+1, {t}) && \\
	\textup{where } (m, {t}) &=  \sn{\term{M'}}^W_{v\{\term{x} \leftarrow \sn{  \term{N}}^W\}} (\sn{S_A}^W), && 
\end{align*}
where $v$ is the least valuation. 
Applying the Substitution Lemma \ref{lem:weak-substitution}, we get $(m, {t}) = \sn{ \term{\{N/x\}M'}}^W (\sn{S_A}^W)$
and so altogether we have $(m,t) = (n, \sn{T_A}^W)$ and thus the required result
\begin{align*}
\sn{ \term{a<x>.M'}}^W (\sn{S_A}^W, a(\sn{ \term{N}})) &= (n+1, \sn{T_A}^W) .
\qedhere
\end{align*}
\end{itemize}
\end{proof}

\begin{remark}
Note that reduction is monotonic with respect to the weak interpretation: that is, $\term{G |- M} \rw \term{M': t}$ implies $\snterm{M}^W \geq_{\type{t}} \snterm{M'}^W$, using a similar proof to Lemma \ref{non-increasing}. 
\end{remark}
This proof confirms the intuition that we have made explicit latent operational intuitions in Gandy's proof of strong normalisation.

\chapter{Related Literature}\label{chapter:related-lit}

Landin was the first to show that the $\lambda$-calculus could be used as a tool for studying programming languages by using it as the target of a translation from Algol \cite{Landin-1964,Landin-1965}. Since then, there has been a huge amount of research into applications of the $\lambda$-calculus to the theory of programming languages. One major obstacle to its application to \textit{real-world} programming languages is finding the right way to incorporate computational effects. 

Something common to the various approaches to effectful $\lambda$-calculi is that {they provide some way to express the \textit{sequencing} of computations within syntax}\footnote{As in the introduction, the notion of ``sequencing'' referred to here is a semantic one, rather than in the sense of reduction order.}. Approaches such as  Moggi's computational metalanguage and Call-by-Push-Value, to be discussed -- 
are thus similar in spirit to the Sequential $\lambda$-calculus. Note, however, that the various approaches presented here introduce new constructors at term and type level to achieve this, as in Figure \ref{fig:approaches-to-sequencing}. In contrast, with SLC \textit{decomposes} and \textit{generalizes} the existing variable construct.
At type level, there are no new constructors and the typed SLC (and FMC) can be considered a novel computational interpretation of
intuitionistic logic, as shown in Chapter \ref{chapter:CCC-eqns}. This has the advantage of parsimony.

However, the approach of the FMC differs \textit{significantly} from other approaches when adding effects.
The major contribution of the FMC is to encode reader/writer effects using the same syntax and operational mechanisms as for higher-order functions. Confluence is retained because the beta-law then captures \textit{both} the semantics of higher-order functions \textit{and} the semantics of these effects. 

We now review some of the large amount of related literature on effects, drawing further comparisons as they become relevant. 

\begin{figure}
\[
	\begin{array}{cc}
	\infer[]
	  {{ \Gamma \vdash \eta(M): TA }}
	  {{\Gamma \vdash  M: A}
	  }
	&
	\infer[]	 
	  {{\Gamma \vdash \textsf{let}_T ~ x \Leftarrow N ~\textsf{in}~ M: TB} 
	  }
	 {{\Gamma \vdash N: TA}\qquad{\Gamma, x:A \vdash M:TB}}
	\\
	\\
	\infer[]
	  {{\Gamma \vdash_\textsf{c} \textsf{return}~ M: FA }}
	  {{ \Gamma \vdash_\textsf{v} M: A}
	  }
	&
	\infer[]	 
	  {{\Gamma \vdash_\textsf{c} N ~\textsf{to}~x .\, M: \underline B} 
	  }
	 {{\Gamma \vdash_\textsf{c} N: FA}\qquad{\Gamma, x:A \vdash_\textsf{c} M:\underline B}}	
	\\ \\
	\infer[]
	  {{\Gamma \vdash_\textsf{c} \textsf{force}~ M: \underline B }}
	  {{ \Gamma \vdash_\textsf{v} M: U\underline B}
	  }
	&	
	\infer[]
	  {{\Gamma \vdash_\textsf{v} \textsf{thunk}~ M: U\underline B }}
	  {{ \Gamma \vdash_\textsf{c} M: \underline B}}
	\\ \\
	\infer[]
	  {{\Gamma, A\vdash B}}
	  {{ \Gamma, !A \vdash B}
	  }
	&
	\infer[]
	  {{! \Gamma \vdash \,!B}}
	  {{! \Gamma \vdash B}
	  }
	\\
	\\
%
\end{array}
\]
\[
\begin{array}{l@{\quad}r@{~}l@{\quad}r@{~}l}
\textsf{Metalanguage:}	&&&(A \to B)_v &= A_v \to TB_v \\
\textsf{Call-by-Push-Value:}	&(A \to B)_n &= UA_n \to B_n & (A \to B)_v &= U(A_v \to FB_v) \\
\textsf{Linear Logic:}&	(A \to B)_n &= ~!A_n \multimap B_n & (A \to B)_v &= ~!(A_v \multimap B_v)\\
\end{array}
\]
\caption{Typing rules for other approaches to sequencing computation}
\label{fig:approaches-to-sequencing}
\end{figure}

\section{Effect-Passing Style}
So-called \textit{effect-passing} style is perhaps the most primitive form of dealing with effects, as a programming style available in the pure $\lambda$-calculus. 
We lead with an example: \textit{state-passing} style. To model global state in the $\lambda$-calculus, it suffices to extend every function with an extra argument and an extra output, modelling the state at the time the function is called, and the updated state at the time the function returns. That is, a procedure of type $A \to B$ making use of state with type $S$ is modelled as a function of type 
\[
	S \times A \to S \times B\, ,
\]
where the state is then threaded through every computation. 

Denotationally, what happens here is in some ways similar to the FMC: stateful processes become pure functions between appropriately enlarged types, as discussed in the introduction to Chapter \ref{chapter:CCC-eqns} (and this in part accounts for the Cartesian closed semantics of the FMC given later in that chapter).  However, the approach of the FMC is much less heavy-handed than simple state-passing style: {permutation equivalence} and the admissibility of type {expansion} means that the programmer need not explicitly record extra input and output types representing, say, initial and final state for terms where it is not used. Yet, these pure terms may still be composed with stateful terms.  Further,  whereas in state-passing style a type $S$ of the state is fixed, in the FMC the type held at any given memory cell (modelled by a given location) can vary. 

A more complicated, but common, example is that of \textit{continuation-passing} style 
\cite{DBLP:journals/lisp/Reynolds93, StreicherT:clacam}. Here, every function takes an extra argument, its continuation, which is called with the result of that function, thus making control flow explicit. Thus, a function of type $A \to B$ becomes a function 
\[
	(B \to \bot) \to A \to \bot\, , 
\]
where $B \to \bot$ is the type of continuations of $B$. This gives explicit access to the continuation of a function, allowing for expression of various control structures
Similarly, one can use \textit{exception}-passing style to model programs with exceptions, \textit{etc}.

A major downside of effect-passing style is that it requires a restructuring of the entire program to account for effects.

\section{Monads and Moggi's Computational Metalanguage}
In Moggi's seminal work of `89, he axiomatized the common structure found in effect-passing styles using the categorical structure of a \textit{strong monad} \cite{Moggi-1989-LICS}\cite{Plotkin-Power-2002-FOSSACS}.
 Since then, monads have been widely adopted as a technique for structuring effectful functional programs, most notably in Haskell \cite{DBLP:conf/popl/JonesW93,DBLP:conf/lfp/Wadler90,DBLP:conf/popl/Wadler92}.

Of course,  a monad is just a monoid in the category of endofunctors.
A \textit{strong} monad on a monoidal category is a monad equipped with a natural transformation, as below, 
called the \textit{strength}, satisfying certain coherence conditions. 
\[
	\textsf{str}_{A,B}: A \otimes TB \to T(A \otimes B)
\]

Moggi notes that the usual semantic interpretation of the type $A \to B$ as the set of total functions from $A$ to $B$ is inadequate for modelling programs with side-effects. He instead developed a variant of the CBV $\lambda$-calculus called the \textit{computational metalanguage} ($ML$),  where the type of such programs is given as $A \to TB$,
for some strong monad $T$. Here, $B$ denotes \textit{values} of type $B$, whereas $TB$ denotes \textit{computations} which return values of type $B$. 
The type and term constructs for $T$ in the meta-language are as follows.
\[
	\infer[]
	  {{ \Gamma \vdash \eta(M): TA }}
	  {{\Gamma \vdash  M: A}
	  }
	\qquad
	\infer[]	 
	  {{\Gamma \vdash \textsf{let}_T ~ x \Leftarrow N ~\textsf{in}~ M: TB} 
	  }
	 {{\Gamma \vdash N: TA}\qquad{\Gamma, x:A \vdash M:TB}}
\]
These rules correspond to the \textit{unit} and \textit{Kleisli composition} associated with the monad $T$, respectively.  As expected, the categorical models of the $ML$ are given by Cartesian closed categories equipped with a strong monad.  In this sense,  $ML$ is a \textit{pure} calculus.  

While in the metalanguage side-effecting computations are explicitly marked as having monadic type, many real-world programming languages do not do this. In order to model such languages, Moggi introduced a second language, the \textit{computational $\lambda$-calculus},  often denoted $\lambda_c$, intended as a effectful call-by-value programming language.  Intuitively, this can be thought of as $ML$ with the monadic type constructor $T$ made invisible.  Thus a program $\Gamma \vdash M:A$ is a possibly effectful computation of type $A$.  
As such,  $\lambda_c$ is naturally interpreted in the Kleisli category of the monad $T$. \footnote{To be more precise, the interpetation of  $\Gamma \vdash M:A$  is given by a morphism from $\llbracket \Gamma \rrbracket \to T \llbracket A\rrbracket$.  As such,  $\lambda x.M : \top \to A$ where $x: \top$ is not free in $M: A$ is given type isomorphic to $T\llbracket A \rrbracket$. In particular, the function type is interpreted as possibly effectful. } More precisely, models of $\lambda_c$ are given by a Cartesian category $\mathbb{C}$ equipped with a strong monad $T$ and \textit{Kleisli exponentials}: that is all exponents of the form $A \Rightarrow TB$ in $\mathbb{C}$, which is just enough to ensure an appropriate exponential structure on the Kleisli category of $T$. 

A major strength of monads is that they capture a remarkable variety of effects. We give just a few example of (necessarily strong) monads on the category $\textsf{Set}$:
\[\begin{array}{c|c}
\textup{Effect} & \textup{Monad} \\
\hline
\textup{Input} & I \to - \\
\textup{Output} & - \times O^* \\
\textup{State} & S \to (- \times S) \\
\textup{Non-determinism} & \mathcal{P}(-) \\
\textup{Continuations} & (- \to \bot) \to \bot \\
\end{array}
\]
where we have a set of inputs $I$, a set of outputs $O$, where $O^*$ is the free monoid over $O$, a set of states $S$, $\bot$ is the empty set, and $\mathcal{P}$ is the powerset operator. 

Note how, when taking $T$ to be the state or continuation monad, $A \to TB$ indeed corresponds to the state- or continuation-passing style interpretation of $A \to B$ given in the previous subsection, modulo currying. 

However, monads have a fundamental problem: they do not compose. That is, given two monads $S$ and $T$, $ST$ is not a monad in general. In some specific cases, there is a \textit{distributive law} between $S$ and $T$: a natural transformation $l_A: STA \to TSA$, satisfying certain coherence conditions, which makes $ST$ into a monad. Such laws do not always exist \cite{DBLP:conf/haskell/KiselyovSS13}.  Furthermore, when considering more than two monads, \textit{iterated} distributive laws are required \cite{Cheng2011IteratedDL,DBLP:journals/corr/abs-2205-03640}, and even for the case of three monads, these are required to satisfy large commutative diagrams.

In practice, multiple effects are combined by building a stack of \textit{monad transformers} (which take a monad as an argument and return a monad as a result), with each layer adding both computational overhead and mental overhead for the programmer. In fact, some practical situations require effects to be interleaved in such a way as for no fixed stack of transformers to suffice \cite{DBLP:conf/haskell/KiselyovSS13}.

The translation of the metalanguage into the SLC extends the encoding of the CBV $\lambda$-calculus $(-)_v$, given in Section \ref{sec:intro:translations} of the introduction, as follows. Recall that evaluation of the translation of a CBV $\lambda$-term returns a value, which is pushed to the stack. Corresponding to the intution given above, we translate $T(A)$ as the type of computations which return values of type $A$ to the stack, with the $\textsf{let}_T$ construct as sequencing of computations. 
\[
T(A)_v ~\defeq~ \type{>} A_v \qquad \eta(M)_v ~\defeq~ \term{[}M_v\term{]} \qquad (\textsf{let}_T ~ x \Leftarrow N ~\textsf{in}~ M)_v ~\defeq~ N_v\term{;<x>.}M_v
\]
Related, we mention also Moggi's \textit{computational $\lambda$-calculus}, which gives the internal language for the Kleisli category of a strong monad on category with products.

In comparison with the FMC, the monadic approach has the undeniable benefit of generality: it accounts for strictly more effects than the FMC (for example, continuations).  
However, for the effects the FMC does deal with, it promises a less problematic approach to composition. See Section 4 of \cite{Barrett-Heijltjes-McCusker-2022} for an example comparing the two.  

In the introduction of \cite{DBLP:conf/fossacs/PlotkinP01},  Plotkin and Power argue that Moggi's original work does not give a satisfactory account of the operational semantics of effects.  They indicate that \textit{operations},  such as \textit{raise} for exceptions, \textit{read} and \textit{write} for input and output ,  and \textit{lookup}, \textit{update} for state,  are the source of effects and they give an operational semantics for a subset of \textit{algebraic} effects (Section 2, \textit{loc. cit.}), which we discuss later, in Section \ref{sec:algebraic-effects}.  However, the FMC gives a alternative account of the operational semantics of the relevant operations for reader/writer effects; in fact, an account in  which operational semantics is \textit{primary}, as detailed in Section \ref{sec:sequencing} of the introduction to this thesis.

Other related work includes \cite{DBLP:journals/entcs/UustaluV08}, which develops the study of comonads with applications to modelling stream-based computation.
 Filinski's PhD thesis introduces a notion of \textit{monadic reflection} to deal with effects \cite{Filinski-1996}.

\section{Pre-monoidal and Freyd Categories}

Pre-monoidal categories, introduced by Power and Robinson \cite{power-robinson-1997,premonoidal-cats-alg-structure,Power2000ModelsFT}, are like monoidal categories, except where the \textit{interchange} law may fail: that is, unlike in a monoidal category, we may have the following inequation, also given in string diagrams below.
\[
	\begin{array}{c@{\quad}c@{\quad}c}
	(f \otimes \textsf{id}) ; (\textsf{id} \otimes g) &\neq& (\textsf{id} \otimes g ) ; (f \otimes \textsf{id})
\\  \\
	\vc
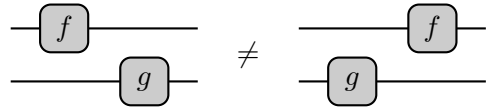
{\begin{tikzpicture}[x=1pt,y=1pt,thick, scale=2]
		\draw (0,20) -- (70,20);
		\draw (0, 0) -- (70, 0);
		\node[draw,fill=black!20,rounded corners,minimum size=18pt] at (20,20) {$f$};
		\node[draw,fill=black!20,rounded corners,minimum size=18pt] at (50, 0) {$g$};
	\end{tikzpicture}}
	&\neq&
	\vc{\begin{tikzpicture}[x=1pt,y=1pt,thick, scale=2]
		\draw (0,20) -- (70,20) ;
		\draw (0, 0) -- (70, 0) ;
		\node[draw,fill=black!20,rounded corners,minimum size=18pt] at (50,20) {$f$};
		\node[draw,fill=black!20,rounded corners,minimum size=18pt] at (20, 0) {$g$};
	\end{tikzpicture}}
	\end{array}
\]
Equivalently stated, the pre-monoidal tensor $\otimes$ need not be \textit{bi}functorial, although it must still be functorial in each of its arguments separately. 
The failure of interchange enforces a strong notion of sequentiality. 
That is, the pre-monoidal tensor specifies the order of evaluation of its arguments, which in general will affect semantics in the presence of effects.
For example, consider $f$ and $g$ above to be processes with some side-effect, such as printing to output. Then it matters whether $f$ or $g$ is evaluated first, necessitating the inequation given above.

Indeed, the Kleisli category of a strong monad $T$ on some Cartesian category has a pre-monoidal tensor defined as follows. It inherits the Cartesian product's structure on objects and, for a morphism $f: A \to TB$, has left- and right-actions given by
\begin{align*}
f \otimes C &\defeq A \times C \xrightarrow{f \times C} TB \times C \xrightarrow{\textsf{str}} T(B \times C)\\
C \otimes f &\defeq  C \times A \xrightarrow{C \times f} C \times TB \xrightarrow{\textsf{str}} T(C \times B)
\end{align*}
The pre-monoidal product given above is monoidal (\textit{i.e.}, satisfies interchange) if and only if the monad $T$ is \textit{commutative}, \textit{i.e.}, the two canonical morphism of type
\[
	TA \otimes TB \to T(A \otimes B)\, ,
\]
formed from the strength of $T$ and its multiplication, coincide. This is \textit{not} the case for many common effects. 

Building on this simple idea, Power and Robinson then introduced (closed) Freyd categories \cite{Power2000ModelsFT}\cite{Levy-Power-Thielecke-2003}, axiomatizing the Kleisli category associated with a strong monad over a Cartesian category. 
These are (higher-order analogues of) symmetric pre-monoidal categories which are additionally equipped with a wide Cartesian subcategory representing \textit{pure} functions, which are thus duplicable, discardable, and satisfy interchange with respect to every other process. \textit{Impure} processes, which live outside of this subcategory, however, cannot be expected to satisfy any of these properties in general. 

More formally, a Freyd category is given as a pre-monoidal functor $J: \mathbb{C} \to \mathbb{P}$, where $\mathbb{C}$ is a Cartesian category and a $\mathbb{P}$ is a pre-monoidal category with the same objects as $\mathbb{C}$, such that $J$ is identity-on-objects.  This models the scenario we have described above where, given a Cartesian catgeory $\mathbb{C}$ and a strong monad $T$ on $\mathbb{C}$, there is a canonical free functor from $\mathbb{C}$ to the Kleisli category of $T$. 
If one further asks that the functor $J(- \times A )$ has a right adjoint, one recovers the notion of a \textit{closed} Freyd category which is equivalent to the notion of a $\lambda_c$ model described in the previous section. 

The strict form of sequentiality imposed by pre-monoidality is only necessary if  processes interact through side-effects which are hidden from the type system. 
An example of this is detailed in the opening of this thesis, in Subsection \ref{subsec:pmc-example}.
{Because the FMC makes these explicit, they are immediately accounted for in the semantics, which is then a standard Cartesian (closed) category.}

Other related work includes Hughes' \textit{Arrows} \cite{Hughes-2000,Atkey-2011,lindley_wadler_yallop_2010,conf/mpc/0001H06,DBLP:journals/entcs/HeunenJ06,DBLP:journals/jfp/JacobsHH09}, which are a generalization of strong monads, \textit{Cartesian effect categories} \cite{DBLP:journals/corr/abs-0903-3311} , and Schewimer and Jeffrey's work on graphical models of effectful programming languages based on a version of string diagrams for pre-monoidal categories \cite{Schweimeier2001CategoricalAG,JEFFREY199851}, and the more recent work of Roman \cite{DBLP:journals/corr/abs-2205-07664}.

\section{Call-by-Push-Value}

Levy, in his PhD thesis, introduced Call-by-Push Value (CBPV) \cite{Levy-2003,Levy-2006} with the slogan a ``synthesis of functional and imperative programming".  
 Here, a strict distinction is drawn between values and computations, with explicit coercions between the two.
 Following Levy, we will underline computation types. Two new type constrctors are introduced: $U$, taking computations to values, and $F$, taking values to computations, so that  $U\underline{B}$ is the value type of \textit{thunks} of $B$, \textit{i.e.}, frozen computations,  and $FA$ is the type of computations which return values of type $A$. 

CBPV may be seen as arising from a decomposition of Moggi's strong monad into a (strong) adjunction between a category of \textit{values} and a category of \textit{stacks}, a third natural judgement of CBPV \cite{Levy-2003-adjunction}.\footnote{Recall that every monad arises from an adjunction. In fact, it may arise from many adjunctions in general. Given a monad $T$ on $\mathbb{C}$, there are \textit{two} canonical such adjunctions: one between $\mathbb{C}$ and the Kleisli category of $T$ and one between $\mathbb{C}$ and the Eilenberg-Moore category of $T$.} 
Concrete examples of the decomposition of Moggi's monads into adjunctons are as follows, where Moggi's monad $T$ is formed from the composite $UF$. Again, note these correspond to the given monads (and to state- and continuation-passing transformations) up to the currying isomorphism. 
\[
\begin{array}{c|c|c}
\textup{Monad} & U & F \\
\hline
\textup{State} 				& S \to - & S \times - \\
\textup{Continuations}			& (- \to \bot) & (- \to \bot) 
\end{array}
\]
The typing rules extend the $\lambda$-calculus with rules associated to $U$ and $F$, given below. Note, there are two different judgements, $\vdash_\textsf{c}$ for computations and $\vdash_\textsf{v}$ for values. {The rule for  sequencing computations, $N ~\textsf{to}~x .\, M$}, takes a similar form to Moggi's $\textsf{let}_T ~ x \Leftarrow N ~\textsf{in}~ M$.
\[
\begin{array}{cc}
	\infer[]
	  {{\Gamma \vdash_\textsf{c} \textsf{return}~ M: FA }}
	  {{ \Gamma \vdash_\textsf{v} M: A}
	  }
	&
	\infer[]	 
	  {{\Gamma \vdash_\textsf{c} N ~\textsf{to}~x .\, M: \underline B} 
	  }
	 {{\Gamma \vdash_\textsf{c} N: FA}\qquad{\Gamma, x:A \vdash_\textsf{c} M:\underline B}}
	\\ \\
	\infer[]
	  {{\Gamma \vdash_\textsf{c} \textsf{force}~ M: \underline B }}
	  {{ \Gamma \vdash_\textsf{v} M: U\underline B}
	  }
	&
	\infer[]
	  {{\Gamma \vdash_\textsf{v} \textsf{thunk}~ M: U\underline B }}
	  {{ \Gamma \vdash_\textsf{c} M: \underline B}}
\end{array}
\]

One of the main aims of CBPV is to allow the study of the operational and denotational semantics of CBN and CBV in a unified framework. 
Indeed, it allows to capture both the CBN and CBV evaluation strategies of the $\lambda$-calculus, by thunking arguments or functions, respectively, giving rise to two different translations of the $\lambda$-calculus into CBPV. In this respect, it is very similar to the SLC. The two translations of the function type of the $\lambda$-calculus are given below. 
\[
	(A \to B)_n = U \underline A_n \to \underline B_n \qquad (A \to B)_v = U(A_v \to FB_v) 
\]
Note that, in CBPV, unlike  CBV, abstractions are considered \textit{computations} and not values, and the computation function type is of the form $A \to \underline B$. 
Levy designed CBPV so that its CBN and CBV translations preserve ``all known'' semantics of (potentially effectful) CBN and CBV $\lambda$-calculi.  Note, this is an empirical claim, and indeed CBPV arose as a result of a search for commonalities in existing CBN and CBV semantics. 
The two translations from the CBN and CBV $\lambda$-calculus are as follows (Section 7, \cite{Levy-2006}). 
\begin{align*}
	x _n ~&\defeq~ \textsf{force } x & x_v ~&\defeq~ \textsf{return } x \\
	 (MN)_n ~&\defeq~ M_n@(\textsf{thunk } N_n) &  MN ~&\defeq~ M_v \textsf{ to }x. \,(N_v \textsf{ to } z.\, (\textsf{force z}@x)) \\
	(\lambda x.M)_n ~&\defeq~ \lambda x.M_n & (\lambda x.M)_v ~&\defeq~ \textsf{return thunk } \lambda x.M_v
\end{align*}

The distinction between computations and values in the \FMCV,  detailed in Section \ref{sec:fmc-values} of Chapter \ref{chapter:fmc-prelims}, was inspired by the same in CBPV, and similarly for the coercions force and thunk. 
It is also of interest to compare the CBN and CBV translations for the FMC, as given in  Section \ref{sec:intro:translations} of the introduction, to the CBN and CBV translations for CBPV above. 

The following translation extends our translation $(-)_n$ of the CBN $\lambda$-calculus, referenced above.\footnote{It is clear how to map that translation into the FMC with values.}
Note how the \textsf{return} and  {sequencing constructs} mirror our translation of the unit and Kleisli composition constructs in $ML$. The value functor $U$ is made redundant by the structure of the SLC's stack types.
\begin{align*}
	F(A)_n ~&\defeq~ \type{>} A_n & \type{!t} ~&\defeq~ \type{`Ut_1 \ldots `Ut_n} \\
N\textsf{ to } x.\, M ~&\defeq~ \term{N;<x>.M} & 
\textsf{return } V ~&\defeq~ \term{[V]} & \\
\textsf{thunk } M ~&\defeq~ \term{`!M} &
\textsf{force } V ~&\defeq~ \term{`?V} 
\end{align*}
Another similarity between CBPV and the SLC is that they can both be considered to be based the operational semantics of a stack machine (Section 5, \cite{Levy-2006}).  
The CCC semantics of the SLC can be recovered as a trivial CBPV model based on the identity monad. 
Other work includes extending CBPV with effects to a dependently typed setting \cite{DBLP:journals/corr/Vakar15a}

\section{Linear Logic and the Bang Calculus}\label{sec:bang-calc}

While discussing decompositions of  CBN and CBV, it would be remiss not to mention Linear Logic \cite{Gira:87:Linear-L:wm}. Linear Logic allows control over the structural rules of contraction and weakening via an \textit{exponential} modality $(!)$. For formulae tagged with this modality, these rules are permissible, whereas they are forbidden for formulae in general (which are \textit{linear}). We present the intuitionistic rules of LL involving the \textit{exponential} $!$ below. 
\[
	\infer[]
	  {{! \Gamma \vdash \,!B}}
	  {{! \Gamma \vdash B}
	  }
	\qquad
	\infer[]
	  {{\Gamma, A\vdash B}}
	  {{ \Gamma, !A \vdash B}
	  }
	\qquad
	\infer[]
	  {{\Gamma, !A, !A\vdash B}}
	  {{ \Gamma, !A \vdash B}
	  }
	\qquad
	\infer[]
	  {{\Gamma \vdash B}}
	  {{ \Gamma, !A \vdash B}
	  }
\]
Note how the first two rules give $!$ the structure of a \textit{co-monad}, while the latter two give the structural rules of contraction and weakening. 
 
Girard developed two translations of intuitionistic logic into Linear Logic \cite{Gira:87:Linear-L:wm}, $(-)_n$ and $(-)_v$, corresponding to encodings of the CBN and CBV $\lambda$-calculus, respectively. These two translations decompose the intutitionistic implication differently, similarly to CBPV. 
\[
	(A \to B)_n ~=~ !A_n \multimap B_n \qquad (A \to B)_v ~=~ !(A_v \multimap B_v) 
\]
Note that $!$ has the structure of a \textit{co-monad}, and the CBN-translation corresponds to its co-Kleisli category.

The Bang calculus \cite{Ehrhard-Guerrieri-2016,Guerrieri-Manzonetto-2018, DBLP:conf/flops/BucciarelliKRV20} can be considered an untyped version of CBPV, however it was developed by Ehrhard and Guerrieri from the distinct perspective of Linear Logic. Their syntax similarly makes the a distinction between computations and values and introduces the Linear Logic operators of \textit{box} $(!)$, introducing the exponential comonad , which marks terms which may be duplicated or erased, and \textit{dereliction} (eliminating the exponential).
\begin{align*}
	V,W \quad &\defeq \quad x ~\mid~ !M\\
	M,N \quad &\defeq \quad V ~\mid~ MN ~\mid~ \lambda x.M ~\mid \textsf{der}\,M
\end{align*}
Indeed, the calculus has the expected reduction rule: $\textsf{der}\, !M \to M$. However, beta reduction is restricted to $(\lambda x.M)V \to \{V/x\}M$, \textit{i.e.}, redexes can only be fired when their argument has been reduced to a value. 

It is shown that these two operators correspond to the \textsf{force} and \textsf{thunk} operators of CBPV, respectively, giving a useful operational intuition to these Linear Logic constructs, and a relatioship between CBPV and Linerar Logic. That is, viewing the type/term constructor $!$ as $U$/\textsf{thunk}, we can see that the CBN translation delays evaluation of its argument, whereas in the CBV translation (dubbed by Girard the``boring" translation), evaluation of the function itself is delayed. The CBPV functor $F$ is, however, silent in this presentation. 

The Bang calculus factorizes Girard's translations, giving an intermediate language which can encode both the CBN and CBV $\lambda$-calculus.
The original idea of CBN and CBV corresponding to Girard's translations dates back to \cite{Maraist-Odersky-Turner-Wadler-1999}.
These translations to the Bang calculus are given as follows. 
\begin{align*}
	x _n ~&\defeq~  \textsf{der}\,x & x_v ~&\defeq~ x \\
	 (MN)_n ~&\defeq~ M_n(!N_n)  &  MN ~&\defeq~ !(\lambda x.M_v) \\
	(\lambda x.M)_n ~&\defeq~ \lambda x.M_n & (\lambda x.M)_v ~&\defeq~ (\textsf{der}\,M_v)N_v
\end{align*}
In the SLC with values, we translate their values in the obvious way, and for computations have the following. 
\begin{align*}
	V &\defeq~ \term{[V]} \qquad
	 (MN)_n ~&\defeq~ \term{N;M} \qquad
	(\lambda x.M)_n ~&\defeq~ \term{[`!<x>.M]} \qquad
	\textsf{der}\,M  = \term{M;<x>.`?x}
\end{align*}
Note how their coercion from values to computations reflects the silence of their $F$ with respect to CBPV, which in our translation is no longer silent: it mirrors our translation of the \textsf{return} of CBPV. 
%

In a similar spirit, We mention here also the (enriched) effect calculus \cite{DBLP:conf/csl/EggerMS09,Egger-Mogelberg-Simpson-2014}, which augments $ML$ with constructs from linear logic. Other links between CBPV and Linear Logic are given by Benton's decomposition of the co-monad $(!)$ into an adjunction between a Cartesian category of values and a category of (linear) computations \cite{Benton-Wadler-1996}. This is in some sense dual to CBPV's decomposition of the monad in the metalanguage into an adjunction. 

\section{Universal Algebra and Algebraic Effects}
\label{sec:algebraic-effects}

Perhaps lacking in Moggi's original work on effectful computation is a treatment of the \textit{operations} which generate effects, and of their \textit{operational semantics}. Examples of operations are given by \textit{raise} for exceptions,  \textit{read} and \textit{write} for input and output, \textit{choose} for probabilistic or non-deterministic choice,  and \textit{lookup} and \textit{update} for state, all familiar from real-world programming languages 
\cite{DBLP:journals/corr/abs-1807-05923}.  

Universal algebra is the study of \textit{algebraic theories},  which are generated by a signature of \textit{operations} $\Sigma$,  like those given above,  and a set of \textit{equations} $\mathcal{E}$.  
Linton was first to formalize a general correspondence between universal algebra and monads: 	any algebraic theory $(\Sigma, \mathcal{E})$ gives rise to a ``free model'' monad $T_\Sigma$ on \textsf{Set} whose Eilenberg-Moore category is equivalent to the category of models of that theory \cite{HYLAND2007437}.  
For example, the theory of monoids has a corresponding ``free monoid monad'' whose Eilenberg-Moore category is equivalent to the usual category of monoids.  

Plotkin and Power  \cite{DBLP:journals/acs/PlotkinP03, DBLP:journals/entcs/PlotkinP01,PLOTKIN2004149,DBLP:conf/fossacs/PlotkinP01,Plotkin-Power-2002-FOSSACS,DBLP:journals/entcs/PlotkinP01}
later worked to generalize this connection by replacing the category \textsf{Set} with any $\lambda_c$ model, in order to give a perspective on computational effects with \textit{operations} at the centre. 
In order to do this, it was natural to work with \textit{Lawvere theories}, which are a categorical interpretation of the notion of algebraic theory which naturally allows models of a theory to be taken in any category with products.  

The correspondence mentioned above can be restated in these terms: the category of Lawvere theories is equivalent to the category of \textit{finitary} monads on \textsf{Set}. \footnote{Note,  although most monads corresponding to computational effects are finitary, there are notable exceptions such that of the continuation monad. }
To ease presentation, we work with an intermediate generality between that of the original correspondence and that of Plotkin and Power: we will work with \textit{infinitary} algebraic theories.  The corresponding result is that the category of  \textit{infinitary} Lawvere theories is equivalent to the category of \textit{arbitrary} monads on \textsf{Set}.  In particular, we work on \textsf{Set} throughout,  rather than an arbitrary category with products.  

We will now discuss in detail the three perspectives mentioned above: universal algebra, Lawvere theories, and monads.  Fixing an algebraic theory $(\Sigma, \mathcal{E})$, each perspective gives a path from left to right in the diagram below,  and we will detail how they are equivalent. \footnote{We  write simply $\Sigma$ instead of $(\Sigma, \mathcal{E})$ in subscripts, leaving the equations implicit,  and elide subscripts altogether in the labels of the arrows of the diagram so as to avoid overly heavy notation. }
In particular, we will  explain how a Lawvere theory and a monad can be generated from $(\Sigma, \mathcal{E})$ and how the models of the theory correspond to the Eilenberg-Moore algebras of the monad.  
Finally, we relate this story back to computational effects and describe how the new perspective of universal algebra accounts for operational considerations. 

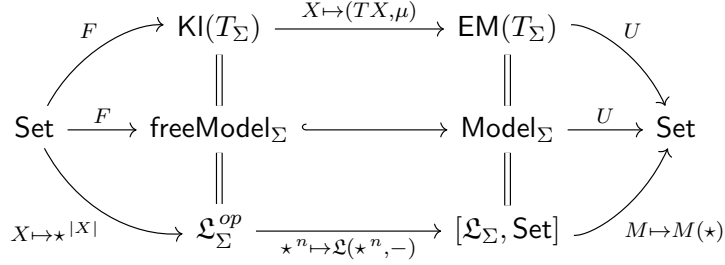
\[\begin{tikzcd}
	& {\textsf{Kl}(T_{\Sigma})} && {\textsf{EM}(T_{\Sigma})} \\
	{\textsf{Set}} & {\textsf{freeModel}_{\Sigma}} && {\textsf{Model}_{\Sigma}} & {\textsf{Set}} \\
	& {\mathfrak{L}_{\Sigma}^{\,op}} && {[\mathfrak{L}_\Sigma,\textsf{Set}]}
	\arrow["F", curve={height=-36pt}, from=2-1, to=1-2]
	\arrow["{X \mapsto \star^{\,|X|}}"', curve={height=36pt}, from=2-1, to=3-2]
	\arrow["F", from=2-1, to=2-2]
	\arrow["{X \mapsto (TX,\mu)}", from=1-2, to=1-4]
	\arrow["", hook, from=2-2, to=2-4]
	\arrow["{\star^{\,n} \mapsto \mathfrak{L}(\star^{\,n},-)}"', from=3-2, to=3-4]
	\arrow["U", curve={height=-36pt}, from=1-4, to=2-5]
	\arrow["{M \mapsto M(\star)}"', curve={height=36pt}, from=3-4, to=2-5]
	\arrow["U", from=2-4, to=2-5]
	\arrow[Rightarrow, no head, from=1-4, to=2-4]
	\arrow[Rightarrow, no head, from=2-4, to=3-4]
	\arrow[Rightarrow, no head, from=1-2, to=2-2]
	\arrow[Rightarrow, no head, from=2-2, to=3-2]
\end{tikzcd}\]

\textbf{Universal Algebra.} 
Algebraic theories are traditionally generated by a signature $\Sigma$ of operations, each with a given \textit{arity} $n \in \mathbb{N}$, and set of equations $\mathcal{E}$ over terms generated by these operations and a set of free variables. 
For example, the theory of a monoid is given by two operations: a nullary unit $u$ and a binary multiplication $(\cdot)$, subject to the following equations:\footnote{More formally, one would write \textit{e.g.}, $x \cdot u$ as $x \vdash \cdot(u(), x)$: we leave implicit the free variables,  use infix notation for multiplication and omit the brackets for the nullary unit operation. } 
\begin{align*}
	u \cdot x = x \qquad x \cdot u = x \qquad (x \cdot y) \cdot z = x \cdot (y \cdot z) \ .
\end{align*}
A second example is the theory of a pointed set, generated by a single nullary operation and no equations. Other examples are given by groups, semi-lattices, and rings. 

A \textit{model} of a theory $(\Sigma, \mathcal{E})$ is given by a choice of set $X$ and for each $n$-ary operation $op$ in $\Sigma$ an interpreting function $f: X^n \to X$ such that the resulting interpretation of the equations in $\mathcal{E}$ hold.  
Given an appropriate notion of model homomorphism, we can thus form a category of models of $(\Sigma, \mathcal{E})$, which we denote $\textsf{Model}_\Sigma$.  This is the expected category of structures: the category of models of the theory of monoids is the usual category of monoids, and so on. 

There is a forgetful functor $U: \textsf{Model}_\Sigma \to \textsf{Set}$ sending a model to its underlying set.  This functor has a left adjoint $F: \textsf{Set} \to \textsf{Model}_\Sigma$ which sends a set $X$ to the freely generated model of $(\Sigma, \mathcal{E})$ over $X$.  
For example,  for the theory of monoids above, $F$ sends $X$ to terms (or trees) inductively generated by operations of $\Sigma$, with free variables (or leaves) taken from $X$:
\[
	t, t' \Coloneqq x \in X ~\mid~ u ~\mid~ (t \cdot t')\ ,
\]
which are finally quotiented by the equations in $\mathcal{E}$. \footnote{The operation $\cdot$ is unimaginatively interpreted as the map $(t,t') \mapsto t \cdot t'$.} The full subcategory of \textit{free} models generated in this way is denoted $\textsf{freeModel}_\Sigma$ in the diagram above.  

While universal algebra traditionally deals with \textit{finitary} theories, some generalization was required in order to model computational effects of interest. 
Reasonably straightforwardly, one can consider \textit{infinitary} theories -- that is, where operations may have arity given by (the cardinality of) any set.  

\textbf{Lawvere Theories.} The definition of a theory as given above is very syntactic: there may be many possible choices of generating operations and equations which result in the same category of models.  This can be a benefit if one cares about operational semantics, but for a more abstract perspective one might consider Lawvere theories.  These are a categorical reformulation of the notion of algebraic theory which is invariant, \textit{i.e.}, independent of any particular choice of generators. 

A Lawvere theory $\mathfrak{L}$ is a category  with  products $(\mathbb{C}, \times, 1)$ with a distinguished object $\star$ such that every object of $\mathbb{C}$ is isomorphic to $\star^{\, n}$ for some $n \in \mathbb{N}$.  Its objects can be considered as abstract \textit{arities}, with \textit{infinitary} Lawvere theories allowing arity $n$ to be (the cardinality of) any set.  Given an algebraic theory $(\Sigma, \mathcal{E})$, a corresponding Lawvere theory $\mathfrak{L}_{\Sigma}$ can be generated by taking hom-sets $\mathfrak{L}_{\Sigma}(\star^{\, n}, \star)$ to be the set of freely generated terms of a theory (modulo equations) with $n$ free variables. \footnote{Hom-sets $\mathfrak{L}_{\Sigma}(\star^{\, n}, \star^{\, m}) \cong \mathfrak{L}_{\Sigma}(\star^{\, n}, \star)^m$ are thus $m$-tuples of terms with $n$ variables and composition is given by substitution; thus the category is determined by this choice.}
Indeed,  every Lawvere theory arises from some algebraic theory in this way. 
 
A model of a Lawvere theory $\mathfrak{L}$ is a (product preserving) functor $M: \mathfrak{L} \to \textsf{Set}$, and the functor category $[\mathfrak{L}, \textsf{Set}]$ is thus the category of models. 
This category has an obvious forgetful functor from $U: [\mathfrak{L}, \textsf{Set}] \to \textsf{Set}$ which sends a model $M: \mathfrak{L} \to \textsf{Set}$ to $M(\star)$.  This functor has a left adjoint $F: \textsf{Set} \to [\mathfrak{L},\textsf{Set}]$ which, as in the diagram above, factorizes as the composition of a functor sending sets $X$ to $\star^{|X|}$ in $\mathfrak{L}^{op}_\Sigma$ and the Yoneda embedding
of $\mathfrak{L}^{op}$ into $[\mathfrak{L}, \textsf{Set}]$.

It can be seen that this perspective is equivalent to the standard perspective of universal algebra: in particular that $\mathfrak{L}_{\Sigma}^{op}$ is equivalent to $\textsf{freeModel}_\Sigma$.\footnote{Let $\underline{n}$ denote a set with $n$ elements, for the finitary case. The infinitary case generalizes easily. 
Let the equivalence send $\star^{n}$ to the free model over $\underline{n}$, $F(\underline{n})$, which is surjective on objects. 
Model homomorphisms $\textsf{freeModel}_\Sigma(F\underline{1}, F\underline{n})$ are determined by where they send the variables of $F\underline{1}$ (by freeness of $F\underline{1}$), of which there is one.  So the set of model morphisms is given by each possible choice of $F\underline{n}$, \textit{i.e.}, each element of the free model of $(\Sigma, \mathcal{E})$ on $n$ variables.  Recall that $\mathfrak{L}(\star^{n}, \star)$ is the set of terms generated by $\Sigma$ with $n$ free variables,  modulo $\mathcal{E}$.  Therefore there is a bijection between $\mathfrak{L}(\star^{n}, \star)$ and $\textsf{freeModel}_\Sigma(F\underline{1}, F\underline{n})$.}
Given this, note that a choice of functor $M: \mathfrak{L}_\Sigma \to \textsf{Set}$ packages the same data as a model of $(\Sigma, \mathcal{E})$: the functor $M$ is determined by where it sends the  generating object $\star$ and generating operations $op \in \mathfrak{L}(\star^{\, n}, \star)$,  and this information constitutes a model. 

Unlike monads,  Lawvere theories can be composed, \textit{e.g.}, by the taking tensor or sum of theories,  and so this avenue presents a potential inroad into dealing with the combination of effects \cite{DBLP:conf/ifipTCS/HylandPP02}.

\textbf{Monads.}
The correspondence between monads on \textsf{Set} and universal algebra can be formulated as follows. 
Given a theory $(\Sigma, \mathcal{E})$,  we can package the data of the signature $\Sigma$ as a \textit{signature endofunctor} $S: \textsf{Set} \to \textsf{Set}$
\[
	S_\Sigma(X) \defeq \bigoplus_{c \in \Sigma}X^{\textsf{ar}(c)}, 
\]
where $\textsf{ar}(c)$ is the arity of $c$. 
For example, for the theory of monoids,   we have $S_\Sigma(X) = X^0 + X^2$,  corresponding to the signature having one \textit{nullary} and one \textit{binary} operation.  

{The \textit{algebras} for $S_\Sigma$ consist of a set $Y$ and morphism $\alpha: S_\Sigma(Y) \to Y$. For the example of the theory of monoids, this means a choice of function $\alpha: 1 + Y^2 \to Y$, or equivalently a choice of a pair of functions $\alpha_l: 1 \to Y$ and and $\alpha_r: Y^2 \to Y$.  
That is,  the algebras of $S_\Sigma$ consist of a set with the operations of a monoid, but nothing to specify that the operations must satisfy the equations of a monoid. 
In fact, the category of $S_\Sigma$-algebras is the category of models of the algebraic theory given by the \textit{signature} of monoids \textit{without} any equations.\footnote{For $S_\Sigma$ above, this is the category of pointed magmas.} For example, the theory of a pointed set, which has no equations associated, generates a signature endofunctor $S_\Sigma(X) = X^0$ and its category of $S_\Sigma$-algebras is in fact equivalent to the category of pointed sets. 

In order to account for the presence of equations we must consider monads. 
Given a signature endofunctor $S_\Sigma$, we can define the \textit{free monad} $T_\Sigma: \textsf{Set} \to \textsf{Set}$ over it as follows: 
\[
	T_{\Sigma}(A) \defeq \mu X. A + S_{\Sigma}(X)\, 
\]
where $\mu$ is the least fixpoint operator.  
In particular, the Eilenberg-Moore category $\textsf{EM}(T_\Sigma)$ of $T_\Sigma$-algebras is equivalent to the category of $S_\Sigma$-algebras. 
Intuitively, this is the free model monad: it takes a set to the set of freely generated terms over $\Sigma$ with variables from $A$. \footnote{Its multiplication takes a set of terms whose variables are taken from a set of terms to a set of terms in the obvious way, and its unit considers variables as terms in the trivial way.} 
Again, consider the example given by the theory of pointed sets, which has a signature endofunctor $S_\Sigma(X) = X^0$ and thus $T_\Sigma$ is exception monad $A \to A + 1$.  

We can now consider equations in this setting: an equation in $n$ elements can be given as a parallel pair of natural transformations in \textsf{Set} (natural in $X$)
\[
	X^n \rightrightarrows T_\Sigma (X)
\]
corresponding to a pair of terms $x_1, \ldots, x_n \vdash t_1, t_2$ each with $n$ free variables. 
Naturality in $X$ reflects that an equation is specified independent of the domain it is instantiated to.
Generally, then, a set of equations can be considered by copairing a number of such transformations together (one for each equation). For example, we can specify the algebraic theory of monoids as a pair of natural transformations
\[
	X^1 + X^1 + X^3 \rightrightarrows T_\Sigma (X)
\]
since we have one free variable for the left and right unit laws, and three free variables in the associativity law.  By considering the free monad $G$ over the endofunctor $X \mapsto X^1 + X^1 + X^3$ we can equivalently (by freeness) give a parallel pair of monad morphisms $
	G \rightrightarrows T_\Sigma $
which specifies the same laws.  In general we can find a monad $G_\mathcal{E}$ and morphisms $G_\mathcal{E} \rightrightarrows T_\Sigma $ to represent any equational theory $\mathcal{E}$ over $\Sigma$. An algebra $(X, \alpha: T_\Sigma (X) \to X)$ for $T_\Sigma X$ satisfies the equations $G_\mathcal{E} \rightrightarrows T_\Sigma (X)$ precisely when the two parallel morphisms 
\[
	G_\mathcal{E}(X) \rightrightarrows T_\Sigma (X) \xrightarrow{\alpha} X
\]
are equal.  Thus we obtain the monad corresponding to the algebraic theory $(\Sigma, \mathcal{E})$ as the coequalizer of $G_\mathcal{E} \rightrightarrows T_\Sigma $ in the category of monads. Indeed,  the (Eilenberg-Moore) category of algebras of the monad so obtained is equivalent to the expected category of models of $(\Sigma, \mathcal{E})$. 
Calling this monad $T_{\Sigma, \mathcal{E}}$, we thus have an equivalence $\textsf{EM}(T_{\Sigma, \mathcal{E}}) \cong \textsf{Model}_{\Sigma, \mathcal{E}}$, as expressed in our original diagram.\footnote{Recalling the omission of equations $\mathcal{E}$ from the notation  used there.}

It is well-known that the Kleisli category of any monad $T$, $\textsf{Kl}(T)$, is the full subcategory of $\textsf{EM}(T)$ consisting of free algebras. 
Intuitively,  $\textsf{Kl}(T_{\Sigma})$ is thus equivalent to $\textsf{freeModel}_{\Sigma}$ since they are both the full subcategories of $\textsf{EM}(T_{\Sigma})$ and $\textsf{Model}_\Sigma$, respectively, containing freely generated objects. 
To complete the picture,  recall that by construction there are canonical free and forgetful functors $F$ and $U$ between $\textsf{EM}(T)$ and $\textsf{Set}$ such that $UF = T$, and similarly for $\textsf{Kl}(T)$. \footnote{In the diagram $U$ sends algebras to their underlying carrier, $F$ considers plain functions as trivially monadic functions, and the embedding of $\textsf{Kl}(T)$ into $\textsf{EM}(T)$ sends objects $X$ to the free algebra on $TX$. } 
Indeed, in each of the three perspectives described above, a free/forgetful adjunction arises,  and each adjunction gives rise to the same monad: the \textit{free model monad},  appearing in the diagram as any path from \textsf{Set} to \textsf{Set}.


\textbf{Algebraic Effects.}
To relate computational effects to algebraic theories, we now give some concrete examples of \textit{algebraic effects} and show how they are included in Moggi's computational metalanguage $ML$. One could similarly incorporate algebraic effects into CBPV.  We consider infinitary theories but continue to work with the category \textsf{Set}. 

A particularly nice example is that of \textit{non-deterministic choice}.  This corresponds to the algebraic theory of a semi-lattice (without unit). That is, we have $\Sigma = \{\vee: 2\}$, together with equations
\[
	x \vee x = x \qquad x \vee y = y \vee x \qquad (x \vee y) \vee z = x \vee (y \vee z) \ . 
\]
Considering $T(X)$ as the set of  free semi-lattice terms,  the operation $\vee$ gives rise to a  family of functions (in \textsf{Set}), $choose_X: T(X)^2 \to T(X)$, natural in $X$, given by $(t_1, t_2) \mapsto t_1 \vee t_2$. 
Another perspective is given by considering $T$ as a free model monad: in fact, it is the finite (non-empty) powerset monad.  This can be seen since the free semi-lattice over $X$ can be considered as the set of all finite subsets of $X$, with $\vee$ corresponding to the union of subsets. This interpretation explains how we are modelling non-determinism.


Another example is given by the theory of pointed sets, which we saw before corresponds to the exception monad $T(X) = X+1$. This has signature $\Sigma = \{{\bot}: 0\}$, \textit{i.e.}, a nullary operation with no equations. This operation  thus gives rise to a family of functions $raise_X: T(X)^0 \to T(X)$,  \textit{i.e.}, $raise_X: 1 \to X+1$, natural in $X$.  Clearly we must thus have that $raise_X$ is given by the right injection. 

Finitary operations without parameters are introduced into $ML$ with the following typing rule 
\[
	\infer[]
	  {{\Gamma \vdash op(M_1, \ldots, M_n): TA}}
	  {{\forall i \in \{1, \ldots, n\}. \  \Gamma \vdash M_i: TA}& { op: n \in \Sigma}
	  }
\]
Operationally, algebraicity is reflected in evaluation contexts commuting with operations,  which reflects a certain naturality property characterizing equationally specified operations. An operational semantics can then be given by `floating' operations to the top of the syntax tree.  
For example,  we would have the following rewrite. 
\[
	\textsf{let } x \Leftarrow op(M_1, \ldots, M_n) \textsf{ in } N\quad  \rightsquigarrow \quad op(\textsf{let } x \Leftarrow M_1\textsf{ in } N, \ldots, \textsf{let } x \Leftarrow M_n\textsf{ in } N)
\]
There exist more sophisticated type systems which track more closely the usage of effects based on these ideas \cite{Plotkin-Pretnar-2009}.  

%

\textbf{State and relationship with the FMC.}
As a leading example from the study of computational effects, we now consider  operations and equations for global state.  
First we will need to generalize our theory so that one can introduce (effectively) an infinite number of operations via \textit{parameterization}: for example, in the theory of vector spaces, there is an operation corresponding to multiplication of a vector by a scalar, say in $\mathbb{R}$. One could introduce $\mathbb{R}$-many unary operations to model this, or one could introduce a single unary operation with a \textit{parameter} $r \in \mathbb{R}$.  
Opting for the latter case, we generalize signatures so that each operation has an arity and an associated set from which it can draw parameters.  Then for a signature with generalized operations of the form $\textit{op}: (P, A)$, with parameter $P$ and arity $A$ (here, again, a set), we can give a model as a set $X$ and for each operation a map $f: P \times X^A \to X$.  
%

We now introduce the operations for global state. We work with a single memory cell, but one can consider multiple memory cells by giving a signature where operations take an extra parameter telling which cell is being referenced. 
One natural choice of operations is as follows: fix a set of states $S$ and let the signature $\Sigma$ consist of the pair of operations
\[
	\textit{lookup}: (1, S) \qquad \textit{update}: (S, 1)\, 
\]
where we write $1$ for the single-element set.  The lookup operation is an $S$-ary operation, and the update operation is a unary operation which takes a parameter in $S$. 
We leave discussion of the associated equations until later. 
However, given these equations, the corresponding monad is of course the state monad (although it takes some work to show this: see Section 3, \cite{Plotkin-Power-2002-FOSSACS}).
The theory gives rise to two families of functions, natural in $X$, as follows. 
\[
	l_X: T(X)^S \to T(X) \qquad u_X: S \times T(X)^1 \to T(X)
\]


Recalling the isomorphism $T(X)^S \cong S \to T(X)$, we can read the intended meaning of $l_X$ as taking an input {continuation} which tells us how to return an element of $T(X)$ given the state $S$; it then \textit{looks up} the current value of the state, and returns the appropriate value of $T(X)$. 

We give the following examples, due to 
\cite{DBLP:journals/corr/abs-1807-05923}, using $\lambda$-notation for set functions.  
Let $S = \mathbb{N}$. 
 The first function below takes an input $x:\mathbb{N}$,  updates the state to $x+1$ and then returns the original value $x$.  
Now, recall that \textit{lookup} takes a continuation $M: \mathbb{N} \to \mathbb{N}$ as input. 
The second function below continues as $M(x):\mathbb{N} $ with $x$ bound to the value held in state. 
\[
	\lambda x. update_\mathbb{N}(x+1,\lambda \_.x): \mathbb{N} \to \mathbb{N} \qquad lookup_\mathbb{N}(\lambda x. M(x)) : \mathbb{N}
\]
Thus, the following function increments the number held in memory by one, and returns the original contents: $	lookup_\mathbb{N}(\lambda x. update_\mathbb{N}(x+1, x)).$ 

We can introduce such infinitary operations into $ML$ using the same syntax, as follows\footnote{Technically,  the binder $\lambda x$ is part of the operation syntax here. We have such a binder in our notation for $update$, whose continuation takes only trivial input. }
The operational semantics then follows that of the previous section.
\[
	\infer[]
	  {{\Gamma \vdash op(N, \lambda x.M): TB}}
	  {\Gamma \vdash N:P & { \Gamma, x:A \vdash M: TB}& { op: (P,A) \in \Sigma}
	  }
\]
\[
	\textsf{let } x \Leftarrow op(N, \lambda y.M) \textsf{ in } M' \quad 
	\rightsquigarrow \quad
	op(N, \lambda y. \textsf{let } x \Leftarrow M \textsf{ in } M')
\]

%

The main result currently connecting the FMC to algebraic effects, due to Heijltjes \cite{Barrett-Heijltjes-McCusker-2022}, is that the algebraic laws for reader/writer effects arise from beta and permutation reduction in the FMC. 
We now give interpretations of \textit{lookup} and \textit{update} in the FMC and 
finally introduce the equations between \textit{lookup} and \textit{update} needed to complete the story.\footnote{
Note, these equations between operations could also naturally be expressed via their corresponding \textit{generic effects}, denoted here by $!a: TS$ and $a := v: T1$ and obtained by passing the trivial continuation in to $l$ and $u$ \cite{DBLP:journals/acs/PlotkinP03, DBLP:journals/entcs/PlotkinP01,PLOTKIN2004149}. By the Yoneda lemma, these are no less general. Further writing $N;M$ as sugar for $\textsf{let} \ () \Leftarrow N \textsf{ in } M$, we can equivalently view the lookup and update equations as the program equations below. 
\[
\begin{array}{lll}
  \textsf{let } y \Leftarrow \ !a \textsf{ in } a := y ; x &=& x \\
\textsf{let } y \Leftarrow \ !a \textsf{ in let } x \Leftarrow \ !a \textsf{ in } M &=& \textsf{let } y \Leftarrow \ !a \textsf{ in } M\{y/x\}\\
 a := v ; a := v' ; x &=& a := v' ; x \\
 a := v ; \textsf{ let } x \Leftarrow\  !a \textsf{ in } M &=& a := v ; \{v/x\}M \\
\end{array}
\]
}
\begin{align*}
	lookup(\lambda x. M)  &\defeq \term{a<x>.[x]a.M} \\
	update(v, M) &\defeq \term{a<\_>.[v]a.M} 
\end{align*} 
Here, the location $\term{a}$ is the single memory cell. 
In the following, we abbreviate the $lookup$ and $update$ operations to $l$ and $u$, respectively. 
%
%
\[
\begin{array}{llll}
l(\lambda y.u(y,x)) &= \term{a<y>.[y]a.a<\_>.[y]a.x} &\rw_\beta \term{a<y>.[y]a.x} &=_\eta x \\
l(\lambda y.l(\lambda x. M)) &= \term{a<y>.[y]a.a<x>[x].M} &\rw_\beta \term{a<y>.[y]a.\{y/x\}M} &= l(\lambda y. \{y/x\}M) \\
u(v, u(v',x)) &= \term{a<\_>.[v]a.a<\_>.[v']a.M} &\rw_\beta \term{a<\_>.[v']a.x} &= u({v'},x) \\
 u(v, l(\lambda x. M)) &= \term{a<\_>.[v]a.a<x>.[x]a.M} &\rw_\beta \term{a<\_>.[v]a.\{v/x\}M} &= u({v},\{v/x\}M) \\
\end{array}
\]
Note the algebraic equations for state are Hilbert-Post complete. 

For simplicity, we have kept to working with models over $\textsf{Set}$, but it is worth emphasizing that 
Plotkin and Power worked to generalize from $\textsf{Set}$ to any $\lambda_c$ model, so as to give the most proper setting for interpreting programs. 

There are at least two significant advantages to taking the (universal) algebraic rather than monadic perspective: first, finding a good presentation of an effect in terms of operations allows for easy incorporation into a programming language and gives the oppurtunity for an account of its operational semantics. Second,  as mentioned briefly before, the Lawvere theoretic perspective allows for natural notions of composition of theories (unlike monads, which do not compose in general). 
Since the original work on algebraic effects, Plotkin and Pretnar also developed a notion of \textit{effect handlers}, which in some sense are \textit{deconstructors} of effects, contra algebraic operations which \textit{construct} effects \cite{Plotkin-Pretnar-2009}. This development has given rise to a large body of research which we do not give an account of here \cite{DBLP:journals/entcs/Pretnar15, DBLP:journals/dagstuhl-reports/ChandrasekaranL18, Ahman-Staton-2013-ENTCS}.

\section{Concatenative Programming and $\kappa$-calculus}

Concatenative programming is a relatively uncommon style of programming which is natural for the Sequential $\lambda$-calculus: sequentially composing functions which act on a stack.
In one paper, concatenative programming is called ``an overlooked paradigm'' \cite{Herzberg-Reichert-2009}.  
There are, however, several language expressing a higher-order variant of this style: for example, $\lambda$-FORTH \cite{Lynas-Stoddart-2006},  Factor \cite{Pestov-Ehrenberg-Groff-2010} and Joy \cite{vonThun-2001}. 
For first investigations into type inference for such systems, see \cite{Diggins-2008} and \cite{Stoddart-Knaggs-1993}.

The $\kappa$-calculus is perhaps the only one that is remotely well-studied from a theoretical perspective \cite{Hasegawa-1995}. It was introduced by Hasegawa as part of a semantic decomposition of the $\lambda$-calculus into the first-order $\kappa$-calculus, as a language for Cartesian categories, and the higher-order $\zeta$-calculus, dealing with continuations.  Power and Thielecke developed the \textit{higher-order} $\kappa$-calculus (Section 3.2, \cite{Power-Thielecke-1999}), similar to \cite{Douence-Fradet-1998} and different from the $\zeta$-calculus, and applied it to the study of effects. 
Similar to the SLC and CBPV, the $\kappa$-calculus comes equipped with a stack-machine based operational intuition. 
Indeed, the type system for the sequential $\lambda$-calculus follows that of the $\kappa$-calculus~\cite{Power-Thielecke-1999}. However, since the 90's, and until recent work on the SLC, this calculus seems to have all but been forgotten.

This chapter has not come close to providing an exhaustive list of approaches to effects. 
We briefly mention the following approaches: \textit{uniqueness types} \cite{Smetsers-Barendsen-vanEekelen-Plasmeijer-1993}, 
and representing references in the $\pi$-calculus \cite{Hirschkoff-Prebet-Sangiorgi-2020}.

\chapter{Conclusion}\label{chapter:conclusion}
The results contained in this thesis, and the approach of the FMC itself, suggest many new and promising avenues of research. A handful of these are listed below.
\section{Further Research}
\textbf{Connection with deep inference:}
The three branches of the Curry-Howard-Lambek correspondence are core to the study of the simply-typed $\lambda$-calculus. 
In this thesis, we have presented in detail the link between the simply-typed FMC and Cartesian closed categories. 
This implies that the type system of the FMC, which is given in natural deduction style, is thus a certain presentation of intuitionistic logic -- although, to achieve this, one must quotient type derivations by the equational theory given in Chapter \ref{chapter:CCC-eqns}. Although this equational theory appears reasonable as an equivalence of terms, it does not appear to be a particularly natural (that is, independently justified) quotient on type derivations. So, it would seem that the link between the FMC and intuitionistic logic is not as direct as might be hoped. 

A promising solution to this is to consider a presentation of intuitionistic logic in a \textit{deep inference} proof system \cite{guglielmi2010proof, Tiu2006,Hugh:04:Deep-Inf:fc}. In such a system, the given equational theory appears extremely natural. In this way, we might achieve a second, convincing perspective on the Curry-Howard-Lambek correspondence. Previous work on the \textit{atomic $\lambda$-calculus} has investigated computational interpretations of inference rules peculiar to deep inference (that is, the \textit{distributor} and \textit{medials}), and it might be interesting to investigate a computational interpretation of these rules in the setting of the FMC \cite{Gundersen2013}.

\textbf{String diagrams:}
There has been much activity in the applied category theory community making use of string diagrams recently \cite{PAVLOVIC201394,Coecke-2015,Bonchi-Holland-Piedelue-Sobocinski-Zanasi-2019}.
It follows trivially from the results of this thesis that the first-order fragment of the FMC gives a sound and complete term language for string diagrams.
 Because we take sequencing as the primitive form of composition, the language thus given is arguably more natural as a representation of string diagrams than the $\lambda$-calculus. 
One source describes a ``dearth of computational tools for working with SMCs'', circa 2020 \cite{DBLP:journals/corr/abs-2101-12046}. 
Another source bemoans the lack of general results about the complexity of decision procedures for string diagrams \cite{DBLP:journals/lmcs/DelpeuchV22}, and perhaps, by giving the right term language, the FMC could help with  this. 
It further appears that the FMC gives a natural language for the recently investigated \textit{higher-order} string diagrams \cite{DBLP:journals/corr/abs-2107-13433}.
These give a way to embed graphs into a higher-order calculus, combining graph and term structure by allowing the ``boxing" of subgraphs, exposing only an interface. 
The higher-order machinery of the FMC gives a natural extension of the first-order fragment allowing for exactly this. One can thus factorize a string diagram by extracting sub-graphs which occur multiple times. 

\textbf{Frobenius algebras:}
One common structure in applied category theory, especially as it relates to string diagrams, is the Frobenius algebra. Early investigations suggest a way to capture this structure via an extension of the first-order fragment of the FMC, by taking seriously the symmetries revealed in its syntax. 
This would provide a term language for a larger class of string diagrams, and put the FMC in contact with recent work on signal flow graphs, Graphical Linear Algebra, and Categorical Quantum Mechanics (\textit{loc.\,cit.}).
Operationally, such a construct could be used to model \textit{conditioning}, as in probabilistic programming languages, as well as provide links with quantum and reversible computation.

\textbf{Linear Logic:}
It would be easy enough to extend the results of Chapter \ref{chapter:CCC-equiv} to give a characterization of the \textit{linear} FMC as the free symmetric monoidal closed category.
Term calculi for linear logic have typically been variants of the $\lambda$-calculus which additionally require a number of \textit{commuting conversions} in the equational theory. 
The FMC, by contrast, restricts nicely to a linear calculus (modulo Remark \ref{rem:sym-in-linear}). 
From here, it is natural to ask about adding exponentials. The discussion in Section \ref{sec:bang-calc} of Related Literature, suggests that the \textit{thunk} construct $(\term{`!})$ may be closely related to Girard's exponential. 
This, along with the question of how Girard's translations might relate to our translations of CBN and CBV into the SLC, would make a good starting point for further investigations.

\textbf{Weaker type systems: forgetting locations and stream types:}
The type system of the FMC is currently \textit{too strong} to be practically useful for modelling effects in many circumstances. For example, when reading from a probabilistic generator, the programmer does not typically want to know \textit{how many times} the generator was consulted, which is currently information recorded in the type system. Indeed, such information is \textit{intensional} and not even observable in general. One obvious way to weaken the type system, recovering something more akin the the effectful $\lambda$-calculus, is to \textit{forget} the type information at certain locations (\textit{e.g.}, the random input stream). 
Initial investigations indicate that, in the first-order setting, ignoring the types at the location of the random stream and suitably generalizating the definition of machine equivalence results not in a Cartesian category, but in a Markov catgeory ---  each term can `secretly' consume random inputs, making it a stochastic process in general, rather than a function. Markov categories are a natural abstract setting for studying probabilistic processes,  where we can express Bayesian networks and even do causal reasoning \cite{Jacobs-Kissinger-Zanasi-2019}. Of course, one also wants to maintain all the good properties of the FMC, so, given this, the question is how far this approach can be usefully pushed. A more sophisticated approach to types for the random stream could be given by a \textit{stream type} $\type{t}^*$, which represents a stream of $\type{t}$'s. 

\textbf{Imperative Programming:}
One promising avenue for immediate investigation is the application of the FMC to imperative programming. 
Perhaps primary is the investigation of \textit{local} state, probably via the inclusion of a \textit{new} location construct. 
Given the multiple arrow types occurring in the semantics of the FMC given in Chapter \ref{chapter:CCC-eqns}, it might further be interesting to investigate a computational interpretation of the logic of Bunched Implications, alternate to the $\alpha\lambda$-calculus, where the linear and intuitionistic arrows correspond to linear and non-linear locations \cite{DBLP:journals/bsl/OHearnP99, DBLP:conf/tlca/OHearn99}. Further, one might attempt to find links between the FMC and type systems for imperative programming such as Reynolds' Syntactic Control of Interference, or Seperation Logic \cite{DBLP:conf/popl/Reynolds78, DBLP:conf/icalp/Reynolds89, DBLP:conf/lics/Reynolds02}.


\textbf{Other:}
One, perhaps obvious, direction is to include sum types, datatypes, and error handling, and it looks possible to capture all three in a uniform way.
Another is to investigate alternate, perhaps more elegant, equivalent equational theories the FMC, and in particular to find one that serves as a well-behaved rewriting system. 
Extending the FMC to incorporate computational effects beyond reader/writer effects is of interest, perhaps especially aiming to be able to express \textit{continuations}. 
One might consider extending the FMC in such a way as to capture the $\lambda\mu$-calculus \cite{Parigot-1992}.
Further, the use of \textit{locations} in the FMC is reminiscent of \textit{channels} in the $\pi$-calculus, except composition in the FMC is \textit{sequential} rather than \textit{parallel}\cite{Milner-1992,Milner-Parrow-Walker-1992}. This could prove a promising avenue of investigation. 

\section{Summary}

Recall that we aim to extend reasoning techniques familiar from the $\lambda$-calculus to real-world, effectful and imperative programming languages. 
We thus began by asking, how can we extend the $\lambda$-calculus to incorporate effects, without losing the good properties which make it a central object of study in the first place?
The approach of the FMC towards modelling effects, while maintaining confluence, appears remarkable, yet raises many questions. One might worry that such an innovation would come at a cost.
This thesis contributes doubly towards allaying these concerns: it is proved that two further fundamental properties of the $\lambda$-calculus (beyond confluence) -- its categorical semantics, and the ability of its type system to guarantee strong normalisation -- are preserved. The results serve as a sanity check, and it appears that the calculus is remarkably well-behaved. The results herein suggest that it has a type system and a denotational semantics as strong and natural (respectively) as the simply-typed $\lambda$-calculus itself.

 The proofs contained in this thesis often have an operational flavour -- especially that of the proof of strong normalisation. It would appear that the FMC naturally leads to clearer links between the operational and the semantic perspectives. The work of a mathematician is not just to prove new theorems, but also to find the right definitions and proofs which simplify, bring new perspectives to, and new \textit{understanding} to, existing ones. 
It is hoped that the reader is convinced that the novel definition of the FMC is indeed a \textit{natural} extension of the $\lambda$-calculus, and that the reader may have found some new perspectives on old ideas. 

\addcontentsline{toc}{chapter}{Bibliography}

\nocite{Winskel-1993}
\nocite{DeBruijn-1993}
\nocite{GiraLafoTayl:90:Proofs-a:lf}

\bibliographystyle{plain}
\bibliography{citations}
\end{document}